\documentclass[a4paper,11pt]{article}

\catcode`\@=11

\newcount\printerresolution
\global\printerresolution=300

\newcount\headsize
\global\headsize=450

\font\dotfont = lcircle10 at 3pt

\newcount\epihead
\global\epihead=200

\newcount\defaultscale
\def\setdefaultscale#1{\global\defaultscale=#1}
\setdefaultscale{100}

\newcount\actualtextarrowspace
\newcount\actualtextarrowlength

\newcommand{\computetextparameters}%
{\global\actualtextarrowspace=\textarrowlength%
\global\advance\actualtextarrowspace by 3%
\global\actualtextarrowlength=\textarrowlength%
\global\multiply\actualtextarrowlength by 100}

\newcount\textarrowlength
\def\settextarrowlength#1{\global\textarrowlength=#1%
\computetextparameters} \settextarrowlength{20}

\newcount\actualdisplayarrowspace
\newcount\actualdisplayarrowlength

\newcommand{\computedisplayparameters}%
{\global\actualdisplayarrowspace=\displayarrowlength%
\global\advance\actualdisplayarrowspace by 3%
\global\actualdisplayarrowlength=\displayarrowlength%
\global\multiply\actualdisplayarrowlength by 100}

\newcount\displayarrowlength
\def\setdisplayarrowlength#1{\global\displayarrowlength=#1%
\computedisplayparameters} \setdisplayarrowlength{30}

\def\@ifnexttok#1#2#3{\let\@tempe #1\def\@tempa{#2}\def\@tempb{#3}%
\futurelet\@tempc\@ifntok}
\def\@ifntok{\ifx \@tempc \@tempe\let\@tempd\@tempa\else\let\@tempd\@tempb\fi%
\@tempd}

\def\@diagramerror#1#2{%
\edef\@@tempc{#2}\expandafter\errhelp\expandafter{\@@tempc}%
\typeout{Diagram error. \space See User's guide for
explanation.^^J
 \space\@spaces\@spaces\@spaces Type \space H <return> \space for
 immediate help.}\errmessage{#1}}

\newif\ifdiagram

\def\testtextmode{%
\ifdiagram\@diagramerror{Text arrows are not allowed in
diagrams}{Here you should use east or west diagram arrows, not
forward or backward text arrows. Try proceeding now, typeset could
succeed but with unpredictable output.}
\else\ifmmode\relax\else%
\@diagramerror{Missing \string$}{Text arrows should be introduced
in math mode. Try proceeding now, typeset could succeed but output
could not be what you expected.}\fi\fi}

\def\testdiagrammode{\ifdiagram\relax\else
\@diagramerror{Diagram arrows are not allowed in formulas}{Here
you should use forward or backward text arrows, not diagram
arrows. Proceeding could work with unpredictable output, but
overflow arithmetic could also occur.}\fi}

\def\checkmode{\ifmmode\@diagramerror{Wrong mode: no diagrams
allowed in math mode.}{You should leave math mode before
introducing your diagram. All items in the diagram will
automatically be processed in math
mode.}\else\relax\fi\global\diagramtrue}

\global\diagramfalse

\def\DOT{{\dotfont q}}

\newcount\basicstep
\newcount\numberofsteps
\numberofsteps=\headsize \global\divide\numberofsteps by
\basicstep

\newcommand{\makehead}[3]{%
\begin{picture}(0,0)%
\multiput(0,0)(#1,#2){#3}{\DOT}%
\multiput(0,0)(-#2,#1){#3}{\DOT}%
\end{picture}}

\newcount\xstep
\newcount\ystep

\newsavebox{\northhead}
\savebox{\northhead}{%
\xstep=-\basicstep%
\multiply\xstep by 7071%
\divide\xstep by 10000%
\ystep=\xstep%
\makehead{\xstep}{\ystep}{\numberofsteps}}
\newcommand{\nhead}{\usebox{\northhead}}

\newsavebox{\easthead}
\savebox{\easthead}{%
\xstep=-\basicstep%
\multiply\xstep by 7071%
\divide\xstep by 10000%
\ystep=-\xstep%
\makehead{\xstep}{\ystep}{\numberofsteps}}
\newcommand{\ehead}{\usebox{\easthead}}

\newsavebox{\southhead}
\savebox{\southhead}{%
\xstep=\basicstep%
\multiply\xstep by 7071%
\divide\xstep by 10000%
\ystep=\xstep%
\makehead{\xstep}{\ystep}{\numberofsteps}}
\newcommand{\shead}{\usebox{\southhead}}

\newsavebox{\westhead}
\savebox{\westhead}{%
\xstep=\basicstep%
\multiply\xstep by 7071%
\divide\xstep by 10000%
\ystep=-\xstep%
\makehead{\xstep}{\ystep}{\numberofsteps}}
\newcommand{\whead}{\usebox{\westhead}}

\newsavebox{\northwesthead}
\savebox{\northwesthead}{%
\makehead{0}{-\basicstep}{\numberofsteps}}
\newcommand{\nwhead}{\usebox{\northwesthead}}

\newsavebox{\northeasthead}
\savebox{\northeasthead}{%
\makehead{-\basicstep}{0}{\numberofsteps}}
\newcommand{\nehead}{\usebox{\northeasthead}}

\newsavebox{\southwesthead}
\savebox{\southwesthead}{%
\makehead{\basicstep}{0}{\numberofsteps}}
\newcommand{\swhead}{\usebox{\southwesthead}}

\newsavebox{\southeasthead}
\savebox{\southeasthead}{%
\makehead{0}{\basicstep}{\numberofsteps}}
\newcommand{\sehead}{\usebox{\southeasthead}}

\newsavebox{\eastnortheasthead}
\savebox{\eastnortheasthead}{%
\xstep=-\basicstep%
\multiply\xstep by 9486%
\divide\xstep by 10000%
\ystep=\xstep%
\divide\ystep by -3%
\makehead{\xstep}{\ystep}{\numberofsteps}}
\newcommand{\enehead}{\usebox{\eastnortheasthead}}

\newsavebox{\northnortheasthead}
\savebox{\northnortheasthead}{%
\xstep=-\basicstep%
\multiply\xstep by 9486%
\divide\xstep by 10000%
\ystep=\xstep%
\divide\ystep by 3%
\makehead{\xstep}{\ystep}{\numberofsteps}}
\newcommand{\nnehead}{\usebox{\northnortheasthead}}

\newsavebox{\southsouthwesthead}
\savebox{\southsouthwesthead}{%
\xstep=\basicstep%
\multiply\xstep by 9486%
\divide\xstep by 10000%
\ystep=\xstep%
\divide\ystep by 3%
\makehead{\xstep}{\ystep}{\numberofsteps}}
\newcommand{\sswhead}{\usebox{\southsouthwesthead}}

\newsavebox{\westsouthwesthead}
\savebox{\westsouthwesthead}{%
\xstep=\basicstep%
\multiply\xstep by 9486%
\divide\xstep by 10000%
\ystep=\xstep%
\divide\ystep by -3%
\makehead{\xstep}{\ystep}{\numberofsteps}}
\newcommand{\wswhead}{\usebox{\westsouthwesthead}}

\newsavebox{\westnorthwesthead}
\savebox{\westnorthwesthead}{%
\xstep=\basicstep%
\multiply\xstep by 3162%
\divide\xstep by 10000%
\ystep=\xstep%
\multiply\ystep by -3%
\makehead{\xstep}{\ystep}{\numberofsteps}}
\newcommand{\wnwhead}{\usebox{\westnorthwesthead}}

\newsavebox{\eastsoutheasthead}
\savebox{\eastsoutheasthead}{%
\xstep=-\basicstep%
\multiply\xstep by 3162%
\divide\xstep by 10000%
\ystep=\xstep%
\multiply\ystep by -3%
\makehead{\xstep}{\ystep}{\numberofsteps}}
\newcommand{\esehead}{\usebox{\eastsoutheasthead}}

\newsavebox{\northnorthwesthead}
\savebox{\northnorthwesthead}{%
\xstep=-\basicstep%
\multiply\xstep by 3162%
\divide\xstep by 10000%
\ystep=\xstep%
\multiply\ystep by 3%
\makehead{\xstep}{\ystep}{\numberofsteps}}
\newcommand{\nnwhead}{\usebox{\northnorthwesthead}}

\newsavebox{\southsoutheasthead}
\savebox{\southsoutheasthead}{%
\xstep=\basicstep%
\multiply\xstep by 3162%
\divide\xstep by 10000%
\ystep=\xstep%
\multiply\ystep by 3%
\makehead{\xstep}{\ystep}{\numberofsteps}}
\newcommand{\ssehead}{\usebox{\southsoutheasthead}}

\newsavebox{\easteastnortheasthead}
\savebox{\easteastnortheasthead}{%
\xstep=-\basicstep%
\multiply\xstep by 8944%
\divide\xstep by 10000%
\ystep=\xstep%
\divide\ystep by -2%
\makehead{\xstep}{\ystep}{\numberofsteps}}
\newcommand{\eenehead}{\usebox{\easteastnortheasthead}}

\newsavebox{\northnorthnortheasthead}
\savebox{\northnorthnortheasthead}{%
\xstep=-\basicstep%
\multiply\xstep by 8944%
\divide\xstep by 10000%
\ystep=\xstep%
\divide\ystep by 2%
\makehead{\xstep}{\ystep}{\numberofsteps}}
\newcommand{\nnnehead}{\usebox{\northnorthnortheasthead}}

\newsavebox{\southsouthsouthwesthead}
\savebox{\southsouthsouthwesthead}{%
\xstep=\basicstep%
\multiply\xstep by 8944%
\divide\xstep by 10000%
\ystep=\xstep%
\divide\ystep by 2%
\makehead{\xstep}{\ystep}{\numberofsteps}}
\newcommand{\ssswhead}{\usebox{\southsouthsouthwesthead}}

\newsavebox{\westwestsouthwesthead}
\savebox{\westwestsouthwesthead}{%
\xstep=\basicstep%
\multiply\xstep by 8944%
\divide\xstep by 10000%
\ystep=\xstep%
\divide\ystep by -2%
\makehead{\xstep}{\ystep}{\numberofsteps}}
\newcommand{\wwswhead}{\usebox{\westwestsouthwesthead}}

\newsavebox{\westwestnorthwesthead}
\savebox{\westwestnorthwesthead}{%
\xstep=\basicstep%
\multiply\xstep by 4472%
\divide\xstep by 10000%
\ystep=\xstep%
\multiply\ystep by -2%
\makehead{\xstep}{\ystep}{\numberofsteps}}
\newcommand{\wwnwhead}{\usebox{\westwestnorthwesthead}}

\newsavebox{\easteastsoutheasthead}
\savebox{\easteastsoutheasthead}{%
\xstep=-\basicstep%
\multiply\xstep by 4472%
\divide\xstep by 10000%
\ystep=\xstep%
\multiply\ystep by -2%
\makehead{\xstep}{\ystep}{\numberofsteps}}
\newcommand{\eesehead}{\usebox{\easteastsoutheasthead}}

\newsavebox{\northnorthnorthwesthead}
\savebox{\northnorthnorthwesthead}{%
\xstep=-\basicstep%
\multiply\xstep by 4472%
\divide\xstep by 10000%
\ystep=\xstep%
\multiply\ystep by 2%
\makehead{\xstep}{\ystep}{\numberofsteps}}
\newcommand{\nnnwhead}{\usebox{\northnorthnorthwesthead}}

\newsavebox{\southsouthsoutheasthead}
\savebox{\southsouthsoutheasthead}{%
\xstep=\basicstep%
\multiply\xstep by 4472%
\divide\xstep by 10000%
\ystep=\xstep%
\multiply\ystep by 2%
\makehead{\xstep}{\ystep}{\numberofsteps}}
\newcommand{\sssehead}{\usebox{\southsouthsoutheasthead}}

\newsavebox{\northeasteastnortheasthead}
\savebox{\northeasteastnortheasthead}{%
\xstep=-\basicstep%
\multiply\xstep by 9806%
\divide\xstep by 10000%
\ystep=\xstep%
\divide\ystep by -5%
\makehead{\xstep}{\ystep}{\numberofsteps}}
\newcommand{\neenehead}{\usebox{\northeasteastnortheasthead}}

\newsavebox{\northeastnorthnortheasthead}
\savebox{\northeastnorthnortheasthead}{%
\xstep=-\basicstep%
\multiply\xstep by 9806%
\divide\xstep by 10000%
\ystep=\xstep%
\divide\ystep by 5%
\makehead{\xstep}{\ystep}{\numberofsteps}}
\newcommand{\nennehead}{\usebox{\northeastnorthnortheasthead}}

\newsavebox{\southwestsouthsouthwesthead}
\savebox{\southwestsouthsouthwesthead}{%
\xstep=\basicstep%
\multiply\xstep by 9806%
\divide\xstep by 10000%
\ystep=\xstep%
\divide\ystep by 5%
\makehead{\xstep}{\ystep}{\numberofsteps}}
\newcommand{\swsswhead}{\usebox{\southwestsouthsouthwesthead}}

\newsavebox{\southwestwestsouthwesthead}
\savebox{\southwestwestsouthwesthead}{%
\xstep=\basicstep%
\multiply\xstep by 9806%
\divide\xstep by 10000%
\ystep=\xstep%
\divide\ystep by -5%
\makehead{\xstep}{\ystep}{\numberofsteps}}
\newcommand{\swwswhead}{\usebox{\southwestwestsouthwesthead}}

\newsavebox{\northwestwestnorthwesthead}
\savebox{\northwestwestnorthwesthead}{%
\xstep=\basicstep%
\multiply\xstep by 1961%
\divide\xstep by 10000%
\ystep=\xstep%
\multiply\ystep by -5%
\makehead{\xstep}{\ystep}{\numberofsteps}}
\newcommand{\nwwnwhead}{\usebox{\northwestwestnorthwesthead}}

\newsavebox{\southeasteastsoutheasthead}
\savebox{\southeasteastsoutheasthead}{%
\xstep=-\basicstep%
\multiply\xstep by 1961%
\divide\xstep by 10000%
\ystep=\xstep%
\multiply\ystep by -5%
\makehead{\xstep}{\ystep}{\numberofsteps}}
\newcommand{\seesehead}{\usebox{\southeasteastsoutheasthead}}

\newsavebox{\northwestnorthnorthwesthead}
\savebox{\northwestnorthnorthwesthead}{%
\xstep=-\basicstep%
\multiply\xstep by 1961%
\divide\xstep by 10000%
\ystep=\xstep%
\multiply\ystep by 5%
\makehead{\xstep}{\ystep}{\numberofsteps}}
\newcommand{\nwnnwhead}{\usebox{\northwestnorthnorthwesthead}}

\newsavebox{\southeastsouthsoutheasthead}
\savebox{\southeastsouthsoutheasthead}{%
\xstep=\basicstep%
\multiply\xstep by 1961%
\divide\xstep by 10000%
\ystep=\xstep%
\multiply\ystep by 5%
\makehead{\xstep}{\ystep}{\numberofsteps}}
\newcommand{\sessehead}{\usebox{\southeastsouthsoutheasthead}}

\newsavebox{\isomorphismmark}
\newcommand{\isomark}[1]{\savebox{\isomorphismmark}{#1}}
\isomark{$\cong$}

\newif\ifuserdist
\newsavebox{\distributormark}
\newcommand{\distmark}[1]{\ifx#1\distcircle\userdistfalse\else%
\userdisttrue\savebox{\distributormark}{#1}\fi}
\newsavebox{\distributorcircle}
\savebox{\distributorcircle}{\begin{picture}(0,0)%
\put(0,0){\circle{4}}\end{picture}}

\userdistfalse

\newcount\monolength
\newcount\epilength
\newcount\bimolength
\newcount\secondmonolength
\newcount\secondepilength

\newcount\monotail%
\global\monotail=\headsize%
\global\multiply\monotail by 7070%
\global\divide\monotail by 10000%

\newcount\truemonotail%
\newcount\trueepihead%

\def\truetail{\truemonotail=\monotail%
\multiply\truemonotail by 100%
\divide\truemonotail by \SCALE}

\def\truehead{\trueepihead=\epihead%
\multiply\trueepihead by 100%
\divide\trueepihead by \SCALE}

\newcount\Monotail%
\global\Monotail=\headsize%
\global\divide\Monotail by 2%

\newcount\Epihead%
\global\Epihead=\epihead%
\global\multiply\Epihead by 7070%
\global\divide\Epihead by 10000%

\newcount\Truemonotail%
\newcount\Trueepihead%

\def\Truetail{\Truemonotail=\Monotail%
\multiply\Truemonotail by 100%
\divide\Truemonotail by \SCALE}%

\def\Truehead{\Trueepihead=\Epihead%
\multiply\Trueepihead by 100%
\divide\Trueepihead by \SCALE}

\newcount\MonoTail%
\global\MonoTail=\headsize%
\global\multiply\MonoTail by 6324%
\global\divide\MonoTail by 10000%

\newcount\EpiHead%
\global\EpiHead=\epihead%
\global\multiply\EpiHead by 8944%
\global\divide\EpiHead by 10000%

\newcount\TrueMonoTail%
\newcount\TrueEpiHead%

\def\TrueTail{\TrueMonoTail=\MonoTail%
\multiply\TrueMonoTail by 100%
\divide\TrueMonoTail by \SCALE}%

\def\TrueHead{\TrueEpiHead=\EpiHead%
\multiply\TrueEpiHead by 100%
\divide\TrueEpiHead by \SCALE}

\newcount\monotaiL%
\global\monotaiL=\headsize%
\global\multiply\monotaiL by 3162%
\global\divide\monotaiL by 10000%

\newcount\epiheaD%
\global\epiheaD=\epihead%
\global\multiply\epiheaD by 4472%
\global\divide\epiheaD by 10000%

\newcount\truemonotaiL%
\newcount\trueepiheaD%

\def\truetaiL{\truemonotaiL=\monotaiL%
\multiply\truemonotaiL by 100%
\divide\truemonotaiL by \SCALE}%

\def\trueheaD{\trueepiheaD=\epiheaD%
\multiply\trueepiheaD by 100%
\divide\trueepiheaD by \SCALE}

\newcount\SCALE%

\newcount\NUMBER%

\newcount\LINE%

\newcount\COLUMN%

\newcount\WIDTH%

\newcount\SOURCE%

\newcount\ARROW%

\newcount\TARGET%

\newcount\ARROWLENGTH%

\newcount\NUMBEROFDOTS%

\newcounter{x}%

\newcounter{y}%

\newcounter{z}%

\newcounter{horizontal}%

\newcounter{vertical}%

\newskip\itemlength%

\newskip\firstitem%

\newskip\seconditem%

\newcommand{\printarrow}{}%

\newcount\X%

\newcount\Y%

\newcount\Z%

\newcommand{\truex}[1]{%
\NUMBER=#1%
\multiply\NUMBER by 100%
\divide\NUMBER by \SCALE%
\setcounter{x}{\NUMBER}}%

\newcommand{\truey}[1]{%
\NUMBER=#1%
\multiply\NUMBER by 100%
\divide\NUMBER by \SCALE%
\setcounter{y}{\NUMBER}}%

\newcommand{\truez}[1]{%
\NUMBER=#1%
\multiply\NUMBER by 100%
\divide\NUMBER by \SCALE%
\setcounter{z}{\NUMBER}}%

\newcommand{\changecounters}[1]{%
\SOURCE=\ARROW%
\ARROW=\TARGET%
\settowidth{\itemlength}{#1}%
\ifdim \itemlength > 2800\unitlength%
\addtolength{\itemlength}{-2800\unitlength}%
\TARGET=\itemlength%
\divide\TARGET by 1310%
\multiply\TARGET by 100%
\divide\TARGET by \SCALE%
\else%
\TARGET=0%
\fi%
\ARROWLENGTH=5000%
\advance\ARROWLENGTH by -\SOURCE%
\advance\ARROWLENGTH by -\TARGET%
\divide\ARROWLENGTH by 100%
\advance\SOURCE by -\TARGET}%

\newcommand{\initialize}[1]{%
\LINE=0%
\COLUMN=0%
\WIDTH=0%
\ARROW=0%
\TARGET=0%
\changecounters{#1}%
\renewcommand{\printarrow}{#1}%
\begin{center}%
\vspace{2pt}%
\begin{picture}(0,0)}%

\newcommand{\DIAGV}[2]{%
\checkmode%
\SCALE=#1%
\setlength{\unitlength}{655sp}%
\multiply\unitlength by \SCALE%
\divide\unitlength by 100%
\initialize{\mbox{$#2$}}}%

\newcommand{\n}[1]{%
\changecounters{\mbox{$#1$}}%
\put(\COLUMN,\LINE){\makebox(0,0){\printarrow}}%
\thinlines%
\renewcommand{\printarrow}{\mbox{$#1$}}%
\advance\COLUMN by 4000}%

\newcommand{\nn}[1]{%
\put(\COLUMN,\LINE){\makebox(0,0){\printarrow}}%
\thinlines%
\ifnum \WIDTH < \COLUMN%
\WIDTH=\COLUMN%
\else%
\fi%
\advance\LINE by -4000%
\COLUMN=0%
\ARROW=0%
\TARGET=0%
\changecounters{\mbox{$#1$}}%
\renewcommand{\printarrow}{\mbox{$#1$}}}%

\newcommand{\conclude}{%
\put(\COLUMN,\LINE){\makebox(0,0){\printarrow}}%
\thinlines%
\ifnum \WIDTH < \COLUMN%
\WIDTH=\COLUMN%
\else%
\fi%
\setcounter{horizontal}{\WIDTH}%
\setcounter{vertical}{-\LINE}%
\end{picture}}%

\newcommand{\diag}{%
\conclude%
\raisebox{0pt}[0pt][\value{vertical}\unitlength]{}%
\hspace*{\value{horizontal}\unitlength}%
\vspace{12pt}%
\end{center}%
\setlength{\unitlength}{1pt}%
\global\diagramfalse}%

\newcommand{\diagv}[3]{%
\conclude%
\NUMBER=#1%
\rule{0pt}{\NUMBER pt}%
\hspace*{-#2pt}%
\raisebox{0pt}[0pt][\value{vertical}\unitlength]{}%
\hspace*{\value{horizontal}\unitlength}
\NUMBER=#3%
\advance\NUMBER by 12%
\vspace*{\NUMBER pt}%
\end{center}%
\setlength{\unitlength}{1pt}%
\global\diagramfalse}%

\def\movename(#1,#2)#3{%
\hspace{#1pt}%
\raisebox{#2pt}[5pt][2pt]{\raisebox{#2pt}{$#3$}}%
\hspace{-#1pt}}%

\def\movearrow(#1,#2)#3{%
\makebox[0pt]{%
\hspace{#1pt}\hspace{#1pt}%
\raisebox{#2pt}[0pt][0pt]{\raisebox{#2pt}{$#3$}}}}%

\def\movevertex(#1,#2)#3{%
\mbox{\hspace{#1pt}%
\raisebox{#2pt}{\raisebox{#2pt}{$#3$}}%
\hspace{-#1pt}}}%

\newcommand{\crosslength}[2]{%
\settowidth{\firstitem}{#1}%
\settowidth{\seconditem}{#2}%
\ifdim\firstitem < \seconditem%
\itemlength=\seconditem%
\else%
\itemlength=\firstitem%
\fi%
\divide\itemlength by 2%
\hspace{\itemlength}}%

\newcommand{\bold}{\ifdiagram\thicklines\else\typeout{Sorry: command
\string\bold does not apply to text arrows; I am ignoring it.}\fi}

\def\basicDIAG#1¤{\DIAGV{\defaultscale}{#1}\@ifnexttok¤{\finishline}{\basicn}}
\def\basicDIAGV[#1]#2¤{\DIAGV{#1}{#2}\@ifnexttok¤{\finishline}{\basicn}}
\def\basicn#1¤{\n{#1}\@ifnexttok¤{\finishline}{\basicn}}
\def\basicnn#1¤{\nn{#1}\@ifnexttok¤{\finishline}{\basicn}}
\def\finishline#1{\@ifnextchar\end{\diag}%
{\@ifnextchar\spacing{\relax}{\basicnn}}}
\def\spacing(#1,#2,#3){\diagv{#1}{#2}{#3}}

\newif\ifcaption%

\newenvironment{diagram}{%
\iffloatdiag\relax\else
\global\def\diagramcaption##1{%
\global\captiontrue%
\global\def\@diagcaption{##1}}%
\global\def\@diagcaption{}\fi%
\@ifnextchar[{\basicDIAGV}{\basicDIAG}}%
{\iffloatdiag\relax\else%
\ifcaption
\begin{center}\mbox{}\@diagcaption\end{center}%
\else\relax\fi\fi\global\captionfalse}

\global\captionfalse

\gdef\@diaglabel{Diagram}
\gdef\diagramlabel#1{\gdef\@diaglabel{#1}}

\newcounter{Diagram}
\def\theDiagram{\@arabic\c@Diagram}
\def\fps@Diagram{tbp}
\def\ftype@Diagram{1}
\def\ext@Diagram{lof}
\def\fnum@Diagram{\@diaglabel\ \theDiagram}
\def\Diagram{\@float{Diagram}}
\let\endDiagram\end@float
\@namedef{Diagram*}{\@dblfloat{Diagram}}
\@namedef{endDiagram*}{\end@dblfloat}

\def\setdiagramcounter#1{\@addtoreset{Diagram}{#1}%
\def\theDiagram{\arabic{#1}.\@arabic\c@Diagram}}

\newif\iffloatdiag

{\end{diagram}\caption{\@diagcaption}\end{Diagram}%
\global\floatdiagfalse}

\global\floatdiagfalse

\newcommand{\TUP}[1]{\raisebox{0pt}[0pt][3pt]{}#1}

\newcommand{\TDOWN}[1]{\raisebox{0pt}[6pt][0pt]{}#1}

\newcommand{\tlowername}[2]%
{$\stackrel{\makebox[1pt]{#1}}%
{\begin{picture}(0,0)%
\put(0,0){\makebox(0,6)[t]{\makebox[1pt]{$\scriptstyle#2$}}}%
\end{picture}}$}%

\newcommand{\tcase}[1]{%
\testtextmode%
\setlength{\unitlength}{0.01pt}%
\makebox[\actualtextarrowspace pt]%
{\raisebox{2.5pt}{#1{\actualtextarrowlength}}}%
\setlength{\unitlength}{1pt}}%

\newcommand{\Tcase}[2]{%
\testtextmode%
\setlength{\unitlength}{0.01pt}%
\makebox[\actualtextarrowspace pt]%
{\raisebox{2.5pt}{$\stackrel{\scriptstyle #2}{#1{\actualtextarrowlength}}$}}%
\setlength{\unitlength}{1pt}}%

\newcommand{\tbicase}[1]{%
\testtextmode%
\setlength{\unitlength}{0.01pt}%
\makebox[\actualtextarrowspace pt]%
{\raisebox{1pt}{#1{\actualtextarrowlength}}}%
\setlength{\unitlength}{1pt}}%

\newcommand{\Tbicase}[3]{%
\testtextmode%
\setlength{\unitlength}{0.01pt}%
\makebox[\actualtextarrowspace pt]%
{\raisebox{-1pt}%
{$\stackrel{\scriptstyle #2}%
{\mbox{\tlowername{#1{\actualtextarrowlength}}%
{\scriptstyle #3}}}$}}%
\setlength{\unitlength}{1pt}}%

\newcommand{\DUP}[1]{\raisebox{0pt}[0pt][4pt]{}#1}

\newcommand{\DDOWN}[1]{\raisebox{0pt}[9pt][0pt]{}#1}

\newcommand{\dlowername}[2]%
{$\stackrel{\makebox[1pt]{#1}}%
{\begin{picture}(0,0)%
\put(0,0){\makebox(0,6)[t]{\makebox[1pt]{$\textstyle#2$}}}%
\end{picture}}$}%

\newcommand{\dcase}[1]{%
\testtextmode%
\setlength{\unitlength}{0.01pt}%
\makebox[\actualdisplayarrowspace pt]%
{\raisebox{2.5pt}{#1{\actualdisplayarrowlength}}}%
\setlength{\unitlength}{1pt}}%

\newcommand{\Dcase}[2]{%
\testtextmode%
\setlength{\unitlength}{0.01pt}%
\makebox[\actualdisplayarrowspace pt]%
{\raisebox{2.5pt}{$\stackrel{\textstyle #2}{#1{\actualdisplayarrowlength}}$}}%
\setlength{\unitlength}{1pt}}%

\newcommand{\dbicase}[1]{%
\testtextmode%
\setlength{\unitlength}{0.01pt}%
\makebox[\actualdisplayarrowspace pt]%
{\raisebox{1pt}{#1{\actualdisplayarrowlength}}}%
\setlength{\unitlength}{1pt}}%

\newcommand{\Dbicase}[3]{%
\testtextmode%
\setlength{\unitlength}{0.01pt}%
\makebox[\actualdisplayarrowspace pt]%
{\raisebox{-1pt}%
{$\stackrel{\textstyle #2}%
{\mbox{\tlowername{#1{\actualdisplayarrowlength}}%
{\textstyle #3}}}$}}%
\setlength{\unitlength}{1pt}}%

\newcommand{\AR}[1]%
{\begin{picture}(#1,0)%
\put(0,0){\line(1,0){#1}}%
\put(#1,0){\ehead}%
\end{picture}}%

\newcommand{\DIST}[1]%
{\begin{picture}(#1,0)%
\put(0,0){\line(1,0){#1}}%
\put(#1,0){\ehead}%
\NUMBER=#1%
\divide\NUMBER by 2%
\put(\NUMBER,0){\circle{400}}%
\end{picture}}%

\newcommand{\DOTAR}[1]%
{\NUMBEROFDOTS=#1%
\divide\NUMBEROFDOTS by 300%
\advance\NUMBEROFDOTS by 1%
\begin{picture}(#1,0)%
\multiput(0,0)(300,0){\NUMBEROFDOTS}{\circle*{100}}%
\put(#1,0){\ehead}%
\end{picture}}%

\newcommand{\MONO}[1]%
{\monolength=#1%
\advance\monolength by -\monotail%
\begin{picture}(#1,0)%
\put(\monotail,0){\line(1,0){\monolength}}%
\put(#1,0){\ehead}%
\put(\monotail,0){\ehead}%
\end{picture}}%

\newcommand{\EPI}[1]%
{\epilength=#1%
\advance\epilength by -\epihead%
\begin{picture}(#1,0)(-#1,0)%
\put(-#1,0){\line(1,0){\epilength}}%
\put(-\epihead,0){\ehead}%
\put(0,0){\ehead}%
\end{picture}}%

\newcommand{\BIMO}[1]%
{\monolength=#1%
\advance\monolength by -\monotail%
\epilength=\monolength%
\advance\epilength by -\epihead%
\begin{picture}(#1,0)(-#1,0)%
\put(-\monolength,0){\line(1,0){\epilength}}%
\put(-\monolength,0){\ehead}%
\put(-\epihead,0){\ehead}%
\put(0,0){\ehead}%
\end{picture}}%

\newcommand{\BIAR}[1]%
{\begin{picture}(#1,700)%
\put(0,0){\line(1,0){#1}}%
\put(#1,0){\ehead}%
\put(0,700){\line(1,0){#1}}%
\put(#1,700){\ehead}%
\end{picture}}%

\newcommand{\BIDIST}[1]%
{\begin{picture}(#1,700)%
\put(0,0){\line(1,0){#1}}%
\put(#1,0){\ehead}%
\put(0,700){\line(1,0){#1}}%
\put(#1,700){\ehead}%
\NUMBER=#1%
\divide\NUMBER by 2%
\put(\NUMBER,0){\circle{400}}%
\put(\NUMBER,700){\circle{400}}%
\end{picture}}%

\newcommand{\EQL}[1]%
{\begin{picture}(#1,0)%
\put(0,100){\line(1,0){#1}}%
\put(0,-100){\line(1,0){#1}}%
\end{picture}}%

\newcommand{\ADJAR}[1]%
{\begin{picture}(#1,700)%
\put(0,0){\line(1,0){#1}}%
\put(#1,0){\ehead}%
\put(#1,700){\line(-1,0){#1}}%
\put(0,700){\whead}
\end{picture}}%

\newcommand{\ADJDIST}[1]%
{\begin{picture}(#1,700)%
\put(0,0){\line(1,0){#1}}%
\put(#1,0){\ehead}%
\put(#1,700){\line(-1,0){#1}}%
\put(0,700){\whead}
\NUMBER=#1%
\divide\NUMBER by 2%
\put(\NUMBER,0){\circle{400}}%
\put(\NUMBER,700){\circle{400}}%
\end{picture}}%

\newcommand{\ar}{\ifinner\tcase{\AR}\else\dcase{\AR}\fi}%

\newcommand{\Ar}[1]{\ifinner\Tcase{\AR}{#1}\else\Dcase{\AR}{#1}\fi}%

\newcommand{\dist}{\ifinner\tcase{\DIST}\else\dcase{\DIST}\fi}%

\newcommand{\Dist}[1]{\ifinner\Tcase{\DIST}{\TUP{#1}}%
\else\Dcase{\DIST}{\TUP{#1}}\fi}%

\newcommand{\dotar}{\ifinner\tcase{\DOTAR}\else\dcase{\DOTAR}\fi}%

\newcommand{\Dotar}[1]{\ifinner\Tcase{\DOTAR}{#1}%
\else\Dcase{\DOTAR}{#1}\fi}%

\newcommand{\mono}{\ifinner\tcase{\MONO}\else\dcase{\MONO}\fi}%

\newcommand{\Mono}[1]{\ifinner\Tcase{\MONO}{#1}\else\Dcase{\MONO}{#1}\fi}%

\newcommand{\epi}{\ifinner\tcase{\EPI}\else\dcase{\EPI}\fi}%

\newcommand{\Epi}[1]{\ifinner\Tcase{\EPI}{#1}\else\Dcase{\EPI}{#1}\fi}%

\newcommand{\bimo}{\ifinner\tcase{\BIMO}\else\dcase{\BIMO}\fi}%

\newcommand{\Bimo}[1]{\ifinner\Tcase{\BIMO}{#1}%
\else\Dcase{\BIMO}{#1}\fi}%

\newcommand{\iso}{\ifinner\Tcase{\AR}{\cong}\else\Dcase{\AR}{\cong}\fi}%

\newcommand{\Iso}[1]{\ifinner\Tcase{\AR}{\cong{#1}}%
\else\Dcase{\AR}{\cong{#1}}\fi}%

\newcommand{\biar}{\ifinner\tbicase{\BIAR}\else\dbicase{\BIAR}\fi}%

\newcommand{\Biar}[2]{\ifinner\Tbicase{\BIAR}{#1}{#2}%
\else\Dbicase{\BIAR}{#1}{#2}\fi}%

\newcommand{\bidist}{\ifinner\tbicase{\BIDIST}\else\dbicase{\BIDIST}\fi}%

\newcommand{\Bidist}[2]{\ifinner\Tbicase{\BIDIST}{\TUP{#1}}{\TDOWN{#2}}%
\else\Dbicase{\BIDIST}{\DUP{#1}}{\DDOWN{#2}}\fi}%

\newcommand{\eql}{\ifinner\tcase{\EQL}\else\dcase{\EQL}\fi}%

\newcommand{\Eql}[1]{\ifinner\Tcase{\EQL}{\TUP{#1}}%
\else\Dcase{\EQL}{\DUP{#1}}\fi}%

\newcommand{\adjar}{\ifinner\tbicase{\ADJAR}\else\dbicase{\ADJAR}\fi}%

\newcommand{\Adjar}[2]{\ifinner\Tbicase{\ADJAR}{#1}{#2}%
\else\Dbicase{\ADJAR}{#1}{#2}\fi}%

\newcommand{\adjdist}{\ifinner\tbicase{\ADJDIST}\else\dbicase{\ADJDIST}\fi}%

\newcommand{\Adjdist}[2]{\ifinner\Tbicase{\ADJDIST}{\TUP{#1}}{\TDOWN{#2}}%
\else\Dbicase{\ADJDIST}{\DUP{#1}}{\DDOWN{#2}}\fi}%

\newcommand{\BKAR}[1]%
{\begin{picture}(#1,0)%
\put(#1,0){\line(-1,0){#1}}%
\put(0,0){\whead}%
\end{picture}}%

\newcommand{\BKDIST}[1]%
{\begin{picture}(#1,0)%
\put(#1,0){\line(-1,0){#1}}%
\put(0,0){\whead}%
\NUMBER=#1%
\divide\NUMBER by 2%
\put(\NUMBER,0){\circle{400}}%
\end{picture}}%

\newcommand{\BKDOTAR}[1]%
{\NUMBEROFDOTS=#1%
\divide\NUMBEROFDOTS by 300%
\advance\NUMBEROFDOTS by 1%
\begin{picture}(#1,0)%
\multiput(#1,0)(-300,0){\NUMBEROFDOTS}{\circle*{100}}%
\put(0,0){\whead}%
\end{picture}}%

\newcommand{\BKMONO}[1]%
{\monolength=#1%
\advance\monolength by -\monotail%
\begin{picture}(#1,0)(-#1,0)%
\put(-\monotail,0){\line(-1,0){\monolength}}%
\put(-\monotail,0){\whead}%
\put(-#1,0){\whead}%
\end{picture}}%

\newcommand{\BKEPI}[1]%
{\epilength=#1%
\advance\epilength by -\epihead%
\begin{picture}(#1,0)%
\put(#1,0){\line(-1,0){\epilength}}%
\put(\epihead,0){\whead}%
\put(0,0){\whead}%
\end{picture}}%

\newcommand{\BKBIMO}[1]%
{\monolength=#1%
\advance\monolength by -\monotail%
\epilength=\monolength%
\advance\epilength by -\epihead%
\begin{picture}(#1,0)%
\put(\monolength,0){\line(-1,0){\epilength}}%
\put(\monolength,0){\whead}%
\put(\epihead,0){\whead}%
\put(0,0){\whead}%
\end{picture}}%

\newcommand{\BKBIAR}[1]%
{\begin{picture}(#1,700)%
\put(#1,0){\line(-1,0){#1}}%
\put(0,0){\whead}%
\put(#1,700){\line(-1,0){#1}}%
\put(0,700){\whead}%
\end{picture}}%

\newcommand{\BKBIDIST}[1]%
{\begin{picture}(#1,700)%
\put(#1,0){\line(-1,0){#1}}%
\put(0,0){\whead}%
\put(#1,700){\line(-1,0){#1}}%
\put(0,700){\whead}%
\NUMBER=#1%
\divide\NUMBER by 2%
\put(\NUMBER,0){\circle{400}}%
\put(\NUMBER,700){\circle{400}}%
\end{picture}}%

\newcommand{\BKADJAR}[1]%
{\begin{picture}(#1,700)%
\put(0,700){\line(1,0){#1}}%
\put(#1,700){\ehead}%
\put(#1,0){\line(-1,0){#1}}%
\put(0,0){\whead}%
\end{picture}}%

\newcommand{\BKADJDIST}[1]%
{\begin{picture}(#1,700)%
\put(0,700){\line(1,0){#1}}%
\put(#1,700){\ehead}%
\put(#1,0){\line(-1,0){#1}}%
\put(0,0){\whead}%
\NUMBER=#1%
\divide\NUMBER by 2%
\put(\NUMBER,0){\circle{400}}%
\put(\NUMBER,700){\circle{400}}%
\end{picture}}%

\newcommand{\bkar}{\ifinner\tcase{\BKAR}\else\dcase{\BKAR}\fi}%

\newcommand{\Bkar}[1]{\ifinner\Tcase{\BKAR}{#1}\else\Dcase{\BKAR}{#1}\fi}%

\newcommand{\bkdist}{\ifinner\tcase{\BKDIST}\else\dcase{\BKDIST}\fi}%

\newcommand{\Bkdist}[1]{\ifinner\Tcase{\BKDIST}{\TUP{#1}}%
\else\Dcase{\BKDIST}{\TUP{#1}}\fi}%

\newcommand{\bkdotar}{\ifinner\tcase{\BKDOTAR}\else\dcase{\BKDOTAR}\fi}%

\newcommand{\Bkdotar}[1]{\ifinner\Tcase{\BKDOTAR}{#1}%
\else\Dcase{\BKDOTAR}{#1}\fi}%

\newcommand{\bkmono}{\ifinner\tcase{\BKMONO}\else\dcase{\BKMONO}\fi}%

\newcommand{\Bkmono}[1]{\ifinner\Tcase{\BKMONO}{#1}%
\else\Dcase{\BKMONO}{#1}\fi}%

\newcommand{\bkepi}{\ifinner\tcase{\BKEPI}\else\dcase{\BKEPI}\fi}%

\newcommand{\Bkepi}[1]{\ifinner\Tcase{\BKEPI}{#1}%
\else\Dcase{\BKEPI}{#1}\fi}%

\newcommand{\bkbimo}{\ifinner\tcase{\BKBIMO}\else\dcase{\BKBIMO}\fi}%

\newcommand{\Bkbimo}[1]{\ifinner\Tcase{\BKBIMO}{\hspace{9pt}#1}%
\else\Dcase{\BKBIMO}{\hspace{9pt}#1}\fi}%

\newcommand{\bkiso}{\ifinner\Tcase{\BKAR}{\cong}%
\else\Dcase{\BKAR}{\cong}\fi}%

\newcommand{\Bkiso}[1]{\ifinner\Tcase{\BKAR}{\cong{#1}}%
\else\Dcase{\BKAR}{\cong{#1}}\fi}%

\newcommand{\bkbiar}{\ifinner\tbicase{\BKBIAR}\else\dbicase{\BKBIAR}\fi}%

\newcommand{\Bkbiar}[2]{\ifinner\Tbicase{\BKBIAR}{#1}{#2}%
\else\Dbicase{\BKBIAR}{#1}{#2}\fi}%

\newcommand{\bkbidist}{\ifinner\tbicase{\BKBIDIST}%
\else\dbicase{\BKBIDIST}\fi}%

\newcommand{\Bkbidist}[2]{\ifinner\Tbicase{\BKBIDIST}{\TUP{#1}}{\TDOWN{#2}}%
\else\Tbicase{\BKBIDIST}{\DUP{#1}}{\DDOWN{#2}}\fi}%

\newcommand{\bkadjar}{\ifinner\tbicase{\BKADJAR}%
\else\dbicase{\BKADJAR}\fi}%

\newcommand{\Bkadjar}[2]{\ifinner\Tbicase{\BKADJAR}{#1}{#2}%
\else\Dbicase{\BKADJAR}{#1}{#2}\fi}%

\newcommand{\bkadjdist}{\ifinner\tbicase{\BKADJDIST}%
\else\dbicase{\BKADJDIST}\fi}%

\newcommand{\Bkadjdist}[2]{\ifinner\Tbicase{\BKADJDIST}{\TUP{#1}}{\TDOWN{#2}}%
\else\Dbicase{\BKADJDIST}{\TUP{#1}}{\TDOWN{#2}}\fi}%

\newcommand{\lowername}[2]%
{$\stackrel{\makebox[1pt]{#1}}%
{\begin{picture}(0,0)%
\truex{600}%
\put(0,0){\makebox(0,\value{x})[t]{\makebox[1pt]{$#2$}}}%
\end{picture}}$}%

\newcommand{\hcase}[2]%
{\testdiagrammode\makebox[0pt]%
{\raisebox{0pt}[0pt][0pt]{#1{#2}}}}%

\newcommand{\Hcase}[3]%
{\testdiagrammode\makebox[0pt]
{\raisebox{0pt}[0pt][0pt]%
{$\stackrel{\makebox[0pt]{$\textstyle{#2}$}}{#1{#3}}$}}}%

\newcommand{\hcasE}[3]%
{\testdiagrammode\makebox[0pt]%
{\raisebox{-8pt}[0pt][0pt]%
{\lowername{#1{#3}}{#2}}}}%

\newcommand{\Hisocase}[4]%
{\testdiagrammode\makebox[0pt]
{\raisebox{-8pt}[0pt][0pt]%
{$\stackrel{\makebox[0pt]{$\textstyle{#2}$}}%
{\mbox{\lowername{#1{#4}}{#3}}}$}}}%

\newcommand{\hbicase}[2]%
{\testdiagrammode\makebox[0pt]%
{\raisebox{-2.4pt}[0pt][0pt]{#1{#2}}}}%

\newcommand{\Hbicase}[4]%
{\testdiagrammode\makebox[0pt]
{\raisebox{-10.4pt}[0pt][0pt]%
{$\stackrel{\makebox[0pt]{$\textstyle{#2}$}}%
{\mbox{\lowername{#1{#4}}{#3}}}$}}}%

\newcommand{\EAR}[1]%
{\begin{picture}(#1,0)%
\put(0,0){\line(1,0){#1}}%
\put(#1,0){\ehead}%
\end{picture}}%

\newcommand{\EDIST}[1]%
{\begin{picture}(#1,0)%
\put(0,0){\line(1,0){#1}}%
\put(#1,0){\ehead}%
\truex{400}
\NUMBER=#1%
\divide\NUMBER by 2%
\put(\NUMBER,0){\circle{\value{x}}}
\end{picture}}%

\newcommand{\EDOTAR}[1]%
{\truex{100}\truey{300}%
\NUMBEROFDOTS=#1%
\divide\NUMBEROFDOTS by \value{y}%
\advance\NUMBEROFDOTS by 1%
\begin{picture}(#1,0)%
\multiput(0,0)(\value{y},0){\NUMBEROFDOTS}%
{\circle*{\value{x}}}%
\put(#1,0){\ehead}%
\end{picture}}%

\newcommand{\EMONO}[1]%
{\truetail
\monolength=#1%
\advance\monolength by -\truemonotail%
\begin{picture}(#1,0)%
\put(\truemonotail,0){\line(1,0){\monolength}}%
\put(#1,0){\ehead}%
\put(\truemonotail,0){\ehead}%
\end{picture}}%

\newcommand{\EEPI}[1]%
{\truehead%
\epilength=#1%
\advance\epilength by -\trueepihead%
\begin{picture}(#1,0)(-#1,0)%
\put(-#1,0){\line(1,0){\epilength}}%
\put(-\trueepihead,0){\ehead}%
\put(0,0){\ehead}%
\end{picture}}%

\newcommand{\EBIMO}[1]%
{\truehead\truetail%
\monolength=#1%
\advance\monolength by -\truemonotail%
\epilength=\monolength%
\advance\epilength by -\trueepihead%
\begin{picture}(#1,0)(-#1,0)%
\put(-\monolength,0){\line(1,0){\epilength}}%
\put(-\monolength,0){\ehead}%
\put(-\trueepihead,0){\ehead}%
\put(0,0){\ehead}%
\end{picture}}%

\newcommand{\EBIAR}[1]%
{\truex{700}%
\begin{picture}(#1,\value{x})%
\put(0,0){\line(1,0){#1}}%
\put(#1,0){\ehead}%
\put(0,\value{x}){\line(1,0){#1}}%
\put(#1,\value{x}){\ehead}%
\end{picture}}%

\newcommand{\EBIDIST}[1]%
{\truex{700}%
\begin{picture}(#1,\value{x})%
\put(0,0){\line(1,0){#1}}%
\put(#1,0){\ehead}%
\put(0,\value{x}){\line(1,0){#1}}%
\put(#1,\value{x}){\ehead}%
\truey{400}%
\NUMBER=#1%
\divide\NUMBER by 2%
\put(\NUMBER,0){\circle{\value{y}}}
\put(\NUMBER,\value{x}){\circle{\value{y}}}%
\end{picture}}%

\newcommand{\EEQL}[1]%
{\begin{picture}(#1,0)%
\truex{200}%
\put(0,\value{x}){\line(1,0){#1}}%
\put(0,0){\line(1,0){#1}}%
\end{picture}}%

\newcommand{\EADJAR}[1]%
{\truex{700}%
\begin{picture}(#1,\value{x})%
\put(0,0){\line(1,0){#1}}%
\put(#1,0){\ehead}%
\put(#1,\value{x}){\line(-1,0){#1}}%
\put(0,\value{x}){\whead}%
\end{picture}}%

\newcommand{\EADJDIST}[1]%
{\truex{700}%
\begin{picture}(#1,\value{x})%
\put(0,0){\line(1,0){#1}}%
\put(#1,0){\ehead}%
\put(#1,\value{x}){\line(-1,0){#1}}%
\put(0,\value{x}){\whead}%
\truey{400}%
\NUMBER=#1%
\divide\NUMBER by 2%
\put(\NUMBER,0){\circle{\value{y}}}
\put(\NUMBER,\value{x}){\circle{\value{y}}}%
\end{picture}}%

\def\basicear[#1]{%
\Z=#1%
\multiply \Z by 100%
\hcase{\EAR}{\Z}}%

\newcommand{\ear}{\@ifnextchar[{\basicear}%
{\hspace{\SOURCE\unitlength}\basicear[\ARROWLENGTH]}}%

\def\basicEar[#1]#2{%
\Z=#1%
\multiply \Z by 100%
\Hcase{\EAR}{#2}{\Z}}%

\newcommand{\Ear}{\@ifnextchar[{\basicEar}%
{\hspace{\SOURCE\unitlength}\basicEar[\ARROWLENGTH]}}%

\def\basiceaR[#1]#2{%
\Z=#1%
\multiply \Z by 100%
\hcasE{\EAR}{#2}{\Z}}%

\newcommand{\eaR}{\@ifnextchar[{\basiceaR}%
{\hspace{\SOURCE\unitlength}\basiceaR[\ARROWLENGTH]}}%

\def\basicedist[#1]{%
\Z=#1%
\multiply \Z by 100%
\hcase{\EDIST}{\Z}}%

\newcommand{\edist}{\@ifnextchar[{\basicedist}%
{\hspace{\SOURCE\unitlength}\basicedist[\ARROWLENGTH]}}%

\def\basicEdist[#1]#2{%
\Z=#1%
\multiply \Z by 100%
\Hcase{\EDIST}{\DUP{#2}}{\Z}}%

\newcommand{\Edist}{\@ifnextchar[{\basicEdist}%
{\hspace{\SOURCE\unitlength}\basicEdist[\ARROWLENGTH]}}%

\def\basicedisT[#1]#2{%
\Z=#1%
\multiply \Z by 100%
\hcasE{\EDIST}{\DDOWN{#2}}{\Z}}%

\newcommand{\edisT}{\@ifnextchar[{\basicedisT}%
{\hspace{\SOURCE\unitlength}\basicedisT[\ARROWLENGTH]}}%

\def\basicedotar[#1]{%
\Z=#1%
\multiply \Z by 100%
\hcase{\EDOTAR}{\Z}}%

\newcommand{\edotar}{\@ifnextchar[{\basicedotar}%
{\hspace{\SOURCE\unitlength}\basicedotar[\ARROWLENGTH]}}%

\def\basicEdotar[#1]#2{%
\Z=#1%
\multiply \Z by 100%
\Hcase{\EDOTAR}{#2}{\Z}}%

\newcommand{\Edotar}{\@ifnextchar[{\basicEdotar}%
{\hspace{\SOURCE\unitlength}\basicEdotar[\ARROWLENGTH]}}%

\def\basicedotaR[#1]#2{%
\Z=#1%
\multiply \Z by 100%
\hcasE{\EDOTAR}{#2}{\Z}}%

\newcommand{\edotaR}{\@ifnextchar[{\basicedotaR}%
{\hspace{\SOURCE\unitlength}\basicedotaR[\ARROWLENGTH]}}%

\def\basicemono[#1]{%
\Z=#1%
\multiply \Z by 100%
\hcase{\EMONO}{\Z}}%

\newcommand{\emono}{\@ifnextchar[{\basicemono}%
{\hspace{\SOURCE\unitlength}\basicemono[\ARROWLENGTH]}}%

\def\basicEmono[#1]#2{%
\Z=#1%
\multiply \Z by 100%
\Hcase{\EMONO}{#2}{\Z}}%

\newcommand{\Emono}{\@ifnextchar[{\basicEmono}%
{\hspace{\SOURCE\unitlength}\basicEmono[\ARROWLENGTH]}}%

\def\basicemonO[#1]#2{%
\Z=#1%
\multiply \Z by 100%
\hcasE{\EMONO}{#2}{\Z}}%

\newcommand{\emonO}{\@ifnextchar[{\basicemonO}%
{\hspace{\SOURCE\unitlength}\basicemonO[\ARROWLENGTH]}}%

\def\basiceepi[#1]{%
\Z=#1%
\multiply \Z by 100%
\hcase{\EEPI}{\Z}}%

\newcommand{\eepi}{\@ifnextchar[{\basiceepi}%
{\hspace{\SOURCE\unitlength}\basiceepi[\ARROWLENGTH]}}%

\def\basicEepi[#1]#2{%
\Z=#1%
\multiply \Z by 100%
\Hcase{\EEPI}{#2}{\Z}}%

\newcommand{\Eepi}{\@ifnextchar[{\basicEepi}%
{\hspace{\SOURCE\unitlength}\basicEepi[\ARROWLENGTH]}}%

\def\basiceepI[#1]#2{%
\Z=#1%
\multiply \Z by 100%
\hcasE{\EEPI}{#2}{\Z}}%

\newcommand{\eepI}{\@ifnextchar[{\basiceepI}%
{\hspace{\SOURCE\unitlength}\basiceepI[\ARROWLENGTH]}}%

\def\basicebimo[#1]{%
\Z=#1%
\multiply \Z by 100%
\hcase{\EBIMO}{\Z}}%

\newcommand{\ebimo}{\@ifnextchar[{\basicebimo}%
{\hspace{\SOURCE\unitlength}\basicebimo[\ARROWLENGTH]}}%

\def\basicEbimo[#1]#2{%
\Z=#1%
\multiply \Z by 100%
\Hcase{\EBIMO}{#2}{\Z}}%

\newcommand{\Ebimo}{\@ifnextchar[{\basicEbimo}%
{\hspace{\SOURCE\unitlength}\basicEbimo[\ARROWLENGTH]}}%

\def\basicebimO[#1]#2{%
\Z=#1%
\multiply \Z by 100%
\hcasE{\EBIMO}{#2}{\Z}}%

\newcommand{\ebimO}{\@ifnextchar[{\basicebimO}%
{\hspace{\SOURCE\unitlength}\basicebimO[\ARROWLENGTH]}}%

\def\basiceiso[#1]{%
\Z=#1%
\multiply \Z by 100%
\Hisocase{\EAR}{\cong}{}{\Z}}%

\newcommand{\eiso}{\@ifnextchar[{\basiceiso}%
{\hspace{\SOURCE\unitlength}\basiceiso[\ARROWLENGTH]}}%

\def\basicEiso[#1]#2{%
\Z=#1%
\multiply \Z by 100%
\Hisocase{\EAR}{#2}{\cong}{\Z}}%

\newcommand{\Eiso}{\@ifnextchar[{\basicEiso}%
{\hspace{\SOURCE\unitlength}\basicEiso[\ARROWLENGTH]}}%

\def\basiceisO[#1]#2{%
\Z=#1%
\multiply \Z by 100%
\Hisocase{\EAR}{\cong}{#2}{\Z}}%

\newcommand{\eisO}{\@ifnextchar[{\basiceisO}%
{\hspace{\SOURCE\unitlength}\basiceisO[\ARROWLENGTH]}}%

\def\basiceeql[#1]{%
\Z=#1%
\multiply \Z by 100%
\hcase{\EEQL}{\Z}}%

\newcommand{\eeql}{\@ifnextchar[{\basiceeql}%
{\hspace{\SOURCE\unitlength}\basiceeql[\ARROWLENGTH]}}%

\def\basicEeql[#1]#2{%
\Z=#1%
\multiply \Z by 100%
\Hcase{\EEQL}{\DUP{#2}}{\Z}}%

\newcommand{\Eeql}{\@ifnextchar[{\basicEeql}%
{\hspace{\SOURCE\unitlength}\basicEeql[\ARROWLENGTH]}}%

\def\basiceeqL[#1]#2{%
\Z=#1%
\multiply \Z by 100%
\hcasE{\EEQL}{#2}{\Z}}%

\newcommand{\eeqL}{\@ifnextchar[{\basiceeqL}%
{\hspace{\SOURCE\unitlength}\basiceeqL[\ARROWLENGTH]}}%

\def\basicebiar[#1]{%
\Z=#1%
\multiply \Z by 100%
\hbicase{\EBIAR}{\Z}}%

\newcommand{\ebiar}{\@ifnextchar[{\basicebiar}%
{\hspace{\SOURCE\unitlength}\basicebiar[\ARROWLENGTH]}}%

\def\basicEbiar[#1]#2#3{%
\Z=#1%
\multiply \Z by 100%
\Hbicase{\EBIAR}{#2}{#3}{\Z}}%

\newcommand{\Ebiar}{\@ifnextchar[{\basicEbiar}%
{\hspace{\SOURCE\unitlength}\basicEbiar[\ARROWLENGTH]}}%

\def\basicebidist[#1]{%
\Z=#1%
\multiply \Z by 100%
\hbicase{\EBIDIST}{\Z}}%

\newcommand{\ebidist}{\@ifnextchar[{\basicebidist}%
{\hspace{\SOURCE\unitlength}\basicebidist[\ARROWLENGTH]}}%

\def\basicEbidist[#1]#2#3{%
\Z=#1%
\multiply \Z by 100%
\Hbicase{\EBIDIST}{\DUP{#2}}{\DDOWN{#3}}{\Z}}%

\newcommand{\Ebidist}{\@ifnextchar[{\basicEbidist}%
{\hspace{\SOURCE\unitlength}\basicEbidist[\ARROWLENGTH]}}%

\def\basiceadjar[#1]{%
\Z=#1%
\multiply \Z by 100%
\hbicase{\EADJAR}{\Z}}%

\newcommand{\eadjar}{\@ifnextchar[{\basiceadjar}%
{\hspace{\SOURCE\unitlength}\basiceadjar[\ARROWLENGTH]}}%

\def\basicEadjar[#1]#2#3{%
\Z=#1%
\multiply \Z by 100%
\Hbicase{\EADJAR}{#2}{#3}{\Z}}%

\newcommand{\Eadjar}{\@ifnextchar[{\basicEadjar}%
{\hspace{\SOURCE\unitlength}\basicEadjar[\ARROWLENGTH]}}%

\def\basiceadjdist[#1]{%
\Z=#1%
\multiply \Z by 100%
\hbicase{\EADJDIST}{\Z}}%

\newcommand{\eadjdist}{\@ifnextchar[{\basiceadjdist}%
{\hspace{\SOURCE\unitlength}\basiceadjdist[\ARROWLENGTH]}}%

\def\basicEadjdist[#1]#2#3{%
\Z=#1%
\multiply \Z by 100%
\Hbicase{\EADJDIST}{\DUP{#2}}{\DDOWN{#3}}{\Z}}%

\newcommand{\Eadjdist}{\@ifnextchar[{\basicEadjdist}%
{\hspace{\SOURCE\unitlength}\basicEadjdist[\ARROWLENGTH]}}%

\newcommand{\WAR}[1]%
{\begin{picture}(#1,0)%
\put(#1,0){\line(-1,0){#1}}%
\put(0,0){\whead}%
\end{picture}}%

\newcommand{\WDIST}[1]%
{\begin{picture}(#1,0)%
\put(#1,0){\line(-1,0){#1}}%
\put(0,0){\whead}%
\truex{400}%
\NUMBER=#1%
\divide\NUMBER by 2%
\put(\NUMBER,0){\circle{\value{x}}}%
\end{picture}}%

\newcommand{\WDOTAR}[1]%
{\truex{100}\truey{300}%
\NUMBEROFDOTS=#1%
\divide\NUMBEROFDOTS by \value{y}%
\advance\NUMBEROFDOTS by 1%
\begin{picture}(#1,0)%
\multiput(#1,0)(-\value{y},0){\NUMBEROFDOTS}%
{\circle*{\value{x}}}%
\put(0,0){\whead}%
\end{picture}}%

\newcommand{\WMONO}[1]%
{\truetail%
\monolength=#1%
\advance\monolength by -\truemonotail%
\begin{picture}(#1,0)(-#1,0)%
\put(-\truemonotail,0){\line(-1,0){\monolength}}%
\put(-\truemonotail,0){\whead}%
\put(-#1,0){\whead}%
\end{picture}}%

\newcommand{\WEPI}[1]%
{\truehead%
\epilength=#1%
\advance\epilength by -\trueepihead%
\begin{picture}(#1,0)%
\put(#1,0){\line(-1,0){\epilength}}%
\put(\trueepihead,0){\whead}%
\put(0,0){\whead}%
\end{picture}}%

\newcommand{\WBIMO}[1]%
{\truehead\truetail%
\monolength=#1
\advance\monolength by -\truemonotail%
\epilength=\monolength%
\advance\epilength by -\trueepihead%
\begin{picture}(#1,0)%
\put(\monolength,0){\line(-1,0){\epilength}}%
\put(\monolength,0){\whead}%
\put(\trueepihead,0){\whead}%
\put(0,0){\whead}%
\end{picture}}%

\newcommand{\WBIAR}[1]%
{\truex{700}%
\begin{picture}(#1,\value{x})%
\put(#1,0){\line(-1,0){#1}}%
\put(0,0){\whead}%
\put(#1,\value{x}){\line(-1,0){#1}}%
\put(0,\value{x}){\whead}%
\end{picture}}%

\newcommand{\WBIDIST}[1]%
{\truex{700}%
\begin{picture}(#1,\value{x})%
\put(#1,0){\line(-1,0){#1}}%
\put(0,0){\whead}%
\put(#1,\value{x}){\line(-1,0){#1}}%
\put(0,\value{x}){\whead}%
\truey{400}%
\NUMBER=#1%
\divide\NUMBER by 2%
\put(\NUMBER,0){\circle{\value{y}}}%
\put(\NUMBER,\value{x}){\circle{\value{y}}}%
\end{picture}}%

\newcommand{\WADJAR}[1]%
{\truex{700}%
\begin{picture}(#1,\value{x})%
\put(0,\value{x}){\line(1,0){#1}}%
\put(#1,\value{x}){\ehead}%
\put(#1,0){\line(-1,0){#1}}%
\put(0,0){\whead}%
\end{picture}}%

\newcommand{\WADJDIST}[1]%
{\truex{700}%
\begin{picture}(#1,\value{x})%
\put(0,\value{x}){\line(1,0){#1}}%
\put(#1,\value{x}){\ehead}%
\put(#1,0){\line(-1,0){#1}}%
\put(0,0){\whead}%
\truey{400}%
\NUMBER=#1%
\divide\NUMBER by 2%
\put(\NUMBER,0){\circle{\value{y}}}%
\put(\NUMBER,\value{x}){\circle{\value{y}}}%
\end{picture}}%

\def\basicwar[#1]{%
\Z=#1%
\multiply \Z by 100%
\hcase{\WAR}{\Z}}%

\newcommand{\war}{\@ifnextchar[{\basicwar}%
{\hspace{\SOURCE\unitlength}\basicwar[\ARROWLENGTH]}}%

\def\basicWar[#1]#2{%
\Z=#1%
\multiply \Z by 100%
\Hcase{\WAR}{#2}{\Z}}%

\newcommand{\War}{\@ifnextchar[{\basicWar}%
{\hspace{\SOURCE\unitlength}\basicWar[\ARROWLENGTH]}}%

\def\basicwaR[#1]#2{%
\Z=#1%
\multiply \Z by 100%
\hcasE{\WAR}{#2}{\Z}}%

\newcommand{\waR}{\@ifnextchar[{\basicwaR}%
{\hspace{\SOURCE\unitlength}\basicwaR[\ARROWLENGTH]}}%

\def\basicwdist[#1]{%
\Z=#1%
\multiply \Z by 100%
\hcase{\WDIST}{\Z}}%

\newcommand{\wdist}{\@ifnextchar[{\basicwdist}%
{\hspace{\SOURCE\unitlength}\basicwdist[\ARROWLENGTH]}}%

\def\basicWdist[#1]#2{%
\Z=#1%
\multiply \Z by 100%
\Hcase{\WDIST}{\DUP{#2}}{\Z}}%

\newcommand{\Wdist}{\@ifnextchar[{\basicWdist}%
{\hspace{\SOURCE\unitlength}\basicWdist[\ARROWLENGTH]}}%

\def\basicwdisT[#1]#2{%
\Z=#1%
\multiply \Z by 100%
\hcasE{\WDIST}{\DDOWN{#2}}{\Z}}%

\newcommand{\wdisT}{\@ifnextchar[{\basicwdisT}%
{\hspace{\SOURCE\unitlength}\basicwdisT[\ARROWLENGTH]}}%

\def\basicwdotar[#1]{%
\Z=#1%
\multiply \Z by 100%
\hcase{\WDOTAR}{\Z}}%

\newcommand{\wdotar}{\@ifnextchar[{\basicwdotar}%
{\hspace{\SOURCE\unitlength}\basicwdotar[\ARROWLENGTH]}}%

\def\basicWdotar[#1]#2{%
\Z=#1%
\multiply \Z by 100%
\Hcase{\WDOTAR}{#2}{\Z}}%

\newcommand{\Wdotar}{\@ifnextchar[{\basicWdotar}%
{\hspace{\SOURCE\unitlength}\basicWdotar[\ARROWLENGTH]}}%

\def\basicwdotaR[#1]#2{%
\Z=#1%
\multiply \Z by 100%
\hcasE{\WDOTAR}{#2}{\Z}}%

\newcommand{\wdotaR}{\@ifnextchar[{\basicwdotaR}%
{\hspace{\SOURCE\unitlength}\basicwdotaR[\ARROWLENGTH]}}%

\def\basicwmono[#1]{%
\Z=#1%
\multiply \Z by 100%
\hcase{\WMONO}{\Z}}%

\newcommand{\wmono}{\@ifnextchar[{\basicwmono}%
{\hspace{\SOURCE\unitlength}\basicwmono[\ARROWLENGTH]}}%

\def\basicWmono[#1]#2{%
\Z=#1%
\multiply \Z by 100%
\Hcase{\WMONO}{#2}{\Z}}%

\newcommand{\Wmono}{\@ifnextchar[{\basicWmono}%
{\hspace{\SOURCE\unitlength}\basicWmono[\ARROWLENGTH]}}%

\def\basicwmonO[#1]#2{%
\Z=#1%
\multiply \Z by 100%
\hcasE{\WMONO}{#2}{\Z}}%

\newcommand{\wmonO}{\@ifnextchar[{\basicwmonO}%
{\hspace{\SOURCE\unitlength}\basicwmonO[\ARROWLENGTH]}}%

\def\basicwepi[#1]{%
\Z=#1%
\multiply \Z by 100%
\hcase{\WEPI}{\Z}}%

\newcommand{\wepi}{\@ifnextchar[{\basicwepi}%
{\hspace{\SOURCE\unitlength}\basicwepi[\ARROWLENGTH]}}%

\def\basicWepi[#1]#2{%
\Z=#1%
\multiply \Z by 100%
\Hcase{\WEPI}{#2}{\Z}}%

\newcommand{\Wepi}{\@ifnextchar[{\basicWepi}%
{\hspace{\SOURCE\unitlength}\basicWepi[\ARROWLENGTH]}}%

\def\basicwepI[#1]#2{%
\Z=#1%
\multiply \Z by 100%
\hcasE{\WEPI}{#2}{\Z}}%

\newcommand{\wepI}{\@ifnextchar[{\basicwepI}%
{\hspace{\SOURCE\unitlength}\basicwepI[\ARROWLENGTH]}}%

\def\basicwbimo[#1]{%
\Z=#1%
\multiply \Z by 100%
\hcase{\WBIMO}{\Z}}%

\newcommand{\wbimo}{\@ifnextchar[{\basicwbimo}%
{\hspace{\SOURCE\unitlength}\basicwbimo[\ARROWLENGTH]}}%

\def\basicWbimo[#1]#2{%
\Z=#1%
\multiply \Z by 100%
\Hcase{\WBIMO}{#2}{\Z}}%

\newcommand{\Wbimo}{\@ifnextchar[{\basicWbimo}%
{\hspace{\SOURCE\unitlength}\basicWbimo[\ARROWLENGTH]}}%

\def\basicwbimO[#1]#2{%
\Z=#1%
\multiply \Z by 100%
\hcasE{\WBIMO}{#2}{\Z}}%

\newcommand{\wbimO}{\@ifnextchar[{\basicwbimO}%
{\hspace{\SOURCE\unitlength}\basicwbimO[\ARROWLENGTH]}}%

\def\basicwiso[#1]{%
\Z=#1%
\multiply \Z by 100%
\Hisocase{\WAR}{\cong}{}{\Z}}%

\newcommand{\wiso}{\@ifnextchar[{\basicwiso}%
{\hspace{\SOURCE\unitlength}\basicwiso[\ARROWLENGTH]}}%

\def\basicWiso[#1]#2{%
\Z=#1%
\multiply \Z by 100%
\Hisocase{\WAR}{#2}{\cong}{\Z}}%

\newcommand{\Wiso}{\@ifnextchar[{\basicWiso}%
{\hspace{\SOURCE\unitlength}\basicWiso[\ARROWLENGTH]}}%

\def\basicwisO[#1]#2{%
\Z=#1%
\multiply \Z by 100%
\Hisocase{\WAR}{\cong}{#2}{\Z}}%

\newcommand{\wisO}{\@ifnextchar[{\basicwisO}%
{\hspace{\SOURCE\unitlength}\basicwisO[\ARROWLENGTH]}}%

\def\basicwbiar[#1]{%
\Z=#1%
\multiply \Z by 100%
\hbicase{\WBIAR}{\Z}}%

\newcommand{\wbiar}{\@ifnextchar[{\basicwbiar}%
{\hspace{\SOURCE\unitlength}\basicwbiar[\ARROWLENGTH]}}%

\def\basicWbiar[#1]#2#3{%
\Z=#1%
\multiply \Z by 100%
\Hbicase{\WBIAR}{#2}{#3}{\Z}}%

\newcommand{\Wbiar}{\@ifnextchar[{\basicWbiar}%
{\hspace{\SOURCE\unitlength}\basicWbiar[\ARROWLENGTH]}}%

\def\basicwbidist[#1]{%
\Z=#1%
\multiply \Z by 100%
\hbicase{\WBIDIST}{\Z}}%

\newcommand{\wbidist}{\@ifnextchar[{\basicwbidist}%
{\hspace{\SOURCE\unitlength}\basicwbidist[\ARROWLENGTH]}}%

\def\basicWbidist[#1]#2#3{%
\Z=#1%
\multiply \Z by 100%
\Hbicase{\WBIDIST}{\DUP{#2}}{\DDOWN{#3}}{\Z}}%

\newcommand{\Wbidist}{\@ifnextchar[{\basicWbidist}%
{\hspace{\SOURCE\unitlength}\basicWbidist[\ARROWLENGTH]}}%

\def\basicwadjar[#1]{%
\Z=#1%
\multiply \Z by 100%
\hbicase{\WADJAR}{\Z}}%

\newcommand{\wadjar}{\@ifnextchar[{\basicwadjar}%
{\hspace{\SOURCE\unitlength}\basicwadjar[\ARROWLENGTH]}}%

\def\basicWadjar[#1]#2#3{%
\Z=#1%
\multiply \Z by 100%
\Hbicase{\WADJAR}{#2}{#3}{\Z}}%

\newcommand{\Wadjar}{\@ifnextchar[{\basicWadjar}%
{\hspace{\SOURCE\unitlength}\basicWadjar[\ARROWLENGTH]}}%

\def\basicwadjdist[#1]{%
\Z=#1%
\multiply \Z by 100%
\hbicase{\WADJDIST}{\Z}}%

\newcommand{\wadjdist}{\@ifnextchar[{\basicwadjdist}%
{\hspace{\SOURCE\unitlength}\basicwadjdist[\ARROWLENGTH]}}%

\def\basicWadjdist[#1]#2#3{%
\Z=#1%
\multiply \Z by 100%
\Hbicase{\WADJDIST}{\DUP{#2}}{\DDOWN{#3}}{\Z}}%

\newcommand{\Wadjdist}{\@ifnextchar[{\basicWadjdist}%
{\hspace{\SOURCE\unitlength}\basicWadjdist[\ARROWLENGTH]}}%

\newcommand{\vcase}[2]{\testdiagrammode#1{#2}}%

\newcommand{\Vcase}[3]{\testdiagrammode\makebox[0pt]%
{\makebox[0pt][r]{\raisebox{0pt}[0pt][0pt]{${#2}\hspace{2pt}$}}}#1{#3}}%

\newcommand{\vcasE}[3]{\testdiagrammode\makebox[0pt]%
{#1{#3}\makebox[0pt][l]{\raisebox{0pt}[0pt][0pt]{\hspace{2pt}$#2$}}}}%

\newcommand{\Visocase}[4]{\testdiagrammode\makebox[0pt]%
{\makebox[0pt][r]{\raisebox{0pt}[0pt][0pt]{$#2$\hspace{2pt}}}#1{#4}%
\makebox[0pt][l]{\raisebox{0pt}[0pt][0pt]{\hspace{2pt}$#3$}}}}%
\newcommand{\vbicase}[2]{\testdiagrammode\makebox[0pt]{{#1{#2}}}}%

\newcommand{\Vbicase}[4]{\testdiagrammode\makebox[0pt]%
{\makebox[0pt][r]{\raisebox{0pt}[0pt][0pt]{$#2$\hspace{5.5pt}}}#1{#4}%
\makebox[0pt][l]{\raisebox{0pt}[0pt][0pt]{\hspace{6.5pt}$#3$}}}}%

\newcommand{\SAR}[1]%
{\begin{picture}(0,0)%
\put(0,0){\makebox(0,0)%
{\begin{picture}(0,#1)%
\put(0,#1){\line(0,-1){#1}}%
\put(0,0){\shead}%
\end{picture}}}\end{picture}}%

\newcommand{\SDIST}[1]%
{\begin{picture}(0,0)%
\put(0,0){\makebox(0,0)%
{\begin{picture}(0,#1)%
\put(0,#1){\line(0,-1){#1}}%
\put(0,0){\shead}%
\end{picture}}}%
\truex{400}%
\put(0,0){\circle{\value{x}}}%
\end{picture}}%

\newcommand{\SDOTAR}[1]%
{\truex{100}\truey{300}%
\NUMBEROFDOTS=#1%
\divide\NUMBEROFDOTS by \value{y}%
\advance\NUMBEROFDOTS by 1%
\begin{picture}(0,0)%
\put(0,0){\makebox(0,0)%
{\begin{picture}(0,#1)%
\multiput(0,#1)(0,-\value{y}){\NUMBEROFDOTS}%
{\circle*{\value{x}}}%
\put(0,0){\shead}%
\end{picture}}}\end{picture}}%

\newcommand{\SMONO}[1]%
{\truetail%
\monolength=#1%
\advance\monolength by -\truemonotail%
\begin{picture}(0,0)%
\put(0,0){\makebox(0,0)%
{\begin{picture}(0,#1)%
\put(0,\monolength){\line(0,-1){\monolength}}%
\put(0,\monolength){\shead}%
\put(0,0){\shead}%
\end{picture}}}\end{picture}}%

\newcommand{\SEPI}[1]%
{\truehead%
\epilength=#1%
\advance\epilength by -\trueepihead%
\begin{picture}(0,0)%
\put(0,0){\makebox(0,0)%
{\begin{picture}(0,#1)%
\put(0,#1){\line(0,-1){\epilength}}%
\put(0,\trueepihead){\shead}%
\put(0,0){\shead}%
\end{picture}}}\end{picture}}%

\newcommand{\SBIMO}[1]%
{\truehead\truetail%
\monolength=#1%
\advance\monolength by -\truemonotail%
\epilength=\monolength%
\advance\epilength by -\trueepihead%
\begin{picture}(0,0)%
\put(0,0){\makebox(0,0)%
{\begin{picture}(0,#1)%
\put(0,\monolength){\line(0,-1){\epilength}}%
\put(0,\monolength){\shead}%
\put(0,\trueepihead){\shead}%
\put(0,0){\shead}%
\end{picture}}}\end{picture}}%

\newcommand{\SBIAR}[1]%
{\begin{picture}(0,0)%
\truex{350}%
\put(0,0){\makebox(0,0)%
{\begin{picture}(0,#1)%
\put(-\value{x},#1){\line(0,-1){#1}}%
\put(-\value{x},0){\shead}%
\put(\value{x},#1){\line(0,-1){#1}}%
\put(\value{x},0){\shead}%
\end{picture}}}\end{picture}}%

\newcommand{\SBIDIST}[1]%
{\begin{picture}(0,0)%
\truex{350}%
\put(0,0){\makebox(0,0)%
{\begin{picture}(0,#1)%
\put(-\value{x},#1){\line(0,-1){#1}}%
\put(-\value{x},0){\shead}%
\put(\value{x},#1){\line(0,-1){#1}}%
\put(\value{x},0){\shead}%
\end{picture}}}%
\truey{400}%
\put(-\value{x},0){\circle{\value{y}}}%
\put(\value{x},0){\circle{\value{y}}}%
\end{picture}}%

\newcommand{\SEQL}[1]%
{\begin{picture}(0,0)%
\truex{100}%
\put(0,0){\makebox(0,0)%
{\begin{picture}(0,#1)\put(-\value{x},#1){\line(0,-1){#1}}%
\put(\value{x},#1){\line(0,-1){#1}}%
\end{picture}}}\end{picture}}%

\newcommand{\SADJAR}[1]{\begin{picture}(0,0)%
\truex{350}%
\put(0,0){\makebox(0,0)%
{\begin{picture}(0,#1)%
\put(-\value{x},#1){\line(0,-1){#1}}%
\put(-\value{x},0){\shead}%
\put(\value{x},0){\line(0,1){#1}}%
\put(\value{x},#1){\nhead}%
\end{picture}}}\end{picture}}%

\newcommand{\SADJDIST}[1]{\begin{picture}(0,0)%
\truex{350}%
\put(0,0){\makebox(0,0)%
{\begin{picture}(0,#1)%
\put(-\value{x},#1){\line(0,-1){#1}}%
\put(-\value{x},0){\shead}%
\put(\value{x},0){\line(0,1){#1}}%
\put(\value{x},#1){\nhead}%
\end{picture}}}%
\truey{400}%
\put(-\value{x},0){\circle{\value{y}}}%
\put(\value{x},0){\circle{\value{y}}}%
\end{picture}}%

\def\basicsar[#1]{\vcase{\SAR}{#100}}%

\newcommand{\sar}{\@ifnextchar[{\basicsar}{\basicsar[50]}}%

\def\basicSar[#1]#2{\Vcase{\SAR}{#2}{#100}}%

\newcommand{\Sar}{\@ifnextchar[{\basicSar}{\basicSar[50]}}%

\def\basicsaR[#1]#2{\vcasE{\SAR}{#2}{#100}}%

\newcommand{\saR}{\@ifnextchar[{\basicsaR}{\basicsaR[50]}}%

\def\basicsdist[#1]{\vcase{\SDIST}{#100}}%

\newcommand{\sdist}{\@ifnextchar[{\basicsdist}{\basicsdist[50]}}%

\def\basicSdist[#1]#2{\Vcase{\SDIST}{#2\hspace*{2pt}}{#100}}%

\newcommand{\Sdist}{\@ifnextchar[{\basicSdist}{\basicSdist[50]}}%

\def\basicsdisT[#1]#2{\vcasE{\SDIST}{\hspace*{2pt}#2}{#100}}%

\newcommand{\sdisT}{\@ifnextchar[{\basicsdisT}{\basicsdisT[50]}}%

\def\basicsdotar[#1]{\vcase{\SDOTAR}{#100}}%

\newcommand{\sdotar}{\@ifnextchar[{\basicsdotar}{\basicsdotar[50]}}%

\def\basicSdotar[#1]#2{\Vcase{\SDOTAR}{#2}{#100}}%

\newcommand{\Sdotar}{\@ifnextchar[{\basicSdotar}{\basicSdotar[50]}}%

\def\basicsdotaR[#1]#2{\vcasE{\SDOTAR}{#2}{#100}}%

\newcommand{\sdotaR}{\@ifnextchar[{\basicsdotaR}{\basicsdotaR[50]}}%

\def\basicsmono[#1]{\vcase{\SMONO}{#100}}%

\newcommand{\smono}{\@ifnextchar[{\basicsmono}{\basicsmono[50]}}%

\def\basicSmono[#1]#2{\Vcase{\SMONO}{#2}{#100}}%

\newcommand{\Smono}{\@ifnextchar[{\basicSmono}{\basicSmono[50]}}%

\def\basicsmonO[#1]#2{\vcasE{\SMONO}{#2}{#100}}%

\newcommand{\smonO}{\@ifnextchar[{\basicsmonO}{\basicsmonO[50]}}%

\def\basicsepi[#1]{\vcase{\SEPI}{#100}}%

\newcommand{\sepi}{\@ifnextchar[{\basicsepi}{\basicsepi[50]}}%

\def\basicSepi[#1]#2{\Vcase{\SEPI}{#2}{#100}}%

\newcommand{\Sepi}{\@ifnextchar[{\basicSepi}{\basicSepi[50]}}%

\def\basicsepI[#1]#2{\vcasE{\SEPI}{#2}{#100}}%

\newcommand{\sepI}{\@ifnextchar[{\basicsepI}{\basicsepI[50]}}%

\def\basicsbimo[#1]{\vcase{\SBIMO}{#100}}%

\newcommand{\sbimo}{\@ifnextchar[{\basicsbimo}{\basicsbimo[50]}}%

\def\basicSbimo[#1]#2{\Vcase{\SBIMO}{#2}{#100}}%

\newcommand{\Sbimo}{\@ifnextchar[{\basicSbimo}{\basicSbimo[50]}}%

\def\basicsbimO[#1]#2{\vcasE{\SBIMO}{#2}{#100}}%

\newcommand{\sbimO}{\@ifnextchar[{\basicsbimO}{\basicsbimO[50]}}%

\def\basicsiso[#1]{\Visocase{\SAR}{\cong}{}{#100}}%

\newcommand{\siso}{\@ifnextchar[{\basicsiso}{\basicsiso[50]}}%

\def\basicSiso[#1]#2{\Visocase{\SAR}{#2}{\cong}{#100}}%

\newcommand{\Siso}{\@ifnextchar[{\basicSiso}{\basicSiso[50]}}%

\def\basicsisO[#1]#2{\Visocase{\SAR}{\cong}{#2}{#100}}%

\newcommand{\sisO}{\@ifnextchar[{\basicsisO}{\basicsisO[50]}}%

\def\basicseql[#1]{\vcase{\SEQL}{#100}}%

\newcommand{\seql}{\@ifnextchar[{\basicseql}{\basicseql[50]}}%

\def\basicSeql[#1]#2{\Vcase{\SEQL}{#2\hspace*{2pt}}{#100}}%

\newcommand{\Seql}{\@ifnextchar[{\basicSeql}{\basicSeql[50]}}%

\def\basicseqL[#1]#2{\vcasE{\SEQL}{\hspace*{2pt}#2}{#100}}%

\newcommand{\seqL}{\@ifnextchar[{\basicseqL}{\basicseqL[50]}}%

\def\basicsbiar[#1]{\vbicase{\SBIAR}{#100}}%

\newcommand{\sbiar}{\@ifnextchar[{\basicsbiar}{\basicsbiar[50]}}%

\def\basicSbiar[#1]#2#3{\Vbicase{\SBIAR}{#2}{#3}{#100}}%

\newcommand{\Sbiar}{\@ifnextchar[{\basicSbiar}{\basicSbiar[50]}}%

\def\basicsbidist[#1]{\vbicase{\SBIDIST}{#100}}%

\newcommand{\sbidist}{\@ifnextchar[{\basicsbidist}{\basicsbidist[50]}}%

\def\basicSbidist[#1]#2#3%
{\Vbicase{\SBIDIST}{#2\hspace*{2pt}}{\hspace*{2pt}#3}{#100}}%

\newcommand{\Sbidist}{\@ifnextchar[{\basicSbidist}{\basicSbidist[50]}}%

\def\basicsadjar[#1]{\vbicase{\SADJAR}{#100}}%

\newcommand{\sadjar}{\@ifnextchar[{\basicsadjar}{\basicsadjar[50]}}%

\def\basicSadjar[#1]#2#3{\Vbicase{\SADJAR}{#2}{#3}{#100}}%

\newcommand{\Sadjar}{\@ifnextchar[{\basicSadjar}{\basicSadjar[50]}}%

\def\basicsadjdist[#1]{\vbicase{\SADJDIST}{#100}}%

\newcommand{\sadjdist}{\@ifnextchar[{\basicsadjdist}{\basicsadjdist[50]}}%

\def\basicSadjdist[#1]#2#3%
{\Vbicase{\SADJDIST}{#2\hspace*{2pt}}{\hspace*{2pt}#3}{#100}}%

\newcommand{\Sadjdist}{\@ifnextchar[{\basicSadjdist}{\basicSadjdist[50]}}%

\newcommand{\NAR}[1]%
{\begin{picture}(0,0)%
\put(0,0){\makebox(0,0)%
{\begin{picture}(0,#1)%
\put(0,0){\line(0,1){#1}}%
\put(0,#1){\nhead}%
\end{picture}}}\end{picture}}%

\newcommand{\NDIST}[1]%
{\begin{picture}(0,0)%
\put(0,0){\makebox(0,0)%
{\begin{picture}(0,#1)%
\put(0,0){\line(0,1){#1}}%
\put(0,#1){\nhead}%
\end{picture}}}
\truex{400}%
\put(0,0){\circle{\value{x}}}%
\end{picture}}%

\newcommand{\NDOTAR}[1]%
{\truex{100}\truey{300}%
\NUMBEROFDOTS=#1%
\divide\NUMBEROFDOTS by \value{y}%
\advance\NUMBEROFDOTS by 1%
\begin{picture}(0,0)%
\put(0,0){\makebox(0,0)%
{\begin{picture}(0,#1)%
\multiput(0,0)(0,\value{y}){\NUMBEROFDOTS}%
{\circle*{\value{x}}}%
\put(0,#1){\nhead}%
\end{picture}}}\end{picture}}%

\newcommand{\NMONO}[1]%
{\truetail%
\monolength=#1%
\advance\monolength by -\truemonotail%
\begin{picture}(0,0)%
\put(0,0){\makebox(0,0)%
{\begin{picture}(0,#1)%
\put(0,\truemonotail){\line(0,1){\monolength}}%
\put(0,#1){\nhead}%
\put(0,\truemonotail){\nhead}%
\end{picture}}}\end{picture}}%

\newcommand{\NEPI}[1]%
{\truehead%
\epilength=#1%
\advance\epilength by -\trueepihead%
\begin{picture}(0,0)%
\put(0,0){\makebox(0,0)%
{\begin{picture}(0,#1)%
\put(0,0){\line(0,1){\epilength}}%
\put(0,#1){\nhead}%
\put(0,\epilength){\nhead}%
\end{picture}}}\end{picture}}%

\newcommand{\NBIMO}[1]%
{\truehead\truetail%
\epilength=#1%
\advance\epilength by -\trueepihead%
\monolength=\epilength%
\advance\monolength by -\truemonotail%
\begin{picture}(0,0)%
\put(0,0){\makebox(0,0)%
{\begin{picture}(0,#1)%
\put(0,\truemonotail){\line(0,1){\monolength}}%
\put(0,#1){\nhead}%
\put(0,\truemonotail){\nhead}%
\put(0,\epilength){\nhead}%
\end{picture}}}\end{picture}}%

\newcommand{\NBIAR}[1]%
{\begin{picture}(0,0)%
\truex{350}%
\put(0,0){\makebox(0,0)%
{\begin{picture}(0,#1)%
\put(-\value{x},0){\line(0,1){#1}}%
\put(-\value{x},#1){\nhead}%
\put(\value{x},0){\line(0,1){#1}}%
\put(\value{x},#1){\nhead}%
\end{picture}}}\end{picture}}%

\newcommand{\NBIDIST}[1]%
{\begin{picture}(0,0)%
\truex{350}%
\put(0,0){\makebox(0,0)%
{\begin{picture}(0,#1)%
\put(-\value{x},0){\line(0,1){#1}}%
\put(-\value{x},#1){\nhead}%
\put(\value{x},0){\line(0,1){#1}}%
\put(\value{x},#1){\nhead}%
\end{picture}}}
\truey{400}%
\put(-\value{x},0){\circle{\value{y}}}%
\put(\value{x},0){\circle{\value{y}}}%
\end{picture}}%

\newcommand{\NADJAR}[1]{\begin{picture}(0,0)%
\truex{350}%
\put(0,0){\makebox(0,0)%
{\begin{picture}(0,#1)%
\put(\value{x},#1){\line(0,-1){#1}}%
\put(\value{x},0){\shead}%
\put(-\value{x},0){\line(0,1){#1}}%
\put(-\value{x},#1){\nhead}%
\end{picture}}}\end{picture}}%

\newcommand{\NADJDIST}[1]{\begin{picture}(0,0)%
\truex{350}%
\put(0,0){\makebox(0,0)%
{\begin{picture}(0,#1)%
\put(\value{x},#1){\line(0,-1){#1}}%
\put(\value{x},0){\shead}%
\put(-\value{x},0){\line(0,1){#1}}%
\put(-\value{x},#1){\nhead}%
\end{picture}}}
\truey{400}%
\put(-\value{x},0){\circle{\value{y}}}%
\put(\value{x},0){\circle{\value{y}}}%
\end{picture}}%

\def\basicnar[#1]{\vcase{\NAR}{#100}}%

\newcommand{\nar}{\@ifnextchar[{\basicnar}{\basicnar[50]}}%

\def\basicNar[#1]#2{\Vcase{\NAR}{#2}{#100}}%

\newcommand{\Nar}{\@ifnextchar[{\basicNar}{\basicNar[50]}}%

\def\basicnaR[#1]#2{\vcasE{\NAR}{#2}{#100}}%

\newcommand{\naR}{\@ifnextchar[{\basicnaR}{\basicnaR[50]}}%

\def\basicndist[#1]{\vcase{\NDIST}{#100}}%

\newcommand{\ndist}{\@ifnextchar[{\basicndist}{\basicndist[50]}}%

\def\basicNdist[#1]#2{\Vcase{\NDIST}{#2\hspace*{2pt}}{#100}}%

\newcommand{\Ndist}{\@ifnextchar[{\basicNdist}{\basicNdist[50]}}%

\def\basicndisT[#1]#2{\vcasE{\NDIST}{\hspace*{2pt}#2}{#100}}%

\newcommand{\ndisT}{\@ifnextchar[{\basicndisT}{\basicndisT[50]}}%

\def\basicndotar[#1]{\vcase{\NDOTAR}{#100}}%

\newcommand{\ndotar}{\@ifnextchar[{\basicndotar}{\basicndotar[50]}}%

\def\basicNdotar[#1]#2{\Vcase{\NDOTAR}{#2}{#100}}%

\newcommand{\Ndotar}{\@ifnextchar[{\basicNdotar}{\basicNdotar[50]}}%

\def\basicndotaR[#1]#2{\vcasE{\NDOTAR}{#2}{#100}}%

\newcommand{\ndotaR}{\@ifnextchar[{\basicndotaR}{\basicndotaR[50]}}%

\def\basicnmono[#1]{\vcase{\NMONO}{#100}}%

\newcommand{\nmono}{\@ifnextchar[{\basicnmono}%
{\basicnmono[50]}}%

\def\basicNmono[#1]#2{\Vcase{\NMONO}{#2}{#100}}%

\newcommand{\Nmono}{\@ifnextchar[{\basicNmono}{\basicNmono[50]}}%

\def\basicnmonO[#1]#2{\vcasE{\NMONO}{#2}{#100}}%

\newcommand{\nmonO}{\@ifnextchar[{\basicnmonO}{\basicnmonO[50]}}%

\def\basicnepi[#1]{\vcase{\NEPI}{#100}}%

\newcommand{\nepi}{\@ifnextchar[{\basicnepi}{\basicnepi[50]}}%

\def\basicNepi[#1]#2{\Vcase{\NEPI}{#2}{#100}}%

\newcommand{\Nepi}{\@ifnextchar[{\basicNepi}{\basicNepi[50]}}%

\def\basicnepI[#1]#2{\vcasE{\NEPI}{#2}{#100}}%

\newcommand{\nepI}{\@ifnextchar[{\basicnepI}{\basicnepI[50]}}%

\def\basicnbimo[#1]{\vcase{\NBIMO}{#100}}%

\newcommand{\nbimo}{\@ifnextchar[{\basicnbimo}{\basicnbimo[50]}}%

\def\basicNbimo[#1]#2{\Vcase{\NBIMO}{#2}{#100}}%

\newcommand{\Nbimo}{\@ifnextchar[{\basicNbimo}{\basicNbimo[50]}}%

\def\basicnbimO[#1]#2{\vcasE{\NBIMO}{#2}{#100}}%

\newcommand{\nbimO}{\@ifnextchar[{\basicnbimO}{\basicnbimO[50]}}%

\def\basicniso[#1]{\Visocase{\NAR}{\cong}{}{#100}}%

\newcommand{\niso}{\@ifnextchar[{\basicniso}{\basicniso[50]}}%

\def\basicNiso[#1]#2{\Visocase{\NAR}{#2}{\cong}{#100}}%

\newcommand{\Niso}{\@ifnextchar[{\basicNiso}{\basicNiso[50]}}%

\def\basicnisO[#1]#2{\Visocase{\NAR}{\cong}{#2}{#100}}%

\newcommand{\nisO}{\@ifnextchar[{\basicnisO}{\basicnisO[50]}}%

\def\basicnbiar[#1]{\vbicase{\NBIAR}{#100}}%

\newcommand{\nbiar}{\@ifnextchar[{\basicnbiar}{\basicnbiar[50]}}%

\def\basicNbiar[#1]#2#3{\Vbicase{\NBIAR}{#2}{#3}{#100}}%

\newcommand{\Nbiar}{\@ifnextchar[{\basicNbiar}{\basicNbiar[50]}}%

\def\basicnbidist[#1]{\vbicase{\NBIDIST}{#100}}%

\newcommand{\nbidist}{\@ifnextchar[{\basicnbidist}{\basicnbidist[50]}}%

\def\basicNbidist[#1]#2#3%
{\Vbicase{\NBIDIST}{#2\hspace*{2pt}}{\hspace*{2pt}#3}{#100}}%

\newcommand{\Nbidist}{\@ifnextchar[{\basicNbidist}{\basicNbidist[50]}}%

\def\basicnadjar[#1]{\vbicase{\NADJAR}{#100}}%

\newcommand{\nadjar}{\@ifnextchar[{\basicnadjar}{\basicnadjar[50]}}%

\def\basicNadjar[#1]#2#3{\Vbicase{\NADJAR}{#2}{#3}{#100}}%

\newcommand{\Nadjar}{\@ifnextchar[{\basicNadjar}{\basicNadjar[50]}}%

\def\basicnadjdist[#1]{\vbicase{\NADJDIST}{#100}}%

\newcommand{\nadjdist}{\@ifnextchar[{\basicnadjdist}{\basicnadjdist[50]}}%

\def\basicNadjdist[#1]#2#3%
{\Vbicase{\NADJDIST}{#2\hspace*{2pt}}{\hspace*{2pt}#3}{#100}}%

\newcommand{\Nadjdist}{\@ifnextchar[{\basicNadjdist}{\basicNadjdist[50]}}%

\newcommand{\fdcase}[4]{\testdiagrammode\begin{picture}(0,0)%
\put(0,0){#1{#4}}%
\truex{200}\truey{600}\truez{600}%
\put(-\value{x},-\value{x}){\makebox(0,\value{z})[r]{${#2}$}}%
\put(\value{x},-\value{y}){\makebox(0,\value{z})[l]{${#3}$}}%
\end{picture}}%

\newcommand{\fdbicase}[4]{\testdiagrammode\begin{picture}(0,0)%
\put(0,0){#1{#4}}%
\truex{900}\truey{150}%
\put(-\value{x},\value{y}){${#2}$}%
\truex{300}\truey{1050}%
\put(\value{x},-\value{y}){${#3}$}%
\end{picture}}%

\newcommand{\NEAR}[1]{%
\Y=#1%
\divide\Y by 2%
\begin{picture}(0,0)%
\put(-\Y,-\Y){\line(1,1){#1}}%
\put(\Y,\Y){\nehead}%
\end{picture}}%

\newcommand{\NEDIST}[1]{%
\Y=#1%
\divide\Y by 2%
\begin{picture}(0,0)%
\put(-\Y,-\Y){\line(1,1){#1}}%
\put(\Y,\Y){\nehead}%
\truex{400}%
\put(0,0){\circle{\value{x}}}%
\end{picture}}%

\newcommand{\NEDOTAR}[1]%
{\truex{100}\truey{212}%
\Y=#1%
\divide\Y by 2%
\NUMBEROFDOTS=#1%
\divide\NUMBEROFDOTS by \value{y}%
\advance\NUMBEROFDOTS by 1%
\begin{picture}(0,0)%
\multiput(-\Y,-\Y)(\value{y},\value{y}){\NUMBEROFDOTS}%
{\circle*{\value{x}}}%
\put(\Y,\Y){\nehead}%
\end{picture}}%

\newcommand{\NEMONO}[1]{%
\Y=#1%
\divide \Y by 2%
\Truetail%
\bimolength=#1%
\advance\bimolength by -\Truemonotail%
\monolength=\bimolength%
\advance\monolength by -\Y%
\begin{picture}(0,0)%
\put(-\monolength,-\monolength){\line(1,1){\bimolength}}%
\put(-\monolength,-\monolength){\nehead}%
\put(\Y,\Y){\nehead}%
\end{picture}}%

\newcommand{\NEEPI}[1]{%
\Y=#1%
\divide\Y by 2%
\Truehead%
\bimolength=#1%
\advance\bimolength by -\Trueepihead%
\epilength=\bimolength%
\advance\epilength by -\Y%
\begin{picture}(0,0)%
\put(-\Y,-\Y){\line(1,1){\bimolength}}%
\put(\epilength,\epilength){\nehead}%
\put(\Y,\Y){\nehead}%
\end{picture}}%

\newcommand{\NEBIMO}[1]{%
\Y=#1%
\divide\Y by 2%
\Truetail\Truehead%
\bimolength=#1%
\advance\bimolength by -\Truemonotail%
\monolength=\bimolength%
\advance\monolength by -\Y%
\advance\bimolength by -\Trueepihead%
\epilength=\bimolength%
\advance\epilength by -\monolength%
\begin{picture}(0,0)%
\put(-\monolength,-\monolength){\line(1,1){\bimolength}}%
\put(-\monolength,-\monolength){\nehead}%
\put(\epilength,\epilength){\nehead}%
\put(\Y,\Y){\nehead}%
\end{picture}}%

\newcommand{\NEBIAR}[1]{%
\Y=#1%
\divide\Y by 2%
\begin{picture}(0,0)%
\put(-\Y,-\Y){\begin{picture}(0,0)%
\truex{247}%
\put(-\value{x},\value{x}){\line(1,1){#1}}%
\put(\value{x},-\value{x}){\line(1,1){#1}}%
\monolength=#1%
\advance\monolength by -\value{x}%
\epilength=#1%
\advance\epilength by \value{x}%
\put(\monolength,\epilength){\nehead}%
\put(\epilength,\monolength){\nehead}%
\end{picture}}\end{picture}}%

\newcommand{\NEBIDIST}[1]{%
\Y=#1%
\divide\Y by 2%
\truey{400}%
\begin{picture}(0,0)%
\put(-\Y,-\Y){\begin{picture}(0,0)%
\truex{247}%
\monolength=#1%
\advance\monolength by -\value{x}%
\epilength=#1%
\advance\epilength by \value{x}%
\put(\value{x},-\value{x}){\line(1,1){#1}}%
\put(\epilength,\monolength){\nehead}%
\end{picture}}%
\put(-\Y,-\Y){\begin{picture}(0,0)%
\truex{247}%
\monolength=#1%
\advance\monolength by \value{x}%
\epilength=#1%
\advance\epilength by -\value{x}%
\put(-\value{x},\value{x}){\line(1,1){#1}}%
\put(\epilength,\monolength){\nehead}%
\end{picture}}%
\put(-\value{x},\value{x}){\circle{\value{y}}}%
\put(\value{x},-\value{x}){\circle{\value{y}}}%
\end{picture}}%

\newcommand{\NEEQL}[1]{%
\Y=#1%
\divide\Y by 2%
\begin{picture}(0,0)%
\put(-\Y,-\Y){\begin{picture}(0,0)%
\truex{70}%
\put(-\value{x},\value{x}){\line(1,1){#1}}%
\put(\value{x},-\value{x}){\line(1,1){#1}}%
\end{picture}}\end{picture}}%

\newcommand{\NEADJAR}[1]{%
\Y=#1%
\divide\Y by 2%
\begin{picture}(0,0)%
\put(-\Y,-\Y){\begin{picture}(0,0)%
\truex{247}%
\monolength=#1%
\advance\monolength by -\value{x}%
\epilength=#1%
\advance\epilength by \value{x}%
\put(\value{x},-\value{x}){\line(1,1){#1}}%
\put(\epilength,\monolength){\nehead}%
\end{picture}}%
\put(\Y,\Y){\begin{picture}(0,0)%
\truex{247}%
\monolength=#1%
\advance\monolength by -\value{x}%
\epilength=#1%
\advance\epilength by \value{x}%
\put(-\value{x},\value{x}){\line(-1,-1){#1}}%
\put(-\epilength,-\monolength){\swhead}%
\end{picture}}\end{picture}}%

\newcommand{\NEADJDIST}[1]{%
\Y=#1%
\divide\Y by 2%
\truey{400}%
\begin{picture}(0,0)%
\put(-\Y,-\Y){\begin{picture}(0,0)%
\truex{247}%
\monolength=#1%
\advance\monolength by -\value{x}%
\epilength=#1%
\advance\epilength by \value{x}%
\put(\value{x},-\value{x}){\line(1,1){#1}}%
\put(\epilength,\monolength){\nehead}%
\end{picture}}%
\put(\Y,\Y){\begin{picture}(0,0)%
\truex{247}%
\monolength=#1%
\advance\monolength by -\value{x}%
\epilength=#1%
\advance\epilength by \value{x}%
\put(-\value{x},\value{x}){\line(-1,-1){#1}}%
\put(-\epilength,-\monolength){\swhead}%
\end{picture}}%
\put(-\value{x},\value{x}){\circle{\value{y}}}%
\put(\value{x},-\value{x}){\circle{\value{y}}}%
\end{picture}}%

\def\basicnear[#1]{\fdcase{\NEAR}{}{}{#100}}%

\newcommand{\near}{\@ifnextchar[{\basicnear}{\basicnear[59]}}%

\def\basicNear[#1]#2{\fdcase{\NEAR}{#2}{}{#100}}%

\newcommand{\Near}{\@ifnextchar[{\basicNear}{\basicNear[59]}}%

\def\basicneaR[#1]#2{\fdcase{\NEAR}{}{#2}{#100}}%

\newcommand{\neaR}{\@ifnextchar[{\basicneaR}{\basicneaR[59]}}%

\def\basicnedist[#1]{\fdcase{\NEDIST}{}{}{#100}}%

\newcommand{\nedist}{\@ifnextchar[{\basicnedist}{\basicnedist[59]}}%

\def\basicNedist[#1]#2{\fdcase{\NEDIST}{#2}{}{#100}}%

\newcommand{\Nedist}{\@ifnextchar[{\basicNedist}{\basicNedist[59]}}%

\def\basicnedisT[#1]#2{\fdcase{\NEDIST}{}{#2}{#100}}%

\newcommand{\nedisT}{\@ifnextchar[{\basicnedisT}{\basicnedisT[59]}}%

\def\basicnedotar[#1]{\fdcase{\NEDOTAR}{}{}{#100}}%

\newcommand{\nedotar}{\@ifnextchar[{\basicnedotar}{\basicnedotar[59]}}%

\def\basicNedotar[#1]#2{\fdcase{\NEDOTAR}{#2}{}{#100}}%

\newcommand{\Nedotar}{\@ifnextchar[{\basicNedotar}{\basicNedotar[59]}}%

\def\basicnedotaR[#1]#2{\fdcase{\NEDOTAR}{}{#2}{#100}}%

\newcommand{\nedotaR}{\@ifnextchar[{\basicnedotaR}{\basicnedotaR[59]}}%

\def\basicnemono[#1]{\fdcase{\NEMONO}{}{}{#100}}%

\newcommand{\nemono}{\@ifnextchar[{\basicnemono}{\basicnemono[59]}}%

\def\basicNemono[#1]#2{\fdcase{\NEMONO}{#2}{}{#100}}%

\newcommand{\Nemono}{\@ifnextchar[{\basicNemono}{\basicNemono[59]}}%

\def\basicnemonO[#1]#2{\fdcase{\NEMONO}{}{#2}{#100}}%

\newcommand{\nemonO}{\@ifnextchar[{\basicnemonO}{\basicnemonO[59]}}%

\def\basicneepi[#1]{\fdcase{\NEEPI}{}{}{#100}}%

\newcommand{\neepi}{\@ifnextchar[{\basicneepi}{\basicneepi[59]}}%

\def\basicNeepi[#1]#2{\fdcase{\NEEPI}{#2}{}{#100}}%

\newcommand{\Neepi}{\@ifnextchar[{\basicNeepi}{\basicNeepi[59]}}%

\def\basicneepI[#1]#2{\fdcase{\NEEPI}{}{#2}{#100}}%

\newcommand{\neepI}{\@ifnextchar[{\basicneepI}{\basicneepI[59]}}%

\def\basicnebimo[#1]{\fdcase{\NEBIMO}{}{}{#100}}%

\newcommand{\nebimo}{\@ifnextchar[{\basicnebimo}{\basicnebimo[59]}}%

\def\basicNebimo[#1]#2{\fdcase{\NEBIMO}{#2}{}{#100}}%

\newcommand{\Nebimo}{\@ifnextchar[{\basicNebimo}{\basicNebimo[59]}}%

\def\basicnebimO[#1]#2{\fdcase{\NEBIMO}{}{#2}{#100}}%

\newcommand{\nebimO}{\@ifnextchar[{\basicnebimO}{\basicnebimO[59]}}%

\def\basicneiso[#1]{\fdcase{\NEAR}{\hspace{-2pt}\cong}{}{#100}}%

\newcommand{\neiso}{\@ifnextchar[{\basicneiso}{\basicneiso[59]}}%

\def\basicNeiso[#1]#2{\fdcase{\NEAR}{#2}{\cong}{#100}}%

\newcommand{\Neiso}{\@ifnextchar[{\basicNeiso}{\basicNeiso[59]}}%

\def\basicneisO[#1]#2{\fdcase{\NEAR}{\hspace{-2pt}\cong}{#2}{#100}}%

\newcommand{\neisO}{\@ifnextchar[{\basicneisO}{\basicneisO[59]}}%

\def\basicneeql[#1]{\fdcase{\NEEQL}{}{}{#100}}%

\newcommand{\neeql}{\@ifnextchar[{\basicneeql}{\basicneeql[59]}}%

\def\basicNeeql[#1]#2{\fdcase{\NEEQL}{#2}{}{#100}}%

\newcommand{\Neeql}{\@ifnextchar[{\basicNeeql}{\basicNeeql[59]}}%

\def\basicneeqL[#1]#2{\fdcase{\NEEQL}{}{#2}{#100}}%

\newcommand{\neeqL}{\@ifnextchar[{\basicneeqL}{\basicneeqL[59]}}%

\def\basicnebiar[#1]{\fdbicase{\NEBIAR}{}{}{#100}}%

\newcommand{\nebiar}{\@ifnextchar[{\basicnebiar}{\basicnebiar[59]}}%

\def\basicNebiar[#1]#2#3{\fdbicase{\NEBIAR}{#2}{#3}{#100}}%

\newcommand{\Nebiar}{\@ifnextchar[{\basicNebiar}{\basicNebiar[59]}}%

\def\basicneadjar[#1]{\fdbicase{\NEADJAR}{}{}{#100}}%

\newcommand{\neadjar}{\@ifnextchar[{\basicneadjar}{\basicneadjar[59]}}%

\def\basicNeadjar[#1]#2#3{\fdbicase{\NEADJAR}{#2}{#3}{#100}}%

\newcommand{\Neadjar}{\@ifnextchar[{\basicNeadjar}{\basicNeadjar[59]}}%

\def\basicnebidist[#1]{\fdbicase{\NEBIDIST}{}{}{#100}}%

\newcommand{\nebidist}{\@ifnextchar[{\basicnebidist}{\basicnebidist[59]}}%

\def\basicNebidist[#1]#2#3{\fdbicase{\NEBIDIST}{#2}{#3}{#100}}%

\newcommand{\Nebidist}{\@ifnextchar[{\basicNebidist}{\basicNebidist[59]}}%

\def\basicneadjdist[#1]{\fdbicase{\NEADJDIST}{}{}{#100}}%

\newcommand{\neadjdist}{\@ifnextchar[{\basicneadjdist}{\basicneadjdist[59]}}%

\def\basicNeadjdist[#1]#2#3{\fdbicase{\NEADJDIST}{#2}{#3}{#100}}%

\newcommand{\Neadjdist}{\@ifnextchar[{\basicNeadjdist}{\basicNeadjdist[59]}}%

\newcommand{\SWAR}[1]{%
\Y=#1%
\divide\Y by 2%
\begin{picture}(0,0)%
\put(\Y,\Y){\line(-1,-1){#1}}%
\put(-\Y,-\Y){\swhead}%
\end{picture}}%

\newcommand{\SWDIST}[1]{%
\Y=#1%
\divide\Y by 2%
\begin{picture}(0,0)%
\put(\Y,\Y){\line(-1,-1){#1}}%
\put(-\Y,-\Y){\swhead}%
\truex{400}%
\put(0,0){\circle{\value{x}}}%
\end{picture}}%

\newcommand{\SWDOTAR}[1]%
{\truex{100}\truey{212}%
\Y=#1%
\divide\Y by 2%
\NUMBEROFDOTS=#1%
\divide\NUMBEROFDOTS by \value{y}%
\advance\NUMBEROFDOTS by 1%
\begin{picture}(0,0)%
\multiput(\Y,\Y)(-\value{y},-\value{y}){\NUMBEROFDOTS}%
{\circle*{\value{x}}}%
\put(-\Y,-\Y){\swhead}%
\end{picture}}%

\newcommand{\SWMONO}[1]{%
\Y=#1%
\divide \Y by 2%
\Truetail%
\bimolength=#1%
\advance\bimolength by -\Truemonotail%
\monolength=\bimolength%
\advance\monolength by -\Y%
\begin{picture}(0,0)%
\put(\monolength,\monolength){\line(-1,-1){\bimolength}}%
\put(\monolength,\monolength){\swhead}%
\put(-\Y,-\Y){\swhead}%
\end{picture}}%

\newcommand{\SWEPI}[1]{%
\Y=#1%
\divide\Y by 2%
\Truehead%
\bimolength=#1%
\advance\bimolength by -\Trueepihead%
\epilength=\bimolength%
\advance\epilength by -\Y%
\begin{picture}(0,0)%
\put(\Y,\Y){\line(-1,-1){\bimolength}}%
\put(-\epilength,-\epilength){\swhead}%
\put(-\Y,-\Y){\swhead}%
\end{picture}}%

\newcommand{\SWBIMO}[1]{%
\Y=#1%
\divide\Y by 2%
\Truetail\Truehead%
\bimolength=#1%
\advance\bimolength by -\Truemonotail%
\monolength=\bimolength%
\advance\monolength by -\Y%
\advance\bimolength by -\Trueepihead%
\epilength=\bimolength%
\advance\epilength by -\monolength%
\begin{picture}(0,0)%
\put(\monolength,\monolength){\line(-1,-1){\bimolength}}%
\put(\monolength,\monolength){\swhead}%
\put(-\epilength,-\epilength){\swhead}%
\put(-\Y,-\Y){\swhead}%
\end{picture}}%

\newcommand{\SWBIAR}[1]{%
\Y=#1%
\divide\Y by 2%
\begin{picture}(0,0)%
\put(\Y,\Y){\begin{picture}(0,0)%
\truex{247}%
\put(\value{x},-\value{x}){\line(-1,-1){#1}}%
\put(-\value{x},\value{x}){\line(-1,-1){#1}}%
\monolength=#1%
\advance\monolength by -\value{x}%
\epilength=#1%
\advance\epilength by \value{x}%
\put(-\monolength,-\epilength){\swhead}%
\put(-\epilength,-\monolength){\swhead}%
\end{picture}}\end{picture}}%

\newcommand{\SWBIDIST}[1]{%
\Y=#1%
\divide\Y by 2%
\truey{400}%
\begin{picture}(0,0)%
\put(\Y,\Y){\begin{picture}(0,0)%
\truex{247}%
\monolength=#1%
\advance\monolength by -\value{x}%
\epilength=#1%
\advance\epilength by \value{x}%
\put(-\value{x},\value{x}){\line(-1,-1){#1}}%
\put(-\epilength,-\monolength){\swhead}%
\end{picture}}%
\put(\Y,\Y){\begin{picture}(0,0)%
\truex{247}%
\monolength=#1%
\advance\monolength by \value{x}%
\epilength=#1%
\advance\epilength by -\value{x}%
\put(\value{x},-\value{x}){\line(-1,-1){#1}}%
\put(-\epilength,-\monolength){\swhead}%
\end{picture}}%
\put(\value{x},-\value{x}){\circle{\value{y}}}%
\put(-\value{x},\value{x}){\circle{\value{y}}}%
\end{picture}}%

\newcommand{\SWADJAR}[1]{%
\Y=#1%
\divide\Y by 2%
\begin{picture}(0,0)%
\put(\Y,\Y){\begin{picture}(0,0)%
\truex{247}%
\monolength=#1%
\advance\monolength by -\value{x}%
\epilength=#1%
\advance\epilength by \value{x}%
\put(\value{x},-\value{x}){\line(-1,-1){#1}}%
\put(-\monolength,-\epilength){\swhead}%
\end{picture}}%
\put(-\Y,-\Y){\begin{picture}(0,0)%
\truex{247}%
\monolength=#1%
\advance\monolength by -\value{x}%
\epilength=#1%
\advance\epilength by \value{x}%
\put(-\value{x},\value{x}){\line(1,1){#1}}%
\put(\monolength,\epilength){\nehead}%
\end{picture}}\end{picture}}%

\newcommand{\SWADJDIST}[1]{%
\Y=#1%
\divide\Y by 2%
\truey{400}%
\begin{picture}(0,0)%
\put(\Y,\Y){\begin{picture}(0,0)%
\truex{247}%
\monolength=#1%
\advance\monolength by -\value{x}%
\epilength=#1%
\advance\epilength by \value{x}%
\put(\value{x},-\value{x}){\line(-1,-1){#1}}%
\put(-\monolength,-\epilength){\swhead}%
\end{picture}}%
\put(-\Y,-\Y){\begin{picture}(0,0)%
\truex{247}%
\monolength=#1%
\advance\monolength by -\value{x}%
\epilength=#1%
\advance\epilength by \value{x}%
\put(-\value{x},\value{x}){\line(1,1){#1}}%
\put(\monolength,\epilength){\nehead}%
\end{picture}}%
\put(-\value{x},\value{x}){\circle{\value{y}}}%
\put(\value{x},-\value{x}){\circle{\value{y}}}%
\end{picture}}%

\def\basicswar[#1]{\fdcase{\SWAR}{}{}{#100}}%

\newcommand{\swar}{\@ifnextchar[{\basicswar}{\basicswar[59]}}%

\def\basicSwar[#1]#2{\fdcase{\SWAR}{#2}{}{#100}}%

\newcommand{\Swar}{\@ifnextchar[{\basicSwar}{\basicSwar[59]}}%

\def\basicswaR[#1]#2{\fdcase{\SWAR}{}{#2}{#100}}%

\newcommand{\swaR}{\@ifnextchar[{\basicswaR}{\basicswaR[59]}}%

\def\basicswdist[#1]{\fdcase{\SWDIST}{}{}{#100}}%

\newcommand{\swdist}{\@ifnextchar[{\basicswdist}{\basicswdist[59]}}%

\def\basicSwdist[#1]#2{\fdcase{\SWDIST}{#2}{}{#100}}%

\newcommand{\Swdist}{\@ifnextchar[{\basicSwdist}{\basicSwdist[59]}}%

\def\basicswdisT[#1]#2{\fdcase{\SWDIST}{}{#2}{#100}}%

\newcommand{\swdisT}{\@ifnextchar[{\basicswdisT}{\basicswdisT[59]}}%

\def\basicswdotar[#1]{\fdcase{\SWDOTAR}{}{}{#100}}%

\newcommand{\swdotar}{\@ifnextchar[{\basicswdotar}{\basicswdotar[59]}}%

\def\basicSwdotar[#1]#2{\fdcase{\SWDOTAR}{#2}{}{#100}}%

\newcommand{\Swdotar}{\@ifnextchar[{\basicSwdotar}{\basicSwdotar[59]}}%

\def\basicswdotaR[#1]#2{\fdcase{\SWDOTAR}{}{#2}{#100}}%

\newcommand{\swdotaR}{\@ifnextchar[{\basicswdotaR}{\basicswdotaR[59]}}%

\def\basicswmono[#1]{\fdcase{\SWMONO}{}{}{#100}}%

\newcommand{\swmono}{\@ifnextchar[{\basicswmono}{\basicswmono[59]}}%

\def\basicSwmono[#1]#2{\fdcase{\SWMONO}{#2}{}{#100}}%

\newcommand{\Swmono}{\@ifnextchar[{\basicSwmono}{\basicSwmono[59]}}%

\def\basicswmonO[#1]#2{\fdcase{\SWMONO}{}{#2}{#100}}%

\newcommand{\swmonO}{\@ifnextchar[{\basicswmonO}{\basicswmonO[59]}}%

\def\basicswepi[#1]{\fdcase{\SWEPI}{}{}{#100}}%

\newcommand{\swepi}{\@ifnextchar[{\basicswepi}{\basicswepi[59]}}%

\def\basicSwepi[#1]#2{\fdcase{\SWEPI}{#2}{}{#100}}%

\newcommand{\Swepi}{\@ifnextchar[{\basicSwepi}{\basicSwepi[59]}}%

\def\basicswepI[#1]#2{\fdcase{\SWEPI}{}{#2}{#100}}%

\newcommand{\swepI}{\@ifnextchar[{\basicswepI}{\basicswepI[59]}}%

\def\basicswbimo[#1]{\fdcase{\SWBIMO}{}{}{#100}}%

\newcommand{\swbimo}{\@ifnextchar[{\basicswbimo}{\basicswbimo[59]}}%

\def\basicSwbimo[#1]#2{\fdcase{\SWBIMO}{#2}{}{#100}}%

\newcommand{\Swbimo}{\@ifnextchar[{\basicSwbimo}{\basicSwbimo[59]}}%

\def\basicswbimO[#1]#2{\fdcase{\SWBIMO}{}{#2}{#100}}%

\newcommand{\swbimO}{\@ifnextchar[{\basicswbimO}{\basicswbimO[59]}}%

\def\basicswiso[#1]{\fdcase{\SWAR}{\hspace{-2pt}\cong}{}{#100}}%

\newcommand{\swiso}{\@ifnextchar[{\basicswiso}{\basicswiso[59]}}%

\def\basicSwiso[#1]#2{\fdcase{\SWAR}{#2}{\cong}{#100}}%

\newcommand{\Swiso}{\@ifnextchar[{\basicSwiso}{\basicSwiso[59]}}%

\def\basicswisO[#1]#2{\fdcase{\SWAR}{\hspace{-2pt}\cong}{#2}{#100}}%

\newcommand{\swisO}{\@ifnextchar[{\basicswisO}{\basicswisO[59]}}%

\def\basicswbiar[#1]{\fdbicase{\SWBIAR}{}{}{#100}}%

\newcommand{\swbiar}{\@ifnextchar[{\basicswbiar}{\basicswbiar[59]}}%

\def\basicSwbiar[#1]#2#3{\fdbicase{\SWBIAR}{#2}{#3}{#100}}%

\newcommand{\Swbiar}{\@ifnextchar[{\basicSwbiar}{\basicSwbiar[59]}}%

\def\basicswadjar[#1]{\fdbicase{\SWADJAR}{}{}{#100}}%

\newcommand{\swadjar}{\@ifnextchar[{\basicswadjar}{\basicswadjar[59]}}%

\def\basicSwadjar[#1]#2#3{\fdbicase{\SWADJAR}{#2}{#3}{#100}}%

\newcommand{\Swadjar}{\@ifnextchar[{\basicSwadjar}{\basicSwadjar[59]}}%

\def\basicswbidist[#1]{\fdbicase{\SWBIDIST}{}{}{#100}}%

\newcommand{\swbidist}{\@ifnextchar[{\basicswbidist}{\basicswbidist[59]}}%

\def\basicSwbidist[#1]#2#3{\fdbicase{\SWBIDIST}{#2}{#3}{#100}}%

\newcommand{\Swbidist}{\@ifnextchar[{\basicSwbidist}{\basicSwbidist[59]}}%

\def\basicswadjdist[#1]{\fdbicase{\SWADJDIST}{}{}{#100}}%

\newcommand{\swadjdist}{\@ifnextchar[{\basicswadjdist}{\basicswadjdist[59]}}%

\def\basicSwadjdist[#1]#2#3{\fdbicase{\SWADJDIST}{#2}{#3}{#100}}%

\newcommand{\Swadjdist}{\@ifnextchar[{\basicSwadjdist}{\basicSwadjdist[59]}}%

\newcommand{\sdcase}[4]{\testdiagrammode\begin{picture}(0,0)%
\put(0,0){#1{#4}}%
\truex{100}\truez{600}%
\put(\value{x},\value{x}){\makebox(0,\value{z})[l]{${#2}$}}%
\truex{300}\truey{800}%
\put(-\value{x},-\value{y}){\makebox(0,\value{z})[r]{${#3}$}}%
\end{picture}}%

\newcommand{\sdbicase}[4]{\testdiagrammode\begin{picture}(0,0)%
\put(0,0){#1{#4}}%
\truex{350}\truey{600}\truez{950}%
\put(\value{x},\value{x}){\makebox(0,\value{y})[l]{${#2}$}}%
\truex{450}\truey{600}\truez{1050}%
\put(-\value{x},-\value{z}){\makebox(0,\value{y})[r]{${#3}$}}%
\end{picture}}%

\newcommand{\SEAR}[1]{%
\Y=#1%
\divide\Y by 2%
\begin{picture}(0,0)%
\put(-\Y,\Y){\line(1,-1){#1}}%
\put(\Y,-\Y){\sehead}%
\end{picture}}%

\newcommand{\SEDIST}[1]{%
\Y=#1%
\divide\Y by 2%
\begin{picture}(0,0)%
\put(-\Y,\Y){\line(1,-1){#1}}%
\put(\Y,-\Y){\sehead}%
\truex{400}%
\put(0,0){\circle{\value{x}}}%
\end{picture}}%

\newcommand{\SEDOTAR}[1]%
{\truex{100}\truey{212}%
\Y=#1%
\divide\Y by 2%
\NUMBEROFDOTS=#1%
\divide\NUMBEROFDOTS by \value{y}%
\advance\NUMBEROFDOTS by 1%
\begin{picture}(0,0)%
\multiput(-\Y,\Y)(\value{y},-\value{y}){\NUMBEROFDOTS}%
{\circle*{\value{x}}}%
\put(\Y,-\Y){\sehead}%
\end{picture}}%

\newcommand{\SEMONO}[1]{%
\Y=#1%
\divide \Y by 2%
\Truetail%
\bimolength=#1%
\advance\bimolength by -\Truemonotail%
\monolength=\bimolength%
\advance\monolength by -\Y%
\begin{picture}(0,0)%
\put(-\monolength,\monolength){\line(1,-1){\bimolength}}%
\put(-\monolength,\monolength){\sehead}%
\put(\Y,-\Y){\sehead}%
\end{picture}}%

\newcommand{\SEEPI}[1]{%
\Y=#1%
\divide\Y by 2%
\Truehead%
\bimolength=#1%
\advance\bimolength by -\Trueepihead%
\epilength=\bimolength%
\advance\epilength by -\Y%
\begin{picture}(0,0)%
\put(-\Y,\Y){\line(1,-1){\bimolength}}%
\put(\epilength,-\epilength){\sehead}%
\put(\Y,-\Y){\sehead}%
\end{picture}}%

\newcommand{\SEBIMO}[1]{%
\Y=#1%
\divide\Y by 2%
\Truetail\Truehead%
\bimolength=#1%
\advance\bimolength by -\Truemonotail%
\monolength=\bimolength%
\advance\monolength by -\Y%
\advance\bimolength by -\Trueepihead%
\epilength=\bimolength%
\advance\epilength by -\monolength%
\begin{picture}(0,0)%
\put(-\monolength,\monolength){\line(1,-1){\bimolength}}%
\put(-\monolength,\monolength){\sehead}%
\put(\epilength,-\epilength){\sehead}%
\put(\Y,-\Y){\sehead}%
\end{picture}}%

\newcommand{\SEBIAR}[1]{%
\Y=#1%
\divide\Y by 2%
\begin{picture}(0,0)%
\put(-\Y,\Y){\begin{picture}(0,0)%
\truex{247}%
\put(-\value{x},-\value{x}){\line(1,-1){#1}}%
\put(\value{x},\value{x}){\line(1,-1){#1}}%
\monolength=#1%
\advance\monolength by -\value{x}%
\epilength=#1%
\advance\epilength by \value{x}%
\put(\monolength,-\epilength){\sehead}%
\put(\epilength,-\monolength){\sehead}%
\end{picture}}\end{picture}}%

\newcommand{\SEBIDIST}[1]{%
\Y=#1%
\divide\Y by 2%
\truey{400}%
\begin{picture}(0,0)%
\put(-\Y,\Y){\begin{picture}(0,0)%
\truex{247}%
\monolength=#1%
\advance\monolength by -\value{x}%
\epilength=#1%
\advance\epilength by \value{x}%
\put(\value{x},\value{x}){\line(1,-1){#1}}%
\put(\epilength,-\monolength){\sehead}%
\end{picture}}%
\put(-\Y,\Y){\begin{picture}(0,0)%
\truex{247}%
\monolength=#1%
\advance\monolength by \value{x}%
\epilength=#1%
\advance\epilength by -\value{x}%
\put(-\value{x},-\value{x}){\line(1,-1){#1}}%
\put(\epilength,-\monolength){\sehead}%
\end{picture}}%
\put(-\value{x},-\value{x}){\circle{\value{y}}}%
\put(\value{x},\value{x}){\circle{\value{y}}}%
\end{picture}}%

\newcommand{\SEEQL}[1]{%
\Y=#1%
\divide\Y by 2%
\begin{picture}(0,0)%
\put(-\Y,\Y){\begin{picture}(0,0)%
\truex{70}%
\put(-\value{x},-\value{x}){\line(1,-1){#1}}%
\put(\value{x},\value{x}){\line(1,-1){#1}}%
\end{picture}}\end{picture}}%

\newcommand{\SEADJAR}[1]{%
\Y=#1%
\divide\Y by 2%
\begin{picture}(0,0)%
\put(-\Y,\Y){\begin{picture}(0,0)%
\truex{247}%
\monolength=#1%
\advance\monolength by -\value{x}%
\epilength=#1%
\advance\epilength by \value{x}%
\put(-\value{x},-\value{x}){\line(1,-1){#1}}%
\put(\monolength,-\epilength){\sehead}%
\end{picture}}%
\put(\Y,-\Y){\begin{picture}(0,0)%
\truex{247}%
\monolength=#1%
\advance\monolength by -\value{x}%
\epilength=#1%
\advance\epilength by \value{x}%
\put(\value{x},\value{x}){\line(-1,1){#1}}%
\put(-\monolength,\epilength){\nwhead}%
\end{picture}}\end{picture}}%

\newcommand{\SEADJDIST}[1]{%
\Y=#1%
\divide\Y by 2%
\truey{400}%
\begin{picture}(0,0)%
\put(-\Y,\Y){\begin{picture}(0,0)%
\truex{247}%
\monolength=#1%
\advance\monolength by -\value{x}%
\epilength=#1%
\advance\epilength by \value{x}%
\put(-\value{x},-\value{x}){\line(1,-1){#1}}%
\put(\monolength,-\epilength){\sehead}%
\end{picture}}%
\put(\Y,-\Y){\begin{picture}(0,0)%
\truex{247}%
\monolength=#1%
\advance\monolength by -\value{x}%
\epilength=#1%
\advance\epilength by \value{x}%
\put(\value{x},\value{x}){\line(-1,1){#1}}%
\put(-\monolength,\epilength){\nwhead}%
\end{picture}}%
\put(-\value{x},-\value{x}){\circle{\value{y}}}%
\put(\value{x},\value{x}){\circle{\value{y}}}%
\end{picture}}%

\def\basicsear[#1]{\sdcase{\SEAR}{}{}{#100}}%

\newcommand{\sear}{\@ifnextchar[{\basicsear}{\basicsear[59]}}%

\def\basicSear[#1]#2{\sdcase{\SEAR}{#2}{}{#100}}%

\newcommand{\Sear}{\@ifnextchar[{\basicSear}{\basicSear[59]}}%

\def\basicseaR[#1]#2{\sdcase{\SEAR}{}{#2}{#100}}%

\newcommand{\seaR}{\@ifnextchar[{\basicseaR}{\basicseaR[59]}}%

\def\basicsedist[#1]{\sdcase{\SEDIST}{}{}{#100}}%

\newcommand{\sedist}{\@ifnextchar[{\basicsedist}{\basicsedist[59]}}%

\def\basicSedist[#1]#2{\sdcase{\SEDIST}{#2}{}{#100}}%

\newcommand{\Sedist}{\@ifnextchar[{\basicSedist}{\basicSedist[59]}}%

\def\basicsedisT[#1]#2{\sdcase{\SEDIST}{}{#2}{#100}}%

\newcommand{\sedisT}{\@ifnextchar[{\basicsedisT}{\basicsedisT[59]}}%

\def\basicsedotar[#1]{\sdcase{\SEDOTAR}{}{}{#100}}%

\newcommand{\sedotar}{\@ifnextchar[{\basicsedotar}{\basicsedotar[59]}}%

\def\basicSedotar[#1]#2{\sdcase{\SEDOTAR}{#2}{}{#100}}%

\newcommand{\Sedotar}{\@ifnextchar[{\basicSedotar}{\basicSedotar[59]}}%

\def\basicsedotaR[#1]#2{\sdcase{\SEDOTAR}{}{#2}{#100}}%

\newcommand{\sedotaR}{\@ifnextchar[{\basicsedotaR}{\basicsedotaR[59]}}%

\def\basicsemono[#1]{\sdcase{\SEMONO}{}{}{#100}}%

\newcommand{\semono}{\@ifnextchar[{\basicsemono}{\basicsemono[59]}}%

\def\basicSemono[#1]#2{\sdcase{\SEMONO}{#2}{}{#100}}%

\newcommand{\Semono}{\@ifnextchar[{\basicSemono}{\basicSemono[59]}}%

\def\basicsemonO[#1]#2{\sdcase{\SEMONO}{}{#2}{#100}}%

\newcommand{\semonO}{\@ifnextchar[{\basicsemonO}{\basicsemonO[59]}}%

\def\basicseepi[#1]{\sdcase{\SEEPI}{}{}{#100}}%

\newcommand{\seepi}{\@ifnextchar[{\basicseepi}{\basicseepi[59]}}%

\def\basicSeepi[#1]#2{\sdcase{\SEEPI}{#2}{}{#100}}%

\newcommand{\Seepi}{\@ifnextchar[{\basicSeepi}{\basicSeepi[59]}}%

\def\basicseepI[#1]#2{\sdcase{\SEEPI}{}{#2}{#100}}%

\newcommand{\seepI}{\@ifnextchar[{\basicseepI}{\basicseepI[59]}}%

\def\basicsebimo[#1]{\sdcase{\SEBIMO}{}{}{#100}}%

\newcommand{\sebimo}{\@ifnextchar[{\basicsebimo}{\basicsebimo[59]}}%

\def\basicSebimo[#1]#2{\sdcase{\SEBIMO}{#2}{}{#100}}%

\newcommand{\Sebimo}{\@ifnextchar[{\basicSebimo}{\basicSebimo[59]}}%

\def\basicsebimO[#1]#2{\sdcase{\SEBIMO}{}{#2}{#100}}%

\newcommand{\sebimO}{\@ifnextchar[{\basicsebimO}{\basicsebimO[59]}}%

\def\basicseiso[#1]{\sdcase{\SEAR}{\hspace{-2pt}\cong}{}{#100}}%

\newcommand{\seiso}{\@ifnextchar[{\basicseiso}{\basicseiso[59]}}%

\def\basicSeiso[#1]#2{\sdcase{\SEAR}{#2}{\cong}{#100}}%

\newcommand{\Seiso}{\@ifnextchar[{\basicSeiso}{\basicSeiso[59]}}%

\def\basicseisO[#1]#2{\sdcase{\SEAR}{\hspace{-2pt}\cong}{#2}{#100}}%

\newcommand{\seisO}{\@ifnextchar[{\basicseisO}{\basicseisO[59]}}%

\def\basicseeql[#1]{\sdcase{\SEEQL}{}{}{#100}}%

\newcommand{\seeql}{\@ifnextchar[{\basicseeql}{\basicseeql[59]}}%

\def\basicSeeql[#1]#2{\sdcase{\SEEQL}{#2}{}{#100}}%

\newcommand{\Seeql}{\@ifnextchar[{\basicSeeql}{\basicSeeql[59]}}%

\def\basicseeqL[#1]#2{\sdcase{\SEEQL}{}{#2}{#100}}%

\newcommand{\seeqL}{\@ifnextchar[{\basicseeqL}{\basicseeqL[59]}}%

\def\basicsebiar[#1]{\sdbicase{\SEBIAR}{}{}{#100}}%

\newcommand{\sebiar}{\@ifnextchar[{\basicsebiar}{\basicsebiar[59]}}%

\def\basicSebiar[#1]#2#3{\sdbicase{\SEBIAR}{#2}{#3}{#100}}%

\newcommand{\Sebiar}{\@ifnextchar[{\basicSebiar}{\basicSebiar[59]}}%

\def\basicseadjar[#1]{\sdbicase{\SEADJAR}{}{}{#100}}%

\newcommand{\seadjar}{\@ifnextchar[{\basicseadjar}{\basicseadjar[59]}}%

\def\basicSeadjar[#1]#2#3{\sdbicase{\SEADJAR}{#2}{#3}{#100}}%

\newcommand{\Seadjar}{\@ifnextchar[{\basicSeadjar}{\basicSeadjar[59]}}%

\def\basicsebidist[#1]{\sdbicase{\SEBIDIST}{}{}{#100}}%

\newcommand{\sebidist}{\@ifnextchar[{\basicsebidist}{\basicsebidist[59]}}%

\def\basicSebidist[#1]#2#3{\sdbicase{\SEBIDIST}{#2}{#3}{#100}}%

\newcommand{\Sebidist}{\@ifnextchar[{\basicSebidist}{\basicSebidist[59]}}%

\def\basicseadjdist[#1]{\sdbicase{\SEADJDIST}{}{}{#100}}%

\newcommand{\seadjdist}{\@ifnextchar[{\basicseadjdist}{\basicseadjdist[59]}}%

\def\basicSeadjdist[#1]#2#3{\sdbicase{\SEADJDIST}{#2}{#3}{#100}}%

\newcommand{\Seadjdist}{\@ifnextchar[{\basicSeadjdist}{\basicSeadjdist[59]}}%

\newcommand{\NWAR}[1]{%
\Y=#1%
\divide\Y by 2%
\begin{picture}(0,0)%
\put(\Y,-\Y){\line(-1,1){#1}}%
\put(-\Y,\Y){\nwhead}%
\end{picture}}%

\newcommand{\NWDIST}[1]{%
\Y=#1%
\divide\Y by 2%
\begin{picture}(0,0)%
\put(\Y,-\Y){\line(-1,1){#1}}%
\put(-\Y,\Y){\nwhead}%
\truex{400}%
\put(0,0){\circle{\value{x}}}%
\end{picture}}%

\newcommand{\NWDOTAR}[1]%
{\truex{100}\truey{212}%
\Y=#1%
\divide\Y by 2%
\NUMBEROFDOTS=#1%
\divide\NUMBEROFDOTS by \value{y}%
\advance\NUMBEROFDOTS by 1%
\begin{picture}(0,0)%
\multiput(\Y,-\Y)(-\value{y},\value{y}){\NUMBEROFDOTS}%
{\circle*{\value{x}}}%
\put(-\Y,\Y){\nwhead}%
\end{picture}}%

\newcommand{\NWMONO}[1]{%
\Y=#1%
\divide \Y by 2%
\Truetail%
\bimolength=#1%
\advance\bimolength by -\Truemonotail%
\monolength=\bimolength%
\advance\monolength by -\Y%
\begin{picture}(0,0)%
\put(\monolength,-\monolength){\line(-1,1){\bimolength}}%
\put(\monolength,-\monolength){\nwhead}%
\put(-\Y,\Y){\nwhead}%
\end{picture}}%

\newcommand{\NWEPI}[1]{%
\Y=#1%
\divide\Y by 2%
\Truehead%
\bimolength=#1%
\advance\bimolength by -\Trueepihead%
\epilength=\bimolength%
\advance\epilength by -\Y%
\begin{picture}(0,0)%
\put(\Y,-\Y){\line(-1,1){\bimolength}}%
\put(-\epilength,\epilength){\nwhead}%
\put(-\Y,\Y){\nwhead}%
\end{picture}}%

\newcommand{\NWBIMO}[1]{%
\Y=#1%
\divide\Y by 2%
\Truetail\Truehead%
\bimolength=#1%
\advance\bimolength by -\Truemonotail%
\monolength=\bimolength%
\advance\monolength by -\Y%
\advance\bimolength by -\Trueepihead%
\epilength=\bimolength%
\advance\epilength by -\monolength%
\begin{picture}(0,0)%
\put(\monolength,-\monolength){\line(-1,1){\bimolength}}%
\put(\monolength,-\monolength){\nwhead}%
\put(-\epilength,\epilength){\nwhead}%
\put(-\Y,\Y){\nwhead}%
\end{picture}}%

\newcommand{\NWBIAR}[1]{%
\Y=#1%
\divide\Y by 2%
\begin{picture}(0,0)%
\put(\Y,-\Y){\begin{picture}(0,0)%
\truex{247}%
\put(-\value{x},-\value{x}){\line(-1,1){#1}}%
\put(\value{x},\value{x}){\line(-1,1){#1}}%
\monolength=#1%
\advance\monolength by -\value{x}%
\epilength=#1%
\advance\epilength by \value{x}%
\put(-\monolength,\epilength){\nwhead}%
\put(-\epilength,\monolength){\nwhead}%
\end{picture}}\end{picture}}%

\newcommand{\NWBIDIST}[1]{%
\Y=#1%
\divide\Y by 2%
\truey{400}%
\begin{picture}(0,0)%
\put(\Y,-\Y){\begin{picture}(0,0)%
\truex{247}%
\monolength=#1%
\advance\monolength by -\value{x}%
\epilength=#1%
\advance\epilength by \value{x}%
\put(-\value{x},-\value{x}){\line(-1,1){#1}}%
\put(-\epilength,\monolength){\nwhead}%
\end{picture}}%
\put(\Y,-\Y){\begin{picture}(0,0)%
\truex{247}%
\monolength=#1%
\advance\monolength by \value{x}%
\epilength=#1%
\advance\epilength by -\value{x}%
\put(\value{x},\value{x}){\line(-1,1){#1}}%
\put(-\epilength,\monolength){\nwhead}%
\end{picture}}%
\put(-\value{x},-\value{x}){\circle{\value{y}}}%
\put(\value{x},\value{x}){\circle{\value{y}}}%
\end{picture}}%

\newcommand{\NWADJAR}[1]{%
\Y=#1%
\divide\Y by 2%
\begin{picture}(0,0)%
\put(\Y,-\Y){\begin{picture}(0,0)%
\truex{247}%
\monolength=#1%
\advance\monolength by -\value{x}%
\epilength=#1%
\advance\epilength by \value{x}%
\put(-\value{x},-\value{x}){\line(-1,1){#1}}%
\put(-\epilength,\monolength){\nwhead}%
\end{picture}}%
\put(-\Y,\Y){\begin{picture}(0,0)%
\truex{247}%
\monolength=#1%
\advance\monolength by -\value{x}%
\epilength=#1%
\advance\epilength by \value{x}%
\put(\value{x},\value{x}){\line(1,-1){#1}}%
\put(\epilength,-\monolength){\sehead}%
\end{picture}}\end{picture}}%

\newcommand{\NWADJDIST}[1]{%
\Y=#1%
\divide\Y by 2%
\truey{400}%
\begin{picture}(0,0)%
\put(\Y,-\Y){\begin{picture}(0,0)%
\truex{247}%
\monolength=#1%
\advance\monolength by -\value{x}%
\epilength=#1%
\advance\epilength by \value{x}%
\put(-\value{x},-\value{x}){\line(-1,1){#1}}%
\put(-\epilength,\monolength){\nwhead}%
\end{picture}}%
\put(-\Y,\Y){\begin{picture}(0,0)%
\truex{247}%
\monolength=#1%
\advance\monolength by -\value{x}%
\epilength=#1%
\advance\epilength by \value{x}%
\put(\value{x},\value{x}){\line(1,-1){#1}}%
\put(\epilength,-\monolength){\sehead}%
\end{picture}}%
\put(-\value{x},-\value{x}){\circle{\value{y}}}%
\put(\value{x},\value{x}){\circle{\value{y}}}%
\end{picture}}%

\def\basicnwar[#1]{\sdcase{\NWAR}{}{}{#100}}%

\newcommand{\nwar}{\@ifnextchar[{\basicnwar}{\basicnwar[59]}}%

\def\basicNwar[#1]#2{\sdcase{\NWAR}{#2}{}{#100}}%

\newcommand{\Nwar}{\@ifnextchar[{\basicNwar}{\basicNwar[59]}}%

\def\basicnwaR[#1]#2{\sdcase{\NWAR}{}{#2}{#100}}%

\newcommand{\nwaR}{\@ifnextchar[{\basicnwaR}{\basicnwaR[59]}}%

\def\basicnwdist[#1]{\sdcase{\NWDIST}{}{}{#100}}%

\newcommand{\nwdist}{\@ifnextchar[{\basicnwdist}{\basicnwdist[59]}}%

\def\basicNwdist[#1]#2{\sdcase{\NWDIST}{#2}{}{#100}}%

\newcommand{\Nwdist}{\@ifnextchar[{\basicNwdist}{\basicNwdist[59]}}%

\def\basicnwdisT[#1]#2{\sdcase{\NWDIST}{}{#2}{#100}}%

\newcommand{\nwdisT}{\@ifnextchar[{\basicnwdisT}{\basicnwdisT[59]}}%

\def\basicnwdotar[#1]{\sdcase{\NWDOTAR}{}{}{#100}}%

\newcommand{\nwdotar}{\@ifnextchar[{\basicnwdotar}{\basicnwdotar[59]}}%

\def\basicNwdotar[#1]#2{\sdcase{\NWDOTAR}{#2}{}{#100}}%

\newcommand{\Nwdotar}{\@ifnextchar[{\basicNwdotar}{\basicNwdotar[59]}}%

\def\basicnwdotaR[#1]#2{\sdcase{\NWDOTAR}{}{#2}{#100}}%

\newcommand{\nwdotaR}{\@ifnextchar[{\basicnwdotaR}{\basicnwdotaR[59]}}%

\def\basicnwmono[#1]{\sdcase{\NWMONO}{}{}{#100}}%

\newcommand{\nwmono}{\@ifnextchar[{\basicnwmono}{\basicnwmono[59]}}%

\def\basicNwmono[#1]#2{\sdcase{\NWMONO}{#2}{}{#100}}%

\newcommand{\Nwmono}{\@ifnextchar[{\basicNwmono}{\basicNwmono[59]}}%

\def\basicnwmonO[#1]#2{\sdcase{\NWMONO}{}{#2}{#100}}%

\newcommand{\nwmonO}{\@ifnextchar[{\basicnwmonO}{\basicnwmonO[59]}}%

\def\basicnwepi[#1]{\sdcase{\NWEPI}{}{}{#100}}%

\newcommand{\nwepi}{\@ifnextchar[{\basicnwepi}{\basicnwepi[59]}}%

\def\basicNwepi[#1]#2{\sdcase{\NWEPI}{#2}{}{#100}}%

\newcommand{\Nwepi}{\@ifnextchar[{\basicNwepi}{\basicNwepi[59]}}%

\def\basicnwepI[#1]#2{\sdcase{\NWEPI}{}{#2}{#100}}%

\newcommand{\nwepI}{\@ifnextchar[{\basicnwepI}{\basicnwepI[59]}}%

\def\basicnwbimo[#1]{\sdcase{\NWBIMO}{}{}{#100}}%

\newcommand{\nwbimo}{\@ifnextchar[{\basicnwbimo}{\basicnwbimo[59]}}%

\def\basicNwbimo[#1]#2{\sdcase{\NWBIMO}{#2}{}{#100}}%

\newcommand{\Nwbimo}{\@ifnextchar[{\basicNwbimo}{\basicNwbimo[59]}}%

\def\basicnwbimO[#1]#2{\sdcase{\NWBIMO}{}{#2}{#100}}%

\newcommand{\nwbimO}{\@ifnextchar[{\basicnwbimO}{\basicnwbimO[59]}}%

\def\basicnwiso[#1]{\sdcase{\NWAR}{\hspace{-2pt}\cong}{}{#100}}%

\newcommand{\nwiso}{\@ifnextchar[{\basicnwiso}{\basicnwiso[59]}}%

\def\basicNwiso[#1]#2{\sdcase{\NWAR}{#2}{\cong}{#100}}%

\newcommand{\Nwiso}{\@ifnextchar[{\basicNwiso}{\basicNwiso[59]}}%

\def\basicnwisO[#1]#2{\sdcase{\NWAR}{\hspace{-2pt}\cong}{#2}{#100}}%

\newcommand{\nwisO}{\@ifnextchar[{\basicnwisO}{\basicnwisO[59]}}%

\let\nweql=\seeql%

\def\basicnwbiar[#1]{\sdbicase{\NWBIAR}{}{}{#100}}%

\newcommand{\nwbiar}{\@ifnextchar[{\basicnwbiar}{\basicnwbiar[59]}}%

\def\basicNwbiar[#1]#2#3{\sdbicase{\NWBIAR}{#2}{#3}{#100}}%

\newcommand{\Nwbiar}{\@ifnextchar[{\basicNwbiar}{\basicNwbiar[59]}}%

\def\basicnwadjar[#1]{\sdbicase{\NWADJAR}{}{}{#100}}%

\newcommand{\nwadjar}{\@ifnextchar[{\basicnwadjar}{\basicnwadjar[59]}}%

\def\basicNwadjar[#1]#2#3{\sdbicase{\NWADJAR}{#2}{#3}{#100}}%

\newcommand{\Nwadjar}{\@ifnextchar[{\basicNwadjar}{\basicNwadjar[59]}}%

\def\basicnwbidist[#1]{\sdbicase{\NWBIDIST}{}{}{#100}}%

\newcommand{\nwbidist}{\@ifnextchar[{\basicnwbidist}{\basicnwbidist[59]}}%

\def\basicNwbidist[#1]#2#3{\sdbicase{\NWBIDIST}{#2}{#3}{#100}}%

\newcommand{\Nwbidist}{\@ifnextchar[{\basicNwbidist}{\basicNwbidist[59]}}%

\def\basicnwadjdist[#1]{\sdbicase{\NWADJDIST}{}{}{#100}}%

\newcommand{\nwadjdist}{\@ifnextchar[{\basicnwadjdist}{\basicnwadjdist[59]}}%

\def\basicNwadjdist[#1]#2#3{\sdbicase{\NWADJDIST}{#2}{#3}{#100}}%

\newcommand{\Nwadjdist}{\@ifnextchar[{\basicNwadjdist}{\basicNwadjdist[59]}}%

\newcommand{\ENEAR}[3]{\testdiagrammode%
\Y=#3%
\divide\Y by 2%
\Z=\Y%
\divide\Z by 2%
\begin{picture}(0,0)%
\put(-\Y,-\Z){\line(2,1){#3}}%
\put(\Y,\Z){\enehead}%
\truex{200}\truey{800}\truez{600}%
\put(-\value{x},\value{x}){\makebox(0,\value{z})[r]{${#1}$}}%
\put(\value{x},-\value{y}){\makebox(0,\value{z})[l]{${#2}$}}%
\end{picture}}%

\newcommand{\ENEDIST}[3]{\testdiagrammode%
\Y=#3%
\divide\Y by 2%
\Z=\Y%
\divide\Z by 2%
\begin{picture}(0,0)%
\put(-\Y,-\Z){\line(2,1){#3}}%
\put(\Y,\Z){\enehead}%
\truex{400}%
\put(0,0){\circle{\value{x}}}%
\truex{200}\truey{800}\truez{600}%
\put(-\value{x},\value{x}){\makebox(0,\value{z})[r]{${#1}$}}%
\put(\value{x},-\value{y}){\makebox(0,\value{z})[l]{${#2}$}}%
\end{picture}}%

\newcommand{\ENEDOTAR}[3]{\testdiagrammode%
\truex{100}\truey{268}\truez{134}%
\Y=#3%
\divide\Y by 2%
\Z=\Y%
\divide\Z by 2%
\NUMBEROFDOTS=#3%
\divide\NUMBEROFDOTS by \value{y}%
\advance\NUMBEROFDOTS by 1%
\begin{picture}(0,0)%
\multiput(-\Y,-\Z)(\value{y},\value{z}){\NUMBEROFDOTS}%
{\circle*{\value{x}}}%
\put(\Y,\Z){\enehead}%
\truex{200}\truey{800}\truez{600}%
\put(-\value{x},\value{x}){\makebox(0,\value{z})[r]{${#1}$}}%
\put(\value{x},-\value{y}){\makebox(0,\value{z})[l]{${#2}$}}%
\end{picture}}%

\newcommand{\ENEMONO}[3]{\testdiagrammode%
\Y=#3%
\divide\Y by 2%
\Z=\Y%
\divide\Z by 2%
\TrueTail%
\bimolength=#3%
\advance\bimolength by -\TrueMonoTail%
\monolength=\bimolength%
\advance\monolength by -\Y%
\secondmonolength=\monolength%
\divide\secondmonolength by 2%
\begin{picture}(0,0)%
\put(-\monolength,-\secondmonolength){\line(2,1){\bimolength}}%
\put(-\monolength,-\secondmonolength){\enehead}%
\put(\Y,\Z){\enehead}%
\truex{200}\truey{800}\truez{600}%
\put(-\value{x},\value{x}){\makebox(0,\value{z})[r]{${#1}$}}%
\put(\value{x},-\value{y}){\makebox(0,\value{z})[l]{${#2}$}}%
\end{picture}}%

\newcommand{\ENEEPI}[3]{\testdiagrammode%
\Y=#3%
\divide\Y by 2%
\Z=\Y%
\divide\Z by 2%
\TrueHead%
\bimolength=#3%
\advance\bimolength by -\TrueEpiHead%
\epilength=\bimolength%
\advance\epilength by -\Y%
\secondepilength=\epilength%
\divide\secondepilength by 2%
\begin{picture}(0,0)%
\put(-\Y,-\Z){\line(2,1){\bimolength}}%
\put(\epilength,\secondepilength){\enehead}%
\put(\Y,\Z){\enehead}%
\truex{200}\truey{800}\truez{600}%
\put(-\value{x},\value{x}){\makebox(0,\value{z})[r]{${#1}$}}%
\put(\value{x},-\value{y}){\makebox(0,\value{z})[l]{${#2}$}}%
\end{picture}}%

\newcommand{\ENEBIMO}[3]{\testdiagrammode%
\Y=#3%
\divide\Y by 2%
\Z=\Y%
\divide\Z by 2%
\TrueTail\TrueHead%
\bimolength=#3%
\advance\bimolength by -\TrueMonoTail%
\monolength=\bimolength%
\advance\monolength by -\Y%
\advance\bimolength by -\TrueEpiHead%
\epilength=\bimolength%
\advance\epilength by -\monolength%
\secondmonolength=\monolength%
\divide\secondmonolength by 2%
\secondepilength=\epilength%
\divide\secondepilength by 2%
\begin{picture}(0,0)%
\put(-\monolength,-\secondmonolength){\line(2,1){\bimolength}}%
\put(-\monolength,-\secondmonolength){\enehead}%
\put(\epilength,\secondepilength){\enehead}%
\put(\Y,\Z){\enehead}%
\truex{200}\truey{800}\truez{600}%
\put(-\value{x},\value{x}){\makebox(0,\value{z})[r]{${#1}$}}%
\put(\value{x},-\value{y}){\makebox(0,\value{z})[l]{${#2}$}}%
\end{picture}}%

\newcommand{\ENEEQL}[3]{\testdiagrammode%
\Y=#3%
\divide\Y by 2%
\Z=\Y%
\divide\Z by 2%
\begin{picture}(0,0)%
\put(-\Y,-\Z){\begin{picture}(0,0)%
\truex{44}\truey{89}%
\put(-\value{x},\value{y}){\line(2,1){#3}}%
\put(\value{x},-\value{y}){\line(2,1){#3}}%
\end{picture}}%
\truex{200}\truey{800}\truez{600}%
\put(-\value{x},\value{x}){\makebox(0,\value{z})[r]{${#1}$}}%
\put(\value{x},-\value{y}){\makebox(0,\value{z})[l]{${#2}$}}%
\end{picture}}%

\newcommand{\ENEBIAR}[3]{\testdiagrammode%
\Y=#3%
\divide\Y by 2%
\Z=\Y%
\divide\Z by 2%
\begin{picture}(0,0)%
\put(-\Y,-\Z){\begin{picture}(0,0)%
\truex{156}\truey{313}%
\put(-\value{x},\value{y}){\line(2,1){#3}}%
\put(\value{x},-\value{y}){\line(2,1){#3}}%
\monolength=#3%
\advance\monolength by -\value{x}%
\epilength=#3%
\advance\epilength by \value{x}%
\secondmonolength=\Y%
\advance\secondmonolength by -\value{y}%
\secondepilength=\Y%
\advance\secondepilength by \value{y}%
\put(\monolength,\secondepilength){\enehead}%
\put(\epilength,\secondmonolength){\enehead}%
\end{picture}}
\truex{300}\truey{1000}\truez{600}%
\put(-\value{x},\value{x}){\makebox(0,\value{z})[r]{${#1}$}}%
\put(\value{x},-\value{y}){\makebox(0,\value{z})[l]{${#2}$}}%
\end{picture}}%

\newcommand{\ENEBIDIST}[3]{\testdiagrammode%
\Y=#3%
\divide\Y by 2%
\Z=\Y%
\divide\Z by 2%
\begin{picture}(0,0)%
\truex{156}\truey{313}\truez{400}%
\put(-\Y,-\Z){\begin{picture}(0,0)%
\put(-\value{x},\value{y}){\line(2,1){#3}}%
\put(\value{x},-\value{y}){\line(2,1){#3}}%
\monolength=#3%
\advance\monolength by -\value{x}%
\epilength=#3%
\advance\epilength by \value{x}%
\secondmonolength=\Y%
\advance\secondmonolength by -\value{y}%
\secondepilength=\Y%
\advance\secondepilength by \value{y}%
\put(\monolength,\secondepilength){\enehead}%
\put(\epilength,\secondmonolength){\enehead}%
\end{picture}}
\put(-\value{x},\value{y}){\circle{\value{z}}}%
\put(\value{x},-\value{y}){\circle{\value{z}}}%
\truex{300}\truey{1000}\truez{600}%
\put(-\value{x},\value{x}){\makebox(0,\value{z})[r]{${#1}$}}%
\put(\value{x},-\value{y}){\makebox(0,\value{z})[l]{${#2}$}}%
\end{picture}}%

\newcommand{\ENEADJAR}[3]{\testdiagrammode%
\Y=#3%
\divide\Y by 2%
\Z=\Y%
\divide\Z by 2%
\begin{picture}(0,0)%
\put(-\Y,-\Z){\begin{picture}(0,0)%
\truex{156}\truey{313}%
\monolength=#3%
\advance\monolength by -\value{x}%
\epilength=#3%
\advance\epilength by \value{x}%
\secondmonolength=\Y%
\advance\secondmonolength by -\value{y}%
\secondepilength=\Y%
\advance\secondepilength by \value{y}%
\put(\value{x},-\value{y}){\line(2,1){#3}}%
\put(\epilength,\secondmonolength){\enehead}%
\put(\monolength,\secondepilength){\line(-2,-1){#3}}%
\put(-\value{x},\value{y}){\wswhead}%
\end{picture}}
\truex{300}\truey{1000}\truez{600}%
\put(-\value{x},\value{x}){\makebox(0,\value{z})[r]{${#1}$}}%
\put(\value{x},-\value{y}){\makebox(0,\value{z})[l]{${#2}$}}%
\end{picture}}%

\newcommand{\ENEADJDIST}[3]{\testdiagrammode%
\Y=#3%
\divide\Y by 2%
\Z=\Y%
\divide\Z by 2%
\begin{picture}(0,0)%
\truex{156}\truey{313}\truez{400}%
\put(-\Y,-\Z){\begin{picture}(0,0)%
\monolength=#3%
\advance\monolength by -\value{x}%
\epilength=#3%
\advance\epilength by \value{x}%
\secondmonolength=\Y%
\advance\secondmonolength by -\value{y}%
\secondepilength=\Y%
\advance\secondepilength by \value{y}%
\put(\value{x},-\value{y}){\line(2,1){#3}}%
\put(\epilength,\secondmonolength){\enehead}%
\put(\monolength,\secondepilength){\line(-2,-1){#3}}%
\put(-\value{x},\value{y}){\wswhead}%
\end{picture}}
\put(-\value{x},\value{y}){\circle{\value{z}}}%
\put(\value{x},-\value{y}){\circle{\value{z}}}%
\truex{300}\truey{1000}\truez{600}%
\put(-\value{x},\value{x}){\makebox(0,\value{z})[r]{${#1}$}}%
\put(\value{x},-\value{y}){\makebox(0,\value{z})[l]{${#2}$}}%
\end{picture}}%

\def\basicenear[#1]{\ENEAR{}{}{#100}}%

\newcommand{\enear}{\@ifnextchar[{\basicenear}{\basicenear[133]}}%

\def\basicEnear[#1]#2{\ENEAR{#2}{}{#100}}%

\newcommand{\Enear}{\@ifnextchar[{\basicEnear}{\basicEnear[133]}}%

\def\basiceneaR[#1]#2{\ENEAR{}{#2}{#100}}%

\newcommand{\eneaR}{\@ifnextchar[{\basiceneaR}{\basiceneaR[133]}}%

\def\basicenedist[#1]{\ENEDIST{}{}{#100}}%

\newcommand{\enedist}{\@ifnextchar[{\basicenedist}{\basicenedist[133]}}%

\def\basicEnedist[#1]#2{\ENEDIST{#2}{}{#100}}%

\newcommand{\Enedist}{\@ifnextchar[{\basicEnedist}{\basicEnedist[133]}}%

\def\basicenedisT[#1]#2{\ENEDIST{}{#2}{#100}}%

\newcommand{\enedisT}{\@ifnextchar[{\basicenedisT}{\basicenedisT[133]}}%

\def\basicenedotar[#1]{\ENEDOTAR{}{}{#100}}%

\newcommand{\enedotar}{\@ifnextchar[{\basicenedotar}{\basicenedotar[133]}}%

\def\basicEnedotar[#1]#2{\ENEDOTAR{#2}{}{#100}}%

\newcommand{\Enedotar}{\@ifnextchar[{\basicEnedotar}{\basicEnedotar[133]}}%

\def\basicenedotaR[#1]#2{\ENEDOTAR{}{#2}{#100}}%

\newcommand{\enedotaR}{\@ifnextchar[{\basicenedotaR}{\basicenedotaR[133]}}%

\def\basicenemono[#1]{\ENEMONO{}{}{#100}}%

\newcommand{\enemono}{\@ifnextchar[{\basicenemono}{\basicenemono[133]}}%

\def\basicEnemono[#1]#2{\ENEMONO{#2}{}{#100}}%

\newcommand{\Enemono}{\@ifnextchar[{\basicEnemono}{\basicEnemono[133]}}%

\def\basicenemonO[#1]#2{\ENEMONO{}{#2}{#100}}%

\newcommand{\enemonO}{\@ifnextchar[{\basicenemonO}{\basicenemonO[133]}}%

\def\basiceneepi[#1]{\ENEEPI{}{}{#100}}%

\newcommand{\eneepi}{\@ifnextchar[{\basiceneepi}{\basiceneepi[133]}}%

\def\basicEneepi[#1]#2{\ENEEPI{#2}{}{#100}}%

\newcommand{\Eneepi}{\@ifnextchar[{\basicEneepi}{\basicEneepi[133]}}%

\def\basiceneepI[#1]#2{\ENEEPI{}{#2}{#100}}%

\newcommand{\eneepI}{\@ifnextchar[{\basiceneepI}{\basiceneepI[133]}}%

\def\basicenebimo[#1]{\ENEBIMO{}{}{#100}}%

\newcommand{\enebimo}{\@ifnextchar[{\basicenebimo}{\basicenebimo[133]}}%

\def\basicEnebimo[#1]#2{\ENEBIMO{#2}{}{#100}}%

\newcommand{\Enebimo}{\@ifnextchar[{\basicEnebimo}{\basicEnebimo[133]}}%

\def\basicenebimO[#1]#2{\ENEBIMO{}{#2}{#100}}%

\newcommand{\enebimO}{\@ifnextchar[{\basicenebimO}{\basicenebimO[133]}}%

\def\basiceneiso[#1]{\ENEAR{\cong}{}{#100}}%

\newcommand{\eneiso}{\@ifnextchar[{\basiceneiso}{\basiceneiso[133]}}%

\def\basicEneiso[#1]#2{\ENEAR{#2}{\cong}{#100}}%

\newcommand{\Eneiso}{\@ifnextchar[{\basicEneiso}{\basicEneiso[133]}}%

\def\basiceneisO[#1]#2{\ENEAR{\cong}{#2}{#100}}%

\newcommand{\eneisO}{\@ifnextchar[{\basiceneisO}{\basiceneisO[133]}}%

\def\basiceneeql[#1]{\ENEEQL{}{}{#100}}%

\newcommand{\eneeql}{\@ifnextchar[{\basiceneeql}{\basiceneeql[133]}}%

\def\basicEneeql[#1]#2{\ENEEQL{#2}{}{#100}}%

\newcommand{\Eneeql}{\@ifnextchar[{\basicEneeql}{\basicEneeql[133]}}%

\def\basiceneeqL[#1]#2{\ENEEQL{}{#2}{#100}}%

\newcommand{\eneeqL}{\@ifnextchar[{\basiceneeqL}{\basiceneeqL[133]}}%

\def\basicenebiar[#1]{\ENEBIAR{}{}{#100}}%

\newcommand{\enebiar}{\@ifnextchar[{\basicenebiar}{\basicenebiar[133]}}%

\def\basicEnebiar[#1]#2#3{\ENEBIAR{#2}{#3}{#100}}%

\newcommand{\Enebiar}{\@ifnextchar[{\basicEnebiar}{\basicEnebiar[133]}}%

\def\basicenebidist[#1]{\ENEBIDIST{}{}{#100}}%

\newcommand{\enebidist}{\@ifnextchar[{\basicenebidist}{\basicenebidist[133]}}%

\def\basicEnebidist[#1]#2#3{\ENEBIDIST{#2}{#3}{#100}}%

\newcommand{\Enebidist}{\@ifnextchar[{\basicEnebidist}{\basicEnebidist[133]}}%

\def\basiceneadjar[#1]{\ENEADJAR{}{}{#100}}%

\newcommand{\eneadjar}{\@ifnextchar[{\basiceneadjar}{\basiceneadjar[133]}}%

\def\basicEneadjar[#1]#2#3{\ENEADJAR{#2}{#3}{#100}}%

\newcommand{\Eneadjar}{\@ifnextchar[{\basicEneadjar}{\basicEneadjar[133]}}%

\def\basiceneadjdist[#1]{\ENEADJDIST{}{}{#100}}%

\newcommand{\eneadjdist}{\@ifnextchar[{\basiceneadjdist}{\basiceneadjdist[133]}}%

\def\basicEneadjdist[#1]#2#3{\ENEADJDIST{#2}{#3}{#100}}%

\newcommand{\Eneadjdist}{\@ifnextchar[{\basicEneadjdist}{\basicEneadjdist[133]}}%

\newcommand{\ESEAR}[3]{\testdiagrammode%
\Y=#3%
\divide\Y by 2%
\Z=\Y%
\divide\Z by 2%
\begin{picture}(0,0)%
\put(-\Y,\Z){\line(2,-1){#3}}%
\put(\Y,-\Z){\esehead}%
\truex{200}\truey{800}\truez{600}%
\put(\value{x},\value{x}){\makebox(0,\value{z})[l]{${#1}$}}%
\put(-\value{x},-\value{y}){\makebox(0,\value{z})[r]{${#2}$}}%
\end{picture}}%

\newcommand{\ESEDIST}[3]{\testdiagrammode%
\Y=#3%
\divide\Y by 2%
\Z=\Y%
\divide\Z by 2%
\begin{picture}(0,0)%
\put(-\Y,\Z){\line(2,-1){#3}}%
\put(\Y,-\Z){\esehead}%
\truex{400}%
\put(0,0){\circle{\value{x}}}%
\truex{200}\truey{800}\truez{600}%
\put(\value{x},\value{x}){\makebox(0,\value{z})[l]{${#1}$}}%
\put(-\value{x},-\value{y}){\makebox(0,\value{z})[r]{${#2}$}}%
\end{picture}}%

\newcommand{\ESEDOTAR}[3]{\testdiagrammode%
\truex{100}\truey{268}\truez{134}%
\Y=#3%
\divide\Y by 2%
\Z=\Y%
\divide\Z by 2%
\NUMBEROFDOTS=#3%
\divide\NUMBEROFDOTS by \value{y}%
\advance\NUMBEROFDOTS by 1%
\begin{picture}(0,0)%
\multiput(-\Y,\Z)(\value{y},-\value{z}){\NUMBEROFDOTS}%
{\circle*{\value{x}}}%
\put(\Y,-\Z){\esehead}%
\truex{200}\truey{800}\truez{600}%
\put(\value{x},\value{x}){\makebox(0,\value{z})[l]{${#1}$}}%
\put(-\value{x},-\value{y}){\makebox(0,\value{z})[r]{${#2}$}}%
\end{picture}}%

\newcommand{\ESEMONO}[3]{\testdiagrammode%
\Y=#3%
\divide\Y by 2%
\Z=\Y%
\divide\Z by 2%
\TrueTail%
\bimolength=#3%
\advance\bimolength by -\TrueMonoTail%
\monolength=\bimolength%
\advance\monolength by -\Y%
\secondmonolength=\monolength%
\divide\secondmonolength by 2%
\begin{picture}(0,0)%
\put(-\monolength,\secondmonolength){\line(2,-1){\bimolength}}%
\put(-\monolength,\secondmonolength){\esehead}%
\put(\Y,-\Z){\esehead}%
\truex{200}\truey{800}\truez{600}%
\put(\value{x},\value{x}){\makebox(0,\value{z})[l]{${#1}$}}%
\put(-\value{x},-\value{y}){\makebox(0,\value{z})[r]{${#2}$}}%
\end{picture}}%

\newcommand{\ESEEPI}[3]{\testdiagrammode%
\Y=#3%
\divide\Y by 2%
\Z=\Y%
\divide\Z by 2%
\TrueHead%
\bimolength=#3%
\advance\bimolength by -\TrueEpiHead%
\epilength=\bimolength%
\advance\epilength by -\Y%
\secondepilength=\epilength%
\divide\secondepilength by 2%
\begin{picture}(0,0)%
\put(-\Y,\Z){\line(2,-1){\bimolength}}%
\put(\epilength,-\secondepilength){\esehead}%
\put(\Y,-\Z){\esehead}%
\truex{200}\truey{800}\truez{600}%
\put(\value{x},\value{x}){\makebox(0,\value{z})[l]{${#1}$}}%
\put(-\value{x},-\value{y}){\makebox(0,\value{z})[r]{${#2}$}}%
\end{picture}}%

\newcommand{\ESEBIMO}[3]{\testdiagrammode%
\Y=#3%
\divide\Y by 2%
\Z=\Y%
\divide\Z by 2%
\TrueTail\TrueHead%
\bimolength=#3%
\advance\bimolength by -\TrueMonoTail%
\monolength=\bimolength%
\advance\monolength by -\Y%
\advance\bimolength by -\TrueEpiHead%
\epilength=\bimolength%
\advance\epilength by -\monolength%
\secondmonolength=\monolength%
\divide\secondmonolength by 2%
\secondepilength=\epilength%
\divide\secondepilength by 2%
\begin{picture}(0,0)%
\put(-\monolength,\secondmonolength){\line(2,-1){\bimolength}}%
\put(-\monolength,\secondmonolength){\esehead}%
\put(\epilength,-\secondepilength){\esehead}%
\put(\Y,-\Z){\esehead}%
\truex{200}\truey{800}\truez{600}%
\put(\value{x},\value{x}){\makebox(0,\value{z})[l]{${#1}$}}%
\put(-\value{x},-\value{y}){\makebox(0,\value{z})[r]{${#2}$}}%
\end{picture}}%

\newcommand{\ESEEQL}[3]{\testdiagrammode%
\Y=#3%
\divide\Y by 2%
\Z=\Y%
\divide\Z by 2%
\begin{picture}(0,0)%
\put(-\Y,\Z){\begin{picture}(0,0)%
\truex{44}\truey{89}%
\put(-\value{x},-\value{y}){\line(2,-1){#3}}%
\put(\value{x},\value{y}){\line(2,-1){#3}}%
\end{picture}}%
\truex{200}\truey{800}\truez{600}%
\put(\value{x},\value{x}){\makebox(0,\value{z})[l]{${#1}$}}%
\put(-\value{x},-\value{y}){\makebox(0,\value{z})[r]{${#2}$}}%
\end{picture}}%

\newcommand{\ESEBIAR}[3]{%
\Y=#3%
\divide\Y by 2%
\Z=\Y%
\divide\Z by 2%
\begin{picture}(0,0)%
\put(-\Y,\Z){\begin{picture}(0,0)%
\truex{156}\truey{313}%
\put(-\value{x},-\value{y}){\line(2,-1){#3}}%
\put(\value{x},\value{y}){\line(2,-1){#3}}%
\monolength=#3%
\advance\monolength by -\value{x}%
\epilength=#3%
\advance\epilength by \value{x}%
\secondmonolength=\Y%
\advance\secondmonolength by -\value{y}%
\secondepilength=\Y%
\advance\secondepilength by \value{y}%
\put(\monolength,-\secondepilength){\esehead}%
\put(\epilength,-\secondmonolength){\esehead}%
\end{picture}}
\truex{400}\truey{1000}\truez{600}%
\put(\value{x},\value{x}){\makebox(0,\value{z})[l]{${#1}$}}%
\put(-\value{x},-\value{y}){\makebox(0,\value{z})[r]{${#2}$}}%
\end{picture}}%

\newcommand{\ESEBIDIST}[3]{\testdiagrammode%
\Y=#3%
\divide\Y by 2%
\Z=\Y%
\divide\Z by 2%
\begin{picture}(0,0)%
\truex{156}\truey{313}\truez{400}%
\put(-\Y,\Z){\begin{picture}(0,0)%
\put(-\value{x},-\value{y}){\line(2,-1){#3}}%
\put(\value{x},\value{y}){\line(2,-1){#3}}%
\monolength=#3%
\advance\monolength by -\value{x}%
\epilength=#3%
\advance\epilength by \value{x}%
\secondmonolength=\Y%
\advance\secondmonolength by -\value{y}%
\secondepilength=\Y%
\advance\secondepilength by \value{y}%
\put(\monolength,-\secondepilength){\esehead}%
\put(\epilength,-\secondmonolength){\esehead}%
\end{picture}}
\put(\value{x},\value{y}){\circle{\value{z}}}%
\put(-\value{x},-\value{y}){\circle{\value{z}}}%
\truex{400}\truey{1000}\truez{600}%
\put(\value{x},\value{x}){\makebox(0,\value{z})[l]{${#1}$}}%
\put(-\value{x},-\value{y}){\makebox(0,\value{z})[r]{${#2}$}}%
\end{picture}}%

\newcommand{\ESEADJAR}[3]{\testdiagrammode%
\Y=#3%
\divide\Y by 2%
\Z=\Y%
\divide\Z by 2%
\begin{picture}(0,0)%
\put(-\Y,\Z){\begin{picture}(0,0)%
\truex{156}\truey{313}%
\monolength=#3%
\advance\monolength by -\value{x}%
\epilength=#3%
\advance\epilength by \value{x}%
\secondmonolength=\Y%
\advance\secondmonolength by -\value{y}%
\secondepilength=\Y%
\advance\secondepilength by \value{y}%
\put(-\value{x},-\value{y}){\line(2,-1){#3}}%
\put(\monolength,-\secondepilength){\esehead}%
\put(\epilength,-\secondmonolength){\line(-2,1){#3}}%
\put(\value{x},\value{y}){\wnwhead}%
\end{picture}}
\truex{400}\truey{1000}\truez{600}%
\put(\value{x},\value{x}){\makebox(0,\value{z})[l]{${#1}$}}%
\put(-\value{x},-\value{y}){\makebox(0,\value{z})[r]{${#2}$}}%
\end{picture}}%

\newcommand{\ESEADJDIST}[3]{\testdiagrammode%
\Y=#3%
\divide\Y by 2%
\Z=\Y%
\divide\Z by 2%
\begin{picture}(0,0)%
\truex{156}\truey{313}\truez{400}%
\put(-\Y,\Z){\begin{picture}(0,0)%
\monolength=#3%
\advance\monolength by -\value{x}%
\epilength=#3%
\advance\epilength by \value{x}%
\secondmonolength=\Y%
\advance\secondmonolength by -\value{y}%
\secondepilength=\Y%
\advance\secondepilength by \value{y}%
\put(-\value{x},-\value{y}){\line(2,-1){#3}}%
\put(\monolength,-\secondepilength){\esehead}%
\put(\epilength,-\secondmonolength){\line(-2,1){#3}}%
\put(\value{x},\value{y}){\wnwhead}%
\end{picture}}
\put(\value{x},\value{y}){\circle{\value{z}}}%
\put(-\value{x},-\value{y}){\circle{\value{z}}}%
\truex{400}\truey{1000}\truez{600}%
\put(\value{x},\value{x}){\makebox(0,\value{z})[l]{${#1}$}}%
\put(-\value{x},-\value{y}){\makebox(0,\value{z})[r]{${#2}$}}%
\end{picture}}%

\def\basicesear[#1]{\ESEAR{}{}{#100}}%

\newcommand{\esear}{\@ifnextchar[{\basicesear}{\basicesear[133]}}%

\def\basicEsear[#1]#2{\ESEAR{#2}{}{#100}}%

\newcommand{\Esear}{\@ifnextchar[{\basicEsear}{\basicEsear[133]}}%

\def\basiceseaR[#1]#2{\ESEAR{}{#2}{#100}}%

\newcommand{\eseaR}{\@ifnextchar[{\basiceseaR}{\basiceseaR[133]}}%

\def\basicesedist[#1]{\ESEDIST{}{}{#100}}%

\newcommand{\esedist}{\@ifnextchar[{\basicesedist}{\basicesedist[133]}}%

\def\basicEsedist[#1]#2{\ESEDIST{#2}{}{#100}}%

\newcommand{\Esedist}{\@ifnextchar[{\basicEsedist}{\basicEsedist[133]}}%

\def\basicesedisT[#1]#2{\ESEDIST{}{#2}{#100}}%

\newcommand{\esedisT}{\@ifnextchar[{\basicesedisT}{\basicesedisT[133]}}%

\def\basicesedotar[#1]{\ESEDOTAR{}{}{#100}}%

\newcommand{\esedotar}{\@ifnextchar[{\basicesedotar}{\basicesedotar[133]}}%

\def\basicEsedotar[#1]#2{\ESEDOTAR{#2}{}{#100}}%

\newcommand{\Esedotar}{\@ifnextchar[{\basicEsedotar}{\basicEsedotar[133]}}%

\def\basicesedotaR[#1]#2{\ESEDOTAR{}{#2}{#100}}%

\newcommand{\esedotaR}{\@ifnextchar[{\basicesedotaR}{\basicesedotaR[133]}}%

\def\basicesemono[#1]{\ESEMONO{}{}{#100}}%

\newcommand{\esemono}{\@ifnextchar[{\basicesemono}{\basicesemono[133]}}%

\def\basicEsemono[#1]#2{\ESEMONO{#2}{}{#100}}%

\newcommand{\Esemono}{\@ifnextchar[{\basicEsemono}{\basicEsemono[133]}}%

\def\basicesemonO[#1]#2{\ESEMONO{}{#2}{#100}}%

\newcommand{\esemonO}{\@ifnextchar[{\basicesemonO}{\basicesemonO[133]}}%

\def\basiceseepi[#1]{\ESEEPI{}{}{#100}}%

\newcommand{\eseepi}{\@ifnextchar[{\basiceseepi}{\basiceseepi[133]}}%

\def\basicEseepi[#1]#2{\ESEEPI{#2}{}{#100}}%

\newcommand{\Eseepi}{\@ifnextchar[{\basicEseepi}{\basicEseepi[133]}}%

\def\basiceseepI[#1]#2{\ESEEPI{}{#2}{#100}}%

\newcommand{\eseepI}{\@ifnextchar[{\basiceseepI}{\basiceseepI[133]}}%

\def\basicesebimo[#1]{\ESEBIMO{}{}{#100}}%

\newcommand{\esebimo}{\@ifnextchar[{\basicesebimo}{\basicesebimo[133]}}%

\def\basicEsebimo[#1]#2{\ESEBIMO{#2}{}{#100}}%

\newcommand{\Esebimo}{\@ifnextchar[{\basicEsebimo}{\basicEsebimo[133]}}%

\def\basicesebimO[#1]#2{\ESEBIMO{}{#2}{#100}}%

\newcommand{\esebimO}{\@ifnextchar[{\basicesebimO}{\basicesebimO[133]}}%

\def\basiceseiso[#1]{\ESEAR{\cong}{}{#100}}%

\newcommand{\eseiso}{\@ifnextchar[{\basiceseiso}{\basiceseiso[133]}}%

\def\basicEseiso[#1]#2{\ESEAR{#2}{\cong}{#100}}%

\newcommand{\Eseiso}{\@ifnextchar[{\basicEseiso}{\basicEseiso[133]}}%

\def\basiceseisO[#1]#2{\ESEAR{\cong}{#2}{#100}}%

\newcommand{\eseisO}{\@ifnextchar[{\basiceseisO}{\basiceseisO[133]}}%

\def\basiceseeql[#1]{\ESEEQL{}{}{#100}}%

\newcommand{\eseeql}{\@ifnextchar[{\basiceseeql}{\basiceseeql[133]}}%

\def\basicEseeql[#1]#2{\ESEEQL{#2}{}{#100}}%

\newcommand{\Eseeql}{\@ifnextchar[{\basicEseeql}{\basicEseeql[133]}}%

\def\basiceseeqL[#1]#2{\ESEEQL{}{#2}{#100}}%

\newcommand{\eseeqL}{\@ifnextchar[{\basiceseeqL}{\basiceseeqL[133]}}%

\def\basicesebiar[#1]{\ESEBIAR{}{}{#100}}%

\newcommand{\esebiar}{\@ifnextchar[{\basicesebiar}{\basicesebiar[133]}}%

\def\basicEsebiar[#1]#2#3{\ESEBIAR{#2}{#3}{#100}}%

\newcommand{\Esebiar}{\@ifnextchar[{\basicEsebiar}{\basicEsebiar[133]}}%

\def\basicesebidist[#1]{\ESEBIDIST{}{}{#100}}%

\newcommand{\esebidist}{\@ifnextchar[{\basicesebidist}{\basicesebidist[133]}}%

\def\basicEsebidist[#1]#2#3{\ESEBIDIST{#2}{#3}{#100}}%

\newcommand{\Esebidist}{\@ifnextchar[{\basicEsebidist}{\basicEsebidist[133]}}%

\def\basiceseadjar[#1]{\ESEADJAR{}{}{#100}}%

\newcommand{\eseadjar}{\@ifnextchar[{\basiceseadjar}{\basiceseadjar[133]}}%

\def\basicEseadjar[#1]#2#3{\ESEADJAR{#2}{#3}{#100}}%

\newcommand{\Eseadjar}{\@ifnextchar[{\basicEseadjar}{\basicEseadjar[133]}}%

\def\basiceseadjdist[#1]{\ESEADJDIST{}{}{#100}}%

\newcommand{\eseadjdist}{\@ifnextchar[{\basiceseadjdist}{\basiceseadjdist[133]}}%

\def\basicEseadjdist[#1]#2#3{\ESEADJDIST{#2}{#3}{#100}}%

\newcommand{\Eseadjdist}{\@ifnextchar[{\basicEseadjdist}{\basicEseadjdist[133]}}%

\newcommand{\WSWAR}[3]{\testdiagrammode%
\Y=#3%
\divide\Y by 2%
\Z=\Y%
\divide\Z by 2%
\begin{picture}(0,0)%
\put(\Y,\Z){\line(-2,-1){#3}}%
\put(-\Y,-\Z){\wswhead}%
\truex{200}\truey{800}\truez{600}%
\put(-\value{x},\value{x}){\makebox(0,\value{z})[r]{${#1}$}}%
\put(\value{x},-\value{y}){\makebox(0,\value{z})[l]{${#2}$}}%
\end{picture}}%

\newcommand{\WSWDIST}[3]{\testdiagrammode%
\Y=#3%
\divide\Y by 2%
\Z=\Y%
\divide\Z by 2%
\begin{picture}(0,0)%
\put(\Y,\Z){\line(-2,-1){#3}}%
\put(-\Y,-\Z){\wswhead}%
\truex{400}%
\put(0,0){\circle{\value{x}}}%
\truex{200}\truey{800}\truez{600}%
\put(-\value{x},\value{x}){\makebox(0,\value{z})[r]{${#1}$}}%
\put(\value{x},-\value{y}){\makebox(0,\value{z})[l]{${#2}$}}%
\end{picture}}%

\newcommand{\WSWDOTAR}[3]{\testdiagrammode%
\truex{100}\truey{268}\truez{134}%
\Y=#3%
\divide\Y by 2%
\Z=\Y%
\divide\Z by 2%
\NUMBEROFDOTS=#3%
\divide\NUMBEROFDOTS by \value{y}%
\advance\NUMBEROFDOTS by 1%
\begin{picture}(0,0)%
\multiput(\Y,\Z)(-\value{y},-\value{z}){\NUMBEROFDOTS}%
{\circle*{\value{x}}}%
\put(-\Y,-\Z){\wswhead}%
\truex{200}\truey{800}\truez{600}%
\put(-\value{x},\value{x}){\makebox(0,\value{z})[r]{${#1}$}}%
\put(\value{x},-\value{y}){\makebox(0,\value{z})[l]{${#2}$}}%
\end{picture}}%

\newcommand{\WSWMONO}[3]{\testdiagrammode%
\Y=#3%
\divide\Y by 2%
\Z=\Y%
\divide\Z by 2%
\TrueTail%
\bimolength=#3%
\advance\bimolength by -\TrueMonoTail%
\monolength=\bimolength%
\advance\monolength by -\Y%
\secondmonolength=\monolength%
\divide\secondmonolength by 2%
\begin{picture}(0,0)%
\put(\monolength,\secondmonolength){\line(-2,-1){\bimolength}}%
\put(\monolength,\secondmonolength){\wswhead}%
\put(-\Y,-\Z){\wswhead}%
\truex{200}\truey{800}\truez{600}%
\put(-\value{x},\value{x}){\makebox(0,\value{z})[r]{${#1}$}}%
\put(\value{x},-\value{y}){\makebox(0,\value{z})[l]{${#2}$}}%
\end{picture}}%

\newcommand{\WSWEPI}[3]{\testdiagrammode%
\Y=#3%
\divide\Y by 2%
\Z=\Y%
\divide\Z by 2%
\TrueHead%
\bimolength=#3%
\advance\bimolength by -\TrueEpiHead%
\epilength=\bimolength%
\advance\epilength by -\Y%
\secondepilength=\epilength%
\divide\secondepilength by 2%
\begin{picture}(0,0)%
\put(\Y,\Z){\line(-2,-1){\bimolength}}%
\put(-\epilength,-\secondepilength){\wswhead}%
\put(-\Y,-\Z){\wswhead}%
\truex{200}\truey{800}\truez{600}%
\put(-\value{x},\value{x}){\makebox(0,\value{z})[r]{${#1}$}}%
\put(\value{x},-\value{y}){\makebox(0,\value{z})[l]{${#2}$}}%
\end{picture}}%

\newcommand{\WSWBIMO}[3]{\testdiagrammode%
\Y=#3%
\divide\Y by 2%
\Z=\Y%
\divide\Z by 2%
\TrueTail\TrueHead%
\bimolength=#3%
\advance\bimolength by -\TrueMonoTail%
\monolength=\bimolength%
\advance\monolength by -\Y%
\advance\bimolength by -\TrueEpiHead%
\epilength=\bimolength%
\advance\epilength by -\monolength%
\secondmonolength=\monolength%
\divide\secondmonolength by 2%
\secondepilength=\epilength%
\divide\secondepilength by 2%
\begin{picture}(0,0)%
\put(\monolength,\secondmonolength){\line(-2,-1){\bimolength}}%
\put(\monolength,\secondmonolength){\wswhead}%
\put(-\epilength,-\secondepilength){\wswhead}%
\put(-\Y,-\Z){\wswhead}%
\truex{200}\truey{800}\truez{600}%
\put(-\value{x},\value{x}){\makebox(0,\value{z})[r]{${#1}$}}%
\put(\value{x},-\value{y}){\makebox(0,\value{z})[l]{${#2}$}}%
\end{picture}}%

\newcommand{\WSWBIAR}[3]{\testdiagrammode%
\Y=#3%
\divide\Y by 2%
\Z=\Y%
\divide\Z by 2%
\begin{picture}(0,0)%
\put(\Y,\Z){\begin{picture}(0,0)%
\truex{156}\truey{313}%
\put(-\value{x},\value{y}){\line(-2,-1){#3}}%
\put(\value{x},-\value{y}){\line(-2,-1){#3}}%
\monolength=#3%
\advance\monolength by -\value{x}%
\epilength=#3%
\advance\epilength by \value{x}%
\secondmonolength=\Y%
\advance\secondmonolength by -\value{y}%
\secondepilength=\Y%
\advance\secondepilength by \value{y}%
\put(-\monolength,-\secondepilength){\wswhead}%
\put(-\epilength,-\secondmonolength){\wswhead}%
\end{picture}}
\truex{300}\truey{1000}\truez{600}%
\put(-\value{x},\value{x}){\makebox(0,\value{z})[r]{${#1}$}}%
\put(\value{x},-\value{y}){\makebox(0,\value{z})[l]{${#2}$}}%
\end{picture}}%

\newcommand{\WSWBIDIST}[3]{\testdiagrammode%
\Y=#3%
\divide\Y by 2%
\Z=\Y%
\divide\Z by 2%
\begin{picture}(0,0)%
\truex{156}\truey{313}\truez{400}%
\put(\Y,\Z){\begin{picture}(0,0)%
\put(-\value{x},\value{y}){\line(-2,-1){#3}}%
\put(\value{x},-\value{y}){\line(-2,-1){#3}}%
\monolength=#3%
\advance\monolength by -\value{x}%
\epilength=#3%
\advance\epilength by \value{x}%
\secondmonolength=\Y%
\advance\secondmonolength by -\value{y}%
\secondepilength=\Y%
\advance\secondepilength by \value{y}%
\put(-\monolength,-\secondepilength){\wswhead}%
\put(-\epilength,-\secondmonolength){\wswhead}%
\end{picture}}
\put(-\value{x},\value{y}){\circle{\value{z}}}%
\put(\value{x},-\value{y}){\circle{\value{z}}}%
\truex{300}\truey{1000}\truez{600}%
\put(-\value{x},\value{x}){\makebox(0,\value{z})[r]{${#1}$}}%
\put(\value{x},-\value{y}){\makebox(0,\value{z})[l]{${#2}$}}%
\end{picture}}%

\newcommand{\WSWADJAR}[3]{\testdiagrammode%
\Y=#3%
\divide\Y by 2%
\Z=\Y%
\divide\Z by 2%
\begin{picture}(0,0)%
\put(\Y,\Z){\begin{picture}(0,0)%
\truex{156}\truey{313}%
\monolength=#3%
\advance\monolength by -\value{x}%
\epilength=#3%
\advance\epilength by \value{x}%
\secondmonolength=\Y%
\advance\secondmonolength by -\value{y}%
\secondepilength=\Y%
\advance\secondepilength by \value{y}%
\put(\value{x},-\value{y}){\line(-2,-1){#3}}%
\put(-\monolength,-\secondepilength){\wswhead}%
\put(-\epilength,-\secondmonolength){\line(2,1){#3}}%
\put(-\value{x},\value{y}){\enehead}%
\end{picture}}
\truex{300}\truey{1000}\truez{600}%
\put(-\value{x},\value{x}){\makebox(0,\value{z})[r]{${#1}$}}%
\put(\value{x},-\value{y}){\makebox(0,\value{z})[l]{${#2}$}}%
\end{picture}}%

\newcommand{\WSWADJDIST}[3]{\testdiagrammode%
\Y=#3%
\divide\Y by 2%
\Z=\Y%
\divide\Z by 2%
\begin{picture}(0,0)%
\truex{156}\truey{313}\truez{400}%
\put(\Y,\Z){\begin{picture}(0,0)%
\monolength=#3%
\advance\monolength by -\value{x}%
\epilength=#3%
\advance\epilength by \value{x}%
\secondmonolength=\Y%
\advance\secondmonolength by -\value{y}%
\secondepilength=\Y%
\advance\secondepilength by \value{y}%
\put(\value{x},-\value{y}){\line(-2,-1){#3}}%
\put(-\monolength,-\secondepilength){\wswhead}%
\put(-\epilength,-\secondmonolength){\line(2,1){#3}}%
\put(-\value{x},\value{y}){\enehead}%
\end{picture}}
\put(-\value{x},\value{y}){\circle{\value{z}}}%
\put(\value{x},-\value{y}){\circle{\value{z}}}%
\truex{300}\truey{1000}\truez{600}%
\put(-\value{x},\value{x}){\makebox(0,\value{z})[r]{${#1}$}}%
\put(\value{x},-\value{y}){\makebox(0,\value{z})[l]{${#2}$}}%
\end{picture}}%

\def\basicwswar[#1]{\WSWAR{}{}{#100}}%

\newcommand{\wswar}{\@ifnextchar[{\basicwswar}{\basicwswar[133]}}%

\def\basicWswar[#1]#2{\WSWAR{#2}{}{#100}}%

\newcommand{\Wswar}{\@ifnextchar[{\basicWswar}{\basicWswar[133]}}%

\def\basicwswaR[#1]#2{\WSWAR{}{#2}{#100}}%

\newcommand{\wswaR}{\@ifnextchar[{\basicwswaR}{\basicwswaR[133]}}%

\def\basicwswdist[#1]{\WSWDIST{}{}{#100}}%

\newcommand{\wswdist}{\@ifnextchar[{\basicwswdist}{\basicwswdist[133]}}%

\def\basicWswdist[#1]#2{\WSWDIST{#2}{}{#100}}%

\newcommand{\Wswdist}{\@ifnextchar[{\basicWswdist}{\basicWswdist[133]}}%

\def\basicwswdisT[#1]#2{\WSWDIST{}{#2}{#100}}%

\newcommand{\wswdisT}{\@ifnextchar[{\basicwswdisT}{\basicwswdisT[133]}}%

\def\basicwswdotar[#1]{\WSWDOTAR{}{}{#100}}%

\newcommand{\wswdotar}{\@ifnextchar[{\basicwswdotar}{\basicwswdotar[133]}}%

\def\basicWswdotar[#1]#2{\WSWDOTAR{#2}{}{#100}}%

\newcommand{\Wswdotar}{\@ifnextchar[{\basicWswdotar}{\basicWswdotar[133]}}%

\def\basicwswdotaR[#1]#2{\WSWDOTAR{}{#2}{#100}}%

\newcommand{\wswdotaR}{\@ifnextchar[{\basicwswdotaR}{\basicwswdotaR[133]}}%

\def\basicwswmono[#1]{\WSWMONO{}{}{#100}}%

\newcommand{\wswmono}{\@ifnextchar[{\basicwswmono}{\basicwswmono[133]}}%

\def\basicWswmono[#1]#2{\WSWMONO{#2}{}{#100}}%

\newcommand{\Wswmono}{\@ifnextchar[{\basicWswmono}{\basicWswmono[133]}}%

\def\basicwswmonO[#1]#2{\WSWMONO{}{#2}{#100}}%

\newcommand{\wswmonO}{\@ifnextchar[{\basicwswmonO}{\basicwswmonO[133]}}%

\def\basicwswepi[#1]{\WSWEPI{}{}{#100}}%

\newcommand{\wswepi}{\@ifnextchar[{\basicwswepi}{\basicwswepi[133]}}%

\def\basicWswepi[#1]#2{\WSWEPI{#2}{}{#100}}%

\newcommand{\Wswepi}{\@ifnextchar[{\basicWswepi}{\basicWswepi[133]}}%

\def\basicwswepI[#1]#2{\WSWEPI{}{#2}{#100}}%

\newcommand{\wswepI}{\@ifnextchar[{\basicwswepI}{\basicwswepI[133]}}%

\def\basicwswbimo[#1]{\WSWBIMO{}{}{#100}}%

\newcommand{\wswbimo}{\@ifnextchar[{\basicwswbimo}{\basicwswbimo[133]}}%

\def\basicWswbimo[#1]#2{\WSWBIMO{#2}{}{#100}}%

\newcommand{\Wswbimo}{\@ifnextchar[{\basicWswbimo}{\basicWswbimo[133]}}%

\def\basicwswbimO[#1]#2{\WSWBIMO{}{#2}{#100}}%

\newcommand{\wswbimO}{\@ifnextchar[{\basicwswbimO}{\basicwswbimO[133]}}%

\def\basicwswiso[#1]{\WSWAR{\cong}{}{#100}}%

\newcommand{\wswiso}{\@ifnextchar[{\basicwswiso}{\basicwswiso[133]}}%

\def\basicWswiso[#1]#2{\WSWAR{#2}{\cong}{#100}}%

\newcommand{\Wswiso}{\@ifnextchar[{\basicWswiso}{\basicWswiso[133]}}%

\def\basicwswisO[#1]#2{\WSWAR{\cong}{#2}{#100}}%

\newcommand{\wswisO}{\@ifnextchar[{\basicwswisO}{\basicwswisO[133]}}%

\def\basicwswbiar[#1]{\WSWBIAR{}{}{#100}}%

\newcommand{\wswbiar}{\@ifnextchar[{\basicwswbiar}{\basicwswbiar[133]}}%

\def\basicWswbiar[#1]#2#3{\WSWBIAR{#2}{#3}{#100}}%

\newcommand{\Wswbiar}{\@ifnextchar[{\basicWswbiar}{\basicWswbiar[133]}}%

\def\basicwswbidist[#1]{\WSWBIDIST{}{}{#100}}%

\newcommand{\wswbidist}{\@ifnextchar[{\basicwswbidist}{\basicwswbidist[133]}}%

\def\basicWswbidist[#1]#2#3{\WSWBIDIST{#2}{#3}{#100}}%

\newcommand{\Wswbidist}{\@ifnextchar[{\basicWswbidist}{\basicWswbidist[133]}}%

\def\basicwswadjar[#1]{\WSWADJAR{}{}{#100}}%

\newcommand{\wswadjar}{\@ifnextchar[{\basicwswadjar}{\basicwswadjar[133]}}%

\def\basicWswadjar[#1]#2#3{\WSWADJAR{#2}{#3}{#100}}%

\newcommand{\Wswadjar}{\@ifnextchar[{\basicWswadjar}{\basicWswadjar[133]}}%

\def\basicwswadjdist[#1]{\WSWADJDIST{}{}{#100}}%

\newcommand{\wswadjdist}{\@ifnextchar[{\basicwswadjdist}{\basicwswadjdist[133]}}%

\def\basicWswadjdist[#1]#2#3{\WSWADJDIST{#2}{#3}{#100}}%

\newcommand{\Wswadjdist}{\@ifnextchar[{\basicWswadjdist}{\basicWswadjdist[133]}}%

\newcommand{\WNWAR}[3]{\testdiagrammode%
\Y=#3%
\divide\Y by 2%
\Z=\Y%
\divide\Z by 2%
\begin{picture}(0,0)%
\put(\Y,-\Z){\line(-2,1){#3}}%
\put(-\Y,\Z){\wnwhead}%
\truex{200}\truey{800}\truez{600}%
\put(\value{x},\value{x}){\makebox(0,\value{z})[l]{${#1}$}}%
\put(-\value{x},-\value{y}){\makebox(0,\value{z})[r]{${#2}$}}%
\end{picture}}%

\newcommand{\WNWDIST}[3]{\testdiagrammode%
\Y=#3%
\divide\Y by 2%
\Z=\Y%
\divide\Z by 2%
\begin{picture}(0,0)%
\put(\Y,-\Z){\line(-2,1){#3}}%
\put(-\Y,\Z){\wnwhead}%
\truex{400}%
\put(0,0){\circle{\value{x}}}%
\truex{200}\truey{800}\truez{600}%
\put(\value{x},\value{x}){\makebox(0,\value{z})[l]{${#1}$}}%
\put(-\value{x},-\value{y}){\makebox(0,\value{z})[r]{${#2}$}}%
\end{picture}}%

\newcommand{\WNWDOTAR}[3]{\testdiagrammode%
\truex{100}\truey{268}\truez{134}%
\Y=#3%
\divide\Y by 2%
\Z=\Y%
\divide\Z by 2%
\NUMBEROFDOTS=#3%
\divide\NUMBEROFDOTS by \value{y}%
\advance\NUMBEROFDOTS by 1%
\begin{picture}(0,0)%
\multiput(\Y,-\Z)(-\value{y},\value{z}){\NUMBEROFDOTS}%
{\circle*{\value{x}}}%
\put(-\Y,\Z){\wnwhead}%
\truex{200}\truey{800}\truez{600}%
\put(\value{x},\value{x}){\makebox(0,\value{z})[l]{${#1}$}}%
\put(-\value{x},-\value{y}){\makebox(0,\value{z})[r]{${#2}$}}%
\end{picture}}%

\newcommand{\WNWMONO}[3]{\testdiagrammode%
\Y=#3%
\divide\Y by 2%
\Z=\Y%
\divide\Z by 2%
\TrueTail%
\bimolength=#3%
\advance\bimolength by -\TrueMonoTail%
\monolength=\bimolength%
\advance\monolength by -\Y%
\secondmonolength=\monolength%
\divide\secondmonolength by 2%
\begin{picture}(0,0)%
\put(\monolength,-\secondmonolength){\line(-2,1){\bimolength}}%
\put(\monolength,-\secondmonolength){\wnwhead}%
\put(-\Y,\Z){\wnwhead}%
\truex{200}\truey{800}\truez{600}%
\put(\value{x},\value{x}){\makebox(0,\value{z})[l]{${#1}$}}%
\put(-\value{x},-\value{y}){\makebox(0,\value{z})[r]{${#2}$}}%
\end{picture}}%

\newcommand{\WNWEPI}[3]{\testdiagrammode%
\Y=#3%
\divide\Y by 2%
\Z=\Y%
\divide\Z by 2%
\TrueHead%
\bimolength=#3%
\advance\bimolength by -\TrueEpiHead%
\epilength=\bimolength%
\advance\epilength by -\Y%
\secondepilength=\epilength%
\divide\secondepilength by 2%
\begin{picture}(0,0)%
\put(\Y,-\Z){\line(-2,1){\bimolength}}%
\put(-\epilength,\secondepilength){\wnwhead}%
\put(-\Y,\Z){\wnwhead}%
\truex{200}\truey{800}\truez{600}%
\put(\value{x},\value{x}){\makebox(0,\value{z})[l]{${#1}$}}%
\put(-\value{x},-\value{y}){\makebox(0,\value{z})[r]{${#2}$}}%
\end{picture}}%

\newcommand{\WNWBIMO}[3]{\testdiagrammode%
\Y=#3%
\divide\Y by 2%
\Z=\Y%
\divide\Z by 2%
\TrueTail\TrueHead%
\bimolength=#3%
\advance\bimolength by -\TrueMonoTail%
\monolength=\bimolength%
\advance\monolength by -\Y%
\advance\bimolength by -\TrueEpiHead%
\epilength=\bimolength%
\advance\epilength by -\monolength%
\secondmonolength=\monolength%
\divide\secondmonolength by 2%
\secondepilength=\epilength%
\divide\secondepilength by 2%
\begin{picture}(0,0)%
\put(\monolength,-\secondmonolength){\line(-2,1){\bimolength}}%
\put(\monolength,-\secondmonolength){\wnwhead}%
\put(-\epilength,\secondepilength){\wnwhead}%
\put(-\Y,\Z){\wnwhead}%
\truex{200}\truey{800}\truez{600}%
\put(\value{x},\value{x}){\makebox(0,\value{z})[l]{${#1}$}}%
\put(-\value{x},-\value{y}){\makebox(0,\value{z})[r]{${#2}$}}%
\end{picture}}%

\newcommand{\WNWBIAR}[3]{\testdiagrammode%
\Y=#3%
\divide\Y by 2%
\Z=\Y%
\divide\Z by 2%
\begin{picture}(0,0)%
\put(\Y,-\Z){\begin{picture}(0,0)%
\truex{156}\truey{313}%
\put(-\value{x},-\value{y}){\line(-2,1){#3}}%
\put(\value{x},\value{y}){\line(-2,1){#3}}%
\monolength=#3%
\advance\monolength by -\value{x}%
\epilength=#3%
\advance\epilength by \value{x}%
\secondmonolength=\Y%
\advance\secondmonolength by -\value{y}%
\secondepilength=\Y%
\advance\secondepilength by \value{y}%
\put(-\monolength,\secondepilength){\wnwhead}%
\put(-\epilength,\secondmonolength){\wnwhead}%
\end{picture}}
\truex{400}\truey{1000}\truez{600}%
\put(\value{x},\value{x}){\makebox(0,\value{z})[l]{${#1}$}}%
\put(-\value{x},-\value{y}){\makebox(0,\value{z})[r]{${#2}$}}%
\end{picture}}%

\newcommand{\WNWBIDIST}[3]{\testdiagrammode%
\Y=#3%
\divide\Y by 2%
\Z=\Y%
\divide\Z by 2%
\begin{picture}(0,0)%
\truex{156}\truey{313}\truez{400}%
\put(\Y,-\Z){\begin{picture}(0,0)%
\put(-\value{x},-\value{y}){\line(-2,1){#3}}%
\put(\value{x},\value{y}){\line(-2,1){#3}}%
\monolength=#3%
\advance\monolength by -\value{x}%
\epilength=#3%
\advance\epilength by \value{x}%
\secondmonolength=\Y%
\advance\secondmonolength by -\value{y}%
\secondepilength=\Y%
\advance\secondepilength by \value{y}%
\put(-\monolength,\secondepilength){\wnwhead}%
\put(-\epilength,\secondmonolength){\wnwhead}%
\end{picture}}
\put(\value{x},\value{y}){\circle{\value{z}}}%
\put(-\value{x},-\value{y}){\circle{\value{z}}}%
\truex{400}\truey{1000}\truez{600}%
\put(\value{x},\value{x}){\makebox(0,\value{z})[l]{${#1}$}}%
\put(-\value{x},-\value{y}){\makebox(0,\value{z})[r]{${#2}$}}%
\end{picture}}%

\newcommand{\WNWADJAR}[3]{\testdiagrammode%
\Y=#3%
\divide\Y by 2%
\Z=\Y%
\divide\Z by 2%
\begin{picture}(0,0)%
\put(\Y,-\Z){\begin{picture}(0,0)%
\truex{156}\truey{313}%
\monolength=#3%
\advance\monolength by -\value{x}%
\epilength=#3%
\advance\epilength by \value{x}%
\secondmonolength=\Y%
\advance\secondmonolength by -\value{y}%
\secondepilength=\Y%
\advance\secondepilength by \value{y}%
\put(-\value{x},-\value{y}){\line(-2,1){#3}}%
\put(-\epilength,\secondmonolength){\wnwhead}%
\put(-\monolength,\secondepilength){\line(2,-1){#3}}%
\put(\value{x},\value{y}){\esehead}%
\end{picture}}
\truex{400}\truey{1000}\truez{600}%
\put(\value{x},\value{x}){\makebox(0,\value{z})[l]{${#1}$}}%
\put(-\value{x},-\value{y}){\makebox(0,\value{z})[r]{${#2}$}}%
\end{picture}}%

\newcommand{\WNWADJDIST}[3]{\testdiagrammode%
\Y=#3%
\divide\Y by 2%
\Z=\Y%
\divide\Z by 2%
\begin{picture}(0,0)%
\truex{156}\truey{313}\truez{400}%
\put(\Y,-\Z){\begin{picture}(0,0)%
\monolength=#3%
\advance\monolength by -\value{x}%
\epilength=#3%
\advance\epilength by \value{x}%
\secondmonolength=\Y%
\advance\secondmonolength by -\value{y}%
\secondepilength=\Y%
\advance\secondepilength by \value{y}%
\put(-\value{x},-\value{y}){\line(-2,1){#3}}%
\put(-\epilength,\secondmonolength){\wnwhead}%
\put(-\monolength,\secondepilength){\line(2,-1){#3}}%
\put(\value{x},\value{y}){\esehead}%
\end{picture}}
\put(\value{x},\value{y}){\circle{\value{z}}}%
\put(-\value{x},-\value{y}){\circle{\value{z}}}%
\truex{400}\truey{1000}\truez{600}%
\put(\value{x},\value{x}){\makebox(0,\value{z})[l]{${#1}$}}%
\put(-\value{x},-\value{y}){\makebox(0,\value{z})[r]{${#2}$}}%
\end{picture}}%

\def\basicwnwar[#1]{\WNWAR{}{}{#100}}%

\newcommand{\wnwar}{\@ifnextchar[{\basicwnwar}{\basicwnwar[133]}}%

\def\basicWnwar[#1]#2{\WNWAR{#2}{}{#100}}%

\newcommand{\Wnwar}{\@ifnextchar[{\basicWnwar}{\basicWnwar[133]}}%

\def\basicwnwaR[#1]#2{\WNWAR{}{#2}{#100}}%

\newcommand{\wnwaR}{\@ifnextchar[{\basicwnwaR}{\basicwnwaR[133]}}%

\def\basicwnwdist[#1]{\WNWDIST{}{}{#100}}%

\newcommand{\wnwdist}{\@ifnextchar[{\basicwnwdist}{\basicwnwdist[133]}}%

\def\basicWnwdist[#1]#2{\WNWDIST{#2}{}{#100}}%

\newcommand{\Wnwdist}{\@ifnextchar[{\basicWnwdist}{\basicWnwdist[133]}}%

\def\basicwnwdisT[#1]#2{\WNWDIST{}{#2}{#100}}%

\newcommand{\wnwdisT}{\@ifnextchar[{\basicwnwdisT}{\basicwnwdisT[133]}}%

\def\basicwnwdotar[#1]{\WNWDOTAR{}{}{#100}}%

\newcommand{\wnwdotar}{\@ifnextchar[{\basicwnwdotar}{\basicwnwdotar[133]}}%

\def\basicWnwdotar[#1]#2{\WNWDOTAR{#2}{}{#100}}%

\newcommand{\Wnwdotar}{\@ifnextchar[{\basicWnwdotar}{\basicWnwdotar[133]}}%

\def\basicwnwdotaR[#1]#2{\WNWDOTAR{}{#2}{#100}}%

\newcommand{\wnwdotaR}{\@ifnextchar[{\basicwnwdotaR}{\basicwnwdotaR[133]}}%

\def\basicwnwmono[#1]{\WNWMONO{}{}{#100}}%

\newcommand{\wnwmono}{\@ifnextchar[{\basicwnwmono}{\basicwnwmono[133]}}%

\def\basicWnwmono[#1]#2{\WNWMONO{#2}{}{#100}}%

\newcommand{\Wnwmono}{\@ifnextchar[{\basicWnwmono}{\basicWnwmono[133]}}%

\def\basicwnwmonO[#1]#2{\WNWMONO{}{#2}{#100}}%

\newcommand{\wnwmonO}{\@ifnextchar[{\basicwnwmonO}{\basicwnwmonO[133]}}%

\def\basicwnwepi[#1]{\WNWEPI{}{}{#100}}%

\newcommand{\wnwepi}{\@ifnextchar[{\basicwnwepi}{\basicwnwepi[133]}}%

\def\basicWnwepi[#1]#2{\WNWEPI{#2}{}{#100}}%

\newcommand{\Wnwepi}{\@ifnextchar[{\basicWnwepi}{\basicWnwepi[133]}}%

\def\basicwnwepI[#1]#2{\WNWEPI{}{#2}{#100}}%

\newcommand{\wnwepI}{\@ifnextchar[{\basicwnwepI}{\basicwnwepI[133]}}%

\def\basicwnwbimo[#1]{\WNWBIMO{}{}{#100}}%

\newcommand{\wnwbimo}{\@ifnextchar[{\basicwnwbimo}{\basicwnwbimo[133]}}%

\def\basicWnwbimo[#1]#2{\WNWBIMO{#2}{}{#100}}%

\newcommand{\Wnwbimo}{\@ifnextchar[{\basicWnwbimo}{\basicWnwbimo[133]}}%

\def\basicwnwbimO[#1]#2{\WNWBIMO{}{#2}{#100}}%

\newcommand{\wnwbimO}{\@ifnextchar[{\basicwnwbimO}{\basicwnwbimO[133]}}%

\def\basicwnwiso[#1]{\WNWAR{\cong}{}{#100}}%

\newcommand{\wnwiso}{\@ifnextchar[{\basicwnwiso}{\basicwnwiso[133]}}%

\def\basicWnwiso[#1]#2{\WNWAR{#2}{\cong}{#100}}%

\newcommand{\Wnwiso}{\@ifnextchar[{\basicWnwiso}{\basicWnwiso[133]}}%

\def\basicwnwisO[#1]#2{\WNWAR{\cong}{#2}{#100}}%

\newcommand{\wnwisO}{\@ifnextchar[{\basicwnwisO}{\basicwnwisO[133]}}%

\def\basicwnwbiar[#1]{\WNWBIAR{}{}{#100}}%

\newcommand{\wnwbiar}{\@ifnextchar[{\basicwnwbiar}{\basicwnwbiar[133]}}%

\def\basicWnwbiar[#1]#2#3{\WNWBIAR{#2}{#3}{#100}}%

\newcommand{\Wnwbiar}{\@ifnextchar[{\basicWnwbiar}{\basicWnwbiar[133]}}%

\def\basicwnwbidist[#1]{\WNWBIDIST{}{}{#100}}%

\newcommand{\wnwbidist}{\@ifnextchar[{\basicwnwbidist}{\basicwnwbidist[133]}}%

\def\basicWnwbidist[#1]#2#3{\WNWBIDIST{#2}{#3}{#100}}%

\newcommand{\Wnwbidist}{\@ifnextchar[{\basicWnwbidist}{\basicWnwbidist[133]}}%

\def\basicwnwadjar[#1]{\WNWADJAR{}{}{#100}}%

\newcommand{\wnwadjar}{\@ifnextchar[{\basicwnwadjar}{\basicwnwadjar[133]}}%

\def\basicWnwadjar[#1]#2#3{\WNWADJAR{#2}{#3}{#100}}%

\newcommand{\Wnwadjar}{\@ifnextchar[{\basicWnwadjar}{\basicWnwadjar[133]}}%

\def\basicwnwadjdist[#1]{\WNWADJDIST{}{}{#100}}%

\newcommand{\wnwadjdist}{\@ifnextchar[{\basicwnwadjdist}{\basicwnwadjdist[133]}}%

\def\basicWnwadjdist[#1]#2#3{\WNWADJDIST{#2}{#3}{#100}}%

\newcommand{\Wnwadjdist}{\@ifnextchar[{\basicWnwadjdist}{\basicWnwadjdist[133]}}%

\newcommand{\NNEAR}[3]{\testdiagrammode%
\Z=#3%
\divide\Z by 2%
\begin{picture}(0,0)%
\put(-\Z,-#3){\line(1,2){#3}}%
\put(\Z,#3){\nnehead}%
\truex{200}\truey{800}\truez{600}%
\put(-\value{x},\value{x}){\makebox(0,\value{z})[r]{${#1}$}}%
\put(\value{x},-\value{y}){\makebox(0,\value{z})[l]{${#2}$}}%
\end{picture}}%

\newcommand{\NNEDIST}[3]{\testdiagrammode%
\Z=#3%
\divide\Z by 2%
\begin{picture}(0,0)%
\put(-\Z,-#3){\line(1,2){#3}}%
\put(\Z,#3){\nnehead}%
\truex{400}%
\put(0,0){\circle{\value{x}}}%
\truex{200}\truey{800}\truez{600}%
\put(-\value{x},\value{x}){\makebox(0,\value{z})[r]{${#1}$}}%
\put(\value{x},-\value{y}){\makebox(0,\value{z})[l]{${#2}$}}%
\end{picture}}%

\newcommand{\NNEDOTAR}[3]{\testdiagrammode%
\truex{100}\truey{268}\truez{134}%
\Z=#3%
\divide\Z by 2%
\NUMBEROFDOTS=#3%
\divide\NUMBEROFDOTS by \value{z}%
\advance\NUMBEROFDOTS by 1%
\begin{picture}(0,0)%
\multiput(-\Z,-#3)(\value{z},\value{y}){\NUMBEROFDOTS}%
{\circle*{\value{x}}}%
\put(\Z,#3){\nnehead}%
\truex{200}\truey{800}\truez{600}%
\put(-\value{x},\value{x}){\makebox(0,\value{z})[r]{${#1}$}}%
\put(\value{x},-\value{y}){\makebox(0,\value{z})[l]{${#2}$}}%
\end{picture}}%

\newcommand{\NNEMONO}[3]{\testdiagrammode%
\Z=#3%
\divide\Z by 2%
\truetaiL%
\bimolength=#3%
\advance\bimolength by -\truemonotaiL%
\monolength=\bimolength%
\advance\monolength by -\Z%
\secondmonolength=\monolength%
\multiply\secondmonolength by 2%
\begin{picture}(0,0)%
\put(-\monolength,-\secondmonolength){\line(1,2){\bimolength}}%
\put(-\monolength,-\secondmonolength){\nnehead}%
\put(\Z,#3){\nnehead}%
\truex{200}\truey{800}\truez{600}%
\put(-\value{x},\value{x}){\makebox(0,\value{z})[r]{${#1}$}}%
\put(\value{x},-\value{y}){\makebox(0,\value{z})[l]{${#2}$}}%
\end{picture}}%

\newcommand{\NNEEPI}[3]{\testdiagrammode%
\Z=#3%
\divide\Z by 2%
\trueheaD%
\bimolength=#3%
\advance\bimolength by -\trueepiheaD%
\epilength=\bimolength%
\advance\epilength by -\Z%
\secondepilength=\epilength%
\multiply\secondepilength by 2%
\begin{picture}(0,0)%
\put(-\Z,-#3){\line(1,2){\bimolength}}%
\put(\epilength,\secondepilength){\nnehead}%
\put(\Z,#3){\nnehead}%
\truex{200}\truey{800}\truez{600}%
\put(-\value{x},\value{x}){\makebox(0,\value{z})[r]{${#1}$}}%
\put(\value{x},-\value{y}){\makebox(0,\value{z})[l]{${#2}$}}%
\end{picture}}%

\newcommand{\NNEBIMO}[3]{\testdiagrammode%
\Z=#3%
\divide\Z by 2%
\truetaiL\trueheaD%
\bimolength=#3%
\advance\bimolength by -\truemonotaiL%
\monolength=\bimolength%
\advance\monolength by -\Z%
\advance\bimolength by -\trueepiheaD%
\epilength=\bimolength%
\advance\epilength by -\monolength%
\secondmonolength=\monolength%
\multiply\secondmonolength by 2%
\secondepilength=\epilength%
\multiply\secondepilength by 2%
\begin{picture}(0,0)%
\put(-\monolength,-\secondmonolength){\line(1,2){\bimolength}}%
\put(-\monolength,-\secondmonolength){\nnehead}%
\put(\epilength,\secondepilength){\nnehead}%
\put(\Z,#3){\nnehead}%
\truex{200}\truey{800}\truez{600}%
\put(-\value{x},\value{x}){\makebox(0,\value{z})[r]{${#1}$}}%
\put(\value{x},-\value{y}){\makebox(0,\value{z})[l]{${#2}$}}%
\end{picture}}%

\newcommand{\NNEEQL}[3]{\testdiagrammode%
\Z=#3%
\divide\Z by 2%
\begin{picture}(0,0)%
\put(-\Z,-#3){\begin{picture}(0,0)%
\truex{44}\truey{89}%
\put(-\value{y},\value{x}){\line(1,2){#3}}%
\put(\value{y},-\value{x}){\line(1,2){#3}}%
\end{picture}}%
\truex{200}\truey{800}\truez{600}%
\put(-\value{x},\value{x}){\makebox(0,\value{z})[r]{${#1}$}}%
\put(\value{x},-\value{y}){\makebox(0,\value{z})[l]{${#2}$}}%
\end{picture}}%

\newcommand{\NNEBIAR}[3]{\testdiagrammode%
\Y=#3%
\divide\Y by 2%
\Z=#3%
\multiply \Z by 2%
\begin{picture}(0,0)%
\put(-\Y,-#3){\begin{picture}(0,0)%
\truex{313}\truey{156}%
\put(-\value{x},\value{y}){\line(1,2){#3}}%
\put(\value{x},-\value{y}){\line(1,2){#3}}%
\monolength=#3%
\advance\monolength by -\value{x}%
\epilength=#3%
\advance\epilength by \value{x}%
\secondmonolength=\Z%
\advance\secondmonolength by -\value{y}%
\secondepilength=\Z%
\advance\secondepilength by \value{y}%
\put(\monolength,\secondepilength){\nnehead}%
\put(\epilength,\secondmonolength){\nnehead}%
\end{picture}}
\truex{300}\truey{1000}\truez{600}%
\put(-\value{x},\value{x}){\makebox(0,\value{z})[r]{${#1}$}}%
\put(\value{x},-\value{y}){\makebox(0,\value{z})[l]{${#2}$}}%
\end{picture}}%

\newcommand{\NNEBIDIST}[3]{\testdiagrammode%
\Y=#3%
\divide\Y by 2%
\Z=#3%
\multiply \Z by 2%
\begin{picture}(0,0)%
\truex{313}\truey{156}\truez{400}%
\put(-\Y,-#3){\begin{picture}(0,0)%
\put(-\value{x},\value{y}){\line(1,2){#3}}%
\put(\value{x},-\value{y}){\line(1,2){#3}}%
\monolength=#3%
\advance\monolength by -\value{x}%
\epilength=#3%
\advance\epilength by \value{x}%
\secondmonolength=\Z%
\advance\secondmonolength by -\value{y}%
\secondepilength=\Z%
\advance\secondepilength by \value{y}%
\put(\monolength,\secondepilength){\nnehead}%
\put(\epilength,\secondmonolength){\nnehead}%
\end{picture}}
\put(-\value{x},\value{y}){\circle{\value{z}}}%
\put(\value{x},-\value{y}){\circle{\value{z}}}%
\truex{300}\truey{1000}\truez{600}%
\put(-\value{x},\value{x}){\makebox(0,\value{z})[r]{${#1}$}}%
\put(\value{x},-\value{y}){\makebox(0,\value{z})[l]{${#2}$}}%
\end{picture}}%

\newcommand{\NNEADJAR}[3]{\testdiagrammode%
\Y=#3%
\divide\Y by 2%
\Z=#3%
\multiply \Z by 2%
\begin{picture}(0,0)%
\put(-\Y,-#3){\begin{picture}(0,0)%
\truex{313}\truey{156}%
\monolength=#3%
\advance\monolength by -\value{x}%
\epilength=#3%
\advance\epilength by \value{x}%
\secondmonolength=\Z%
\advance\secondmonolength by -\value{y}%
\secondepilength=\Z%
\advance\secondepilength by \value{y}%
\put(\value{x},-\value{y}){\line(1,2){#3}}%
\put(\epilength,\secondmonolength){\nnehead}%
\put(\monolength,\secondepilength){\line(-1,-2){#3}}%
\put(-\value{x},\value{y}){\sswhead}%
\end{picture}}
\truex{300}\truey{1000}\truez{600}%
\put(-\value{x},\value{x}){\makebox(0,\value{z})[r]{${#1}$}}%
\put(\value{x},-\value{y}){\makebox(0,\value{z})[l]{${#2}$}}%
\end{picture}}%

\newcommand{\NNEADJDIST}[3]{\testdiagrammode%
\Y=#3%
\divide\Y by 2%
\Z=#3%
\multiply \Z by 2%
\begin{picture}(0,0)%
\truex{313}\truey{156}\truez{400}%
\put(-\Y,-#3){\begin{picture}(0,0)%
\monolength=#3%
\advance\monolength by -\value{x}%
\epilength=#3%
\advance\epilength by \value{x}%
\secondmonolength=\Z%
\advance\secondmonolength by -\value{y}%
\secondepilength=\Z%
\advance\secondepilength by \value{y}%
\put(\value{x},-\value{y}){\line(1,2){#3}}%
\put(\epilength,\secondmonolength){\nnehead}%
\put(\monolength,\secondepilength){\line(-1,-2){#3}}%
\put(-\value{x},\value{y}){\sswhead}%
\end{picture}}
\put(-\value{x},\value{y}){\circle{\value{z}}}%
\put(\value{x},-\value{y}){\circle{\value{z}}}%
\truex{300}\truey{1000}\truez{600}%
\put(-\value{x},\value{x}){\makebox(0,\value{z})[r]{${#1}$}}%
\put(\value{x},-\value{y}){\makebox(0,\value{z})[l]{${#2}$}}%
\end{picture}}%

\def\basicnnear[#1]{\NNEAR{}{}{#100}}%

\newcommand{\nnear}{\@ifnextchar[{\basicnnear}{\basicnnear[67]}}%

\def\basicNnear[#1]#2{\NNEAR{#2}{}{#100}}%

\newcommand{\Nnear}{\@ifnextchar[{\basicNnear}{\basicNnear[67]}}%

\def\basicnneaR[#1]#2{\NNEAR{}{#2}{#100}}%

\newcommand{\nneaR}{\@ifnextchar[{\basicnneaR}{\basicnneaR[67]}}%

\def\basicnnedist[#1]{\NNEDIST{}{}{#100}}%

\newcommand{\nnedist}{\@ifnextchar[{\basicnnedist}{\basicnnedist[67]}}%

\def\basicNnedist[#1]#2{\NNEDIST{#2}{}{#100}}%

\newcommand{\Nnedist}{\@ifnextchar[{\basicNnedist}{\basicNnedist[67]}}%

\def\basicnnedisT[#1]#2{\NNEDIST{}{#2}{#100}}%

\newcommand{\nnedisT}{\@ifnextchar[{\basicnnedisT}{\basicnnedisT[67]}}%

\def\basicnnedotar[#1]{\NNEDOTAR{}{}{#100}}%

\newcommand{\nnedotar}{\@ifnextchar[{\basicnnedotar}{\basicnnedotar[67]}}%

\def\basicNnedotar[#1]#2{\NNEDOTAR{#2}{}{#100}}%

\newcommand{\Nnedotar}{\@ifnextchar[{\basicNnedotar}{\basicNnedotar[67]}}%

\def\basicnnedotaR[#1]#2{\NNEDOTAR{}{#2}{#100}}%

\newcommand{\nnedotaR}{\@ifnextchar[{\basicnnedotaR}{\basicnnedotaR[67]}}%

\def\basicnnemono[#1]{\NNEMONO{}{}{#100}}%

\newcommand{\nnemono}{\@ifnextchar[{\basicnnemono}{\basicnnemono[67]}}%

\def\basicNnemono[#1]#2{\NNEMONO{#2}{}{#100}}%

\newcommand{\Nnemono}{\@ifnextchar[{\basicNnemono}{\basicNnemono[67]}}%

\def\basicnnemonO[#1]#2{\NNEMONO{}{#2}{#100}}%

\newcommand{\nnemonO}{\@ifnextchar[{\basicnnemonO}{\basicnnemonO[67]}}%

\def\basicnneepi[#1]{\NNEEPI{}{}{#100}}%

\newcommand{\nneepi}{\@ifnextchar[{\basicnneepi}{\basicnneepi[67]}}%

\def\basicNneepi[#1]#2{\NNEEPI{#2}{}{#100}}%

\newcommand{\Nneepi}{\@ifnextchar[{\basicNneepi}{\basicNneepi[67]}}%

\def\basicnneepI[#1]#2{\NNEEPI{}{#2}{#100}}%

\newcommand{\nneepI}{\@ifnextchar[{\basicnneepI}{\basicnneepI[67]}}%

\def\basicnnebimo[#1]{\NNEBIMO{}{}{#100}}%

\newcommand{\nnebimo}{\@ifnextchar[{\basicnnebimo}{\basicnnebimo[67]}}%

\def\basicNnebimo[#1]#2{\NNEBIMO{#2}{}{#100}}%

\newcommand{\Nnebimo}{\@ifnextchar[{\basicNnebimo}{\basicNnebimo[67]}}%

\def\basicnnebimO[#1]#2{\NNEBIMO{}{#2}{#100}}%

\newcommand{\nnebimO}{\@ifnextchar[{\basicnnebimO}{\basicnnebimO[67]}}%

\def\basicnneiso[#1]{\NNEAR{\cong}{}{#100}}%

\newcommand{\nneiso}{\@ifnextchar[{\basicnneiso}{\basicnneiso[67]}}%

\def\basicNneiso[#1]#2{\NNEAR{#2}{\cong}{#100}}%

\newcommand{\Nneiso}{\@ifnextchar[{\basicNneiso}{\basicNneiso[67]}}%

\def\basicnneisO[#1]#2{\NNEAR{\cong}{#2}{#100}}%

\newcommand{\nneisO}{\@ifnextchar[{\basicnneisO}{\basicnneisO[67]}}%

\def\basicnneeql[#1]{\NNEEQL{}{}{#100}}%

\newcommand{\nneeql}{\@ifnextchar[{\basicnneeql}{\basicnneeql[67]}}%

\def\basicNneeql[#1]#2{\NNEEQL{#2}{}{#100}}%

\newcommand{\Nneeql}{\@ifnextchar[{\basicNneeql}{\basicNneeql[67]}}%

\def\basicnneeqL[#1]#2{\NNEEQL{}{#2}{#100}}%

\newcommand{\nneeqL}{\@ifnextchar[{\basicnneeqL}{\basicnneeqL[67]}}%

\def\basicnnebiar[#1]{\NNEBIAR{}{}{#100}}%

\newcommand{\nnebiar}{\@ifnextchar[{\basicnnebiar}{\basicnnebiar[67]}}%

\def\basicNnebiar[#1]#2#3{\NNEBIAR{#2}{#3}{#100}}%

\newcommand{\Nnebiar}{\@ifnextchar[{\basicNnebiar}{\basicNnebiar[67]}}%

\def\basicnnebidist[#1]{\NNEBIDIST{}{}{#100}}%

\newcommand{\nnebidist}{\@ifnextchar[{\basicnnebidist}{\basicnnebidist[67]}}%

\def\basicNnebidist[#1]#2#3{\NNEBIDIST{#2}{#3}{#100}}%

\newcommand{\Nnebidist}{\@ifnextchar[{\basicNnebidist}{\basicNnebidist[67]}}%

\def\basicnneadjar[#1]{\NNEADJAR{}{}{#100}}%

\newcommand{\nneadjar}{\@ifnextchar[{\basicnneadjar}{\basicnneadjar[67]}}%

\def\basicNneadjar[#1]#2#3{\NNEADJAR{#2}{#3}{#100}}%

\newcommand{\Nneadjar}{\@ifnextchar[{\basicNneadjar}{\basicNneadjar[67]}}%

\def\basicnneadjdist[#1]{\NNEADJDIST{}{}{#100}}%

\newcommand{\nneadjdist}{\@ifnextchar[{\basicnneadjdist}{\basicnneadjdist[67]}}%

\def\basicNneadjdist[#1]#2#3{\NNEADJDIST{#2}{#3}{#100}}%

\newcommand{\Nneadjdist}{\@ifnextchar[{\basicNneadjdist}{\basicNneadjdist[67]}}%

\newcommand{\SSEAR}[3]{\testdiagrammode%
\Z=#3%
\divide\Z by 2%
\begin{picture}(0,0)%
\put(-\Z,#3){\line(1,-2){#3}}%
\put(\Z,-#3){\ssehead}%
\truex{200}\truey{800}\truez{600}%
\put(\value{x},\value{x}){\makebox(0,\value{z})[l]{${#1}$}}%
\put(-\value{x},-\value{y}){\makebox(0,\value{z})[r]{${#2}$}}%
\end{picture}}%

\newcommand{\SSEDIST}[3]{\testdiagrammode%
\Z=#3%
\divide\Z by 2%
\begin{picture}(0,0)%
\put(-\Z,#3){\line(1,-2){#3}}%
\put(\Z,-#3){\ssehead}%
\truex{400}%
\put(0,0){\circle{\value{x}}}%
\truex{200}\truey{800}\truez{600}%
\put(\value{x},\value{x}){\makebox(0,\value{z})[l]{${#1}$}}%
\put(-\value{x},-\value{y}){\makebox(0,\value{z})[r]{${#2}$}}%
\end{picture}}%

\newcommand{\SSEDOTAR}[3]{\testdiagrammode%
\truex{100}\truey{268}\truez{134}%
\Z=#3%
\divide\Z by 2%
\NUMBEROFDOTS=#3%
\divide\NUMBEROFDOTS by \value{z}%
\advance\NUMBEROFDOTS by 1%
\begin{picture}(0,0)%
\multiput(-\Z,#3)(\value{z},-\value{y}){\NUMBEROFDOTS}%
{\circle*{\value{x}}}%
\put(\Z,-#3){\ssehead}%
\truex{200}\truey{800}\truez{600}%
\put(\value{x},\value{x}){\makebox(0,\value{z})[l]{${#1}$}}%
\put(-\value{x},-\value{y}){\makebox(0,\value{z})[r]{${#2}$}}%
\end{picture}}%

\newcommand{\SSEMONO}[3]{\testdiagrammode%
\Z=#3%
\divide\Z by 2%
\truetaiL%
\bimolength=#3%
\advance\bimolength by -\truemonotaiL%
\monolength=\bimolength%
\advance\monolength by -\Z%
\secondmonolength=\monolength%
\multiply\secondmonolength by 2%
\begin{picture}(0,0)%
\put(-\monolength,\secondmonolength){\line(1,-2){\bimolength}}%
\put(-\monolength,\secondmonolength){\ssehead}%
\put(\Z,-#3){\ssehead}%
\truex{200}\truey{800}\truez{600}%
\put(\value{x},\value{x}){\makebox(0,\value{z})[l]{${#1}$}}%
\put(-\value{x},-\value{y}){\makebox(0,\value{z})[r]{${#2}$}}%
\end{picture}}%

\newcommand{\SSEEPI}[3]{\testdiagrammode%
\Z=#3%
\divide\Z by 2%
\trueheaD%
\bimolength=#3%
\advance\bimolength by -\trueepiheaD%
\epilength=\bimolength%
\advance\epilength by -\Z%
\secondepilength=\epilength%
\multiply\secondepilength by 2%
\begin{picture}(0,0)%
\put(-\Z,#3){\line(1,-2){\bimolength}}%
\put(\epilength,-\secondepilength){\ssehead}%
\put(\Z,-#3){\ssehead}%
\truex{200}\truey{800}\truez{600}%
\put(\value{x},\value{x}){\makebox(0,\value{z})[l]{${#1}$}}%
\put(-\value{x},-\value{y}){\makebox(0,\value{z})[r]{${#2}$}}%
\end{picture}}%

\newcommand{\SSEBIMO}[3]{\testdiagrammode%
\Z=#3%
\divide\Z by 2%
\truetaiL\trueheaD%
\bimolength=#3%
\advance\bimolength by -\truemonotaiL%
\monolength=\bimolength%
\advance\monolength by -\Z%
\advance\bimolength by -\trueepiheaD%
\epilength=\bimolength%
\advance\epilength by -\monolength%
\secondmonolength=\monolength%
\multiply\secondmonolength by 2%
\secondepilength=\epilength%
\multiply\secondepilength by 2%
\begin{picture}(0,0)%
\put(-\monolength,\secondmonolength){\line(1,-2){\bimolength}}%
\put(-\monolength,\secondmonolength){\ssehead}%
\put(\epilength,-\secondepilength){\ssehead}%
\put(\Z,-#3){\ssehead}%
\truex{200}\truey{800}\truez{600}%
\put(\value{x},\value{x}){\makebox(0,\value{z})[l]{${#1}$}}%
\put(-\value{x},-\value{y}){\makebox(0,\value{z})[r]{${#2}$}}%
\end{picture}}%

\newcommand{\SSEEQL}[3]{\testdiagrammode%
\Z=#3%
\divide\Z by 2%
\begin{picture}(0,0)%
\put(-\Z,#3){\begin{picture}(0,0)%
\truex{44}\truey{89}%
\put(-\value{y},-\value{x}){\line(1,-2){#3}}%
\put(\value{y},\value{x}){\line(1,-2){#3}}%
\end{picture}}%
\truex{200}\truey{800}\truez{600}%
\put(\value{x},\value{x}){\makebox(0,\value{z})[l]{${#1}$}}%
\put(-\value{x},-\value{y}){\makebox(0,\value{z})[r]{${#2}$}}%
\end{picture}}%

\newcommand{\SSEBIAR}[3]{\testdiagrammode%
\Y=#3%
\divide\Y by 2%
\Z=#3%
\multiply \Z by 2%
\begin{picture}(0,0)%
\put(-\Y,#3){\begin{picture}(0,0)%
\truex{313}\truey{156}%
\put(-\value{x},-\value{y}){\line(1,-2){#3}}%
\put(\value{x},\value{y}){\line(1,-2){#3}}%
\monolength=#3%
\advance\monolength by -\value{x}%
\epilength=#3%
\advance\epilength by \value{x}%
\secondmonolength=\Z%
\advance\secondmonolength by -\value{y}%
\secondepilength=\Z%
\advance\secondepilength by \value{y}%
\put(\monolength,-\secondepilength){\ssehead}%
\put(\epilength,-\secondmonolength){\ssehead}%
\end{picture}}
\truex{400}\truey{1000}\truez{600}%
\put(\value{x},\value{x}){\makebox(0,\value{z})[l]{${#1}$}}%
\put(-\value{x},-\value{y}){\makebox(0,\value{z})[r]{${#2}$}}%
\end{picture}}%

\newcommand{\SSEBIDIST}[3]{\testdiagrammode%
\Y=#3%
\divide\Y by 2%
\Z=#3%
\multiply \Z by 2%
\begin{picture}(0,0)%
\truex{313}\truey{156}\truez{400}%
\put(-\Y,#3){\begin{picture}(0,0)%
\put(-\value{x},-\value{y}){\line(1,-2){#3}}%
\put(\value{x},\value{y}){\line(1,-2){#3}}%
\monolength=#3%
\advance\monolength by -\value{x}%
\epilength=#3%
\advance\epilength by \value{x}%
\secondmonolength=\Z%
\advance\secondmonolength by -\value{y}%
\secondepilength=\Z%
\advance\secondepilength by \value{y}%
\put(\monolength,-\secondepilength){\ssehead}%
\put(\epilength,-\secondmonolength){\ssehead}%
\end{picture}}
\put(-\value{x},-\value{y}){\circle{\value{z}}}%
\put(\value{x},\value{y}){\circle{\value{z}}}%
\truex{500}\truey{1000}\truez{600}%
\put(\value{x},\value{x}){\makebox(0,\value{z})[l]{${#1}$}}%
\put(-\value{x},-\value{y}){\makebox(0,\value{z})[r]{${#2}$}}%
\end{picture}}%

\newcommand{\SSEADJAR}[3]{\testdiagrammode%
\Y=#3%
\divide\Y by 2%
\Z=#3%
\multiply \Z by 2%
\begin{picture}(0,0)%
\put(-\Y,#3){\begin{picture}(0,0)%
\truex{313}\truey{156}%
\monolength=#3%
\advance\monolength by -\value{x}%
\epilength=#3%
\advance\epilength by \value{x}%
\secondmonolength=\Z%
\advance\secondmonolength by -\value{y}%
\secondepilength=\Z%
\advance\secondepilength by \value{y}%
\put(-\value{x},-\value{y}){\line(1,-2){#3}}%
\put(\monolength,-\secondepilength){\ssehead}%
\put(\epilength,-\secondmonolength){\line(-1,2){#3}}%
\put(\value{x},\value{y}){\nnwhead}%
\end{picture}}
\truex{400}\truey{1000}\truez{600}%
\put(\value{x},\value{x}){\makebox(0,\value{z})[l]{${#1}$}}%
\put(-\value{x},-\value{y}){\makebox(0,\value{z})[r]{${#2}$}}%
\end{picture}}%

\newcommand{\SSEADJDIST}[3]{\testdiagrammode%
\Y=#3%
\divide\Y by 2%
\Z=#3%
\multiply \Z by 2%
\begin{picture}(0,0)%
\truex{313}\truey{156}\truez{400}%
\put(-\Y,#3){\begin{picture}(0,0)%
\monolength=#3%
\advance\monolength by -\value{x}%
\epilength=#3%
\advance\epilength by \value{x}%
\secondmonolength=\Z%
\advance\secondmonolength by -\value{y}%
\secondepilength=\Z%
\advance\secondepilength by \value{y}%
\put(-\value{x},-\value{y}){\line(1,-2){#3}}%
\put(\monolength,-\secondepilength){\ssehead}%
\put(\epilength,-\secondmonolength){\line(-1,2){#3}}%
\put(\value{x},\value{y}){\nnwhead}%
\end{picture}}
\put(\value{x},\value{y}){\circle{\value{z}}}%
\put(-\value{x},-\value{y}){\circle{\value{z}}}%
\truex{500}\truey{1000}\truez{600}%
\put(\value{x},\value{x}){\makebox(0,\value{z})[l]{${#1}$}}%
\put(-\value{x},-\value{y}){\makebox(0,\value{z})[r]{${#2}$}}%
\end{picture}}%

\def\basicssear[#1]{\SSEAR{}{}{#100}}%

\newcommand{\ssear}{\@ifnextchar[{\basicssear}{\basicssear[67]}}%

\def\basicSsear[#1]#2{\SSEAR{#2}{}{#100}}%

\newcommand{\Ssear}{\@ifnextchar[{\basicSsear}{\basicSsear[67]}}%

\def\basicsseaR[#1]#2{\SSEAR{}{#2}{#100}}%

\newcommand{\sseaR}{\@ifnextchar[{\basicsseaR}{\basicsseaR[67]}}%

\def\basicssedist[#1]{\SSEDIST{}{}{#100}}%

\newcommand{\ssedist}{\@ifnextchar[{\basicssedist}{\basicssedist[67]}}%

\def\basicSsedist[#1]#2{\SSEDIST{#2}{}{#100}}%

\newcommand{\Ssedist}{\@ifnextchar[{\basicSsedist}{\basicSsedist[67]}}%

\def\basicssedisT[#1]#2{\SSEDIST{}{#2}{#100}}%

\newcommand{\ssedisT}{\@ifnextchar[{\basicssedisT}{\basicssedisT[67]}}%

\def\basicssedotar[#1]{\SSEDOTAR{}{}{#100}}%

\newcommand{\ssedotar}{\@ifnextchar[{\basicssedotar}{\basicssedotar[67]}}%

\def\basicSsedotar[#1]#2{\SSEDOTAR{#2}{}{#100}}%

\newcommand{\Ssedotar}{\@ifnextchar[{\basicSsedotar}{\basicSsedotar[67]}}%

\def\basicssedotaR[#1]#2{\SSEDOTAR{}{#2}{#100}}%

\newcommand{\ssedotaR}{\@ifnextchar[{\basicssedotaR}{\basicssedotaR[67]}}%

\def\basicssemono[#1]{\SSEMONO{}{}{#100}}%

\newcommand{\ssemono}{\@ifnextchar[{\basicssemono}{\basicssemono[67]}}%

\def\basicSsemono[#1]#2{\SSEMONO{#2}{}{#100}}%

\newcommand{\Ssemono}{\@ifnextchar[{\basicSsemono}{\basicSsemono[67]}}%

\def\basicssemonO[#1]#2{\SSEMONO{}{#2}{#100}}%

\newcommand{\ssemonO}{\@ifnextchar[{\basicssemonO}{\basicssemonO[67]}}%

\def\basicsseepi[#1]{\SSEEPI{}{}{#100}}%

\newcommand{\sseepi}{\@ifnextchar[{\basicsseepi}{\basicsseepi[67]}}%

\def\basicSseepi[#1]#2{\SSEEPI{#2}{}{#100}}%

\newcommand{\Sseepi}{\@ifnextchar[{\basicSseepi}{\basicSseepi[67]}}%

\def\basicsseepI[#1]#2{\SSEEPI{}{#2}{#100}}%

\newcommand{\sseepI}{\@ifnextchar[{\basicsseepI}{\basicsseepI[67]}}%

\def\basicssebimo[#1]{\SSEBIMO{}{}{#100}}%

\newcommand{\ssebimo}{\@ifnextchar[{\basicssebimo}{\basicssebimo[67]}}%

\def\basicSsebimo[#1]#2{\SSEBIMO{#2}{}{#100}}%

\newcommand{\Ssebimo}{\@ifnextchar[{\basicSsebimo}{\basicSsebimo[67]}}%

\def\basicssebimO[#1]#2{\SSEBIMO{}{#2}{#100}}%

\newcommand{\ssebimO}{\@ifnextchar[{\basicssebimO}{\basicssebimO[67]}}%

\def\basicsseiso[#1]{\SSEAR{\cong}{}{#100}}%

\newcommand{\sseiso}{\@ifnextchar[{\basicsseiso}{\basicsseiso[67]}}%

\def\basicSseiso[#1]#2{\SSEAR{#2}{\cong}{#100}}%

\newcommand{\Sseiso}{\@ifnextchar[{\basicSseiso}{\basicSseiso[67]}}%

\def\basicsseisO[#1]#2{\SSEAR{\cong}{#2}{#100}}%

\newcommand{\sseisO}{\@ifnextchar[{\basicsseisO}{\basicsseisO[67]}}%

\def\basicsseeql[#1]{\SSEEQL{}{}{#100}}%

\newcommand{\sseeql}{\@ifnextchar[{\basicsseeql}{\basicsseeql[67]}}%

\def\basicSseeql[#1]#2{\SSEEQL{#2}{}{#100}}%

\newcommand{\Sseeql}{\@ifnextchar[{\basicSseeql}{\basicSseeql[67]}}%

\def\basicsseeqL[#1]#2{\SSEEQL{}{#2}{#100}}%

\newcommand{\sseeqL}{\@ifnextchar[{\basicsseeqL}{\basicsseeqL[67]}}%

\def\basicssebiar[#1]{\SSEBIAR{}{}{#100}}%

\newcommand{\ssebiar}{\@ifnextchar[{\basicssebiar}{\basicssebiar[67]}}%

\def\basicSsebiar[#1]#2#3{\SSEBIAR{#2}{#3}{#100}}%

\newcommand{\Ssebiar}{\@ifnextchar[{\basicSsebiar}{\basicSsebiar[67]}}%

\def\basicssebidist[#1]{\SSEBIDIST{}{}{#100}}%

\newcommand{\ssebidist}{\@ifnextchar[{\basicssebidist}{\basicssebidist[67]}}%

\def\basicSsebidist[#1]#2#3{\SSEBIDIST{#2}{#3}{#100}}%

\newcommand{\Ssebidist}{\@ifnextchar[{\basicSsebidist}{\basicSsebidist[67]}}%

\def\basicsseadjar[#1]{\SSEADJAR{}{}{#100}}%

\newcommand{\sseadjar}{\@ifnextchar[{\basicsseadjar}{\basicsseadjar[67]}}%

\def\basicSseadjar[#1]#2#3{\SSEADJAR{#2}{#3}{#100}}%

\newcommand{\Sseadjar}{\@ifnextchar[{\basicSseadjar}{\basicSseadjar[67]}}%

\def\basicsseadjdist[#1]{\SSEADJDIST{}{}{#100}}%

\newcommand{\sseadjdist}{\@ifnextchar[{\basicsseadjdist}{\basicsseadjdist[67]}}%

\def\basicSseadjdist[#1]#2#3{\SSEADJDIST{#2}{#3}{#100}}%

\newcommand{\Sseadjdist}{\@ifnextchar[{\basicSseadjdist}{\basicSseadjdist[67]}}%

\newcommand{\SSWAR}[3]{\testdiagrammode%
\Z=#3%
\divide\Z by 2%
\begin{picture}(0,0)%
\put(\Z,#3){\line(-1,-2){#3}}%
\put(-\Z,-#3){\sswhead}%
\truex{200}\truey{800}\truez{600}%
\put(-\value{x},\value{x}){\makebox(0,\value{z})[r]{${#1}$}}%
\put(\value{x},-\value{y}){\makebox(0,\value{z})[l]{${#2}$}}%
\end{picture}}%

\newcommand{\SSWDIST}[3]{\testdiagrammode%
\Z=#3%
\divide\Z by 2%
\begin{picture}(0,0)%
\put(\Z,#3){\line(-1,-2){#3}}%
\put(-\Z,-#3){\sswhead}%
\truex{400}%
\put(0,0){\circle{\value{x}}}%
\truex{200}\truey{800}\truez{600}%
\put(-\value{x},\value{x}){\makebox(0,\value{z})[r]{${#1}$}}%
\put(\value{x},-\value{y}){\makebox(0,\value{z})[l]{${#2}$}}%
\end{picture}}%

\newcommand{\SSWDOTAR}[3]{\testdiagrammode%
\truex{100}\truey{268}\truez{134}%
\Z=#3%
\divide\Z by 2%
\NUMBEROFDOTS=#3%
\divide\NUMBEROFDOTS by \value{z}%
\advance\NUMBEROFDOTS by 1%
\begin{picture}(0,0)%
\multiput(\Z,#3)(-\value{z},-\value{y}){\NUMBEROFDOTS}%
{\circle*{\value{x}}}%
\put(-\Z,-#3){\sswhead}%
\truex{200}\truey{800}\truez{600}%
\put(-\value{x},\value{x}){\makebox(0,\value{z})[r]{${#1}$}}%
\put(\value{x},-\value{y}){\makebox(0,\value{z})[l]{${#2}$}}%
\end{picture}}%

\newcommand{\SSWMONO}[3]{\testdiagrammode%
\Z=#3%
\divide\Z by 2%
\truetaiL%
\bimolength=#3%
\advance\bimolength by -\truemonotaiL%
\monolength=\bimolength%
\advance\monolength by -\Z%
\secondmonolength=\monolength%
\multiply\secondmonolength by 2%
\begin{picture}(0,0)%
\put(\monolength,\secondmonolength){\line(-1,-2){\bimolength}}%
\put(\monolength,\secondmonolength){\sswhead}%
\put(-\Z,-#3){\sswhead}%
\truex{200}\truey{800}\truez{600}%
\put(-\value{x},\value{x}){\makebox(0,\value{z})[r]{${#1}$}}%
\put(\value{x},-\value{y}){\makebox(0,\value{z})[l]{${#2}$}}%
\end{picture}}%

\newcommand{\SSWEPI}[3]{\testdiagrammode%
\Z=#3%
\divide\Z by 2%
\trueheaD%
\bimolength=#3%
\advance\bimolength by -\trueepiheaD%
\epilength=\bimolength%
\advance\epilength by -\Z%
\secondepilength=\epilength%
\multiply\secondepilength by 2%
\begin{picture}(0,0)%
\put(\Z,#3){\line(-1,-2){\bimolength}}%
\put(-\epilength,-\secondepilength){\sswhead}%
\put(-\Z,-#3){\sswhead}%
\truex{200}\truey{800}\truez{600}%
\put(-\value{x},\value{x}){\makebox(0,\value{z})[r]{${#1}$}}%
\put(\value{x},-\value{y}){\makebox(0,\value{z})[l]{${#2}$}}%
\end{picture}}%

\newcommand{\SSWBIMO}[3]{\testdiagrammode%
\Z=#3%
\divide\Z by 2%
\truetaiL\trueheaD%
\bimolength=#3%
\advance\bimolength by -\truemonotaiL%
\monolength=\bimolength%
\advance\monolength by -\Z%
\advance\bimolength by -\trueepiheaD%
\epilength=\bimolength%
\advance\epilength by -\monolength%
\secondmonolength=\monolength%
\multiply\secondmonolength by 2%
\secondepilength=\epilength%
\multiply\secondepilength by 2%
\begin{picture}(0,0)%
\put(\monolength,\secondmonolength){\line(-1,-2){\bimolength}}%
\put(\monolength,\secondmonolength){\sswhead}%
\put(-\epilength,-\secondepilength){\sswhead}%
\put(-\Z,-#3){\sswhead}%
\truex{200}\truey{800}\truez{600}%
\put(-\value{x},\value{x}){\makebox(0,\value{z})[r]{${#1}$}}%
\put(\value{x},-\value{y}){\makebox(0,\value{z})[l]{${#2}$}}%
\end{picture}}%

\newcommand{\SSWBIAR}[3]{\testdiagrammode%
\Y=#3%
\divide\Y by 2%
\Z=#3%
\multiply \Z by 2%
\begin{picture}(0,0)%
\put(\Y,#3){\begin{picture}(0,0)%
\truex{313}\truey{156}%
\put(-\value{x},\value{y}){\line(-1,-2){#3}}%
\put(\value{x},-\value{y}){\line(-1,-2){#3}}%
\monolength=#3%
\advance\monolength by -\value{x}%
\epilength=#3%
\advance\epilength by \value{x}%
\secondmonolength=\Z%
\advance\secondmonolength by -\value{y}%
\secondepilength=\Z%
\advance\secondepilength by \value{y}%
\put(-\monolength,-\secondepilength){\sswhead}%
\put(-\epilength,-\secondmonolength){\sswhead}%
\end{picture}}
\truex{300}\truey{1000}\truez{600}%
\put(-\value{x},\value{x}){\makebox(0,\value{z})[r]{${#1}$}}%
\put(\value{x},-\value{y}){\makebox(0,\value{z})[l]{${#2}$}}%
\end{picture}}%

\newcommand{\SSWBIDIST}[3]{\testdiagrammode%
\Y=#3%
\divide\Y by 2%
\Z=#3%
\multiply \Z by 2%
\begin{picture}(0,0)%
\truex{313}\truey{156}\truez{400}%
\put(\Y,#3){\begin{picture}(0,0)%
\put(-\value{x},\value{y}){\line(-1,-2){#3}}%
\put(\value{x},-\value{y}){\line(-1,-2){#3}}%
\monolength=#3%
\advance\monolength by -\value{x}%
\epilength=#3%
\advance\epilength by \value{x}%
\secondmonolength=\Z%
\advance\secondmonolength by -\value{y}%
\secondepilength=\Z%
\advance\secondepilength by \value{y}%
\put(-\monolength,-\secondepilength){\sswhead}%
\put(-\epilength,-\secondmonolength){\sswhead}%
\end{picture}}
\put(-\value{x},\value{y}){\circle{\value{z}}}%
\put(\value{x},-\value{y}){\circle{\value{z}}}%
\truex{300}\truey{1000}\truez{600}%
\put(-\value{x},\value{x}){\makebox(0,\value{z})[r]{${#1}$}}%
\put(\value{x},-\value{y}){\makebox(0,\value{z})[l]{${#2}$}}%
\end{picture}}%

\newcommand{\SSWADJAR}[3]{\testdiagrammode%
\Y=#3%
\divide\Y by 2%
\Z=#3%
\multiply \Z by 2%
\begin{picture}(0,0)%
\put(\Y,#3){\begin{picture}(0,0)%
\truex{313}\truey{156}%
\monolength=#3%
\advance\monolength by -\value{x}%
\epilength=#3%
\advance\epilength by \value{x}%
\secondmonolength=\Z%
\advance\secondmonolength by -\value{y}%
\secondepilength=\Z%
\advance\secondepilength by \value{y}%
\put(\value{x},-\value{y}){\line(-1,-2){#3}}%
\put(-\monolength,-\secondepilength){\sswhead}%
\put(-\epilength,-\secondmonolength){\line(1,2){#3}}%
\put(-\value{x},\value{y}){\nnehead}%
\end{picture}}
\truex{300}\truey{1000}\truez{600}%
\put(-\value{x},\value{x}){\makebox(0,\value{z})[r]{${#1}$}}%
\put(\value{x},-\value{y}){\makebox(0,\value{z})[l]{${#2}$}}%
\end{picture}}%

\newcommand{\SSWADJDIST}[3]{\testdiagrammode%
\Y=#3%
\divide\Y by 2%
\Z=#3%
\multiply \Z by 2%
\begin{picture}(0,0)%
\truex{313}\truey{156}\truez{400}%
\put(\Y,#3){\begin{picture}(0,0)%
\monolength=#3%
\advance\monolength by -\value{x}%
\epilength=#3%
\advance\epilength by \value{x}%
\secondmonolength=\Z%
\advance\secondmonolength by -\value{y}%
\secondepilength=\Z%
\advance\secondepilength by \value{y}%
\put(\value{x},-\value{y}){\line(-1,-2){#3}}%
\put(-\monolength,-\secondepilength){\sswhead}%
\put(-\epilength,-\secondmonolength){\line(1,2){#3}}%
\put(-\value{x},\value{y}){\nnehead}%
\end{picture}}
\put(-\value{x},\value{y}){\circle{\value{z}}}%
\put(\value{x},-\value{y}){\circle{\value{z}}}%
\truex{300}\truey{1000}\truez{600}%
\put(-\value{x},\value{x}){\makebox(0,\value{z})[r]{${#1}$}}%
\put(\value{x},-\value{y}){\makebox(0,\value{z})[l]{${#2}$}}%
\end{picture}}%

\def\basicsswar[#1]{\SSWAR{}{}{#100}}%

\newcommand{\sswar}{\@ifnextchar[{\basicsswar}{\basicsswar[67]}}%

\def\basicSswar[#1]#2{\SSWAR{#2}{}{#100}}%

\newcommand{\Sswar}{\@ifnextchar[{\basicSswar}{\basicSswar[67]}}%

\def\basicsswaR[#1]#2{\SSWAR{}{#2}{#100}}%

\newcommand{\sswaR}{\@ifnextchar[{\basicsswaR}{\basicsswaR[67]}}%

\def\basicsswdist[#1]{\SSWDIST{}{}{#100}}%

\newcommand{\sswdist}{\@ifnextchar[{\basicsswdist}{\basicsswdist[67]}}%

\def\basicSswdist[#1]#2{\SSWDIST{#2}{}{#100}}%

\newcommand{\Sswdist}{\@ifnextchar[{\basicSswdist}{\basicSswdist[67]}}%

\def\basicsswdisT[#1]#2{\SSWDIST{}{#2}{#100}}%

\newcommand{\sswdisT}{\@ifnextchar[{\basicsswdisT}{\basicsswdisT[67]}}%

\def\basicsswdotar[#1]{\SSWDOTAR{}{}{#100}}%

\newcommand{\sswdotar}{\@ifnextchar[{\basicsswdotar}{\basicsswdotar[67]}}%

\def\basicSswdotar[#1]#2{\SSWDOTAR{#2}{}{#100}}%

\newcommand{\Sswdotar}{\@ifnextchar[{\basicSswdotar}{\basicSswdotar[67]}}%

\def\basicsswdotaR[#1]#2{\SSWDOTAR{}{#2}{#100}}%

\newcommand{\sswdotaR}{\@ifnextchar[{\basicsswdotaR}{\basicsswdotaR[67]}}%

\def\basicsswmono[#1]{\SSWMONO{}{}{#100}}%

\newcommand{\sswmono}{\@ifnextchar[{\basicsswmono}{\basicsswmono[67]}}%

\def\basicSswmono[#1]#2{\SSWMONO{#2}{}{#100}}%

\newcommand{\Sswmono}{\@ifnextchar[{\basicSswmono}{\basicSswmono[67]}}%

\def\basicsswmonO[#1]#2{\SSWMONO{}{#2}{#100}}%

\newcommand{\sswmonO}{\@ifnextchar[{\basicsswmonO}{\basicsswmonO[67]}}%

\def\basicsswepi[#1]{\SSWEPI{}{}{#100}}%

\newcommand{\sswepi}{\@ifnextchar[{\basicsswepi}{\basicsswepi[67]}}%

\def\basicSswepi[#1]#2{\SSWEPI{#2}{}{#100}}%

\newcommand{\Sswepi}{\@ifnextchar[{\basicSswepi}{\basicSswepi[67]}}%

\def\basicsswepI[#1]#2{\SSWEPI{}{#2}{#100}}%

\newcommand{\sswepI}{\@ifnextchar[{\basicsswepI}{\basicsswepI[67]}}%

\def\basicsswbimo[#1]{\SSWBIMO{}{}{#100}}%

\newcommand{\sswbimo}{\@ifnextchar[{\basicsswbimo}{\basicsswbimo[67]}}%

\def\basicSswbimo[#1]#2{\SSWBIMO{#2}{}{#100}}%

\newcommand{\Sswbimo}{\@ifnextchar[{\basicSswbimo}{\basicSswbimo[67]}}%

\def\basicsswbimO[#1]#2{\SSWBIMO{}{#2}{#100}}%

\newcommand{\sswbimO}{\@ifnextchar[{\basicsswbimO}{\basicsswbimO[67]}}%

\def\basicsswiso[#1]{\SSWAR{\cong}{}{#100}}%

\newcommand{\sswiso}{\@ifnextchar[{\basicsswiso}{\basicsswiso[67]}}%

\def\basicSswiso[#1]#2{\SSWAR{#2}{\cong}{#100}}%

\newcommand{\Sswiso}{\@ifnextchar[{\basicSswiso}{\basicSswiso[67]}}%

\def\basicsswisO[#1]#2{\SSWAR{\cong}{#2}{#100}}%

\newcommand{\sswisO}{\@ifnextchar[{\basicsswisO}{\basicsswisO[67]}}%

\def\basicsswbiar[#1]{\SSWBIAR{}{}{#100}}%

\newcommand{\sswbiar}{\@ifnextchar[{\basicsswbiar}{\basicsswbiar[67]}}%

\def\basicSswbiar[#1]#2#3{\SSWBIAR{#2}{#3}{#100}}%

\newcommand{\Sswbiar}{\@ifnextchar[{\basicSswbiar}{\basicSswbiar[67]}}%

\def\basicsswbidist[#1]{\SSWBIDIST{}{}{#100}}%

\newcommand{\sswbidist}{\@ifnextchar[{\basicsswbidist}{\basicsswbidist[67]}}%

\def\basicSswbidist[#1]#2#3{\SSWBIDIST{#2}{#3}{#100}}%

\newcommand{\Sswbidist}{\@ifnextchar[{\basicSswbidist}{\basicSswbidist[67]}}%

\def\basicsswadjar[#1]{\SSWADJAR{}{}{#100}}%

\newcommand{\sswadjar}{\@ifnextchar[{\basicsswadjar}{\basicsswadjar[67]}}%

\def\basicSswadjar[#1]#2#3{\SSWADJAR{#2}{#3}{#100}}%

\newcommand{\Sswadjar}{\@ifnextchar[{\basicSswadjar}{\basicSswadjar[67]}}%

\def\basicsswadjdist[#1]{\SSWADJDIST{}{}{#100}}%

\newcommand{\sswadjdist}{\@ifnextchar[{\basicsswadjdist}{\basicsswadjdist[67]}}%

\def\basicSswadjdist[#1]#2#3{\SSWADJDIST{#2}{#3}{#100}}%

\newcommand{\Sswadjdist}{\@ifnextchar[{\basicSswadjdist}{\basicSswadjdist[67]}}%

\newcommand{\NNWAR}[3]{\testdiagrammode%
\Z=#3%
\divide\Z by 2%
\begin{picture}(0,0)%
\put(\Z,-#3){\line(-1,2){#3}}%
\put(-\Z,#3){\nnwhead}%
\truex{200}\truey{800}\truez{600}%
\put(\value{x},\value{x}){\makebox(0,\value{z})[l]{${#1}$}}%
\put(-\value{x},-\value{y}){\makebox(0,\value{z})[r]{${#2}$}}%
\end{picture}}%

\newcommand{\NNWDIST}[3]{\testdiagrammode%
\Z=#3%
\divide\Z by 2%
\begin{picture}(0,0)%
\put(\Z,-#3){\line(-1,2){#3}}%
\put(-\Z,#3){\nnwhead}%
\truex{400}%
\put(0,0){\circle{\value{x}}}%
\truex{200}\truey{800}\truez{600}%
\put(\value{x},\value{x}){\makebox(0,\value{z})[l]{${#1}$}}%
\put(-\value{x},-\value{y}){\makebox(0,\value{z})[r]{${#2}$}}%
\end{picture}}%

\newcommand{\NNWDOTAR}[3]{\testdiagrammode%
\truex{100}\truey{268}\truez{134}%
\Z=#3%
\divide\Z by 2%
\NUMBEROFDOTS=#3%
\divide\NUMBEROFDOTS by \value{z}%
\advance\NUMBEROFDOTS by 1%
\begin{picture}(0,0)%
\multiput(\Z,-#3)(-\value{z},\value{y}){\NUMBEROFDOTS}%
{\circle*{\value{x}}}%
\put(-\Z,#3){\nnwhead}%
\truex{200}\truey{800}\truez{600}%
\put(\value{x},\value{x}){\makebox(0,\value{z})[l]{${#1}$}}%
\put(-\value{x},-\value{y}){\makebox(0,\value{z})[r]{${#2}$}}%
\end{picture}}%

\newcommand{\NNWMONO}[3]{\testdiagrammode%
\Z=#3%
\divide\Z by 2%
\truetaiL%
\bimolength=#3%
\advance\bimolength by -\truemonotaiL%
\monolength=\bimolength%
\advance\monolength by -\Z%
\secondmonolength=\monolength%
\multiply\secondmonolength by 2%
\begin{picture}(0,0)%
\put(\monolength,-\secondmonolength){\line(-1,2){\bimolength}}%
\put(\monolength,-\secondmonolength){\nnwhead}%
\put(-\Z,#3){\nnwhead}%
\truex{200}\truey{800}\truez{600}%
\put(\value{x},\value{x}){\makebox(0,\value{z})[l]{${#1}$}}%
\put(-\value{x},-\value{y}){\makebox(0,\value{z})[r]{${#2}$}}%
\end{picture}}%

\newcommand{\NNWEPI}[3]{\testdiagrammode%
\Z=#3%
\divide\Z by 2%
\trueheaD%
\bimolength=#3%
\advance\bimolength by -\trueepiheaD%
\epilength=\bimolength%
\advance\epilength by -\Z%
\secondepilength=\epilength%
\multiply\secondepilength by 2%
\begin{picture}(0,0)%
\put(\Z,-#3){\line(-1,2){\bimolength}}%
\put(-\epilength,\secondepilength){\nnwhead}%
\put(-\Z,#3){\nnwhead}%
\truex{200}\truey{800}\truez{600}%
\put(\value{x},\value{x}){\makebox(0,\value{z})[l]{${#1}$}}%
\put(-\value{x},-\value{y}){\makebox(0,\value{z})[r]{${#2}$}}%
\end{picture}}%

\newcommand{\NNWBIMO}[3]{\testdiagrammode%
\Z=#3%
\divide\Z by 2%
\truetaiL\trueheaD%
\bimolength=#3%
\advance\bimolength by -\truemonotaiL%
\monolength=\bimolength%
\advance\monolength by -\Z%
\advance\bimolength by -\trueepiheaD%
\epilength=\bimolength%
\advance\epilength by -\monolength%
\secondmonolength=\monolength%
\multiply\secondmonolength by 2%
\secondepilength=\epilength%
\multiply\secondepilength by 2%
\begin{picture}(0,0)%
\put(\monolength,-\secondmonolength){\line(-1,2){\bimolength}}%
\put(\monolength,-\secondmonolength){\nnwhead}%
\put(-\epilength,\secondepilength){\nnwhead}%
\put(-\Z,#3){\nnwhead}%
\truex{200}\truey{800}\truez{600}%
\put(\value{x},\value{x}){\makebox(0,\value{z})[l]{${#1}$}}%
\put(-\value{x},-\value{y}){\makebox(0,\value{z})[r]{${#2}$}}%
\end{picture}}%

\newcommand{\NNWBIAR}[3]{\testdiagrammode%
\Y=#3%
\divide\Y by 2%
\Z=#3%
\multiply \Z by 2%
\begin{picture}(0,0)%
\put(\Y,-#3){\begin{picture}(0,0)%
\truex{313}\truey{156}%
\put(-\value{x},-\value{y}){\line(-1,2){#3}}%
\put(\value{x},\value{y}){\line(-1,2){#3}}%
\monolength=#3%
\advance\monolength by -\value{x}%
\epilength=#3%
\advance\epilength by \value{x}%
\secondmonolength=\Z%
\advance\secondmonolength by -\value{y}%
\secondepilength=\Z%
\advance\secondepilength by \value{y}%
\put(-\monolength,\secondepilength){\nnwhead}%
\put(-\epilength,\secondmonolength){\nnwhead}%
\end{picture}}
\truex{400}\truey{1000}\truez{600}%
\put(\value{x},\value{x}){\makebox(0,\value{z})[l]{${#1}$}}%
\put(-\value{x},-\value{y}){\makebox(0,\value{z})[r]{${#2}$}}%
\end{picture}}%

\newcommand{\NNWBIDIST}[3]{\testdiagrammode%
\Y=#3%
\divide\Y by 2%
\Z=#3%
\multiply \Z by 2%
\begin{picture}(0,0)%
\truex{313}\truey{156}\truez{400}%
\put(\Y,-#3){\begin{picture}(0,0)%
\put(-\value{x},-\value{y}){\line(-1,2){#3}}%
\put(\value{x},\value{y}){\line(-1,2){#3}}%
\monolength=#3%
\advance\monolength by -\value{x}%
\epilength=#3%
\advance\epilength by \value{x}%
\secondmonolength=\Z%
\advance\secondmonolength by -\value{y}%
\secondepilength=\Z%
\advance\secondepilength by \value{y}%
\put(-\monolength,\secondepilength){\nnwhead}%
\put(-\epilength,\secondmonolength){\nnwhead}%
\end{picture}}
\put(-\value{x},-\value{y}){\circle{\value{z}}}%
\put(\value{x},\value{y}){\circle{\value{z}}}%
\truex{500}\truey{1000}\truez{600}%
\put(\value{x},\value{x}){\makebox(0,\value{z})[l]{${#1}$}}%
\put(-\value{x},-\value{y}){\makebox(0,\value{z})[r]{${#2}$}}%
\end{picture}}%

\newcommand{\NNWADJAR}[3]{\testdiagrammode%
\Y=#3%
\divide\Y by 2%
\Z=#3%
\multiply \Z by 2%
\begin{picture}(0,0)%
\put(\Y,-#3){\begin{picture}(0,0)%
\truex{313}\truey{156}%
\monolength=#3%
\advance\monolength by -\value{x}%
\epilength=#3%
\advance\epilength by \value{x}%
\secondmonolength=\Z%
\advance\secondmonolength by -\value{y}%
\secondepilength=\Z%
\advance\secondepilength by \value{y}%
\put(-\value{x},-\value{y}){\line(-1,2){#3}}%
\put(-\epilength,\secondmonolength){\nnwhead}%
\put(-\monolength,\secondepilength){\line(1,-2){#3}}%
\put(\value{x},\value{y}){\ssehead}%
\end{picture}}
\truex{400}\truey{1000}\truez{600}%
\put(\value{x},\value{x}){\makebox(0,\value{z})[l]{${#1}$}}%
\put(-\value{x},-\value{y}){\makebox(0,\value{z})[r]{${#2}$}}%
\end{picture}}%

\newcommand{\NNWADJDIST}[3]{\testdiagrammode%
\Y=#3%
\divide\Y by 2%
\Z=#3%
\multiply \Z by 2%
\begin{picture}(0,0)%
\truex{313}\truey{156}\truez{400}%
\put(\Y,-#3){\begin{picture}(0,0)%
\monolength=#3%
\advance\monolength by -\value{x}%
\epilength=#3%
\advance\epilength by \value{x}%
\secondmonolength=\Z%
\advance\secondmonolength by -\value{y}%
\secondepilength=\Z%
\advance\secondepilength by \value{y}%
\put(-\value{x},-\value{y}){\line(-1,2){#3}}%
\put(-\epilength,\secondmonolength){\nnwhead}%
\put(-\monolength,\secondepilength){\line(1,-2){#3}}%
\put(\value{x},\value{y}){\ssehead}%
\end{picture}}
\put(\value{x},\value{y}){\circle{\value{z}}}%
\put(-\value{x},-\value{y}){\circle{\value{z}}}%
\truex{500}\truey{1000}\truez{600}%
\put(\value{x},\value{x}){\makebox(0,\value{z})[l]{${#1}$}}%
\put(-\value{x},-\value{y}){\makebox(0,\value{z})[r]{${#2}$}}%
\end{picture}}%

\def\basicnnwar[#1]{\NNWAR{}{}{#100}}%

\newcommand{\nnwar}{\@ifnextchar[{\basicnnwar}{\basicnnwar[67]}}%

\def\basicNnwar[#1]#2{\NNWAR{#2}{}{#100}}%

\newcommand{\Nnwar}{\@ifnextchar[{\basicNnwar}{\basicNnwar[67]}}%

\def\basicnnwaR[#1]#2{\NNWAR{}{#2}{#100}}%

\newcommand{\nnwaR}{\@ifnextchar[{\basicnnwaR}{\basicnnwaR[67]}}%

\def\basicnnwdist[#1]{\NNWDIST{}{}{#100}}%

\newcommand{\nnwdist}{\@ifnextchar[{\basicnnwdist}{\basicnnwdist[67]}}%

\def\basicNnwdist[#1]#2{\NNWDIST{#2}{}{#100}}%

\newcommand{\Nnwdist}{\@ifnextchar[{\basicNnwdist}{\basicNnwdist[67]}}%

\def\basicnnwdisT[#1]#2{\NNWDIST{}{#2}{#100}}%

\newcommand{\nnwdisT}{\@ifnextchar[{\basicnnwdisT}{\basicnnwdisT[67]}}%

\def\basicnnwdotar[#1]{\NNWDOTAR{}{}{#100}}%

\newcommand{\nnwdotar}{\@ifnextchar[{\basicnnwdotar}{\basicnnwdotar[67]}}%

\def\basicNnwdotar[#1]#2{\NNWDOTAR{#2}{}{#100}}%

\newcommand{\Nnwdotar}{\@ifnextchar[{\basicNnwdotar}{\basicNnwdotar[67]}}%

\def\basicnnwdotaR[#1]#2{\NNWDOTAR{}{#2}{#100}}%

\newcommand{\nnwdotaR}{\@ifnextchar[{\basicnnwdotaR}{\basicnnwdotaR[67]}}%

\def\basicnnwmono[#1]{\NNWMONO{}{}{#100}}%

\newcommand{\nnwmono}{\@ifnextchar[{\basicnnwmono}{\basicnnwmono[67]}}%

\def\basicNnwmono[#1]#2{\NNWMONO{#2}{}{#100}}%

\newcommand{\Nnwmono}{\@ifnextchar[{\basicNnwmono}{\basicNnwmono[67]}}%

\def\basicnnwmonO[#1]#2{\NNWMONO{}{#2}{#100}}%

\newcommand{\nnwmonO}{\@ifnextchar[{\basicnnwmonO}{\basicnnwmonO[67]}}%

\def\basicnnwepi[#1]{\NNWEPI{}{}{#100}}%

\newcommand{\nnwepi}{\@ifnextchar[{\basicnnwepi}{\basicnnwepi[67]}}%

\def\basicNnwepi[#1]#2{\NNWEPI{#2}{}{#100}}%

\newcommand{\Nnwepi}{\@ifnextchar[{\basicNnwepi}{\basicNnwepi[67]}}%

\def\basicnnwepI[#1]#2{\NNWEPI{}{#2}{#100}}%

\newcommand{\nnwepI}{\@ifnextchar[{\basicnnwepI}{\basicnnwepI[67]}}%

\def\basicnnwbimo[#1]{\NNWBIMO{}{}{#100}}%

\newcommand{\nnwbimo}{\@ifnextchar[{\basicnnwbimo}{\basicnnwbimo[67]}}%

\def\basicNnwbimo[#1]#2{\NNWBIMO{#2}{}{#100}}%

\newcommand{\Nnwbimo}{\@ifnextchar[{\basicNnwbimo}{\basicNnwbimo[67]}}%

\def\basicnnwbimO[#1]#2{\NNWBIMO{}{#2}{#100}}%

\newcommand{\nnwbimO}{\@ifnextchar[{\basicnnwbimO}{\basicnnwbimO[67]}}%

\def\basicnnwiso[#1]{\NNWAR{\cong}{}{#100}}%

\newcommand{\nnwiso}{\@ifnextchar[{\basicnnwiso}{\basicnnwiso[67]}}%

\def\basicNnwiso[#1]#2{\NNWAR{#2}{\cong}{#100}}%

\newcommand{\Nnwiso}{\@ifnextchar[{\basicNnwiso}{\basicNnwiso[67]}}%

\def\basicnnwisO[#1]#2{\NNWAR{\cong}{#2}{#100}}%

\newcommand{\nnwisO}{\@ifnextchar[{\basicnnwisO}{\basicnnwisO[67]}}%

\def\basicnnwbiar[#1]{\NNWBIAR{}{}{#100}}%

\newcommand{\nnwbiar}{\@ifnextchar[{\basicnnwbiar}{\basicnnwbiar[67]}}%

\def\basicNnwbiar[#1]#2#3{\NNWBIAR{#2}{#3}{#100}}%

\newcommand{\Nnwbiar}{\@ifnextchar[{\basicNnwbiar}{\basicNnwbiar[67]}}%

\def\basicnnwbidist[#1]{\NNWBIDIST{}{}{#100}}%

\newcommand{\nnwbidist}{\@ifnextchar[{\basicnnwbidist}{\basicnnwbidist[67]}}%

\def\basicNnwbidist[#1]#2#3{\NNWBIDIST{#2}{#3}{#100}}%

\newcommand{\Nnwbidist}{\@ifnextchar[{\basicNnwbidist}{\basicNnwbidist[67]}}%

\def\basicnnwadjar[#1]{\NNWADJAR{}{}{#100}}%

\newcommand{\nnwadjar}{\@ifnextchar[{\basicnnwadjar}{\basicnnwadjar[67]}}%

\def\basicNnwadjar[#1]#2#3{\NNWADJAR{#2}{#3}{#100}}%

\newcommand{\Nnwadjar}{\@ifnextchar[{\basicNnwadjar}{\basicNnwadjar[67]}}%

\def\basicnnwadjdist[#1]{\NNWADJDIST{}{}{#100}}%

\newcommand{\nnwadjdist}{\@ifnextchar[{\basicnnwadjdist}{\basicnnwadjdist[67]}}%

\def\basicNnwadjdist[#1]#2#3{\NNWADJDIST{#2}{#3}{#100}}%

\newcommand{\Nnwadjdist}{\@ifnextchar[{\basicNnwadjdist}{\basicNnwadjdist[67]}}%

\newcommand{\EENEAR}[3]{\testdiagrammode%
\Y=#3%
\divide \Y by 2%
\Z=\Y%
\divide \Z by 3%
\begin{picture}(0,0)%
\put(-\Y,-\Z){\line(3,1){#3}}%
\put(\Y,\Z){\eenehead}%
\truex{200}\truey{800}\truez{600}%
\put(-\value{x},\value{x}){\makebox(0,\value{z})[r]{${#1}$}}%
\put(\value{x},-\value{y}){\makebox(0,\value{z})[l]{${#2}$}}%
\end{picture}}%

\def\basiceenear[#1]{\EENEAR{}{}{#100}}%

\newcommand{\eenear}{\@ifnextchar[{\basiceenear}{\basiceenear[211]}}%

\def\basicEenear[#1]#2{\EENEAR{#2}{}{#100}}%

\newcommand{\Eenear}{\@ifnextchar[{\basicEenear}{\basicEenear[211]}}%

\def\basiceeneaR[#1]#2{\EENEAR{}{#2}{#100}}%

\newcommand{\eeneaR}{\@ifnextchar[{\basiceeneaR}{\basiceeneaR[211]}}%

\newcommand{\EESEAR}[3]{\testdiagrammode%
\Y=#3%
\divide \Y by 2%
\Z=\Y%
\divide \Z by 3%
\begin{picture}(0,0)%
\put(-\Y,\Z){\line(3,-1){#3}}%
\put(\Y,-\Z){\eesehead}%
\truex{200}\truey{800}\truez{600}%
\put(\value{x},\value{x}){\makebox(0,\value{z})[l]{${#1}$}}%
\put(-\value{x},-\value{y}){\makebox(0,\value{z})[r]{${#2}$}}%
\end{picture}}%

\def\basiceesear[#1]{\EESEAR{}{}{#100}}%

\newcommand{\eesear}{\@ifnextchar[{\basiceesear}{\basiceesear[211]}}%

\def\basicEesear[#1]#2{\EESEAR{#2}{}{#100}}%

\newcommand{\Eesear}{\@ifnextchar[{\basicEesear}{\basicEesear[211]}}%

\def\basiceeseaR[#1]#2{\EESEAR{}{#2}{#100}}%

\newcommand{\eeseaR}{\@ifnextchar[{\basiceeseaR}{\basiceeseaR[211]}}%

\newcommand{\WWNWAR}[3]{\testdiagrammode%
\Y=#3%
\divide \Y by 2%
\Z=\Y%
\divide \Z by 3%
\begin{picture}(0,0)%
\put(\Y,-\Z){\line(-3,1){#3}}%
\put(-\Y,\Z){\wwnwhead}%
\truex{200}\truey{800}\truez{600}%
\put(\value{x},\value{x}){\makebox(0,\value{z})[l]{${#1}$}}%
\put(-\value{x},-\value{y}){\makebox(0,\value{z})[r]{${#2}$}}%
\end{picture}}%

\def\basicwwnwar[#1]{\WWNWAR{}{}{#100}}%

\newcommand{\wwnwar}{\@ifnextchar[{\basicwwnwar}{\basicwwnwar[211]}}%

\def\basicWwnwar[#1]#2{\WWNWAR{#2}{}{#100}}%

\newcommand{\Wwnwar}{\@ifnextchar[{\basicWwnwar}{\basicWwnwar[211]}}%

\def\basicwwnwaR[#1]#2{\WWNWAR{}{#2}{#100}}%

\newcommand{\wwnwaR}{\@ifnextchar[{\basicwwnwaR}{\basicwwnwaR[211]}}%

\newcommand{\WWSWAR}[3]{\testdiagrammode%
\Y=#3%
\divide \Y by 2%
\Z=\Y%
\divide \Z by 3%
\begin{picture}(0,0)%
\put(\Y,\Z){\line(-3,-1){#3}}%
\put(-\Y,-\Z){\wwswhead}%
\truex{200}\truey{800}\truez{600}%
\put(-\value{x},\value{x}){\makebox(0,\value{z})[r]{${#1}$}}%
\put(\value{x},-\value{y}){\makebox(0,\value{z})[l]{${#2}$}}%
\end{picture}}%

\def\basicwwswar[#1]{\WWSWAR{}{}{#100}}%

\newcommand{\wwswar}{\@ifnextchar[{\basicwwswar}{\basicwwswar[211]}}%

\def\basicWwswar[#1]#2{\WWSWAR{#2}{}{#100}}%

\newcommand{\Wwswar}{\@ifnextchar[{\basicWwswar}{\basicWwswar[211]}}%

\def\basicwwswaR[#1]#2{\WWSWAR{}{#2}{#100}}%

\newcommand{\wwswaR}{\@ifnextchar[{\basicwwswaR}{\basicwwswaR[211]}}%

\newcommand{\NNNEAR}[3]{\testdiagrammode%
\Y=#3%
\divide \Y by 2%
\Z=\Y%
\multiply \Z by 3%
\begin{picture}(0,0)%
\put(-\Y,-\Z){\line(1,3){#3}}%
\put(\Y,\Z){\nnnehead}%
\truex{100}\truez{600}%
\put(-\value{x},\value{x}){\makebox(0,\value{z})[r]{${#1}$}}%
\put(\value{x},-\value{z}){\makebox(0,\value{z})[l]{${#2}$}}%
\end{picture}}%

\def\basicnnnear[#1]{\NNNEAR{}{}{#100}}%

\newcommand{\nnnear}{\@ifnextchar[{\basicnnnear}{\basicnnnear[71]}}%

\def\basicNnnear[#1]#2{\NNNEAR{#2}{}{#100}}%

\newcommand{\Nnnear}{\@ifnextchar[{\basicNnnear}{\basicNnnear[71]}}%

\def\basicnnneaR[#1]#2{\NNNEAR{}{#2}{#100}}%

\newcommand{\nnneaR}{\@ifnextchar[{\basicnnneaR}{\basicnnneaR[71]}}%

\newcommand{\SSSWAR}[3]{\testdiagrammode%
\Y=#3%
\divide \Y by 2%
\Z=\Y%
\multiply \Z by 3%
\begin{picture}(0,0)%
\put(\Y,\Z){\line(-1,-3){#3}}%
\put(-\Y,-\Z){\ssswhead}%
\truex{100}\truez{600}%
\put(-\value{x},\value{x}){\makebox(0,\value{z})[r]{${#1}$}}%
\put(\value{x},-\value{z}){\makebox(0,\value{z})[l]{${#2}$}}%
\end{picture}}%

\def\basicssswar[#1]{\SSSWAR{}{}{#100}}%

\newcommand{\ssswar}{\@ifnextchar[{\basicssswar}{\basicssswar[71]}}%

\def\basicSsswar[#1]#2{\SSSWAR{#2}{}{#100}}%

\newcommand{\Ssswar}{\@ifnextchar[{\basicSsswar}{\basicSsswar[71]}}%

\def\basicssswaR[#1]#2{\SSSWAR{}{#2}{#100}}%

\newcommand{\ssswaR}{\@ifnextchar[{\basicssswaR}{\basicssswaR[71]}}%

\newcommand{\SSSEAR}[3]{\testdiagrammode%
\Y=#3%
\divide \Y by 2%
\Z=\Y%
\multiply \Z by 3%
\begin{picture}(0,0)%
\put(-\Y,\Z){\line(1,-3){#3}}%
\put(\Y,-\Z){\sssehead}%
\truex{200}\truez{600}%
\put(\value{x},\value{x}){\makebox(0,\value{z})[l]{${#1}$}}%
\put(-\value{x},-\value{z}){\makebox(0,\value{z})[r]{${#2}$}}%
\end{picture}}%

\def\basicsssear[#1]{\SSSEAR{}{}{#100}}%

\newcommand{\sssear}{\@ifnextchar[{\basicsssear}{\basicsssear[71]}}%

\def\basicSssear[#1]#2{\SSSEAR{#2}{}{#100}}%

\newcommand{\Sssear}{\@ifnextchar[{\basicSssear}{\basicSssear[71]}}%

\def\basicssseaR[#1]#2{\SSSEAR{}{#2}{#100}}%

\newcommand{\ssseaR}{\@ifnextchar[{\basicssseaR}{\basicssseaR[71]}}%

\newcommand{\NNNWAR}[3]{\testdiagrammode%
\Y=#3%
\divide \Y by 2%
\Z=\Y%
\multiply \Z by 3%
\begin{picture}(0,0)%
\put(\Y,-\Z){\line(-1,3){#3}}%
\put(-\Y,\Z){\nnnwhead}%
\truex{200}\truez{600}%
\put(\value{x},\value{x}){\makebox(0,\value{z})[l]{${#1}$}}%
\put(-\value{x},-\value{z}){\makebox(0,\value{z})[r]{${#2}$}}%
\end{picture}}%

\def\basicnnnwar[#1]{\NNNWAR{}{}{#100}}%

\newcommand{\nnnwar}{\@ifnextchar[{\basicnnnwar}{\basicnnnwar[71]}}%

\def\basicNnnwar[#1]#2{\NNNWAR{#2}{}{#100}}%

\newcommand{\Nnnwar}{\@ifnextchar[{\basicNnnwar}{\basicNnnwar[71]}}%

\def\basicnnnwaR[#1]#2{\NNNWAR{}{#2}{#100}}%

\newcommand{\nnnwaR}{\@ifnextchar[{\basicnnnwaR}{\basicnnnwaR[71]}}%

\newcommand{\NEENEAR}[3]{\testdiagrammode%
\Y=#3%
\divide \Y by 2%
\Z=#3%
\divide \Z by 3%
\begin{picture}(0,0)%
\put(-\Y,-\Z){\line(3,2){#3}}%
\put(\Y,\Z){\neenehead}%
\truex{200}\truey{800}\truez{600}%
\put(-\value{x},\value{x}){\makebox(0,\value{z})[r]{${#1}$}}%
\put(\value{x},-\value{y}){\makebox(0,\value{z})[l]{${#2}$}}%
\end{picture}}%

\def\basicneenear[#1]{\NEENEAR{}{}{#100}}%

\newcommand{\neenear}{\@ifnextchar[{\basicneenear}{\basicneenear[215]}}%

\def\basicNeenear[#1]#2{\NEENEAR{#2}{}{#100}}%

\newcommand{\Neenear}{\@ifnextchar[{\basicNeenear}{\basicNeenear[215]}}%

\def\basicneeneaR[#1]#2{\NEENEAR{}{#2}{#100}}%

\newcommand{\neeneaR}{\@ifnextchar[{\basicneeneaR}{\basicneeneaR[215]}}%

\newcommand{\SEESEAR}[3]{\testdiagrammode%
\Y=#3%
\divide \Y by 2%
\Z=#3%
\divide \Z by 3%
\begin{picture}(0,0)%
\put(-\Y,\Z){\line(3,-2){#3}}%
\put(\Y,-\Z){\seesehead}%
\truex{200}\truey{800}\truez{600}%
\put(\value{x},\value{x}){\makebox(0,\value{z})[l]{${#1}$}}%
\put(-\value{x},-\value{y}){\makebox(0,\value{z})[r]{${#2}$}}%
\end{picture}}%

\def\basicseesear[#1]{\SEESEAR{}{}{#100}}%

\newcommand{\seesear}{\@ifnextchar[{\basicseesear}{\basicseesear[215]}}%

\def\basicSeesear[#1]#2{\SEESEAR{#2}{}{#100}}%

\newcommand{\Seesear}{\@ifnextchar[{\basicSeesear}{\basicSeesear[215]}}%

\def\basicseeseaR[#1]#2{\SEESEAR{}{#2}{#100}}%

\newcommand{\seeseaR}{\@ifnextchar[{\basicseeseaR}{\basicseeseaR[215]}}%

\newcommand{\NWWNWAR}[3]{\testdiagrammode%
\Y=#3%
\divide \Y by 2%
\Z=#3%
\divide \Z by 3%
\begin{picture}(0,0)%
\put(\Y,-\Z){\line(-3,2){#3}}%
\put(-\Y,\Z){\nwwnwhead}%
\truex{200}\truey{800}\truez{600}%
\put(\value{x},\value{x}){\makebox(0,\value{z})[l]{${#1}$}}%
\put(-\value{x},-\value{y}){\makebox(0,\value{z})[r]{${#2}$}}%
\end{picture}}%

\def\basicnwwnwar[#1]{\NWWNWAR{}{}{#100}}%

\newcommand{\nwwnwar}{\@ifnextchar[{\basicnwwnwar}{\basicnwwnwar[215]}}%

\def\basicNwwnwar[#1]#2{\NWWNWAR{#2}{}{#100}}%

\newcommand{\Nwwnwar}{\@ifnextchar[{\basicNwwnwar}{\basicNwwnwar[215]}}%

\def\basicnwwnwaR[#1]#2{\NWWNWAR{}{#2}{#100}}%

\newcommand{\nwwnwaR}{\@ifnextchar[{\basicnwwnwaR}{\basicnwwnwaR[215]}}%

\newcommand{\SWWSWAR}[3]{\testdiagrammode%
\Y=#3%
\divide \Y by 2%
\Z=#3%
\divide \Z by 3%
\begin{picture}(0,0)%
\put(\Y,\Z){\line(-3,-2){#3}}%
\put(-\Y,-\Z){\swwswhead}%
\truex{200}\truey{800}\truez{600}%
\put(-\value{x},\value{x}){\makebox(0,\value{z})[r]{${#1}$}}%
\put(\value{x},-\value{y}){\makebox(0,\value{z})[l]{${#2}$}}%
\end{picture}}%

\def\basicswwswar[#1]{\SWWSWAR{}{}{#100}}%

\newcommand{\swwswar}{\@ifnextchar[{\basicswwswar}{\basicswwswar[215]}}%

\def\basicSwwswar[#1]#2{\SWWSWAR{#2}{}{#100}}%

\newcommand{\Swwswar}{\@ifnextchar[{\basicSwwswar}{\basicSwwswar[215]}}%

\def\basicswwswaR[#1]#2{\SWWSWAR{}{#2}{#100}}%

\newcommand{\swwswaR}{\@ifnextchar[{\basicswwswaR}{\basicswwswaR[215]}}%

\newcommand{\NENNEAR}[3]{\testdiagrammode%
\Y=#3%
\divide \Y by 2%
\Z=#3%
\multiply \Z by 3%
\divide \Z by 4%
\begin{picture}(0,0)%
\put(-\Y,-\Z){\line(2,3){#3}}%
\put(\Y,\Z){\nennehead}%
\truex{100}\truez{600}%
\put(-\value{x},\value{x}){\makebox(0,\value{z})[r]{${#1}$}}%
\put(\value{x},-\value{z}){\makebox(0,\value{z})[l]{${#2}$}}%
\end{picture}}%

\def\basicnennear[#1]{\NENNEAR{}{}{#100}}%

\newcommand{\nennear}{\@ifnextchar[{\basicnennear}{\basicnennear[143]}}%

\def\basicNennear[#1]#2{\NENNEAR{#2}{}{#100}}%

\newcommand{\Nennear}{\@ifnextchar[{\basicNennear}{\basicNennear[143]}}%

\def\basicnenneaR[#1]#2{\NENNEAR{}{#2}{#100}}%

\newcommand{\nenneaR}{\@ifnextchar[{\basicnenneaR}{\basicnenneaR[143]}}%

\newcommand{\SWSSWAR}[3]{\testdiagrammode%
\Y=#3%
\divide \Y by 2%
\Z=#3%
\multiply \Z by 3%
\divide \Z by 4%
\begin{picture}(0,0)%
\put(\Y,\Z){\line(-2,-3){#3}}%
\put(-\Y,-\Z){\swsswhead}%
\truex{100}\truez{600}%
\put(-\value{x},\value{x}){\makebox(0,\value{z})[r]{${#1}$}}%
\put(\value{x},-\value{z}){\makebox(0,\value{z})[l]{${#2}$}}%
\end{picture}}%

\def\basicswsswar[#1]{\SWSSWAR{}{}{#100}}%

\newcommand{\swsswar}{\@ifnextchar[{\basicswsswar}{\basicswsswar[143]}}%

\def\basicSwsswar[#1]#2{\SWSSWAR{#2}{}{#100}}%

\newcommand{\Swsswar}{\@ifnextchar[{\basicSwsswar}{\basicSwsswar[143]}}%

\def\basicswsswaR[#1]#2{\SWSSWAR{}{#2}{#100}}%

\newcommand{\swsswaR}{\@ifnextchar[{\basicswsswaR}{\basicswsswaR[143]}}%

\newcommand{\SESSEAR}[3]{\testdiagrammode%
\Y=#3%
\divide \Y by 2%
\Z=#3%
\multiply \Z by 3%
\divide \Z by 4%
\begin{picture}(0,0)%
\put(-\Y,\Z){\line(2,-3){#3}}%
\put(\Y,-\Z){\sessehead}%
\truex{200}\truez{600}%
\put(\value{x},\value{x}){\makebox(0,\value{z})[l]{${#1}$}}%
\put(-\value{x},-\value{z}){\makebox(0,\value{z})[r]{${#2}$}}%
\end{picture}}%

\def\basicsessear[#1]{\SESSEAR{}{}{#100}}%

\newcommand{\sessear}{\@ifnextchar[{\basicsessear}{\basicsessear[143]}}%

\def\basicSessear[#1]#2{\SESSEAR{#2}{}{#100}}%

\newcommand{\Sessear}{\@ifnextchar[{\basicSessear}{\basicSessear[143]}}%

\def\basicsesseaR[#1]#2{\SESSEAR{}{#2}{#100}}%

\newcommand{\sesseaR}{\@ifnextchar[{\basicsesseaR}{\basicsesseaR[143]}}%

\newcommand{\NWNNWAR}[3]{\testdiagrammode%
\Y=#3%
\divide \Y by 2%
\Z=#3%
\multiply \Z by 3%
\divide \Z by 4%
\begin{picture}(0,0)%
\put(\Y,-\Z){\line(-2,3){#3}}%
\put(-\Y,\Z){\nwnnwhead}%
\truex{200}\truez{600}%
\put(\value{x},\value{x}){\makebox(0,\value{z})[l]{${#1}$}}%
\put(-\value{x},-\value{z}){\makebox(0,\value{z})[r]{${#2}$}}%
\end{picture}}%

\def\basicnwnnwar[#1]{\NWNNWAR{}{}{#100}}%

\newcommand{\nwnnwar}{\@ifnextchar[{\basicnwnnwar}{\basicnwnnwar[143]}}%

\def\basicNwnnwar[#1]#2{\NWNNWAR{#2}{}{#100}}%

\newcommand{\Nwnnwar}{\@ifnextchar[{\basicNwnnwar}{\basicNwnnwar[143]}}%

\def\basicnwnnwaR[#1]#2{\NWNNWAR{}{#2}{#100}}%

\newcommand{\nwnnwaR}{\@ifnextchar[{\basicnwnnwaR}{\basicnwnnwaR[143]}}%

\newcommand{\Necurve}[2]%
{\testdiagrammode\begin{picture}(0,0)%
\truex{1300}\truey{2000}\truez{200}%
\put(0,\value{x}){\oval(#200,\value{y})[t]}%
\put(0,\value{x}){\makebox(0,0){\begin{picture}(#200,0)%
\put(#200,0){\line(0,-1){\value{z}}}%
\put(#200,-\value{z}){\shead}%
\put(0,0){\line(0,-1){\value{z}}}\end{picture}}}%
\truex{2500}%
\put(0,\value{x}){\makebox(0,0)[b]{${#1}$}}%
\end{picture}}%

\def\basicnecurvar[#1]{\Necurve{}{#1}}

\newcommand{\necurvar}{\@ifnextchar[{\basicnecurvar}{\basicnecurvar[160]}}%

\def\basicNecurvar[#1]#2{\Necurve{#2}{#1}}%

\newcommand{\Necurvar}{\@ifnextchar[{\basicNecurvar}{\basicNecurvar[160]}}%

\newcommand{\Nwcurve}[2]%
{\testdiagrammode\begin{picture}(0,0)%
\truex{1300}\truey{2000}\truez{200}%
\put(0,\value{x}){\oval(#200,\value{y})[t]}%
\put(0,\value{x}){\makebox(0,0){\begin{picture}(#200,0)%
\put(#200,0){\line(0,-1){\value{z}}}%
\put(0,0){\line(0,-1){\value{z}}}%
\put(0,-\value{z}){\shead}%
\end{picture}}}%
\truex{2500}%
\put(0,\value{x}){\makebox(0,0)[b]{${#1}$}}%
\end{picture}}%

\def\basicnwcurvar[#1]{\Nwcurve{}{#1}}

\newcommand{\nwcurvar}{\@ifnextchar[{\basicnwcurvar}{\basicnwcurvar[160]}}%

\def\basicNwcurvar[#1]#2{\Nwcurve{#2}{#1}}%

\newcommand{\Nwcurvar}{\@ifnextchar[{\basicNwcurvar}{\basicNwcurvar[160]}}%

\newcommand{\Securve}[2]%
{\testdiagrammode\begin{picture}(0,0)%
\truex{1300}\truey{2000}\truez{200}%
\put(0,-\value{x}){\oval(#200,\value{y})[b]}%
\put(0,-\value{x}){\makebox(0,0){\begin{picture}(#200,0)%
\put(#200,0){\line(0,1){\value{z}}}%
\put(0,0){\line(0,1){\value{z}}}%
\put(#200,\value{z}){\nhead}%
\end{picture}}}%
\truex{2500}%
\put(0,-\value{x}){\makebox(0,0)[t]{${#1}$}}%
\end{picture}}%

\def\basicsecurvar[#1]{\Securve{}{#1}}

\newcommand{\securvar}{\@ifnextchar[{\basicsecurvar}{\basicsecurvar[160]}}%

\def\basicSecurvar[#1]#2{\Securve{#2}{#1}}%

\newcommand{\Securvar}{\@ifnextchar[{\basicSecurvar}{\basicSecurvar[160]}}%

\newcommand{\Swcurve}[2]%
{\testdiagrammode\begin{picture}(0,0)%
\truex{1300}\truey{2000}\truez{200}%
\put(0,-\value{x}){\oval(#200,\value{y})[b]}%
\put(0,-\value{x}){\makebox(0,0){\begin{picture}(#200,0)%
\put(#200,0){\line(0,1){\value{z}}}%
\put(0,0){\line(0,1){\value{z}}}%
\put(0,\value{z}){\nhead}%
\end{picture}}}%
\truex{2500}%
\put(0,-\value{x}){\makebox(0,0)[t]{${#1}$}}%
\end{picture}}%

\def\basicswcurvar[#1]{\Swcurve{}{#1}}

\newcommand{\swcurvar}{\@ifnextchar[{\basicswcurvar}{\basicswcurvar[160]}}%

\def\basicSwcurvar[#1]#2{\Swcurve{#2}{#1}}%

\newcommand{\Swcurvar}{\@ifnextchar[{\basicSwcurvar}{\basicSwcurvar[160]}}%

\newcommand{\Escurve}[2]%
{\testdiagrammode\begin{picture}(0,0)%
\truex{1400}\truey{2000}\truez{200}%
\put(\value{x},0){\oval(\value{y},#200)[r]}%
\put(\value{x},0){\makebox(0,0){\begin{picture}(0,#200)%
\put(0,0){\line(-1,0){\value{z}}}%
\put(0,#200){\line(-1,0){\value{z}}}%
\put(-\value{z},0){\whead}%
\end{picture}}}%
\truex{2500}%
\put(\value{x},0){\makebox(0,0)[l]{${#1}$}}%
\end{picture}}%

\def\basicescurvar[#1]{\Escurve{}{#1}}

\newcommand{\escurvar}{\@ifnextchar[{\basicescurvar}{\basicescurvar[160]}}%

\def\basicEscurvar[#1]#2{\Escurve{#2}{#1}}%

\newcommand{\Escurvar}{\@ifnextchar[{\basicEscurvar}{\basicEscurvar[160]}}%

\newcommand{\Encurve}[2]%
{\testdiagrammode\begin{picture}(0,0)%
\truex{1400}\truey{2000}\truez{200}%
\put(\value{x},0){\oval(\value{y},#200)[r]}%
\put(\value{x},0){\makebox(0,0){\begin{picture}(0,#200)%
\put(0,0){\line(-1,0){\value{z}}}%
\put(0,#200){\line(-1,0){\value{z}}}%
\put(-\value{z},#200){\whead}%
\end{picture}}}%
\truex{2500}%
\put(\value{x},0){\makebox(0,0)[l]{${#1}$}}%
\end{picture}}%

\def\basicencurvar[#1]{\Encurve{}{#1}}

\newcommand{\encurvar}{\@ifnextchar[{\basicencurvar}{\basicencurvar[160]}}%

\def\basicEncurvar[#1]#2{\Encurve{#2}{#1}}%

\newcommand{\Encurvar}{\@ifnextchar[{\basicEncurvar}{\basicEncurvar[160]}}%

\newcommand{\Wscurve}[2]%
{\testdiagrammode\begin{picture}(0,0)%
\truex{1300}\truey{2000}\truez{200}%
\put(-\value{x},0){\oval(\value{y},#200)[l]}%
\put(-\value{x},0){\makebox(0,0){\begin{picture}(0,#200)%
\put(0,0){\line(1,0){\value{z}}}%
\put(0,#200){\line(1,0){\value{z}}}%
\put(\value{z},0){\ehead}%
\end{picture}}}%
\truex{2400}%
\put(-\value{x},0){\makebox(0,0)[r]{${#1}$}}%
\end{picture}}%

\def\basicwscurvar[#1]{\Wscurve{}{#1}}

\newcommand{\wscurvar}{\@ifnextchar[{\basicwscurvar}{\basicwscurvar[160]}}%

\def\basicWscurvar[#1]#2{\Wscurve{#2}{#1}}%

\newcommand{\Wscurvar}{\@ifnextchar[{\basicWscurvar}{\basicWscurvar[160]}}%

\newcommand{\Wncurve}[2]%
{\testdiagrammode\begin{picture}(0,0)%
\truex{1300}\truey{2000}\truez{200}%
\put(-\value{x},0){\oval(\value{y},#200)[l]}%
\put(-\value{x},0){\makebox(0,0){\begin{picture}(0,#200)%
\put(0,0){\line(1,0){\value{z}}}%
\put(\value{z},#200){\ehead}%
\put(0,#200){\line(1,0){\value{z}}}%
\end{picture}}}%
\truex{2400}%
\put(-\value{x},0){\makebox(0,0)[r]{${#1}$}}%
\end{picture}}%

\def\basicwncurvar[#1]{\Wncurve{}{#1}}

\newcommand{\wncurvar}{\@ifnextchar[{\basicwncurvar}{\basicwncurvar[160]}}%

\def\basicWncurvar[#1]#2{\Wncurve{#2}{#1}}%

\newcommand{\Wncurvar}{\@ifnextchar[{\basicWncurvar}{\basicWncurvar[160]}}%

\catcode`\@=12

\begin{document}

\vbox{\vspace{6mm}}
\begin{center}{ {\large \bf Quantum Observables Algebras and Abstract Differential
Geometry: The Topos-Theoretic Dynamics of Diagrams of Commutative
Algebraic Localizations }
\\[7mm]
ELIAS ZAFIRIS\\ {\it University of Athens \\ Institute of
Mathematics
\\ Panepistimioupolis, 15784 Athens
\\ Greece  \\} \vspace{2mm} }\end{center} \vspace{8mm}

\footnotetext{{\it E-mail }:{ \bf ezafiris@math.uoa.gr}}

\begin{abstract}
We construct a sheaf-theoretic representation of quantum
observables algebras over a base category equipped with a
Grothendieck topology, consisting of epimorphic families of
commutative observables algebras, playing the role of local
arithmetics in measurement situations. This construction makes
possible the adaptation of the methodology of Abstract
Differential Geometry (ADG), {\it $\grave{a}$ la Mallios}, in a
topos-theoretic environment, and hence, the extension of the
``mechanism of differentials" in the quantum regime. The process
of gluing information, within diagrams of commutative algebraic
localizations, generates dynamics, involving the transition from
the classical to the quantum regime, formulated cohomologically in
terms of a functorial quantum connection, and subsequently,
detected via the associated curvature of that connection.
\end{abstract}

\newpage

\section{PROLOGUE}

The working understanding of contemporary physical theories is
grounded on the notion of observables. Observables are associated
with physical quantities that, in principle, can be measured. In
this sense physical systems are completely described by the
collection of all observed data determined by adequate devices in
appropriate measurement situations. The mathematical formalization
of this procedure relies on the idea of expressing the observables
by functions corresponding to measuring devices. Usually it is
also stipulated that quantities admissible as measured results
must be real numbers. It is a common belief that the resort to
real numbers has the advantage of making our empirical access
secure. Hence the underlying assumption on the basis of physical
theories postulates that our form of observation is expressed by
real number representability, and subsequently, observables are
modelled by real-valued functions corresponding to measuring
devices. At a further stage of development of this idea, two
further assumptions are imposed on the structure of observables:
the first of them specifies the algebraic nature of the set of all
observables used for the description of a physical system, by
assuming the structure of a commutative unital algebra $\mathcal
A$ over the real numbers. The second assumption restricts the
content of the set of real-valued functions corresponding to
physical observables to those that admit a mathematical
characterization as measurable, continuous or smooth. Thus,
depending on the means of description of a physical system,
observables are modelled by $\mathcal R$-algebras of measurable,
continuous or smooth functions corresponding to suitably
specifiable in each case measurement environments. Usually the
smoothness assumption is postulated because it is desirable to
consider derivatives of observables and effectively set-up a
kinematical framework of description in terms of differential
equations. Moreover, since we have initially assumed that
real-number representability constitutes our form of observation
in terms of the readings of measuring devices, the set of all
$\mathcal R$-algebra homomorphisms $\mathcal A \rightarrow
\mathcal R$, assigning to each observable in $\mathcal A$, the
reading of a measuring device in $\mathcal R$, encapsulates all
the information collected about a physical system in measurement
situations in terms of algebras of real-valued observables.
Mathematically,  the set of all $\mathcal R$-algebra homomorphisms
$\mathcal A \rightarrow \mathcal R$ is identified as the $\mathcal
R$-spectrum of the unital commutative algebra of observables
$\mathcal A$. The physical semantics of this connotation denotes
the set that can be $\mathcal R$-observed by means of this
algebra. It is well known that, in case $\mathcal A$ stands for a
smooth algebra of real-valued observables, $\mathcal R$-algebra
homomorphisms $\mathcal A \rightarrow \mathcal R$ can be
legitimately identified with the points of a space that can be
observed by means of $\mathcal A$, namely the points of a
differential manifold that, in turn, denote the states of the
observed physical system. From this perspective state spaces in
general are derivative notions referring to sets of points
$\mathcal R$-observed, through the lenses of corresponding
algebras of observables.

An equally important notion referring to the conceptualization of
physical observables is related with the issue of localization.
Usually the operationalizations of  measurement situations assume
their existence locally and the underlying assumption is that the
information gathered about observables in different measurement
situations can be collated together by appropriate means. The
notion of local requires the specification of a topology on an
assumed underlying measurement space over which algebras of
observables may be localized. The net effect of this localization
procedure of algebras of observables together with the requirement
of compatible information collation along localizations are
formalized by the notion of sheaf. A sheaf of commutative unital
$\mathcal R$-algebras of observables incorporates exactly the
conditions for the transition from locally collected observable
data to globally defined ones. In case of smooth observational
procedures the notion of a sheaf of smooth $\mathcal R$-algebras
of observables $\mathcal A$, means that locally $\mathcal A$ is
like the $\mathcal R$-algebra $\mathcal {C^\infty}({{\mathcal
R}^n})$ of infinitely differentiable functions on ${{\mathcal
R}^n}$.

The interpretative power of this modeling scheme, based on sheaves
of algebras of observables, has been recently vastly enhanced by
the development of Abstract Differential Geometry (ADG) {\it
$\grave{a}$ la Mallios} [1-8], which generalizes the differential
geometric mechanism of smooth manifolds. Remarkably, it shows that
most of the usual differential geometric constructions can be
carried out by purely algebraic means without any use of any sort
of $C^\infty$-smoothness or any of the conventional calculus that
goes with it. This conclusion is important because it permits the
legitimate use of any appropriate $\mathcal R$-algebra sheaf of
observables suited to a measurement situation, even Rosinger's
singular algebra sheaf of generalized functions [7, 8], without
loosing the differential mechanism, prior believed to be solely
associated with smooth manifold state-spaces. Most significantly,
ADG has made us realize that the differential geometric mechanism
in its abstract algebraic sheaf theoretical formulation expressing
from a physical viewpoint the kinematics and dynamics of
information propagation through observables, is independent of the
localization method employed for the extraction and subsequent
coordinatization of its content. Thus, algebra sheaves of smooth
real-valued functions together with their associated by
measurement manifold $\mathcal R$-spectrums are by no means unique
coordinatizations of the universal physical mechanism of
qualitative information propagation through observables.

The major foundational difference between classical and quantum
physical systems from the perspective of the modeling scheme by
observables is a consequence of a single principle that can be
termed principle of simultaneous observability. According to this,
in the classical description of physical systems all their
observables are theoretically compatible, or else, they can be
simultaneously specified in a single local measurement context. On
the other side, the quantum description of physical systems is
based on the assertion of incompatibility of all theoretical
observables in a single local measurement context, and as a
consequence quantum-theoretically the simultaneous specification
of all observables is not possible. The conceptual roots of the
violation of the principle of simultaneous observability in the
quantum regime is tied with Heisenberg's uncertainty principle and
Bohr's principle of complementarity of physical descriptions
[9-13]. A natural question that arises in this setting is whether
one could express algebras of quantum observables in terms of
structured families of local commutative algebras of classical
observables capable of carrying all the information encoded in the
former. Of course, the notion of local has to be carefully
redesigned in this formulation, as it will become clear at a later
stage. From a category-theoretic standpoint, the transition from a
classical to a quantum description can be made simply equivalent
to a transition from a category of commutative algebra sheaves of
observables to a category of diagrams of commutative algebra
sheaves of observables. The advantage of this formulation  in
comparison to global non-commutative axiomatizations of operator
algebras of quantum observables is twofold: firstly, it makes
transparent the construction of an algebra of quantum observables
from the interconnection  of locally defined commutative algebras
of classically conceived observables, and secondly, it makes
possible the extension of the differential geometric mechanism of
ADG in the quantum regime, thus, in effect, of the classical one,
as well.

According to the above line of reasoning, we are guided in
expressing a globally non-commutative object, like an algebra of
quantum observables, in terms of structured families of
commutative algebras of observables, which have to satisfy certain
compatibility relations, and also, a closure constraint. Hence,
commutative algebras of real-valued observables are used locally,
in an appropriate manner, accomplishing the task of providing
partial congruent relations with globally non-commutative
observable algebras, the internal structure of which, is being
effectively expressed in terms of the interconnecting machinery
binding the local objects together. This point of view stresses
the contextual character of quantum theory and establishes a
relation with commutative algebras associated with typical
measurement situations. In order to proceed a suitable
mathematical language has to be used. The criterion for choosing
an appropriate language is rather emphasis in the form of the
structures and the universality of the constructions involved. The
ideal candidate for this purpose is provided by category theory
[14-20]. Subsequently, we will see that sheaf theory [21-23],
(yet, see also [1]), is the appropriate mathematical vehicle to
carry out the program implied by the proposed methodology.

The concept of sheaf expresses essentially gluing conditions,
namely, the way by which local data can be collated into global
ones. It is the  mathematical abstraction suited to formalizing
the relations between covering systems and properties, and,
furthermore, provides the means for studying the global
consequences/information from locally defined properties. The
notion of local is characterized ``geometrically", viz. by using a
topology (in the general case a Grothendieck topology on a
category), the axioms of which express closure conditions on the
collection of covers. Essentially, a map which assigns a set to
each object of a topology is called a sheaf if the map is locally
defined, or else the value of the map on an object can be uniquely
obtained from its values on any cover of that object.
Categorically speaking, besides mapping each object to a set, a
sheaf maps each arrow in the topology to a restriction function in
the opposite direction. We stress the point that the transition
from locally defined properties to global consequences/information
happens via a compatible family of elements over a covering system
of the global object. A covering system of the global  object can
be viewed as providing a decomposition of that object into local
objects. The sheaf assigns a set to each element of that system,
or else, to each local piece of the original object. A choice of
elements from these sets, one for each piece, forms a compatible
family if the choice respects the mappings by the restriction
functions and if the elements chosen agree whenever two pieces of
the covering system overlap. If such a locally compatible choice
induces a unique choice for the object being covered, viz. a
global choice, then the condition for being a sheaf is satisfied.
We note that in general, there will be more locally defined or
partial choices than globally defined ones, since not all partial
choices need be extendible to global ones, but a compatible family
of partial choices uniquely extends to a global one, or in other
words, any presheaf uniquely defines a sheaf; thus, see e.g. A.
Mallios [24], for an ``information/choice"-theoretic formulation
of the same.

In the following sections we shall see that a quantum observables
algebra can be understood as a sheaf for a suitable Grothendieck
topology on the category of commutative subalgebras of it. The
idea is based on extension and elaboration of previous works of
the author, communicated, both conceptually and technically, in
the literature [25-29]. In all these papers, the focus has been
shifted from point-set to topological localization models of
quantum algebraic structures, that effectively, induce a
transition in the semantics of observables from a set-theoretic to
a sheaf-theoretic one. The primary physical motivation behind this
strategy, has been generated by investigating the possibility of
mathematically implementing localization processes referring to
physical observation, concerning in particular quantum phenomena,
that is not necessarily based on the existence of an underlying
structure of points on the real line.

It is also instructive to mention that, contextual topos
theoretical approaches to quantum structures have been also
developed, from a different viewpoint in [30-34]. Moreover, the
necessity of implementation of a sheaf-theoretic framework for
overcoming the problems of singularities has been thoroughly
discussed recently in [35]. Finally, the central thesis of [36],
according to which, quantum physics at a fundamental level may
itself be realized as a species of quantum computation, is
strongly embraced by the author.

\section{CATEGORIES OF OBSERVABLES}
Category theory provides a general apparatus for dealing with
mathematical structures and their mutual relations and
transformations.  The basic categorical principles that we adopt
in the subsequent analysis are summarized as follows:

[i]. To each kind of mathematical structure, there corresponds a
{\bf category} whose objects  have that structure, and whose
morphisms preserve it.

[ii].  To any natural construction on structures of one kind,
yielding structures of another kind, there corresponds a {\bf
functor} from the category of the first kind to the category of
the second.

[iii]. To each translation between constructions of the above form
there corresponds a {\bf natural transformation}.

\subsection{Classical and Quantum Observables Structures}

A {\bf Classical Observables structure} is a small category,
denoted by $\mathcal A_C$, and called the {\it category of
Classical Observables algebras}, or of {\it classical
arithmetics}. Its objects are commutative unital $\mathcal
R$-algebras of observables, and its arrows are unit preserving
$\mathcal R$-algebras morphisms. Thus, $\mathcal A_C$ is a
subcategory of that one of commutative unital $\mathcal
R$-algebras and unit preserving $\mathcal R$-algebra morphisms.

\paragraph{Examples:}
[i] We consider the Boolean algebra of events $B$ associated with
the measurement of a physical system. In any experiment performed
by an observer, the propositions that can be made concerning a
physical quantity are of the type, which asserts that, the value
of the physical quantity lies in some Borel set of the real
numbers. The proposition that the value of a physical quantity
lies in a Borel set of the real line corresponds to an event in
the ordered event structure $B$, as it is apprehended by an
observer. Thus we obtain a mapping $A_C : Bor(\mathbf R) \to B$
from the Borel sets of the real line to the event structure which
captures precisely the notion of observable. Most importantly the
above mapping is required to be a homomorphism. In this
representation $Bor(\mathbf R)$ stands for the algebra of events
associated with a measurement device interacting with a physical
system. The homomorphism assigns to every empirical event in
$Bor(\mathbf R)$ a proposition or event in $B$, that states, a
measurement fact about the physical system interacting with the
measuring apparatus. According to  Stone's representation theorem
for Boolean algebras, it is legitimate to replace Boolean algebras
by fields of subsets of a measurement space.  Hence we may replace
the Boolean algebra $B$ by its set-theoretical representation
$[{\mathbf \Sigma}, B_\Sigma]$, consisting of a  measurement space
$\mathbf \Sigma$ and its field of subsets $B_\Sigma$. Then
observables $\xi$ are in injective correspondence with inverses of
random variables $f:{\mathbf \Sigma} \rightarrow \mathcal R$. In
this setting we may also identify a classical observables algebra
with the $\mathcal R$-algebra of measurable functions defined on
the measurement space ${\mathbf \Sigma}$.

[ii] We assume that the measurement space $[{\mathbf \Sigma},
B_\Sigma]$ above, is identified with the $\sigma$-algebra of Borel
subsets of a topological space $X$. In this setting we could
consider as a classical observables algebra the $\mathcal
R$-algebra of continuous functions defined on $X$.

[iii] We assume that the topological space $X$ above is
paracompact and Hausdorff, and furthermore that can be endowed
with the structure of a differential manifold. In this setting we
could consider as a classical observables algebra the $\mathcal
R$-algebra of smooth functions on $X$.

A {\bf Quantum Observables structure} is a small category, denoted
by $\mathcal A_Q$, and called the {\it category of Quantum
Observables algebras}, or of {\it Quantum arithmetics}. Its
objects are thus unital $\mathcal R$-algebras of observables, and
its arrows are unit preserving $\mathcal R$-algebras morphisms.
Hence, in other words, $\mathcal A_Q$ is a subcategory of the
category of unital $\mathcal R$-algebras and unit preserving
algebra morphisms.

\paragraph{Examples:}
[i] We consider the  algebra of events $L$ associated with the
measurement of a quantum system. In this case $L$ is not a Boolean
algebra, but an orthomodular $\sigma$-orthoposet. A quantum
observable $\Xi$ is defined to be an algebra morphism from the
Borel algebra of the real line $Bor(\mathbf{R})$ to the quantum
event algebra $L$ [13, 25-26, 28]. $$ \Xi : Bor(\mathbf{R}) \to
L$$ such that: [i] $ \Xi(\emptyset)=0, \Xi(\mathbf R)=1$, [ii]  $E
\bigcap F=\emptyset \Rightarrow \Xi(E) \perp \Xi(F)$, for  $E, F
\in Bor(\mathbf R)$, [iii] $\Xi({\bigcup}_n E_n)={\bigvee}_n
\Xi(E_n)$,  where $E_1, E_2, \ldots$  sequence of mutually
disjoint Borel sets of the real line. Addition and multiplication
on $\mathcal R$ induce on the set of quantum observables the
structure of a partial commutative algebra over $\mathcal R$. In
most of the cases the stronger assumption of a non-commutative
algebra of quantum observables is adopted.

[ii] If $L$ is isomorphic with the orthocomplemented lattice of
orthogonal projections on a Hilbert space, then it follows from
von Neumann's spectral theorem that the quantum observables are in
1-1 correspondence with the hypermaximal Hermitian operators on
the Hilbert space.

[iii] An algebra of quantum observables can be made isomorphic to
the partial algebra of Hermitian elements in a $C^*$-algebra.

The crucial observation that the development of this paper will be
based on, has to do with the fact that a globally non-commutative
or partial algebra of quantum observables determines an underlying
diagram of commutative subalgebras. Then each commutative
subalgebra can be locally identified, in a sense that will be made
clear later, with an algebra of classical observables. Thus the
information that is contained in an algebra of quantum observabes
can be recovered by a gluing construction referring to its
commutative subalgebras. This construction is also capable of
extending the differential geometric mechanism to the regime of
quantum systems via {\it ADG sheaf-theoretical methodology} and
collating information appropriately.

\subsection{The Notion of Differential Triad}
A Differential Triad is a concept introduced by A. Mallios in an
axiomatic approach to Differential Geometry [1]. The notion of
Differential Triad replaces the assumptions on the local structure
of a topological space $X$ for its specification as a manifold,
namely charts and atlases, with assumptions on the existence of a
derivative (``flat connection") on an arbitrary sheaf of algebras
on $X$, playing the equivalent role of the structure sheaf of
germs of smooth functions on $X$. The major novelty of this notion
relies on the fact that any sheaf of algebras may be regarded as
the structural sheaf of a differential triad capable of providing
a differential geometric mechanism, independent thus of any
manifold concept, analogous, however, to the one supported by
smooth manifolds.

The significance of the notion of Differential Triad for the
purposes of the present work can be made clear in a procedure
consisting of two levels:

The first level considers the localization of a commutative unital
$\mathcal R$-algebra of observables over a topological measurement
space $X$. The general methodology of localization, by means of an
arbitrary topological commutative algebra has been discussed
extensively in [24]. The localization procedure provides a sheaf
of unital, commutative $\mathcal R$-algebras of observables over
$X$. Having at our disposal this localized structure we may set up
a differential triad associated with the sheaf of commutative
algebras of observables $ {A}_C$ as follows: Let $\Omega$ be an $
{A}_C$-module, that is, $\Omega$ stands for a sheaf of $\mathcal
R$-vector spaces over $X$, such that $\Omega(U)$ is an $ {A}_C$(U)
module, for every $U$ in the topology $\tau_X$ of $X$. Besides,
let $\vartheta:=(\vartheta_U): {A}_C \rightarrow \Omega$ be a
sheaf morphism. Then the triplet $ \Delta=({A}_C, \vartheta,
\Omega)$ constitutes a differential triad, if it satisfies the
following conditions:

[i] $\vartheta$ is $\mathcal R$-linear, and

[ii] $\vartheta$ satisfies the Leibniz rule: for every pair
$(\xi_1, \xi_2)$ in ${A}_C {\times}_X {A}_C$ it holds that
$$\vartheta(\xi_1 \cdot \xi_2)=\xi_1 \cdot \vartheta(\xi_2)+
{\xi_2} \cdot \vartheta(\xi_1)$$

In this manner for every localized commutative unital $\mathcal
R$-algebra sheaf of observables suited to a measurement situation
we may associate a differential triad $\Delta=({A}_C, \vartheta,
\Omega)$ as above, that is capable of expressing according to ADG
a generalized differential geometric mechanism referred to the
propagation of information encoded in the sheaf ${\mathbf A}_C$,
that in turn, instantiates a coordinatized arithmetic suited for
the study of a physical system associated with a measurement
environment. It is instructive to emphasize here that classically
speaking, viz. for a classical physical system, all the
observables are theoretically compatible, or simultaneously
detectable, thus a single differential triad is enough for the
complete determination of the former mechanism.

The second level of the proposed scheme has the purpose of
extending the differential geometric mechanism to globally
non-commutative algebras of quantum observables. We may remind
that according to the principle of non-simultaneous observability
in the quantum regime, as above, the observables of quantum
systems are not compatible. Thus, a single differential triad
associated with a commutative sector of an algebra of quantum
observables is not possible to encode the totality of the
information required for the set-up of a generalized differential
geometric mechanism in the quantum case. What is needed is a
procedure of gluing together differential triads attached to local
commutative sectors. In this case the notion of local is
distinguished from the classical case and is naturally provided by
the definition of an appropriate Grothendieck topology over the
opposite category of commutative subalgebras of a quantum algebra
of observables. In this perspective an algebra of quantum
observables, or quantum arithmetic, can be made isomorphic with a
sheaf of locally commutative algebras of observables for this
Grothendieck topology. Thus the differential geometric mechanism,
following ADG, can be now applied locally in the quantum regime,
as well, by referring to the aforementioned sheaf of locally
commutative algebras of observables. In the sequel, it will become
clear that the transition to the quantum regime involves
considering diagrams of differential triads attached to
commutative subalgebras of an algebra of quantum observables,
together with a generalized conception of locality in the
Grothendieck sense, that permits collation of local information in
a sheaf-theoretic manner among these diagrams.

Since a quantum algebra of observables could be theoretically
built up from diagrams of commutative algebras of observables,
each one of them carrying a differential triad, it is necessary to
specify their morphisms in a category-theoretic language. Let us
consider that $\Delta_X=({A^X}_C, \vartheta_X, \Omega_X)$,
$\Delta_Y=({A^Y}_C, \vartheta_Y, \Omega_Y)$ are differential
triads associated with measurement situations that take place over
the topological spaces $X$, $Y$ respectively. A morphism from
$\Delta_X$ to $\Delta_Y$ is a triplet $(z,z_{A_C}, z_\Omega)$ such
that:

[i]: $z: X \rightarrow Y$ is a continuous map,

[ii]: $z_{A_C}: {A^Y}_C \rightarrow {z}_* ({A^X}_C)$ is a unit
preserving morphism of classical sheaves of ${\mathcal
R}$-algebras over Y, where ${z}_*: Sh_X \rightarrow Sh_Y$ denotes
the push-out functor,

[iii]: $z_\Omega: \Omega_Y \rightarrow {z}_* (\Omega_X)$ is a
morphism of sheaves of  ${\mathcal R}$-vector spaces over Y, such
that  $z_\Omega(\xi \omega)= z_{A_C}(\xi) z_\Omega(\omega)$
$\forall (\xi, \omega)$ in ${\mathbf A^Y}_C \times_Y \Omega_Y$,

[iv]: the diagram below, denoting push-out operations of
differential triads, commutes;

\begin{diagram}
¤{{A^Y}_C}   ¤\Ear {z_{A_C}} ¤{{z}_* ({A^X}_C)} ¤¤ ¤\Sar
{\vartheta_Y} ¤ ¤\saR  {{z}_* (\vartheta_X)}¤¤ ¤ {\Omega_Y}¤\Ear
{z_\Omega}¤{{z}_* (\Omega_X)}¤¤
\end{diagram}

In the sequel, the category of differential triads associated with
the subcategory of commutative algebras of a quantum algebra of
observables will be used only when we discuss the extension of the
differential geometric mechanism in the quantum regime. The
description of localization of a quantum algebra of observables
with respect to an appropriate Grothendieck topology on the
opposite subcategory of its commutatives algebras will be based
solely on the definitions provided in Section 2.1 for reasons of
simplicity in the exposition of the method. Of course, it is
obvious that a commutative subalgebra of a quantum observables
algebra once localized itself over a measurement topological space
$X$ becomes a sheaf. Hence, as we shall see in detail in what
follows, a differential triad can be appropriately associated with
it.

\section{FUNCTOR OF POINTS OF A QUANTUM OBSERVABLES ALGEBRA} The
development of the ideas contained in the proposed scheme are
based on the notion of the functor of points of a quantum
observables algebra, so it is worthwhile to explicate its meaning
in detail. The ideology behind this notion has its roots in the
work of Grothendieck in algebraic geometry. If we consider the
opposite of the category of algebras of quantum observables, that
is, the category with the same objects but with arrows reversed
${{\mathcal A}_Q}^{op}$, each object in that context can be
thought of as the locus of a quantum observables algebra, or else
it carries the connotation of space. The crucial observation is
that, any such space is determined, up to canonical isomorphism,
if we know all morphisms into this locus from any other locus in
the category. For instance, the set of morphisms from the
one-point locus to $A_Q$ in the categorial context of ${{\mathcal
A}_Q}^{op}$ determines the set of points of the locus $A_Q$. The
philosophy behind this approach amounts to considering any
morphism in ${{\mathcal A}_Q}^{op}$ with target the locus $A_Q$ as
a generalized point of $A_Q$. For our purposes we consider the
description of a locus $A_Q$ in terms of all possible morphisms
from all other objects of ${{\mathcal A}_Q}^{op}$ as redundant.
For this reason we may restrict the generalized points of $A_Q$ to
all those morphisms in ${{\mathcal A}_Q}^{op}$ having as domains
spaces corresponding to commutative subablgebras of a quantum
observables algebra. Variation of generalized points over all
domain-objects of the subcategory of ${{\mathcal A}_Q}^{op}$
consisting of commutative algebras of observables produces the
functor of points of $A_Q$ restricted to the subcategory of
commutative objects, identified, in what follows, with ${{\mathcal
A}_C}^{op}$. This functor of points of $A_Q$ is made then an
object in the category of presheaves ${{\bf Sets}^{{\mathcal
A_C}^{op}}}$, representing a quantum observables algebra -(in the
sequel for simplicity we talk of an algebra and its associated
locus tautologically)- in the environment of the topos of
presheaves over the category of its commutative subalgebras. This
methodology will prove to be successful if it could be possible to
establish an isomorphic representation of $A_Q$ in terms of its
generalized points $A_C \rightarrow A_Q$, considered as morphisms
in the same category, collated together by sheaf-theoretical
means.

From a physical point of view, the domains of generalized points
of $A_Q$ specify precisely the kind of loci of variation that are
used for individuation of observable events in the physical
continuum in a quantum measurement situation, accomplishing an
instantiation of  Bohr's conception of a {\it phenomenon}, as
referring exclusively to observations obtained under specific
circumstances that constitute a physical descriptive context.
Thus, the methodological underpinning of the introduction of
generalized points $A_C \rightarrow A_Q$ adapt Bohr's concept of
phenomenon as a referent of the assignment of an observable
quantity to a system, in the context of a commutative domain
considered appropriately as a local environment of measurement. In
this sense generalized points  play the equivalent role of {\it
generalized reference frames}, such that reference to concrete
events of the specified kind can be made possible only with
respect to the former.  In the trivial case the only locus is a
point serving as a unique idealized measure of localization, and
moreover, the only kind of reference frame is the one attached to
a point. This kind of reference frames are used in classical
physics, but prove to be insufficient for handling information
related with quantum measurement situations due to the principle
of non-simultaneous observability explicated previously. Hence,
generalized points $A_C \rightarrow A_Q$ constitute reference
frames only in a local sense by means of a {\it Grothendieck
topology}, to be introduced at a latter stage, and information
collected in different or overlapping commutative local domains
$A_C$ can be collated appropriately in the form of sheaf
theoretical localization systems of $A_Q$. The net effect of this
procedure, endowed in the above sense with a solid operational
meaning, is the isomorphic representation of a quantum observable
structure via a {\it Grothendieck topos}, understood as a {\it
sheaf for a Grothendieck topology}. The notion of topos is
essential and indispensable to the comprehension of the whole
scheme, because it engulfs the crucial idea of a well-defined
variable structure, admitting localizations over a multiplicity of
generalized reference domains of coordinatizing coefficients, such
that information about observable attributes collected in
partially overlapping domains can be pasted together in a
meaningful manner. Pictorially, the instantiation of such a topos
theoretical scheme can be represented as a {\it fibered
structure}, which we may call, as we shall see, a quantum
observable structure, that fibers over a base category of varying
reference loci, consisting of locally commutative coefficients,
specified by operational means and standing for physical contexts
of quantum measurement. We may formalize the ideas exposed above
as follows:

We make the basic assumption that, there exists a {\it
coordinatization functor}, ${\mathbf M}:{{\mathcal A}_C} \ar {
{\mathcal A}_Q}$, which assigns to commutative observables
algebras in ${{\mathcal A}_C}$, that instantiates a model
category, {\it the underlying quantum algebras} from ${ {\mathcal
A}_Q}$, and to commutative algebras morphisms the underlying
quantum algebraic morphisms.

If ${{\mathcal A}_C}^{op}$ is the opposite category of ${\mathcal
{A_C}}$, then ${{\bf Sets}^{{{\mathcal A}_C}^{op}}}$ denotes the
{\it functor category of presheaves} of commutative observables
algebras, with objects all functors $ {\mathbf P}: {{\mathcal
A}_C}^{op} \ar {\bf Sets}$, and morphisms all natural
transformations between such functors. Each object ${\mathbf P}$
in this category is a contravariant set-valued functor on
${{\mathcal A}_C}$,  called a {\it presheaf} on ${{\mathcal
A}_C}$. The functor category of presheaves on commutative
observables algebras ${{\bf Sets}^{{{\mathcal A}_C}^{op}}}$,
exemplifies a well defined notion of a universe of {\it variable
sets}, and is characterized as a {\it topos}. We recall that a
topos is a category which has a terminal object, pullbacks,
exponentials, and a subobject classifier, that is conceived as an
object of generalized truth values. In this sense, a topos can be
conceived as a local mathematical framework corresponding to a
generalized model of set theory, or as a generalized space.

For each commutative algebra ${A_C}$ of ${{\mathcal A}_C}$,
${\mathbf P}({A_C})$  is a set, and for each arrow $f : {C}_C \ar
{A_C}$, ${\mathbf P} (f) : {\mathbf P}({A_C}) \ar   {\mathbf
P}({C}_C)$ is a set-function. If ${\mathbf P}$ is a presheaf on
${{\mathcal A}_C}$ and $p \in {\mathbf P}({A_C})$, the value
${\mathbf P}(f) (p)$ for an arrow $f : {C}_C \ar {A_C}$ in
${{\mathcal A}_C}$ is called the restriction of $p$ along $f$ and
is denoted by ${\mathbf P}(f) (p)=p \cdot f$.

Each object ${A_C}$ of ${\mathcal A}_C$ gives rise to a
contravariant Hom-functor ${\mathbf y}[{A_C}]:={Hom_{{\mathcal
A}_C}}(-,{A_C})$. This functor defines a presheaf on ${\mathcal
A}_C$. Its action on an object ${C}_C$ of ${\mathcal A}_C$ is
given by $${\mathbf y}[{A_C}]({C}_C):={Hom_{\mathcal
{A_C}}}({C}_C,{A_C})$$ whereas its action on a morphism ${{D}_C}
\Ar x {C}_C$, for $v : {C}_C \ar {A_C}$ is given by $${\mathbf
y}[{A_C}](x) : {Hom_{{\mathcal A}_C}}({C}_C,{A_C}) \ar
{Hom_{{\mathcal A}_C}}({{D}_C},{A_C})$$ $${\mathbf
y}[{A_C}](x)(v)=v \circ x$$ Furthermore $\mathbf y$ can be made
into a functor from ${\mathcal A}_C$ to the contravariant functors
on ${\mathcal A}_C$ $$ \mathbf y : {\mathcal A}_C \ar {{\bf
Sets}^{{{\mathcal A}_C}^{op}}}$$ such that ${A_C} {\mapsto}
{Hom_{{\mathcal A}_C}}(-,{A_C})$. This is called the {\it Yoneda
embedding} and it is a full and faithful functor.

Next we construct the {\it category of elements} of ${\mathbf P}$,
denoted by $\bf{G}({\mathbf P},{{\mathcal A}_C})$. Its objects are
all pairs  $({A_C},p)$, and its morphisms
${(\acute{{A_C}},\acute{p})} {\rightarrow}({A_C},p)$ are those
morphisms $u : \acute{{A_C}} {\rightarrow} {A_C}$ of ${{\mathcal
A}_C}$ for which $pu=\acute{p}$. Projection on the second
coordinate of $\bf{G}({\mathbf P},{{\mathcal A}_C})$, defines a
functor $\bf{G}({\mathbf P}) : \bf{G}({\mathbf P},{{\mathcal
A}_C}) {\rightarrow}  {\mathcal {A_C}}$. $\bf{G}({\mathbf
P},{{\mathcal A}_C})$ together with the projection functor
$\bf{G}({\mathbf P}) $ is called the {\it split discrete
fibration} induced by ${\mathbf P}$, and ${{\mathcal A}_C}$ is the
base category of the fibration. We note that the fibers are
categories in which the only arrows are identity arrows. If
${A_C}$ is an object of ${{\mathcal A}_C}$, the inverse image
under $\bf{G}({\mathbf P})$ of ${A_C}$ is simply the set ${\mathbf
P}({A_C})$, although its elements are written as pairs so as to
form a disjoint union. The construction of the fibration induced
by ${\mathbf P}$, is called the {\it Grothendieck construction}
[13].

\begin{diagram}
¤{\mathbf G}({\mathbf P}, {\mathcal {A_C}} )¤¤ ¤\Sar {{\mathbf
G}({\mathbf P})}¤¤ ¤{\mathcal {A_C}} ¤\Ear {\mathbf P} ¤ \bf
Sets¤¤
\end{diagram}

Now, if we consider the category of quantum observables algebras
${\mathcal {A_Q}}$ and the coefficient functor ${\mathbf M}$,  we
can define the functor; $${\mathbf R}: { {\mathcal A}_Q}
\rightarrow {{\bf Sets}^{{{\mathcal A}_C}^{op}}}$$ from ${
{\mathcal A}_Q}$ to the category of presheaves of commutative
observables algebras given by:

$${\mathbf R}({A_Q}) : {A_C} {\mapsto} {\mathbf
R}({A_Q})(A_C):={{Hom}_{ {\mathcal A}_Q}({\mathbf M}({A_C}),
{A_Q})}$$

According to the philosophy of the {\it functor of points of a
quantum observables algebra}, the objects of the category of
elements ${\mathbf G}({{\mathbf R}({A_Q})},{{\mathcal A}_C})$
constitute {\it generalized points} of $A_Q$ in the environment of
presheaves of commutative observables algebras $A_C$.

We notice that the set of objects of  ${\mathbf G}({{\mathbf
R}({A_Q})},{{\mathcal A}_C})$, considered as a small category,
consists of all the elements of all the sets ${{\mathbf
R}({A_Q})}({A_C})$, and more concretely, has been constructed from
the disjoint union of all the sets of the above form, by labeling
the elements. The elements of this disjoint union are represented
as pairs $({A_C},{\psi}_{{A_C}} | ({\psi}_{{A_C}}: {\mathbf
M}({A_C}) \ar {A_Q}))$ for all objects ${A_C}$ of ${\mathcal
{A_C}}$ and elements ${\psi}_{{A_C}} \in {{\mathbf
R}({A_Q})}({A_C})$.

It is finally instructive to clarify that the functor of points of
a quantum observables algebra, can be also legitimately  made an
object in the category of presheaves of modules ${{\bf
Mod}^{{\mathcal {A_C}}^{op}}}$,  under the requirement that its
composition  with the forgetful functor ${\mathbf Fr: {\bf Mod}
\rightarrow {\bf Sets}}$ is the presheaf of sets functor of points
as determined above.

\section{THE ADJOINT FUNCTORIAL CLASSICAL-QUANTUM RELATION}

The existence of an adjunctive correspondence between the
commutative and quantum observables algebras, which will be proved
in what follows, provides the conceptual ground, concerning the
{\it representation  of quantum observables algebras in terms of
sheaves of structured families of commutative observables
algebras};  this is based on a categorical construction of
colimits over categories of elements of presheaves of commutative
algebras.

A natural transformation $\tau$ between the presheaves on the
category of commutative algebras ${\mathbf P}$ and ${\mathbf
R}({A_Q})$, $\tau : {\mathbf P} \ar {\mathbf R}({A_Q})$, is a
family ${{\tau}_{A_C}}$ indexed by commutative algebras ${A_C}$ of
${{\mathcal A}_C}$ for which each ${\tau}_{A_C}$ is a map of sets,
$${\tau}_{A_C} : {\mathbf P}({A_C}) {\to} {{Hom}_{ {\mathcal
A}_Q}({\mathbf M}({A_C}), {A_Q})} \equiv {\mathbf
R}({A_Q})({A_C})$$ such that the diagram of sets below commutes
for each commutative algebras morphism $u : {\acute{{A_C}}} \to
{A_C}$ of ${{\mathcal A}_C}$.

\begin{diagram} ¤{\mathbf P}({A_C})  ¤\Ear[50]   {{\tau}_{A_C}} ¤
¤{Hom_{ {\mathcal A}_Q}}({{\mathbf M}({A_C})}, {A_Q})¤¤ ¤\Sar
{{\mathbf P}(u)} ¤                      ¤\saR {{\mathbf M}(u)}^*
¤¤ ¤{\mathbf P}(\acute {A_C})  ¤\Ear[50]   {{\tau}_{\acute {A_C}}}
¤ ¤{Hom_{\mathcal {A_Q}}}({{\mathbf M}(\acute {A_C})}, {A_Q})¤¤
\end{diagram}

From the perspective of the category of elements of the
commutative algebras-variable set $P$ the map ${\tau}_{A_C}$,
defined above, is identical with the map: $${\tau}_{A_C} :
({A_C},p) {\to} {{Hom}_{\mathcal {A_Q}}({{\mathbf M} \circ
{G_{\mathbf P}}}({A_C},p), {A_Q})}$$ Subsequently such a $\tau$
may be represented as a family of arrows of ${ {\mathcal A}_Q}$
which is being indexed by objects $({A_C},p)$ of the category of
elements of the presheaf of commutative algebras ${\mathbf P}$,
namely $${\{{{\tau}_{A_C}}(p) : {\mathbf M}({A_C}) \to
{A_Q}\}}_{({A_C},p)}$$ Thus, according to the point of view
provided by the category of elements of ${\mathbf P}$, the
condition of the commutativity of the previous diagram, is
equivalent to the condition that for each arrow $u$ the following
diagram commutes.

\begin{diagram}
¤{{\mathbf M}({A_C})}   ¤ \eeql ¤ {\mathbf M} \circ {{{\mathbf
G}_{\mathbf P}}({A_C},p)}  ¤¤ ¤        ¤          ¤ ¤\Sear
{{{\tau}_{A_C}} (p)} ¤¤ ¤\Nar[133] {{\mathbf M}(u)}  ¤ ¤\naR[133]
{u_*}      ¤   ¤ {A_Q}    ¤¤ ¤       ¤            ¤ ¤\neaR {{{
\tau}_{{\acute A_C}}} (\acute p)} ¤¤ ¤{{\mathbf M}(\acute {A_C})}
¤ \eeql ¤ {\mathbf M} \circ {{{\mathbf G}_{\mathbf P}}(\acute
{A_C},\acute p)} ¤¤
\end{diagram}

Consequently, according to the diagram above, the arrows
${{\tau}_{A_C}}(p)$ form a cocone from the functor ${{\mathbf M}
\circ {G_{\mathbf P}}}$ to the quantum observables algebra
${A_Q}$. The categorical definition of colimit, points to the
conclusion that each such cocone emerges by the composition of the
colimiting cocone with a unique arrow from the colimit $\mathbf L
\mathbf P$ to the quantum observables algebra object ${A_Q}$.
Equivalently, we conclude that there is a bijection, natural in
$\mathbf P$ and ${A_Q}$, as follows:

$$ Nat({\mathbf P},{\mathbf R}({A_Q})) \cong {{Hom}_{\mathcal
{A_Q}}({\mathbf L \mathbf P}, {A_Q})}$$

The established bijective correspondence, interpreted
functorially, says that the functor of points ${\mathbf R}$ from
${\mathcal {A_Q}}$ to presheaves given by $${\mathbf R}({A_Q}) :
{A_C} {\mapsto} {{Hom}_{ {\mathcal A}_Q}({\mathbf M}({A_C}),
{A_Q})}$$ has a left adjoint $\mathbf L :  {{\bf Sets}^{{{\mathcal
A}_C}^{op}}} \to { {\mathcal A}_Q}$, which is defined for each
presheaf of commutative algebras $\mathbf P$ in ${{\bf
Sets}^{{{\mathcal A}_C}^{op}}}$ as the colimit $${\mathbf
{L}}({\mathbf P})= {\it Colim} \{ \bf{G}({\mathbf P},{{\mathcal
A}_C}) \Ar {{\mathbf G}_{\mathbf P}} {{\mathcal A}_C} \Ar {\mathbf
M} { {\mathcal A}_Q} \}$$

Consequently there is a pair of adjoint functors ${\mathbf L}
\dashv {\mathbf R}$ as follows: $$\mathbf L :  {{\bf
Sets}^{{{\mathcal A}_C}^{op}}}  \adjar  { {\mathcal A}_Q} :
\mathbf R$$

Thus we have constructed an adjunction which consists of the
functors $\mathbf L$ and $\mathbf R$, called left and right
adjoints with respect to each other respectively, as well as, the
natural bijection;

$$ Nat({\mathbf P},{\mathbf R}({A_Q})) \cong {{Hom}_{\mathcal
{A_Q}}({\mathbf L \mathbf P}, {A_Q})}$$

The content of the {\it adjunction between  the topos of
presheaves of commutative observables algebras and the category of
quantum observables algebras} can be further developed, if we make
use of the categorical construction of the colimit defined above,
as a coequalizer of a coproduct. We consider the colimit of any
functor $\mathbf F : I \ar {\mathcal A_Q}$ from some index
category $\mathbf I$ to $\mathcal A_Q$, called a diagram of a
quantum observables algebra. Let ${\mu}_i : {\mathbf F}(i) \to
{\amalg}_i {\mathbf F}(i)$, $i \in I$, be the injections into the
coproduct. A morphism from this coproduct, $\chi : {\amalg}_i
{\mathbf F}(i) \to {A_{Q}}$, is determined uniquely by the set of
its components ${\chi}_i=\chi {\mu}_i$. These components
${\chi}_i$ are going to form a cocone over $\mathbf F$ to the
quantum observable vertex $A{_{Q}}$ only when, for all arrows $v :
i \ar j$ of the index category $I$, the following conditions are
satisfied; $$ (\chi {\mu}_j) {\mathbf F}(v)=\chi {\mu}_i$$

\begin{diagram}
¤{{\mathbf F}(i)}¤¤ ¤\Sar {\mu_i} ¤\Sear {{\chi}{\mu_i}}¤¤
¤\coprod {{\mathbf F}(i)}¤\Edotar {\chi}¤A{_{Q}}¤¤ ¤\Nar {\mu_j}
¤\Near {{\chi}{\mu_j}}¤¤ ¤{{\mathbf F}(j)}¤¤
\end{diagram}

So we consider all ${\mathbf F}(dom v)$ for all arrows $v$ with
its injections ${\nu}_v$ and obtain their coproduct ${\amalg}_{v :
i \to j} {\mathbf F}(dom v)$. Next we construct two arrows $\zeta$
and $\eta$, defined in terms of the injections ${\nu}_v$ and
${\mu}_i$, for each $v : i \ar j$ by the conditions $$\zeta
{\nu}_v={\mu}_i$$ $$\eta {\nu}_v={\mu}_j {\mathbf F}(v)$$ as well
as their coequalizer $\chi$

\begin{diagram}
 ¤{{\mathbf F}(dom v)}                        ¤                                             ¤                ¤               ¤{{\mathbf F}(i)}¤¤
 ¤\Sar {\mu_v}                                   ¤                                    ¤              ¤                  ¤\Sar {\mu_i} ¤\Sedotar {{\chi}{\mu_i}}¤¤
 ¤{{\coprod}_ {v : i \to j}}{{{\mathbf F}(dom v)}}¤ ¤ \Ebiar[70]{\zeta}{\eta}¤  ¤\coprod {{\mathbf F}(i)}¤\Edotar {\chi}¤A{_{Q}}¤¤
\end{diagram}

The coequalizer condition $\chi \zeta=\chi \eta$ tells us that the
arrows $\chi{ {\mu}_i}$ form a cocone over $\mathbf F$ to the
quantum observable vertex $\mathcal A_Q$. We further note that
since $\chi$ is the coequalizer of the arrows $\zeta$ and $\eta$
this cocone is the colimiting cocone for the functor $\mathbf F :
I \to {\mathcal A_Q}$ from some index category $I$ to $\mathcal
A_Q$. Hence the colimit of the functor $\mathbf F$ can be
constructed as a coequalizer of coproducts according to

\begin{diagram}
¤{{\coprod}_ {v : i \to j}}{{{\mathbf F}(dom v)}}¤ ¤
\Ebiar[70]{\zeta}{\eta}¤  ¤\coprod {{\mathbf F}(i)}¤\Ear
{\chi}¤Colim \mathbf F¤¤
\end{diagram}

In the case considered the index category is the category of
elements of the presheaf of commutative observables algebras
$\mathbf P$ and the functor  ${{\mathbf M} \circ {G_{\mathbf P}}}$
plays the role of the diagram of quantum observables algebras
$\mathbf F : I \ar {\mathcal A_Q}$. In the diagram above the
second coproduct is over all the objects $(\xi,p)$ with $p \in
{\mathbf P}({A_{C}})$ of the category of elements, while the first
coproduct is over all the maps $v : ({\acute {A_{C}}},{\acute p})
\ar ({A_{C}},p)$ of that category, so that $v : {\acute {A_{C}}}
\ar {A_{C}}$ and the condition $pv=\acute p$ is satisfied. We
conclude that the colimit ${{\mathbf L}_M}(P)$ can be equivalently
presented as the coequalizer:

\begin{diagram}
¤{{\coprod}_ {v : {\acute {A_{C}}} \to {A_{C}}}}{{{\mathbf
M}(\acute {A_{C}})}}¤        ¤ \Ebiar[70]{\zeta}{\eta}¤
¤{{\coprod}_{({A_{C}},p)}} {{\mathbf M}({A_{C}})} ¤\Ear
{\chi}¤{\mathbf P} {{\otimes}_{\mathcal A_C}} {\mathbf M}¤¤
\end{diagram}

The  coequalizer presentation of the colimit  shows that the
Hom-functor  ${\mathbf R}({A_Q})$ has a left adjoint which can be
characterized categorically as  the  tensor product $-
{\otimes}_{\mathcal A_C} {\mathbf M}$.

In order to clarify the above observation, we forget for the
moment that the discussion concerns the category of quantum
observables $\mathcal A_Q$, and we consider instead the category
$\bf Sets$. Then the coproduct ${{\amalg}_p} {\mathbf M}({A_{C}})$
is a coproduct of sets, which is equivalent to the product
${\mathbf P}({A_{C}})  \times {\mathbf M}({A_{C}})$ for ${A_{C}}
\in \mathcal A_C$. The coequalizer is thus the definition of the
tensor product ${\mathcal P} \otimes {\mathcal A}$ of the set-
valued functors: $$\mathbf P : {\mathcal A_C}^{op} \ar {\bf Sets},
\qquad \mathbf M : {\mathcal A_C} \ar {\bf Sets}$$

\begin{diagram}
¤{{\coprod}_ {{A_{C}}, \acute {A_{C}}}} {{\mathbf P}({A_{C}})}
\times Hom(\acute {A_{C}}, {A_{C}}) \times {{{\mathbf M}(\acute
{A_{C}})}}¤ ¤ ¤ \Ebiar[30]   {\zeta}{\eta}¤ ¤{{\coprod}_{A_{C}}}
{{\mathbf P}({A_{C}})} \times {{\mathbf M}({A_{C}})}¤ ¤\Ear
{\chi}¤{\mathbf P} {{\otimes}_{\mathcal A_C}} {\mathbf M}¤¤
\end{diagram}

According to the diagram above for elements $p \in {\mathbf
P}({A_{C}})$, $v : {\acute {A_{C}}} \to {A_{C}}$ and $\acute q \in
{\mathbf M}({\acute {A_{C}}})$ the following equations hold:
$$\zeta (p,v, \acute q)=(pv, \acute q),  \qquad \eta(p,v, {\acute
q})=(p, v \acute q)$$ symmetric in $\mathbf P$ and $\mathbf M$.
Hence, the elements of the set ${\mathbf P} {\otimes}_{\mathcal
A_C} {\mathbf M}$ are all of the form $\chi (p,q)$. This element
can be written as $$ \chi(p,q)=p \otimes q, \quad  p \in {\mathbf
P}({A_{C}}), q \in {\mathbf M}({A_{C}})$$ Thus if we take into
account the definitions of $\zeta$ and $\eta$ above, we obtain
$$pv \otimes \acute q=p \otimes v \acute q$$

Furthermore, if we define the arrows $$ k_{A_{C}} :  {\mathbf P}
{\otimes}_{\mathcal A_C} {\mathbf M} \ar { A{_{Q}}}, \qquad
l_{A_{C}} : {\mathbf P}({A_{C}}) \ar  {{Hom}_{\mathcal
A_Q}({\mathbf M}({A_{C}}), A{_{Q}})}$$ they are related under the
fundamental adjunction by $$ {k_{A_{C}}}(p,q)={l_{A_{C}}}(p)(q),
\qquad {A_{C}} \in {\mathcal A_C}, p \in {\mathbf P}({A_{C}}), q
\in {\mathbf M}({A_{C}})$$ Here we consider $k$ as a function on
${\amalg}_{A_{C}} {\mathbf P}({A_{C}}) \times {\mathbf
M}({A_{C}})$ with components $ k_{A_{C}} : {{\mathbf P}({A_{C}})}
{\times} {{\mathbf M}({A_{C}})}  \ar { A{_{Q}}}$, satisfying the
relation; $${k_{\acute {A_{C}}}}(pv,q)={k_{A_{C}}}(p,vq)$$ in
agreement with the equivalence relation defined above.

Now we replace the category $\bf Sets$ by the category of quantum
observables $\mathcal A_Q$ under study. The element $q$ in the set
${\mathbf M}({A_{C}})$ is replaced by a generalized element $q
:{\mathbf M}({J_{C}})  \to {\mathbf M}({A_{C}}) $ from some
modelling object ${\mathbf M}({J_{C}})$ of $\mathcal A_Q$. Then we
consider $k$ as a function ${\amalg}_{({A_{C}},p)} {{\mathbf
M}({A_{C}})} \ar A{_{Q}}$ with components $k_{({A_{C}},p)} :
{\mathbf M}({A_{C}}) \to { A{_{Q}}}$ for each $p \in {\mathbf
P}({A_{C}})$, that, for all arrows $v : {\acute {A_{C}}} \ar
{A_{C}}$ satisfy; $$ k_{({\acute {A_{C}}}, pv)}=k_{({A_{C}},p)}
\circ {{\mathbf M}(v)}$$ Then the condition defining the bijection
holding by virtue of the fundamental adjunction is given by
$$k_{{({A_{C}},p)}} \circ q={l_{A_{C}}}(p) \circ q : {\mathbf
M}({J_{C}}) \to A{_{Q}}$$ This argument, being natural in the
object ${\mathbf M}({J_{C}})$, is determined by setting ${\mathbf
M}({J_{C}})={\mathbf M}({A_{C}})$ with $q$ being the identity map.
Hence, the bijection takes the form
$k_{{({A_{C}},p)}}={l_{A_{C}}}(p) $, where
$k:{\amalg}_{({A_{C}},p)} {{\mathbf M}({A_{C}})} \ar A{_{Q}}$, and
$l_{A_{C}} : {\mathbf P}({A_{C}}) \ar {{Hom}_{\mathcal
A_Q}({\mathbf M}({A_{C}}), A{_{Q}})}$.

The  physical meaning of the adjunction between presheaves of
commutative observables algebras and quantum observables algebras
is made transparent  if we consider that the adjointly related
functors are associated with the process of encoding and decoding
information relevant to the structural form of their domain and
codomain categories. If we think of ${\bf Sets}^{{{\mathcal
A}_C}^{op}}$ as the topos of variable commutative algebras
modelled in $\bf Sets$, and of ${\mathcal A}_Q$ as the universe of
quantum observable structures, then the functor $\mathbf L : {\bf
Sets}^{{{\mathcal A}_C}^{op}} \ar {\mathcal A}_Q$ signifies a
translational code of information from the topos of commutative
observables structures to the universe of globally non-commutative
ones, whereas the functor $\mathbf R : {\mathcal A}_Q \ar {\bf
Sets}^{{{\mathcal A}_C}^{op}}$ a translational code in the inverse
direction. In general, the content of the information is not
possible to remain completely invariant with respect to
translating transformations from one universe to another and back.
However, there remain two alternatives for a variable set over
commutative observables algebras $\mathbf P$ to exchange
information with a quantum observables algebra ${A_Q}$. Either the
content of information is exchanged  in non-commutative terms with
$\mathbf P$ translating, represented as the quantum morphism
${\mathbf L \mathbf P} \ar {A_Q}$, or the content of information
is exchanged in commutative terms with ${A_Q}$ translating,
represented correspondingly as the natural transformation
${\mathbf P} \ar {{\mathbf R}({A_Q})}$. In the first case, from
the perspective of ${A_Q}$ information is being received in
quantum terms, while in the second, from the perspective of
$\mathbf P$ information is being sent in commutative algebras
terms. The natural bijection then corresponds to the assertion
that these two distinct ways of communicating are equivalent.
Thus, the fact that these two functors are adjoint, expresses a
relation of variation regulated by two poles, with respect to the
meaning of the information related to observation. We claim that
the totality of the content of information included in quantum
observables structures remains {\it invariant under commutative
algebras encodings}, corresponding to local commutative
observables algebras, if and only if the adjunctive correspondence
can be appropriately restricted to an equivalence of the
functorially correlated categories. In the following sections we
will show that this task can be accomplished by defining an
appropriate Grothendieck topology on the category of commutative
observables algebras, that, essentially permits the comprehension
of {\it a quantum observables  structure, as a sheaf of locally
commutative ones over an appropriately  specified covering
system}. Subsequently, the categorical equivalence that will be
established in the sequel, is going to be interpreted, as the
denotator of  an informational invariance property, referring to
the translational code of communication between variable
commutative observables algebras and globally non-commutative
ones.

\section{TOPOLOGIES ON CATEGORIES}
\subsection{Motivation}

Our purpose is to show that the functor ${\mathbf R}$ from ${
{\mathcal A}_Q}$ to presheaves given by $${\mathbf R}({A_Q}) :
{A_C} {\mapsto} {{Hom}_{ {\mathcal A}_Q}({\mathbf M}({A_C}),
{A_Q})}$$ sends quantum observables algebras ${A_Q}$ in $
{\mathcal A}_Q$ not just into presheaves, but actually into
sheaves for a suitable Grothendieck topology ${\mathbf J}$ on the
category of commutative observables algebras ${\mathcal A}_C$, so
that the fundamental adjunction will restrict to an equivalence of
categories ${\bf Sh}({{\mathcal A}_C}, {\mathbf J}) \cong {
{\mathcal A}_Q}$. From a physical perspective the above can be
understood as a topos theoretical formulation of Bohr's
correspondence, or as a generalized ``complementarity principle".

We note at this point that the usual notion of sheaf, in terms of
coverings, restrictions, and collation, can be defined and used
not just in the spatial sense, namely on the usual topological
spaces, but in a generalized spatial sense, on more general
topologies (Grothendieck topologies). In the usual definition of a
sheaf on a topological space we use the open neighborhoods $U$ of
a point in a space $X$; such neighborhoods are actually monic
topological maps $U \to X$.  The neighborhoods $U$ in topological
spaces can be replaced by maps $V \to X$ not necessarily monic,
and this can be done in any category with pullbacks. In effect, a
covering by open sets can be replaced by a new notion of covering
provided by a family of maps satisfying certain conditions.

For an object ${A_C}$ of ${\mathcal A}_C$, we consider indexed
families $$\mathbf S=\{\psi_i : {A_C}_i \to {A_C}, i \in I \}$$ of
maps to ${A_C}$ (viz. maps with common codomain ${A_C}$), and we
assume that, for each object ${A_C}$ of ${\mathcal A}_C$, we have
a set ${\mathbf \Lambda} ({A_C})$ of certain such families
satisfying conditions to be specified later. These families play
the role of coverings of ${A_C}$, under those conditions. Based on
such coverings, it is possible to construct the analogue of the
topological definition of a sheaf, where as presheaves on
${\mathcal A}_C$ we consider the functors $\mathbf P : {\mathcal
{A_C}}^{op} \to \bf {Sets}$. According to the topological
definition of a sheaf on a space, we demand that for each open
cover $\{U_i, i \in I \}$ of some $U$, every family of elements
$\{p_i \in {\mathbf P}(U_i) \}$ that satisfy the compatibility
condition on the intersections $U_i \cap U_j, \forall i,j$, are
pasted together, as a unique element $p \in {\mathbf P}(U)$.
Imitating the above procedure for any covering $\mathbf S$ of an
object ${A_C}$, and replacing the intersection $U_i \cap U_j$ by
the pullback ${A_C}_i {\times}_{{A_C}} {A_C}_j$ in the general
case, according to the diagram
\begin{diagram}
¤{A_C}_i {\times}_{A_C} {A_C}_j  ¤\Ear {g_{ij}} ¤ {A_C}_j ¤¤ ¤\Sar
{h_{ij}} ¤ ¤\saR {\psi_{j}} ¤¤ ¤{A_C}_i¤\Ear {\psi_i}¤{A_C}¤¤
\end{diagram}
we effectively obtain for a given presheaf $\mathbf P : {\mathcal
{A_C}}^{op} \to \bf {Sets}$ a diagram of sets as follows
\begin{diagram}
¤\mathbf P ({A_C}_i {\times}_{A_C} {A_C}_j) ¤   ¤\Ear[72] {\mathbf
P ({g_{ij}})} ¤ ¤ \mathbf P ({A_C}_j) ¤¤ ¤\Sar {\mathbf P
({h_{ij}})} ¤  ¤  ¤ ¤\saR {\mathbf P ({\psi_{j}})} ¤¤ ¤\mathbf P
({A_C}_i) ¤   ¤\Ear[88] {\mathbf P ({\psi_i})} ¤  ¤\mathbf P
({A_C})¤¤
\end{diagram}

In this case the compatibility condition for a sheaf takes the
form: if $\{p_i \in {\mathbf P}_i, i \in I \}$ is a family of
compatible elements, namely satisfy $p_i h_{ij}=p_j g_{ij},
\forall i,j$, then a unique element  $p \in {\mathbf P}({A_C})$ is
being determined by the family such that $p\cdot \psi_i=p_i,
\forall i \in I$, where the notational convention $p\cdot \psi_i=
{\mathbf P ({\psi_i})}(p)$ has been used . Equivalently, this
condition can be expressed in the categorical terminology by the
requirement that in the diagram
\begin{diagram}
¤\prod_{i,j} {\mathbf P ({A_C}_i {\times}_{A_C} {A_C}_j)}   ¤
¤\wbiar[50] ¤ ¤{\prod}_{i} {\mathbf P ({{A_C}_i})}  ¤ ¤\War [70]
e¤ ¤\mathbf P ({A_C})¤¤
\end{diagram}
the arrow $e$, where $e(p)=(p\cdot \psi_i, i \in I)$  is an
equalizer of the maps $(p_i, i \in I) \to ({p_i} h_{ij};i,j \in I
\times I)$ and $(p_i, i \in I) \to ({p_i} g_{ij};i,j \in I \times
I)$, correspondingly.

The specific conditions that the elements of the set $\Lambda
({A_C})$, or else the coverings of ${A_C}$, have to satisfy,
naturally lead to the notion of a Grothendieck topology on the
category ${{\mathcal A}_C}$.

\subsection{Grothendieck topologies}

We start our discussion by explicating the notion of a pretopology
on the category of commutative observables algebras ${\mathcal
A}_C$.

A {\it pretopology} on ${A_C}$ is a function $\bf {\Lambda}$ where
for each object ${A_C}$ there is a set
 ${\bf {\Lambda}}({A_C})$. Each ${\bf {\Lambda}}({A_C})$ contains indexed families of ${\mathcal A}_C$-morphisms
$$\mathbf S=\{\psi_i : {A_C}_i \to {A_C}, i \in I \}$$ of maps to
${A_C}$ such that the following conditions are satisfied:

(1) For each ${A_C}$ in ${\mathcal A}_C$, $\{{id}_{A_C}  \} \in
{{\bf {\Lambda}}({A_C})}$ ;

(2) If ${C}_C \to {A_C}$ in ${\mathcal A}_C$ and $\mathbf
S=\{\psi_i : {A_C}_i \to {A_C}, i \in I \} \in {\bf
{\Lambda}}({A_C})$ then $\{\psi_1 : {C}_C {\times}_{A_C} {A_C}_i
\to {A_C}, i \in I \}  \in {\bf {\Lambda}}({C}_C)$. Note that
$\psi_1$ is the pullback in ${\mathcal A}_C$ of $\psi_i$ along
${C}_C \to {A_C}$;

(3) If $\mathbf S=\{\psi_i : {A_C}_i \to {A_C}, i \in I \} \in
{\bf {\Lambda}}({A_C})$, and for each $i \in I$, $\{{\psi_{ik}}^i
: {{C}_C}_{ik} \to {A_C}_i, k \in K_i \} \in {\bf
{\Lambda}}({A_C}_i)$, then $\{{\psi_{ik}}^i \circ \psi_i :
{{C}_C}_{ik} \to {A_C}_i \to {A_C}, i \in I;  k \in K_i \} \in
{\bf {\Lambda}}({A_C})$. Note that ${{C}_C}_{ik}$ is an example of
a double indexed object rather than the intersection of
${{C}_C}_i$ and ${{C}_C}_k$.

The notion of a pretopology on the category of commutative
algebras ${\mathcal A}_C$ is a categorical generalization of a
system of set-theoretical covers on a topology $\bf T$, where a
cover for $U \in \bf T$ is a set $\{U_i : U_i \in \bf T, i \in I
\}$ such that ${\cup}_i {U_i}=U$. The generalization is achieved
by noting that the topology ordered by inclusion is a poset
category and that any cover corresponds to a collection of
inclusion arrows $U_i \to U$. Given this fact, any family of
arrows contained in ${\bf {\Lambda}}({A_C})$ of a pretopology is a
cover, as well.

The passage from a pretopology to a categorical or Grothendieck
topology on the category of commutative unital $\mathcal
R$-algebras takes place through the introduction of appropriate
covering devices, called {\it covering sieves}. For an object
${A_C}$ in ${\mathcal A}_C$, an ${A_C}$-sieve is a family $R$ of
${\mathcal A}_C$-morphisms with codomain ${A_C}$, such that if
${{C}_C}\rightarrow {A_C}$ belongs to $R$ and ${{D}_C}\rightarrow
{{C}_C}$ is any ${\mathcal A}_C$-morphism, then the composite
${{D}_C}\rightarrow {{C}_C}\rightarrow {A_C}$ belongs to $R$.

A {\it Grothendieck topology} on the category of commutative
algebras ${\mathcal A}_C$, is a system $J$ of sets, $J({A_C})$,
for each ${A_C}$ in ${\mathcal A}_C$, where each $J({A_C})$
consists of a set of ${A_C}$-sieves, (called the covering sieves),
that satisfy the following conditions:

(1) For any ${A_C}$ in ${\mathcal A}_C$ the maximal sieve
$\{g:cod(g)={A_C}\}$ belongs to $J({A_C})$ (maximality condition).

(2) If $R$ belongs to $J({A_C})$ and $f:{{C}_C}\rightarrow {A_C}$
is an $\mathcal {A_C}$-morphism, then
$f^*(R)=\{h:{{C}_C}\rightarrow {A_C}, f\cdot h \in R\}$ belongs to
$J({{C}_C})$ (stability condition).

(3) If $R$ belongs to $J({A_C})$ and $S$ is a sieve on ${{C}_C}$,
where for each $f:{{C}_C}\rightarrow {A_C}$ belonging to $R$, we
have $f^*(S)$ in $J({{C}_C})$, then $S$ belongs to $J({A_C})$
(transitivity condition).

The small category ${\mathcal A}_C$ together with a Grothendieck
topology ${\mathbf J}$, is called a {\it site}. A {\it sheaf on a
site} $({\mathcal A}_C,{\mathbf J})$ is defined to be any
contravariant functor $\mathbf P : {{\mathcal A}_C}^{op} \to \bf
{Sets}$, satisfying the equalizer condition expressed in terms of
covering sieves $S$ for ${A_C}$, as in the following diagram in
$\mathbf {Sets}$:
\begin{diagram}
¤\prod_{{f\cdot g} \in S} {\mathbf P (dom g)}   ¤ ¤\wbiar[50] ¤
¤{\prod}_{f \in S} {\mathbf P (dom f)}  ¤ ¤\War [70] e¤ ¤\mathbf P
({A_C})¤¤
\end{diagram}
If the above diagram is an equalizer for a particular covering
sieve $S$, we obtain that $\mathbf P$ satisfies the sheaf
condition with  respect to the covering sieve $S$.

A {\bf Grothendieck topos} over the small category ${\mathcal
A}_C$ is a category which is equivalent to the category of sheaves
${\bf Sh}({{\mathcal A}_C}, {\mathbf J})$  on a site $(\mathcal
{A_C},{\mathbf J})$. The site can be conceived as a system of
generators and relations for the topos. We note that a category of
sheaves ${\bf Sh}({{\mathcal A}_C}, {\mathbf J})$ on a site
$(\mathcal {A_C},{\mathbf J})$ is a full subcategory of the
functor category of presheaves ${{\bf Sets}^{{{\mathcal
A}_C}^{op}}}$.

The basic properties of a Grothendieck topos are the following:

(1). It admits finite projective limits; in particular, it has a
terminal object, and it admits fibered products.

(2). If $({B_i})_{i \in I}$ is a family of objects of the topos,
then the sum ${\coprod_{i \in I}} {B_i}$ exists and is disjoint.

(3). There exist quotients by equivalence relations and have the
same good properties as in the category of sets.

\section{GROTHENDIECK TOPOLOGY ON ${{\mathcal A}_C}$ }

\subsection{${\mathcal A}_C$ as a generating subcategory of $\mathcal
{A_Q}$}

We consider ${\mathcal A}_C$ as a {\it full subcategory} of
$\mathcal {A_Q}$, whose set of objects $\{{A_C}_i: i \in I\}$,
with $I$ an index set, {\it generate} $ {\mathcal A}_Q$; that is,
for any diagram in $\mathcal {A_Q}$,

\begin{diagram}
¤{A_C}_i¤\Ear w ¤{A_Q} ¤\Ebiar[50]v u¤{{\acute A}_{Q}}¤¤
\end{diagram}
the identity $v \circ w=u \circ w$, for every arrow
$w:{A_C}_i\rightarrow {A_Q}$, and every ${A_C}_i$, implies that
$v=u$. We notice that, for every pair of different parallel
morphisms of ${A_Q}$, with common domain, there is a separating
morphism of ${A_Q}$, with domain in ${A_C}_i \hookrightarrow
{A_Q}$ and codomain the previous common domain. Equivalently, we
can say that the set of all arrows $w:{A_C}_i\rightarrow {A_Q}$,
constitute an epimorphic family. We may verify this claim, if we
take into account the adjunction and observe that objects of
$\mathcal {A_Q}$ are being constructed as {\it colimits} over
categories of elements of presheaves over ${\mathcal A}_C$. Since
objects of $\mathcal {A_Q}$ are constructed as colimits of this
form, whenever two parallel arrows
\begin{diagram}
¤{A_Q} ¤\Ebiar[50]v u¤{{\acute A}_{Q}}¤¤
\end{diagram}
are different, there is an arrow $l: {A_C}_i \rightarrow {A_Q}$
from some ${A_C}_i$ in ${\mathcal A}_C$, such that $v \circ l \neq
u \circ l$.

Since we assume that ${\mathcal A}_C$ is a full subcategory of $
{\mathcal A}_Q$ we omit the explicit presence of the
coordinatization functor ${\mathbf M}$ in the subsequent
discussion.

The consideration that ${\mathcal A}_C$ is a generating
subcategory of $ {\mathcal A}_Q$ points exactly to the depiction
of the appropriate Grothendieck topology on ${\mathcal A}_C$, that
accomplishes our purpose of comprehending quantum observables
algebras as sheaves on ${\mathcal A}_C$.

We assert that a sieve $S$ on a commutative algebra ${A_C}$ in
$\mathcal {A_C}$ is to be a covering sieve of ${A_C}$, when the
arrows $s:{{C}_C}\rightarrow {A_C}$ belonging to the sieve $S$
together form an epimorphic family in $ {\mathcal A}_Q$. This
requirement may be equivalently expressed in terms of a map $$ G_S
: {\coprod}_{\{s:{{C}_C}\rightarrow {A_C} \} \in S} {{C}_C}
\rightarrow {A_C}$$ being an epi in $ {\mathcal A}_Q$.

\subsection{The Grothendieck topology of Epimorphic Families}

We will show in the sequel, that {\it covering sieves on
commutative algebras in ${\mathcal A}_C$, being epimorphic
families in $ {\mathcal A}_Q$, indeed define, a Grothendieck
topology on ${\mathcal A}_C$}.

First of all we notice that the maximal sieve on each commutative
algebra ${A_C}$, includes the identity ${A_C}\rightarrow {A_C}$,
thus it is a covering sieve. Next, the transitivity property of
the depicted covering sieves is obvious. It remains to demonstrate
that the covering sieves remain stable under pullback. For this
purpose we consider the pullback of such a covering sieve $S$ on
${A_C}$ along any arrow $h:{{A_C}^\prime}\rightarrow {A_C}$ in
${\mathcal A}_C$

\begin{diagram}
¤{\coprod}_{s \in S} {{C}_C} {\times}_{A_C}  \acute {A_C} ¤\ear
¤\acute {A_C} ¤¤ ¤\sar ¤ ¤\saR h ¤¤ ¤{\coprod}_{s \in S} {{C}_C}
¤\Ear G ¤{A_C}¤¤
\end{diagram}

The commutative algebras ${A_C}$ in ${\mathcal A}_C$ generate the
category of quantum observables algebras $ {\mathcal A}_Q$, hence,
there exists for each arrow $s: {{D}_C} \rightarrow {A_C}$ in $S$,
an epimorphic family of arrows $ \coprod [{A_C}]^s \rightarrow
{{D}_C} \times_{A_C} \acute {A_C}$, or equivalently
$\{{[{A_C}]^s}_j \rightarrow {{D}_C} \times_{A_C} \acute
{A_C}\}_j$, with each domain $[{A_C}]^s$ a commutative algebra.
Consequently the collection of all the composites: $${[{A_C}]^s}_j
\rightarrow {{D}_C} \times_{A_C} \acute {A_C} \rightarrow \acute
{A_C}$$ for all $s: {{D}_C} \rightarrow {A_C}$ in $S$, and all
indices $j$ together form an epimorphic family in $ {\mathcal
A}_Q$, that is contained in the sieve $h^*(S)$, being the pullback
of $S$ along $h: {A_C} \rightarrow \acute {A_C}$. Therefore the
sieve $h^*(S)$ is a covering sieve.

It is important to construct the representation of covering sieves
within the category of commutative observables algebras ${\mathcal
A}_C$. This is possible, if we first observe that for an object
${{C}_C}$ of ${\mathcal A}_C$, and a covering sieve for the
defined Grothendieck topology on ${\mathcal A}_C$, the  map $$ G_S
: {\coprod}_{(s:{{C}_C}\rightarrow {A_C}) \in S} {{C}_C}
\rightarrow {A_C}$$ being an epi in $ {\mathcal A}_Q$, can be
equivalently presented as the coequalizer of its kernel pair, or
else the pullback of $G_S$ along itself

\begin{diagram}
¤{\coprod}_{\acute s} \acute {{D}_C} {{\times}_{{{C}_C}}}
{\coprod}_s {{D}_C} ¤\ear ¤{\coprod}_s {{D}_C} ¤¤ ¤\sar ¤ ¤\saR
{G_S} ¤¤ ¤{\coprod}_{\acute s} \acute {{D}_C}¤\Ear {G_S}
¤{{C}_C}¤¤
\end{diagram}
Furthermore, using the fact that pullbacks in $ {\mathcal A}_Q$
preserve coproducts, the epimorhic family associated with a
covering sieve on ${{C}_C}$, admits the following coequalizer
presentation

\begin{diagram}
¤{{\coprod}_{\acute s,s} \acute {{D}_C} {{\times}_{{{C}_C}}}
{{D}_C}} ¤ ¤\Ebiar[70] {q_1} {q_2}¤  ¤{\coprod}_s {{D}_C} ¤\Ear
G¤{{C}_C}¤¤
\end{diagram}
Moreover, since ${\mathcal A}_C$ is a generating subcategory of $
{\mathcal A}_Q$, for each pair of arrows $s: {{D}_C}\rightarrow
{{C}_C}$ and $\acute s: \acute {{D}_C} \rightarrow {{C}_C}$ in the
covering sieve $S$ on the commutative algebra ${{C}_C}$, there
exists an epimorphic family $\{{A_C}\rightarrow {\acute {{D}_C}
{\times}_{{C}_C} {{D}_C}}\}$, such that each domain ${A_C}$ is a
commutative algebra object in ${\mathcal A}_C$.

Consequently, each element of the epimorphic family, associated
with a covering sieve $S$ on a commutative algebra ${{C}_C}$ is
represented by a commutative diagram in ${\mathcal A}_C$ of the
following form:

\begin{diagram}
¤{A_C}  ¤\Ear l ¤ {{D}_C} ¤¤ ¤\saR k ¤ ¤\saR s ¤¤ ¤\acute
{{D}_C}¤\Ear {\acute s}¤{{C}_C}¤¤
\end{diagram}
At a further step we may compose the representation of epimorphic
families by commutative squares in ${\mathcal A}_C$, obtained
previously, with the coequalizer presentation of the same
epimorphic families. The composition results in a new coequalizer
diagram in ${\mathcal A}_C$ of the following form:

\begin{diagram}
¤{\coprod}_{A_C} {A_C} ¤ ¤\Ebiar[70] {y_1} {y_2}¤ ¤{\coprod}_s
{{D}_C} ¤\Ear G¤{{C}_C}¤¤
\end{diagram}
where the first coproduct is indexed by all ${A_C}$ in the
commutative diagrams in ${\mathcal A}_C$, representing elements of
epimorphic families.

\subsection{The $\mathbf{J}$-Sheaf ${\mathbf R}({A_Q})$}

For each quantum observables algebra ${A_Q}$ in $ {\mathcal A}_Q$,
we consider the contravariant $Hom$-functor ${\mathbf
R}({A_Q})={{Hom}_{ {\mathcal A}_Q}(-, {A_Q})}$ in ${{\bf
Sets}^{{{\mathcal A}_C}^{op}}}$. If we apply this representable
functor to the latter coequalizer diagram we obtain an equalizer
diagram in ${\bf Sets}$ as follows:
\begin{diagram}
¤\prod_{A_C} {Hom}_{ {\mathcal A}_Q}({A_C}, {A_Q}) ¤ ¤\wbiar[50] ¤
¤{\prod}_{s \in S} {Hom}_{ {\mathcal A}_Q}({{D}_C}, {A_Q}) ¤ ¤\war
[50] ¤ ¤{Hom}_{ {\mathcal A}_Q}({{C}_C}, {A_Q})¤¤
\end{diagram}
where the first product is indexed by all ${A_C}$ in the
commutative diagrams in ${\mathcal A}_C$, representing elements of
epimorphic families.

The equalizer in ${\bf Sets}$, thus obtained, says explicitly that
the contravariant $Hom$-functor ${\mathbf
R}({A_Q})={{Hom}_{\mathcal {A_Q}}(-, {A_Q})}$ in ${{\bf
Sets}^{{{\mathcal A}_C}^{op}}}$, satisfies the sheaf condition for
the covering sieve $S$. Moreover, the equalizer condition holds,
for every covering sieve in the Grothendieck topology of
epimorphic families.

{\it By rephrasing the above, we conclude that the representable
$Hom$-functor ${\mathbf R}({A_Q})$ is a sheaf for the Grothendieck
topology of epimorphic families on the category of commutative
observables algebras}.

\section{EQUIVALENCE OF THE TOPOS ${\bf
Sh}({{\mathcal A}_C}, {\mathbf J})$ WITH ${ {\mathcal A}_Q}$ }

{\it We claim, that if the functor ${\mathbf R}$ from ${ {\mathcal
A}_Q}$ to presheaves $${\mathbf R}({A_Q}) : {A_C} {\mapsto}
{{Hom}_{ {\mathcal A}_Q}({\mathbf M}({A_C}), {A_Q})}$$ sends
quantum observables algebras ${A_Q}$ in $ {\mathcal A}_Q$ not just
into presheaves, but into sheaves for the Grothendieck topology of
epimorphic families, ${\mathbf J}$, on the category of commutative
observables algebras ${\mathcal A}_C$,  the fundamental adjunction
restricts to an equivalence of categories ${\bf Sh}({{\mathcal
A}_C}, {\mathbf J}) \cong { {\mathcal A}_Q}$. Thus, ${ {\mathcal
A}_Q}$ is, in effect, a Grothendieck topos. Hence, in an
epigrammatic manner we can assert that; appropriately sheafifying
in the Grothendieck sense is equivalent to quantizing,
equivalently, quantizing means, in effect, sheafifying {\it
$\grave{a}$ la Grothendieck!}}.

\subsection{Covering Sieves on Quantum Observables Algebras}

If we consider a quantum observables algebra ${A_Q}$, and all
quantum algebraic morphisms of the form $\psi: {{E}_C} \rightarrow
{A_Q}$, with domains ${{E}_C}$, in the generating subcategory of
commutative observables algebras ${\mathcal A}_C$, then the family
of all these maps $\psi$, constitute an epimorphism: $$ T :
{\coprod}_{({{E}_C} \in \mathcal {A_C},\psi:{{E}_C}\rightarrow
{A_Q})} {{E}_C} \rightarrow {A_Q}$$ We notice that the quantum
algebraic epimorphism $T$ is actually the same as the map, $$ T :
{\coprod}_{({{E}_C} \in \mathcal {A_C},\psi:\mathbf M({{E}_C})
\rightarrow {A_Q})} \mathbf M({{E}_C}) \rightarrow {A_Q}$$ since
the coordinatization functor $\mathbf M$ is,  by the fact that
${\mathcal A}_C$ is a full subcategory of $ {\mathcal A}_Q$, just
the inclusion functor $\mathbf M : {\mathcal A}_C \hookrightarrow
{\mathcal A}_Q$.

Subsequently, we may use the same arguments, as in the discussion
of the Grothendieck topology of epimorphic families of the
previous section, in order to assert that the epimorphism $T$ can
be presented as a coequalizer diagram of the form [DI] in
${\mathcal A}_Q$ as follows:

\begin{diagram}
¤{\coprod}_{\nu} {A_C} ¤ ¤\Ebiar[70] {y_1} {y_2}¤ ¤
¤{\coprod}_{({{E}_C} \in {\mathcal A}_C, \psi:{{E}_C}\rightarrow
{A_Q})} {{E}_C}¤ ¤\Ear T¤{A_Q}¤¤
\end{diagram}
where the first coproduct is indexed by all $\nu$, representing
commutative diagrams in $ {\mathcal A}_Q$, of the form:

\begin{diagram}
¤{A_C}  ¤\Ear l ¤ {{E}_C} ¤¤ ¤\saR k ¤ ¤\saR \psi ¤¤ ¤\acute
{{E}_C}¤\Ear {\acute \psi}¤{A_Q}¤¤
\end{diagram}
where ${A_C}$, ${{E}_C}$, $\acute {{E}_C}$ are objects in the
generating subcategory ${\mathcal A}_C$ of  $ {\mathcal A}_Q$.

We say that a sieve on a quantum observables algebra defines a
covering sieve by objects of its generating subcategory ${\mathcal
A}_C$, when the quantum algebraic morphisms belonging to the sieve
define an epimorphism  $$ T : {\coprod}_{({{E}_C} \in \mathcal
{A_C},\psi:\mathbf A({{E}_C}) \rightarrow {A_Q})} \mathbf
M({{E}_C}) \rightarrow {A_Q}$$ In this case the epimorphic
families of quantum algebraic morphisms constituting covering
sieves of quantum observables algebras fit into coequalizer
diagrams of the latter form [DI].

From the physical point of view covering sieves of the form
defined above, are equivalent with commutative algebras
localization systems of quantum observables algebras. These
localization systems filter the information of quantum observabes
algebras, through commutative algebras domains,  associated with
procedures of measurement of observables.  We will discuss
localizations systems in detail, in order to unravel the physical
meaning of the requirements underlying the notion of Grothendieck
topology, and subsequently, the notion of covering sieves defined
previously. It is instructive to begin with the notion of a system
of prelocalizations for a quantum observables algebra.

A {\bf system of prelocalizations} for a quantum observables
algebra ${A_{Q}}$ in $\mathcal A_Q$ is a {\it subfunctor of the
Hom-functor} ${\mathbf R}({A_{Q}})$ of the form $\mathbf S :
{\mathcal {A_{C}}}^{op} \to \bf Sets$, namely, for all ${A_{C}}$
in $\mathcal {A_{C}}$ it satisfies ${\mathbf S}({A_{C}}) \subseteq
[{\mathbf R}({A_{Q}})]({A_{C}})$. Hence, a system of
prelocalizations for a quantum observables algebra ${A_{Q}}$ in
$\mathcal A_Q$ is an {\it ideal ${\mathbf S}({A_{C}})$ of quantum
algebraic morphisms} from commutative algebras domains of the form
$${\psi}_{A_{C}} : {\mathbf M}({A_{C}}) \ar {A_{Q}}, \qquad
{A_{C}} \in {\mathcal {A_{C}}}$$ such that $\{ {\psi}_{A_{C}} :
{\mathbf M}({A_{C}}) \ar {A_{Q}}$ in ${\mathbf S}({A_{C}})$, and
${\mathbf M}(v) : {\mathbf M}({\acute {A_{C}}}) \rightarrow
{\mathbf M}({A_{C}})$ in $\mathcal A_Q$ for $v :{\acute {A_{C}}}
\rightarrow {A_{C}} $ in ${\mathcal {A_{C}}}$, implies
${\psi}_{{A_{C}}} \circ {\mathbf M}(v) : {\mathbf M}({\acute
{A_{C}}}) \ar \mathcal A_Q$ in ${\mathbf S}({A_{C}}) \}$.

The introduction of the notion of a system of prelocalizations of
a quantum observabes algebra has a sound operational physical
basis: In every concrete {\it experimental context}, the set of
observables that can be observed  in this context forms a {\it
unital commutative algebra}.  The above remark is equivalent to
the statement that a {\it measurement-induced commutative algebra
of observables} serves as a {\it local reference frame, in a
topos-theoretical environment}, relative to which a measurement
result is being coordinatized. Adopting the aforementioned
perspective on quantum observables algebras, the operation of the
Hom-functor ${\mathbf R}({A_{Q}})$ is equivalent to  depicting  an
ideal of algebraic morphisms which are to play the role of local
coverings of a quantum observables algebra, by coordinatizing
commutative algebras related with measurement situations. From a
geometrical viewpoint, we may thus characterize  the maps
${\psi}_{A_{C}} : {\mathbf M}({A_{C}}) \ar {A_{Q}}, \quad {A_{C}}
\in {\mathcal {A_{C}}}$, in a system of prelocalizations for a
quantum observables algebra ${A_{Q}}$, as a cover of ${A_{Q}}$ by
an algebra of commutative observables.

Under these intuitive identifications, we say that a family of
commutative domains covers ${\psi}_{A_{C}} : {\mathbf M}({A_{C}})
\ar {A_{Q}}, \quad {A_{C}} \in {\mathcal {A_{C}}}$, is the
generator of the system of prelocalization $\mathbf S$, iff this
system is the smallest among all that contains that family. It is
evident that a quantum observables algebra can have many systems
of measurement prelocalizations, that, remarkably, form an ordered
structure. More specifically, systems of prelocalizations
constitute a partially ordered set, under inclusion. Furthermore,
the intersection of any number of systems of prelocalization is
again a system of prelocalization.We emphasize that the minimal
system is the empty one, namely ${\mathbf S}({A_{C}}) = \emptyset$
for all ${A_{C}} \in {\mathcal {A_{C}}}$, whereas the maximal
system is the Hom-functor ${\mathbf R}({A_{Q}})$ itself, or
equivalently, all quantum algebraic morphisms  ${\psi}_{A_{C}} :
{\mathbf M}({A_{C}}) \ar {A_{Q}}$, that is the set ${{Hom}_{
{\mathcal A}_Q}({\mathbf M}({A_C}), {A_Q})}$.

The transition from a system of prelocalizations to a system of
localizations for a quantum observables algebra, can be effected
under the restriction that, certain compatibility conditions have
to be satisfied on the overlap of the coordinatizing commutative
domain covers. In order to accomplish this, we use a {\it
pullback} diagram in ${\mathcal A_Q}$ as follows:

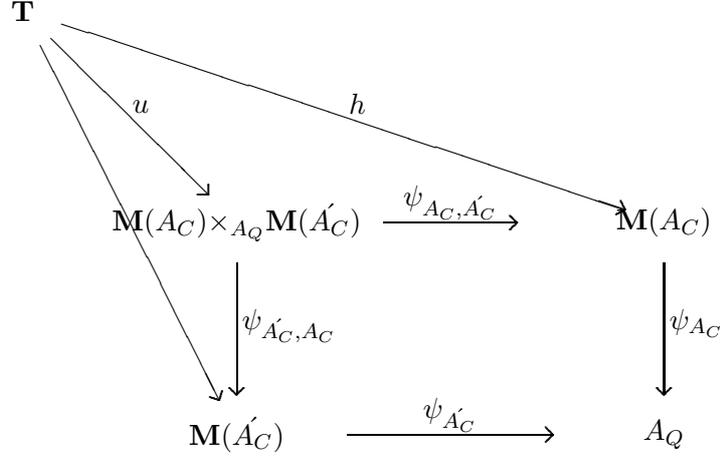
\begin{diagram}
¤{\mathbf T}¤¤ ¤  ¤\Sear u¤ ¤ \Eesear h¤ ¤ ¤¤ ¤ ¤\ssear
¤{{{\mathbf M}({A_{C}})} {\times}_{A_{Q}} {{\mathbf M}(\acute
{A_{C}})}}     ¤ ¤\Ear {{\psi}_{{A_{C}},{\acute {A_{C}}}}} ¤
¤{{\mathbf M}({A_{C}})} ¤¤
 ¤         ¤ ¤\saR {{\psi}_{{\acute {A_{C}}},{A_{C}}}}   ¤     ¤  ¤  ¤\saR  {{\psi}_{A_{C}}}  ¤¤
 ¤         ¤ ¤{{\mathbf M}({\acute {A_{C}}})}   ¤    ¤\Ear[77]  {{\psi}_{\acute {A_{C}}}}¤ ¤{A_{Q}}¤¤
\end{diagram}
The pullback of the commutative domains covers  ${\psi}_{A_{C}} :
{\mathbf M}({A_{C}}) \ar {A_{Q}},  {A_{C}} \in {\mathcal {A_{C}}}$
and ${\psi}_{\acute {A_{C}}} : {\mathbf M}({\acute {A_{C}}}) \ar
{A_{Q}}, {\acute {A_{C}}} \in {\mathcal {A_{C}}}$ with common
codomain the quantum observables algebra ${A_{Q}}$, consists of
the object ${\mathbf M}({A_{C}}) {\times}_{A_{Q}} {\mathbf
M}({\acute {A_{C}}})$ and two arrows $\psi_{{A_{C}} \acute
{A_{C}}}$ and $\psi_{\acute {A_{C}} {A_{C}}}$, called projections,
as shown in the above diagram. The square commutes and, for any
object $T$ and arrows $h$, $g$ that make the outer square commute,
there is a unique $u : T \ar {\mathbf M}({A_{C}}) {\times}_{A_{Q}}
{\mathbf M}({\acute {A_{C}}})$ that makes the whole diagram
commutative. Hence, we obtain  the condition: $${\psi}_{\acute
{A_{C}}} \circ g={\psi}_{A_{C}} \circ h$$ We notice that if
${\psi}_{A_{C}}$ and ${\psi}_{\acute {A_{C}}}$ are 1-1, then the
pullback is isomorphic with the intersection $ {\mathbf
M}({A_{C}}) \cap {\mathbf M}({\acute {A_{C}}})$. Then, we can
define the pasting map, which is an isomorphism, as follows:
$${W}_{{A_{C}}, \acute {A_{C}}} : \psi_{\acute {A_{C}}
{A_{C}}}({\mathbf M}({A_{C}}) {\times}_{A_{Q}} {\mathbf M}({\acute
{A_{C}}})) \ar
 \psi_{{A_{C}} \acute {A_{C}}}({\mathbf M}({A_{C}}) {{\times}_{A_{Q}}} {\mathbf M}({\acute {A_{C}}}))$$ by putting
$${W}_{{A_{C}}, \acute {A_{C}}}=\psi_{{A_{C}} \acute {A_{C}}}
\circ {\psi_{\acute {A_{C}} {A_{C}}}}^{-1}$$ Then we have the
following conditions: (``pull-back compatibility")

$${W}_{{A_{C}}, {A_{C}}}=1_{A_{C}}  \qquad with \qquad \qquad
1_{A_{C}} := id_{A_{C}} $$ $${W}_{{A_{C}}, \acute {A_{C}}} \circ
{W}_{\acute {A_{C}}, \acute{\acute {A_{C}}}}={W}_{{A_{C}},
\acute{\acute {A_{C}}}} \qquad if \quad {\mathbf M}({A_{C}}) \cap
{\mathbf M}({\acute {A_{C}}}) \cap {\mathbf M}({{\acute{\acute
{A_{C}}}}}) \neq 0 $$ $${W}_{{A_{C}}, \acute {A_{C}}}
={{{W}^{-1}}}_{\acute {A_{C}}, {A_{C}}} \qquad if \quad {\mathbf
M}({A_{C}}) \cap {\mathbf M}({\acute {A_{C}}}) \neq 0$$ The
pasting map provides the means to guarantee that $\psi_{\acute
{A_{C}} {A_{C}}}({\mathbf M}({A_{C}}) {{\times}_{A_{Q}}} {\mathbf
M}({\acute {A_{C}}}))$ and $ \psi_{{A_{C}} \acute
{A_{C}}}({\mathbf M}({A_{C}}) {{\times}_{A_{Q}}} {\mathbf
M}({\acute {A_{C}}}))$ are going to cover the same part of a
quantum observables algebra in a compatible way.

{\it Given a system of measurement prelocalizations for a quantum
observables algebra ${A_{Q}} \in {\mathcal A_Q}$, we call it a
{\bf system of localizations} iff the above compatibility
conditions are being satisfied}.

The compatibility conditions established, provide the necessary
relations for understanding a system of measurement localizations
for a quantum observables algebra as a structure sheaf or sheaf of
coefficients from local commutative covering domains of
observables algebras. This is related to the fact that systems of
measurement localizations are actually subfunctors of the
representable Hom-functor ${\mathbf R}({A_{Q}})$ of the form
$\mathbf S : {\mathcal {A_{C}}}^{op} \to \bf Sets$, namely,  all
${A_{C}}$ in $\mathcal {A_{C}}$  satisfy ${\mathbf S}({A_{C}})
\subseteq [{\mathbf R}({A_{Q}})]({A_{C}})$. In this sense the
pullback compatibility conditions express gluing relations on
overlaps of commutative domains covers and convert a presheaf
subfunctor of the Hom-functor (system of prelocalizations) into a
sheaf for the Grothendieck topology specified.

\subsection{Unit and Counit of the Adjunction}

We focus again our attention in the fundamental adjunction and
investigate the unit and the counit of it. For any presheaf
$\mathbf P \in {\bf Sets}^{{{\mathcal A_{C}}}^{op}}$, we deduce
that the unit ${\delta}_{\mathbf P} : \mathbf P \ar
{{Hom}_{\mathcal {A_Q}}}({\mathbf M}(\_), {\mathbf P}
{\otimes}_{{\mathcal A}_C} \mathbf M)$ has components:
$${{\delta}_{\mathbf P}}({{A_{C}}}) : {\mathbf P}({{A_{C}}}) \ar
{{Hom}_{ {\mathcal A}_Q}}({\mathbf M}({{A_{C}}}), {\mathbf P}
{\otimes}_{{\mathcal A}_C} \mathbf M)$$ for each commutative
algebra object ${{A_{C}}}$ of ${\mathcal A}_C$. If we make use of
the representable presheaf $y[{A_C}]$, we obtain:
$${\delta}_{{\mathbf y}[{A_C}]} : {\mathbf y}[{A_C}] \to
{{Hom}_{\mathcal {A_Q}}}({\mathbf M}(\_), {\mathbf y}[{A_C}]
{\otimes}_{{\mathcal A}_C} \mathbf M)$$ Hence, for each object
${A_C}$ of ${\mathcal A}_C$ the unit, in the case considered,
corresponds to a map, $${\mathbf M}({A_C}) \to {\mathbf y}[{A_C}]
{\otimes}_{{\mathcal A}_C} \mathbf M$$ But, since $$ {\mathbf
y}[{A_C}]  {\otimes}_{{\mathcal A}_C} \mathbf M  \cong {\mathbf M
\circ {\mathbf G}_{{\mathbf y}[{A_C}]}}({A_C},1_{A_C})={\mathbf
M}({A_C})$$  the unit for the representable presheaf of
commutative algebras, which is a sheaf for the Grothendieck
topology of epimorphic families, is clearly an isomorphism. By the
preceding discussion we can see that the diagram commutes

\begin{diagram}
¤{{\mathcal A}_C}¤¤ ¤\Smono {\mathbf y} ¤   ¤\Esear[133] {\mathbf
M}¤¤ ¤{\mathbf Sets}^{{{\mathcal A}_C}^{op}}¤    ¤\Ear[100] [-]
{{\otimes}_{ {\mathcal A}_Q}}  {\mathbf M}¤ ¤{ {\mathcal A}_Q}¤¤
\end{diagram}

Thus, the unit of the fundamental adjunction, referring to the
representable sheaf  ${\mathbf y}[{A_C}]$ of the category of
commutative observables algebras, provides a map (quantum
algebraic morphism) ${\mathbf M}({A_C}) \ar {\mathbf y}[{A_C}]
{\otimes}_{{\mathcal A}_C} \mathbf M$,  which is an isomorphism.

On the other side, for each quantum observables algebra object
${A_Q}$ of $ {\mathcal A}_Q$, the counit is $${\epsilon}_{A_Q} :
{{Hom}_{\mathcal {A_Q}}}({\mathbf M}(\_),{A_Q})
{\otimes}_{{\mathcal A}_C} \mathbf M \ar {A_Q}$$ The counit
corresponds to the vertical map in the following coequalizer
diagram [DII]:

\begin{diagram}
¤{{\coprod}_ {v : {{A_C} \to {{E}_C}}}}{{{\mathbf M}({A_C})}}¤ ¤
\Ebiar[70]{\zeta}{\eta}¤ ¤{{\coprod}_{({{E}_C},{\psi })}}
{{\mathbf M} ({{E}_C})}¤ ¤\ear ¤[{{\mathbf R}({A_Q})}](-)
{{\otimes}_{{\mathcal A}_C}} {\mathbf M}¤¤ ¤    ¤    ¤    ¤ ¤ ¤
¤\sear ¤\sdotar {{\epsilon}_{A_Q}} ¤¤ ¤    ¤    ¤     ¤     ¤ ¤ ¤
¤{A_Q} ¤¤
\end{diagram}
where the first coproduct is indexed by all arrows $v : {A_C}
\rightarrow {{E}_C}$, with ${A_C}$, ${{E}_C}$ objects of
${\mathcal A}_C$, whereas the second coproduct is indexed by all
objects ${A_C}$ in ${\mathcal A}_C$ and arrows $\psi : {\mathbf
M({{E}_C})} \rightarrow {A_Q}$, belonging to a covering sieve of
${A_Q}$ by objects of its generating subcategory.

It is important to notice the similarity in form of diagrams [DI]
and [DII]. Based on this observation, it is possible to prove that
if the domain of the counit of the adjunction is restricted to
sheaves for the Grothendieck topology of epimorphic families on
${\mathcal A}_C$, then the counit defines a quantum algebraic
isomorphism; $${\epsilon}_{A_Q} : {{Hom}_{ {\mathcal
A}_Q}}({\mathbf M}(\_),{A_Q}) {\otimes}_{{\mathcal A}_C} \mathbf M
\simeq {A_Q}$$

In order to substantiate our thesis, we inspect diagrams [DI] and
[DII], observing that it is enough to prove that the pairs of
arrows $(\zeta, \eta)$ and $(y_1, y_2)$ have isomorphic
coequalizers, since, then, the counit is obviously an isomorphism.
Thus, we wish to show that a covering sieve of a quantum event
algebra $$ T : {\coprod}_{({{E}_C} \in {\mathcal A}_C,\psi:\mathbf
M({{E}_C}) \rightarrow {A_Q})} \mathbf M({{E}_C}) \rightarrow
{A_Q}$$ is the coequalizer of $(y_1, y_2)$ iff it is the
coequalizer of $(\zeta, \eta)$. In the following discussion, we
may omit the explicit presence of the inclusion functor $\mathbf
M$, for the same reasons stated previously.

We consider a covering sieve of a quantum observables algebra
${A_Q}$, consisting of quantum algebraic morhisms $T_{({{E}_C},
\psi)}$, that together constitute an epimorphic family in ${
{\mathcal A}_Q}$. We observe that the condition $T\cdot y_1=T\cdot
y_2$ is equivalent to the condition [CI] as follows:
$${T_{({{E}_C}, \psi)}} \cdot l={T_{(\acute {{E}_C}, \acute
\psi)}} \cdot k$$ for each commutative square $\nu$. Furthermore,
the condition $T\cdot \zeta=T\cdot \eta$ is equivalent to the
condition [CII] as follows: $${T_{({{E}_C}, \psi)}} \cdot
u={T_{(\acute {{E}_C}, \psi \cdot u)}}$$ for every commutative
algebras morphism
 $u : \acute {{E}_C}
\rightarrow {{E}_C}$, with ${A_C}$, ${{E}_C}$ objects of
${\mathcal A}_C$ and $\psi : {{E}_C} \rightarrow {A_Q}$, belonging
to a covering sieve of ${A_Q}$ by objects of its generating
subcategory. Therefore, our thesis is proved if we show that $[CI]
\Leftrightarrow [CII]$.

On the one hand, $T\cdot \zeta=T\cdot \eta$, implies for every
commutative diagram of the form $\nu$:
\begin{diagram}
¤{A_C}  ¤\Ear l ¤ {{E}_C} ¤¤ ¤\saR k ¤ ¤\saR \psi ¤¤ ¤\acute
{{E}_C}¤\Ear {\acute \psi}¤{A_Q}¤¤
\end{diagram}
the following relations: $${T_{({{E}_C}, \psi)}} \cdot
l={T_{({A_C}, \psi \cdot l )}}={T_{({A_C},\acute \psi \cdot k
)}}={T_{(\acute {{E}_C}, \acute \psi)}} \cdot k$$ Thus $[CI]
\Rightarrow [CII]$

On the other hand, $T\cdot y_1=T\cdot y_2$, implies that for every
commutative algebras morphism
 $u : \acute {{E}_C}
\rightarrow {{E}_C}$, with ${A_C}$, ${{E}_C}$ objects of
${\mathcal A}_C$ and $\psi : {{E}_C} \rightarrow {A_Q}$, the
diagram of the form $\nu$
\begin{diagram}
¤\acute {{E}_C}  ¤\Ear u ¤ {{E}_C} ¤¤ ¤\Sar  {id}¤ ¤\saR \psi ¤¤
¤\acute {{E}_C}¤\Ear { \psi \cdot u}¤{A_Q}¤¤
\end{diagram}
commutes and provides the condition $${T_{({{E}_C}, \psi)}} \cdot
u={T_{(\acute {{E}_C},  \psi \cdot u)}}$$ Thus $[CI] \Leftarrow
[CII]$.

Consequently, the pairs  of arrows $(\zeta, \eta)$ and $(y_1,
y_2)$ have isomorphic coequalizers, proving that {\it the counit
of the fundamental adjunction restricted to sheaves for the
Grothendieck topology of epimorphic families on ${\mathcal A}_C$
is an isomorphism}.

\section{ABSTRACT DIFFERENTIAL GEOMETRY IN THE QUANTUM REGIME}

\subsection{The Quantum Quotient Algebra Sheaf of Coefficients}

Having at our disposal the {\it sheaf theoretical representation
of a quantum observables algebra} through the counit isomorphism
established above, for the Grothendieck topology of epimorphic
families of covers from local commutative algebras domains of
observables, we may attempt to apply the methodology of Abstract
(alias, Modern) Differential Geometry (ADG), in order to set up a
differential geometric mechanism suited to the quantum regime of
observables structures.

First of all we notice that the transition from the classical to
the quantum case is expressed in terms of the relevant arithmetics
used, as a transition from a commutative algebra  of observables
presented as a sheaf, if localized over a measurement topological
space, to a globally non-commutative algebra of observables,
presented correspondingly as a sheaf of locally commutative
algebras of coefficients for the Grothendieck topology specified
over the category of commutative subalgebras of the former. It is
instructive to remind that the latter sheaf theoretical
representation is established according to the counit isomorphism
by  $${\epsilon}_{A_Q} : {\mathbf R}({A_{Q}}) {\otimes}_{{\mathcal
A}_C} \mathbf M \simeq {A_Q}$$ Furthermore, we may give an
explicit form of the elements of ${\mathbf R}({A_{Q}})
{\otimes}_{{\mathcal A}_C} \mathbf M$ according to the coequalizer
of coproduct definition of the above tensor product

\begin{diagram}
¤{{\coprod}_ {v : {\acute {A_{C}}} \to {A_{C}}}}{{{\mathbf
M}(\acute {A_{C}})}}¤        ¤ \Ebiar[70]{\zeta}{\eta}¤
¤{{\coprod}_{({A_{C}},{{\psi}_{A_C}})}} {{\mathbf M}({A_{C}})}
¤\Ear {\chi}¤ ¤{\mathbf R}({A_{Q}}) {{\otimes}_{\mathcal A_C}}
{\mathbf M}¤¤
\end{diagram}

According to the diagram above for elements ${{\psi}_{A_C}} \in
{\mathbf R}({A_{Q}}) ({A_{C}})$, $v : {\acute {A_{C}}} \to
{A_{C}}$ and $\acute \xi \in {\mathbf M}({\acute {A_{C}}})$, the
following equations hold: $$\zeta (\psi_{A_C},v, \acute
\xi)=(\psi_{A_C} v, \acute \xi), \qquad \eta(\psi,v, {\acute
\xi})=(\psi_{A_C}, v \acute \xi)$$ symmetric in ${\mathbf
R}({A_{Q}}) $ and $\mathbf M$. Hence the elements of ${\mathbf
R}({A_{Q}}) {\otimes}_{\mathcal A_C} {\mathbf M}$ are all of the
form $\chi (\psi_{A_C},\xi)$. This element can be written as $$
\chi(\psi_{A_C},\xi)=\psi_{A_C} \otimes \xi, \quad  \psi_{A_C} \in
{\mathbf R}({A_{Q}}) ({A_{C}}), \xi \in {\mathbf M}({A_{C}})$$
Thus if we take into account the definitions of $\zeta$ and $\eta$
above, we obtain $$\psi_{A_C} v \otimes \acute \xi=\psi_{A_C}
\otimes v \acute \xi$$ We conclude that  ${\mathbf R}({A_{Q}})
{{\otimes}_{\mathcal A_C}} {\mathbf M}$ is actually the quotient
of ${{\coprod}_{({A_{C}},{{\psi}_{A_C}})}} {{\mathbf M}({A_{C}})}$
by the smallest equivalence relation generated by the above
equations. Moreover, if there exists $A_{D}$ in ${\mathcal A_C}$
and homomorphisms $w: A_D \rightarrow A_C$, $v: A_D \rightarrow
\acute A_{C}$, such that: $w \bar{\xi}=\xi$, $v \bar{\xi}=\acute
\xi$, and $\psi_{A_C} w=\psi_{\acute {A_{C}}}$, $\xi \in {\mathbf
M}({A_{C}}),\acute \xi \in {\mathbf M}({\acute A_{C}}),\bar{\xi}
\in {\mathbf M}({A_{D}}), \psi_{A_C} \in {\mathbf R}({A_{Q}})
({A_{C}}),\acute  \psi_{A_C} \in {\mathbf R}({A_{Q}}) ({\acute
A_{C}})$ then the identification equations take the form
$$\psi_{A_C} \otimes \xi=\psi_{\acute {A_C}} \otimes \acute \xi$$
If we denote by $l_Q(A_{C})$  the ideal generated by the
equivalence relation, corresponding to the above identification
equations, for each $A_{C}$ in ${\mathcal A_C}$, we conclude that
locally, in the Grothendieck topology defined, an element of
${\mathbf R}({A_{Q}}) {{\otimes}_{\mathcal A_C}} {\mathbf M}$ can
be written in the form: $$\psi_{A_C} \otimes \xi=(\psi_{A_C},
\xi)+ l_Q(A_{C})\equiv [\psi_{A_C} \otimes \xi]$$ Subsequently a
quantum observables algebra admits a sheaf theoretical
representation in terms of an algebra sheaf that, locally, that
is, over a particular cover, has the quotient form; $${\mathbf
R}({A_{Q}}) ({A_{C}}) {{\otimes}_{\mathcal A_C}} {\mathbf M}
({A_{C}})= [{\mathbf R}({A_{Q}}) ({A_{C}}) \times {{\mathbf
M}({A_{C}})}] \slash l_Q(A_{C})$$ In this sense, the quantum
arithmetics  can be described locally, that is, over a particular
cover in a localization system of a quantum observables algebra,
as an algebra ${\mathbf K} (A_C):=[{\mathbf R}({A_{Q}}) ({A_{C}})
\times {{\mathbf M}({A_{C}})}] \slash l_Q(A_{C})$. The latter can
be further localized over a ``topological measurement space",
categorically dual to the commutative observables algebra
${A_{C}}$, that serves as the {\it algebra sheaf of a differential
triad} $\Delta=({\mathbf K} (A_C), \delta_{{\mathbf K} (A_C)},
\Omega({\mathbf K} (A_C)))$, attached to this particular cover.
The appropriate specification of the ${\mathbf K} (A_C)$-module
$\Omega({\mathbf K} (A_C))$ is going to be the subject of a
detailed discussion in what follows: From a physical viewpoint a
reasonable choice would be the identification of $\Omega({\mathbf
K} (A_C))$ with the ${\mathbf K} (A_C)$-module $\Phi({\mathbf K}
(A_C))$ of all localized quotient commutative algebra of
observables sheaf endomorphisms $\nabla_{{\mathbf K} (A_C)}:
{\mathbf K} (A_C) \rightarrow {\mathbf K} (A_C)$, which are
$\mathcal R$-linear and satisfy the Leibniz rule (``derivations").
Thus the differential structure on a local commutative domain
cover, ${\psi}_{A_{C}} : {\mathbf M}({A_{C}}) \hookrightarrow
{A_{Q}}$, $ {A_{C}}$ in ${\mathcal {A_{C}}}$, being an inclusion,
would be naturally defined in the following manner: $$(\psi_{A_C},
\xi)+ l_Q(A_{C}) \mapsto (\psi_{A_C},\nabla_{A_C} \xi)+
l_Q(A_{C})$$ where $\nabla_{A_C}:=\nabla: A_C \rightarrow A_C$ is
an $A_C$-valued derivation of $A_C$, which we call differential
variation of first-order, or equivalently differential
$1$-variation, applied to the observable $\xi$. In the sequel, we
will specify the necessary conditions required for the existence
of $\nabla$ for a general commutative algebra of observables
$A_C$.

In this sense, we may form the conclusion that locally in the
Grothendieck topology specified, there exists a naturally defined
differential operator, that has the following form over a
particular cover for each $A_{C}$ in ${\mathcal A_C}$: $$
\delta_{{\mathbf K} (A_C)}(\psi_{A_C} \otimes
\xi):=(\psi_{A_C},\nabla_{A_C} \xi)+ l_Q(A_{C})$$ At this point we
remind that a covering sieve, or equivalently, localization system
of a quantum observables algebra contains epimorphic families from
local commutative domain covers, such that each element associated
with a covering sieve is represented by a commutative diagram of
the form

\begin{diagram}
¤{A_C}  ¤\Ear l ¤ {{E}_C} ¤¤ ¤\saR k ¤ ¤\saR \psi ¤¤ ¤\acute
{{E}_C}¤\Ear {\acute \psi}¤{A_Q}¤¤
\end{diagram}
where ${A_C}$, ${{E}_C}$, $\acute {{E}_C}$ are objects in the
generating subcategory ${\mathcal A}_C$ of  $ {\mathcal A}_Q$.

Moreover they fit all together in a coequalizer diagram
\begin{diagram}
¤{\coprod}_{\nu} {A_C} ¤ ¤\Ebiar[70] {y_1} {y_2}¤ ¤
¤{\coprod}_{({{E}_C} \in {\mathcal A}_C, \psi:{{E}_C}\rightarrow
{A_Q})} {{E}_C}¤ ¤\Ear T¤{A_Q}¤¤
\end{diagram}
where the first coproduct is indexed by all $\nu$, representing
commutative diagrams in $ {\mathcal A}_Q$ of the form above.

Thus, having specified a differential triad $\Delta=({\mathbf K}
(A_C), \delta_{{\mathbf K} (A_C)}, \Omega({\mathbf K} (A_C)))$
attached to each particular cover, we may specify a diagram of
differential triads that, in turn corresponds to an element
associated with an epimorphic covering sieve in the Grothendieck
topology defined on $ {\mathcal A}_C$. This diagram of
differential triads, together with the corresponding coequalizer
of coproduct diagram, contain all the information necessary for
the set-up of the differential geometric mechanism suited to the
quantum regime. {\it Hence, the transition from the classical to
the quantum case amounts to a change of perspective from a single
differential triad to a diagram of differential triads
interlocking in such a way that information related to observation
in different covering domains is compatible on their overlaps}.

\subsection{Differential $1$-variations}
A derivation $\nabla$ of a commutative observables algebra $A_C$
is an $\mathrm{R}$-linear endomorphism of the
$\mathrm{R}$-commutative arithmetic $A_C$, denoted by $\nabla:A_C
\rightarrow A_C$, that satisfies the Leibniz rule: $$\nabla (\zeta
\xi)=\zeta \nabla(\xi) + \xi \nabla(\zeta)$$ for all $\zeta$,
$\xi$ belonging to $A_C$.

We also define the set of all derivations of $A_C$, denoted by
$G(A_C)$. It is obvious that $G(A_C)$ is a left $A_C$-module.
Remarkably, $G(A_C)$ can be also endowed with a Lie algebra
structure if we define an $\mathrm{R}$-linear skew-symmetric
operator, called commutator of derivations in $G(A_C)$ as follows:

For any two derivations $\nabla_1$, $\nabla_2$ $\in$ $G(A_C)$
their commutator, denoted as $[\nabla_1, \nabla_2]$, is given by;
$$[\nabla_1, \nabla_2]=\nabla_1 \circ \nabla_2 - \nabla_2 \circ
\nabla_1$$ We can easily check that the commutator $[\nabla_1,
\nabla_2]$ is skew-symmetric, and also, it is a derivation
belonging to $G(A_C)$. Furthermore, the commutator derivation
satisfies the Jacobi identity as follows; $$[\nabla_1, [\nabla_2,
\nabla_3]]=[[\nabla_1, \nabla_2], \nabla_3] + [\nabla_2,
[\nabla_1, \nabla_3]]$$ Actually if we consider $\nabla \in
G(A_C)$, and also, $\zeta, \xi \in A_C$, then we define;
$$(\overleftarrow {\zeta} \nabla)(\xi):=\zeta(\nabla (\xi))$$ It
is clear that $\overleftarrow { \zeta} \nabla \in G(A_C)$, thus
$G(A_C)$ is a left $A_C$-module. We notice that we can also define
a right $A_C$-module structure on $G(A_C)$ according to the rule;
$$(\overrightarrow {\zeta} \nabla)(\xi):=\nabla (\zeta \xi)$$ Now,
we may define a commutator as follows; $$[\hat {\zeta},
\nabla](\xi):=(\overrightarrow {\zeta} \nabla - \overleftarrow
{\zeta} \nabla)(\xi)=(\nabla (\zeta)) \xi$$ Thus, for any $\zeta
\in A_C$, we can define the Lie derivative operator;
$${{L}_\zeta}: G(A_C) \rightarrow G(A_C)$$
$${{L}_\zeta}(\nabla):=[\hat {\zeta}, \nabla]$$ Moreover, if we
consider operators ${{L}_\zeta}$, ${{L}_\eta}$, we can easily show
that they commute, and furthermore, the identity below is being
satisfied; $$({{L}_\eta} \circ {{L}_\zeta}) (\nabla)=0$$ for every
$\zeta$, $\eta$ $\in$ $A_C$. Thus we can state the following:

{\it If we consider a commutative observables algebra $A_C$, then
an $\mathrm{R}$-linear morphism $\nabla \in G(A_C)$ is called a
differential $1$-variation if for all $\eta$, $\zeta$ $\in$ $A_C$,
and corresponding commutator operators ${{L}_\eta}$,
${{L}_\zeta}$, the following identity holds:} $$({{L}_\eta} \circ
{{L}_\zeta})(\nabla)=0$$

In the case that the classical commutative arithmetic $A_C$
represents $\mathcal {C^\infty}(X, R)$, then the above identity is
satisfied and differential $1$-variations are tautosemous with the
usual fist-order linear differential operators of the form
$\nabla=\kappa^i \partial_i + \lambda$, where $\kappa^i, \lambda
\in \mathcal {C^\infty}(X, R)$.

The fact that the set of all derivations of $A_C$, say $G(A_C)$,
has an $A_C$-module structure, motivates the definition of an
$M$-valued derivation of an observables algebra $A_C$, for an
arbitrary $A_C$-module $M$ as follows:

An $M$-valued derivation $\nabla_M$ of an observables algebra
$A_C$ is an $\mathrm{R}$-linear morphism, denoted by $\nabla_M:A_C
\rightarrow M$, that satisfies the Leibniz rule: $$\nabla_M (\zeta
\xi)=\zeta \nabla_M (\xi) + \xi \nabla_M (\zeta)$$ for all
$\zeta$, $\xi$ belonging to $A_C$. If we consider $\nabla_M \in
G(M)$, and also, $\zeta, \xi \in A_C$, then we have;
$$(\overleftarrow {\zeta} \nabla_M)(\xi):=\zeta(\nabla_M(\xi))$$
It is clear that $\overleftarrow { \zeta} \nabla_M \in G(M)$, thus
$G(M)$ is a left $A_C$-module. We also notice that we can define a
right $A_C$-module structure on $G(M)$ as follows;
$$(\overrightarrow {\zeta} \nabla_M)(\xi):=\nabla_M(\zeta \xi)$$
Now, we can define a commutator according to; $$[\hat {\zeta},
\nabla_M](\xi):=(\overrightarrow {\zeta} \nabla_M - \overleftarrow
{\zeta} \nabla_M)(\xi)=(\nabla_M(\zeta)) \xi$$ Thus, for any
$\zeta \in A_C$, we can again define the Lie derivative operator,
as follows; $${{L}_\zeta}: G(M) \rightarrow G(M)$$
$${{L}_\zeta}(\nabla_M):=[\hat {\zeta}, \nabla_M]$$  Furthermore,
if we consider operators ${{L}_\zeta}$, ${{L}_\eta}$, for every
$\zeta$, $\eta$ $\in$ $A_C$,  they commute, and also, satisfy the
identity; $$({{L}_\eta} \circ {{L}_\zeta}) (\nabla_M)=0$$ In this
sense, we can state the criterion of identification of $M$-valued
derivations, in an analogous manner as in $8.2$, as follows:

{\it If we consider the set $S(M)$, consisting of
$\mathrm{R}$-linear morphisms of a commutative observables algebra
$A_C$ into an arbitrary $A_C$-module $M$, then the elements of
$S(M)$ are identified as $M$-valued derivations, $\nabla_M$, of
the algebra $A_C$, if for any $\zeta$, $\eta$ $\in$ $A_C$, and
corresponding Lie derivative operators ${{L}_\zeta}$, ${{L}_\eta}$
from $S(M)$ to itself, the following identity holds: $$({{L}_\eta}
\circ {{L}_\zeta}) (\nabla_M)=0$$ }

Furthermore, it is instructive to notice that, for an
$\mathrm{R}$-linear morphism $\theta$ of $A_C$-modules, $M$, $N$
$\in$ $S(M,N)$; where $S(M,N)$ denotes the bimodule of all
$\mathrm{R}$-linear morphisms of $A_C$-modules $M$ and $N$, we can
analogously define a commutator operator ${{\hat L}_\zeta}$, for
every $\zeta$ $\in$ $A_C$, according to; $${{\hat L}_\zeta}:
S(M,N) \rightarrow S(M,N)$$ such that; $${{\hat
L}_\zeta}(\theta):=[\hat \zeta, \theta]=(\overrightarrow {\zeta}
\theta - \overleftarrow {\zeta} \theta)$$ Thus, we can consider
commutator operators ${{\hat L}_\eta}$,  ${{\hat L}_\zeta}$, for
$\eta$, $\zeta$ $\in$ $A_C$, and also, take their composition,
denoted by ${{\hat L}_\eta} \circ {{\hat L}_\zeta}$. Then we can
give the following definition:

{\it We consider an observables algebra $A_C$, and let $M$, $N$ be
$A_C$-modules. An $\mathrm{R}$-linear morphism $\theta \in S(M,N)$
is called a differential $1$-variation induced by the action of
$M$ on $N$ if for all $\eta$, $\zeta$ $\in$ $A_C$, and
corresponding commutator operators ${{\hat L}_\eta}$, ${{\hat
L}_\zeta}$, the following identity holds:} $$({{\hat L}_\eta}
\circ {{\hat L}_\zeta})(\theta)=0$$

Let us denote the set of all differential $1$-variations induced
by the action of $M$ on $N$, by ${{V^1}_{A_C}} (M,N)$. The set
${{V^1}_{A_C}} (M,N)$ can be endowed with a bimodule structure,
where multiplication from the left by elements $\zeta$ of $A_C$ is
denoted by $\overleftarrow {\zeta} \theta$, whereas multiplication
from the right is denoted by $\overrightarrow {\zeta} \theta$,
according to; $$(\overleftarrow {\zeta} \theta)(m):=\zeta \cdot
\theta(m)$$ $$(\overrightarrow {\zeta} \theta)(m):=(\theta \circ
\zeta)(m)$$ for every $\zeta \in A_C$. We also denote by ${{{\hat
V}^1}_{A_C}} (M,N)$ the bimodule structure, whereas by
${{{\overleftarrow V}^1}_{A_C}} (M,N)$ and ${{{\overrightarrow
V}^1}_{A_C}} (M,N)$, the left and right $A_C$-module structures,
respectively.  Moreover, the bimodule of all differential
$1$-variations, induced by the action of $A_C$, being an
$A_C$-module over itself, on $N$, is denoted by ${{{\hat
V}^1}_{A_C}} (N)$. We also denote by
${{\overleftarrow{V}^1}_{A_C}} (N)$ the left $A_C$-module
structure, whereas by ${{\overrightarrow{V}^1}_{A_C}} (N)$ the
right $A_C$-module structure.

\subsection{ Left Modules of $1$-Forms}

From now on, we shall focus our attention to the left module
structure alone. The correspondence $N \mapsto {{{\overleftarrow
V}^1}_{A_C}} (M,N)$ if applied to all objects and arrows of the
category of $A_C$-modules $\mathcal M^{(A_C)}$, specifies a
covariant functor from the category of $A_C$-modules to
themselves;

$${{{\overleftarrow {\mathbf V}}^1}_{A_C}}(M,-): \mathcal
M^{(A_C)} \rightarrow {\mathcal M}^{(A_C)}$$ Furthermore, we
define the $\mathrm{R}$-linear map; $$l: M \rightarrow (A_C)
{\otimes}_\mathrm{R} M$$ by setting; $$l(m)= 1 {\otimes} m$$ where
$m$ $\in$ $M$. The codomain of the map $l$ is called the tensor
product of the left $A_C$-modules $A_C$ and $M$, and most
significantly it is an $A_C$-module itself, where the left
multiplication is specified by; $$\overleftarrow{\zeta}(\xi
\otimes m):=(\zeta \xi) \otimes m$$ where $\zeta$, $\xi$ $\in$
$A_C$, and $m$ $\in$ $M$. The tensor product $(A_C)
{\otimes}_\mathrm{R} M$, can be further endowed with a right
$A_C$-module structure defined by $$\overrightarrow{\zeta} (\xi
\otimes m):=\xi \otimes (\zeta m)$$

Thus, we may form a commutator operator for every $\zeta$ $\in$
$A_C$ defined as follows; $${{\hat L}_\zeta}: (A_C)
{\otimes}_\mathrm{R} M \rightarrow (A_C) {\otimes}_\mathrm{R} M$$
$${{\hat L}_\zeta}(l(m)):=[\hat \zeta, l(m)]=(\overrightarrow
{\zeta} l(m) - \overleftarrow {\zeta} l(m))$$ Subsequently, we can
consider for $\eta$, $\zeta$ $\in$ $A_C$, the corresponding
commutator operators ${{\hat L}_{\eta}}$, ${{\hat L}_{\zeta}}$,
and also, take their composition. Consequently,  the elements
$(({{\hat L}_{\eta}} \circ {{\hat L}_{\zeta}})(l))(m)$ generate a
submodule of the tensor product $(A_C) {\otimes}_\mathrm{R} M$,
denoted by ${\underline{M}}$. Moreover, we may form a quotient
$A_C$-module corresponding to each $A_C$-module $M$, defined as
follows;

$$\pi(M):= \big((A_C) {\otimes}_\mathrm{R} M \big ) /
{\underline{M}}$$ It is straightforward to see, if we take into
account the definition of the quotient $A_C$-module
${\underline{M}}$, that the map ; $$\Pi: M \rightarrow \pi(M)$$
defined by the assignment $$m \mapsto {\Pi}(m):=(l(m))
mod({\underline{M}}):=[l(m)]$$ for each $A_C$-module $M$, is a
differential $1$-variation.

Moreover, the above map for each $A_C$-module $M$, gives rise to a
covariant functor $${\mathbf {\Pi}}: \mathcal M^{(A_C)}
\rightarrow {\mathcal M}^{(A_C)}$$

[i]. Its action on a $A_C$-module in $\mathcal M^{(A_C)}$ is given
by; $${{\mathbf {\Pi}}}(M):=\pi(M)$$

[ii]. Its action on a morphism of $A_C$-modules $\alpha: M
\rightarrow N$, for $[l(m)] \in \pi(M)$ is given by; $${{\mathbf
{\Pi}}}(\alpha): {{\mathbf {\Pi}}}(M) \rightarrow {{\mathbf
{\Pi}}}(N)$$

$${{\mathbf {\Pi}}}(\alpha)([l(m)] )=\alpha \circ [l(m)]$$ Now, we
consider that $\theta$ is a differential $1$-variation, that is
$\theta \in {{{\overleftarrow {\mathbf V}}}_{A_C}}(M,N)$.
Obviously, ${{{\overleftarrow {\mathbf V}}}_{A_C}}(M,N) \subset
{Hom}_{\mathrm{R}}(M,N)$, so we may further consider the morphism;
$$\chi: {Hom}_{A_C}(\big((A_C) {\otimes}_\mathrm{R} M \big ),N)
\rightarrow {Hom}_{\mathrm{R}}(M,N)$$ defined by the relation;
$$\chi(\tau)=\tau \circ l$$ Next, we apply the commutator operator
${{\hat L}_\zeta}$ on $\chi(\tau)$, taking into account that
$\tau$ is an $A_C$-morphism, as follows; $${{\hat
L}_\zeta}(\chi(\tau))={{\hat L}_\zeta}(\tau \circ l)=\tau \circ
{{\hat L}_\zeta}(l)=\chi({{\hat L}_\zeta}(\tau))$$ Consequently,
$\chi(\tau) \in {Hom}_{\mathrm{R}}(M,N)$, is a differential
$1$-variation, iff $${{\hat L}_\zeta}(\chi(\tau))=0$$ or
equivalently, iff; $$\tau({\underline{M}})=0$$ Thus, by
restricting the codomain of $\chi$ to elements being qualified as
differential $1$-variations, we obtain the following isomorphism;
$$\iota: {Hom}_{A_C}({{\mathbf {\Pi}}}(M),N) \rightarrow
{{{\overleftarrow {\mathbf V}}}_{A_C}}(M,N)$$ Its inverse is
denoted by $\varepsilon$ and is subsequently defined as;
$$\varepsilon: {{{\overleftarrow {\mathbf V}}}_{A_C}}(M,N)
\rightarrow {Hom}_{A_C}({{\mathbf {\Pi}}}(M),N)$$ $$\theta \mapsto
\varepsilon_\theta$$ $$\theta=\varepsilon_\theta \circ \Pi$$
according to the diagram below;
\begin{diagram}
¤ ¤ ¤{{{\mathbf {\Pi}}}(M)}¤  ¤ ¤¤ ¤ ¤\Near {\Pi} ¤ ¤\Sear
{\varepsilon_\theta}¤ ¤¤ ¤{M}¤ ¤\Ear[133]{\theta} ¤ ¤{N}   ¤¤
\end{diagram}
{\it Hence, we can draw the conclusion that the covariant functor
corresponding to a left $A_C$-module $M$ in $\mathcal M^{(A_C)}$;
$${{{\overleftarrow {\mathbf V}}}_{A_C}}(M,-): \mathcal M^{(A_C)}
\rightarrow {\mathcal M}^{(A_C)}$$ is being representable by the
left $A_C$-module in $\mathcal M^{(A_C)}$; $${{\mathbf
{\Pi}}}(M):= \big((A_C) {\otimes}_\mathrm{R} M \big ) /
{\underline{M}}$$ according to the established isomorphism;
$${{{\overleftarrow {\mathbf V}}}_{A_C}}(M,N) \cong
{Hom}_{A_C}({{\mathbf {\Pi}}}(M),N)$$} As a consequence, if we
consider the case $M=A_C$, we obtain; $${{{\overleftarrow {\mathbf
V}}}_{A_C}}(N) \cong {Hom}_{A_C}({{\mathbf {\Pi}}}(A_C),N)$$
\begin{diagram}
¤ ¤ ¤{{{\mathbf {\Pi}}}(A_C)}¤  ¤ ¤¤ ¤ ¤\Near {\Pi} ¤ ¤\Sear
{\varepsilon_{\tilde{\theta}}}¤ ¤¤ ¤{A_C}¤
¤\Ear[133]{\tilde{\theta}} ¤ ¤{N} ¤¤
\end{diagram}
where the map, $$\Pi: A_C \rightarrow {{{\mathbf {\Pi}}}(A_C)}$$
is defined by the assignment $$\zeta \mapsto
{\Pi}(\zeta)=[l(\zeta)]=[1 \otimes \zeta]$$ Now, we may form the
quotient left $A_C$-module $\Omega^1(A_C)$ defined as follows;
$$\Omega^1(A_C):={{{\mathbf {\Pi}}}(A_C)} / {Im({\Pi}})$$ where
${Im({\Pi}})$ denotes the submodule of ${{{\mathbf {\Pi}}}(A_C)}$
depicted by the image of the morphism $\Pi$. There exists a
natural projection mapping defined by; $${pr}: {{{\mathbf
{\Pi}}}(A_C)} \rightarrow \Omega^1(A_C)$$ So, we may form the
composition; $$d_{A_C}: A_C \rightarrow \Omega^1(A_C)$$
$$d_{A_C}:={pr} \circ {{\mathbf {\Pi}}}$$ Then, $d_{A_C}$ is
clearly an $\Omega^1(A_C)$-valued derivation of $A_C$.

{\it In a suggestive terminology, $d_{A_C}$ is called a first
order differential of the observables algebra $A_C$, whereas the
left $A_C$-module $\Omega^1(A_C)$ is characterized as the module
of $1$-forms of $A_C$. In this sense, a differential $1$-variation
is tautosemous with a first order differential of $A_C$, evaluated
on $1$-forms in $\Omega^1(A_C)$.}

Consequently, we may further consider the following commutative
diagram;
\begin{diagram}
¤ ¤ ¤\Omega^1(A_C)¤  ¤ ¤¤ ¤ ¤\Near {d_{A_C}} ¤ ¤\Sear
{\varepsilon_{\tilde{[\theta]}}}¤ ¤¤ ¤{A_C}¤
¤\Ear[133]{{\tilde{\theta}}\equiv {\nabla_N}} ¤ ¤{N} ¤¤
\end{diagram}
We conclude that for any $N$-valued derivation
${{\tilde{\theta}}\equiv {\nabla_N}}$ of $A_C$, there exists a
uniquely defined morphism ${\varepsilon_{\tilde{[\theta]}}}:
\Omega^1(A_C) \rightarrow N$ making the diagram above commutative.

{\it In functorial language the statement above means that the
covariant functor of left $A_C$-modules valued derivations of
$A_C$; $$\overleftarrow {\mathcal {{\nabla}}}_{A_C}: \mathcal
M^{(A_C)} \rightarrow \mathcal M^{(A_C)}$$ is being representable
by the left $A_C$-module of $1$-forms in $\mathcal M^{(A_C)}$;
$$\Omega^1(A_C):={{{\mathbf {\Pi}}}(A_C)} / {Im({\Pi)}}$$ for
every commutative arithmetic $A_C$, according to the isomorphism;
$$\overleftarrow {\mathcal {{\nabla}}}_{A_C}(N)\cong
{Hom}_{A_C}(\Omega^1(A_C),N)$$}

Consequently, the conclusion stated above resolves completely the
issue related with the appropriate specification of the ${\mathbf
K} (A_C)$-module $\Omega({\mathbf K} (A_C))$ in 8.1. If we remind
the relevant discussion, it has been initially conjectured that
from a physical viewpoint, a reasonable choice would be the
identification of $\Omega({\mathbf K} (A_C))$ with the ${\mathbf
K} (A_C)$-module $\Phi({\mathbf K} (A_C))$ of all derivations,
that is localized arithmetics endomorphisms $\nabla_{{\mathbf K}
(A_C)}: {\mathbf K} (A_C) \rightarrow {\mathbf K} (A_C)$, which
are $\mathcal R$-linear and satisfy the Leibniz rule. From the
isomorphism established above, the covariant functor of ${\mathbf
K} (A_C)$-modules valued derivations of ${\mathbf K} (A_C)$ is
being representable by the ${\mathbf K} (A_C)$-module of $1$-forms
in $\mathcal M^{{\mathbf K} (A_C)}$; Hence, we finally identify
the ${\mathbf K} (A_C)$-module $\Omega({\mathbf K} (A_C))$ in 8.1
with the ${\mathbf K} (A_C)$-module of $1$-forms
$\Omega^1({\mathbf K} (A_C))$ and from now on we use them
interchangeably.

Summarizing and recapitulating, we state that the differential
structure on a local commutative domain cover, ${\psi}_{A_{C}} :
{\mathbf M}({A_{C}}) \hookrightarrow {A_{Q}}$, $ {A_{C}}$ in
${\mathcal {A_{C}}}$, being an inclusion,  is defined as follows:
$$(\psi_{A_C}, \xi)+ l_Q(A_{C}) \mapsto (\psi_{A_C},d_{A_C} \xi)+
l_Q(A_{C}) \equiv [(\psi_{A_C},d_{A_C} \xi)]$$ Hence, locally in
the Grothendieck topology specified, there exists a naturally
defined differential operator, that has the following form over a
particular cover for each $A_{C}$ in ${\mathcal A_C}$; $$
d_{{\mathbf K} (A_C)}(\psi_{A_C} \otimes \xi):=(\psi_{A_C},d_{A_C}
\xi)+ l_Q(A_{C}) \equiv [(\psi_{A_C},d_{A_C} \xi)]$$

\subsection{Non-Local Information Encoded in Ideals}

If we focus our attention to a localization system of compatible
overlapping commutative domain covers, we can specify accurately
the information encoded in the ideal $l_Q(A_{C})$ in ${\mathbf K}
(A_C)$, where  ${\psi}_{A_{C}} : {\mathbf M}({A_{C}})
\hookrightarrow {A_{Q}}$, $ {A_{C}}$ in ${\mathcal {A_{C}}}$,
stands for a local cover belonging to this system. More concretely
the ideal $l_Q(A_{C})$ contains  information about all the other
covers in the localization system that are compatible in pullback
diagrams over ${A_{Q}}$ with the specified one. This is evident if
we inspect the isomorphism pasting map ${W}_{{A_{C}}, \acute
{A_{C}}}=\psi_{{A_{C}} \acute {A_{C}}} \circ {\psi_{\acute {A_{C}}
{A_{C}}}}^{-1}$ and noticing that its existence guarantees the
satisfaction of the relations needed, as has been explained
previously, for the establishment of the identification equations
$\psi_{A_C} \otimes \xi=\psi_{\acute {A_C}} \otimes \acute \xi$ in
the localization system. Thus, essentially the information encoded
in the ideal $l_Q(A_{C})$ refers to all other local covers that
are compatible with the specified one in the localization system.
This is a unique peculiar characteristic of the quantum
arithmetics as substantiated in the form of the algebras ${\mathbf
K} (A_C)$. Remarkably, each one of them in a covering sieve
contains information about all the others in the same sieve that
can be made compatible, and explicitly, the content of this
information is encoded in the structure of an ideal. This is a
crucial observation and pertains to discussions of non-locality
characterizing the behavior of quantum systems. In our perspective
the assumed paradoxical behaviour of quantum systems exhibiting
non-local correlations  stems  from two factors: The first factor
has to do with the employment of supposedly unrelated classical
arithmetics, while the second stems from the identification of the
general notion of localization in the sense of Grothendieck with
the restricted notion of spatial localization. These two factors,
of course are intimately connected, since if somebody sticks
blindly to the notion of spatial localization, that works nicely
for a space of points but is completely inadequate to function in
a category of points, is not possible to think of a correlation of
arithmetics that are spatially employed far apart from each other
for the description of the observables of the same quantum system,
that can even be the whole universe itself. This is only possible
if the notion of localization is detached from its spatial
connotation, as it is the case with Grothendieck localization in
categories. We have seen in detail how the functioning of covering
sieves permits the conception of localization systems in a
generalized topological sense and subsequently the natural
appearance of commutative local arithmetics correlated by means of
compatible information content.

\subsection{The Abstract De Rham Complex}
We consider the differential triad $\Delta=({\mathbf K} (A_C),
d_{{\mathbf K} (A_C)}, \Omega({\mathbf K} (A_C)))$ that has been
attached to each particular cover in a localization system of a
quantum observables algebra. We further localize over a
topological measurement space $X$, that we may consider as a
nonvoid open subset in $R^n$, or an $n$-dimensional manifold. In
this setting we assume that the classical commutative arithmetic
$A_C$ represents $\mathcal {C^\infty}(X, R)$, whereas its
corresponding module of variations is the respective set of
1-forms.

Now, given the differential triad $\Delta=({\mathbf K} (A_C),
d_{{\mathbf K} (A_C)}, \Omega({\mathbf K} (A_C)))$ localized
sheaf-theoretically over a finite open covering $U=(U_a)$ of $X$
as above, we define algebraically, for each $n \in N$, $n \geq 2$
the $n$-fold exterior product ${\Omega^n({\mathbf K}
(A_C))}={{\bigwedge}^n} {\Omega^1({\mathbf K} (A_C))}$, where
$\Omega({\mathbf K} (A_C)):={\Omega^1({\mathbf K} (A_C))}$.

Furthermore, we assume the existence of an $R$-linear sheaf
morphism $d^1: {\Omega^1({\mathbf K} (A_C))} \rightarrow
{\Omega^2({\mathbf K} (A_C))}$, satisfying the Leibniz rule as
follows: $$d^1(ft)=fd^1(t)+\vartheta (f) \wedge t$$ for every $f
\in {\mathbf K} (A_C)(U)$, $t \in {\Omega^1({\mathbf K}
(A_C))(U)}$, $U \subseteq X$. Moreover, we require that $d^1 \circ
d^0=0$, where $d^0:=d_{{\mathbf K} (A_C)}$.

Based on the above, we can now further construct the $R$-linear
sheaf morphism $d^2: {\Omega^2({\mathbf K} (A_C))} \rightarrow
{\Omega^3({\mathbf K} (A_C))}$, satisfying: $$d^2(t \wedge r)=t
\wedge d^1 (r) + d^1 (t) \wedge r$$ where $t, r \in
{\Omega^1(A_C)(U)}$, $U \subseteq X$. Finally, we may assume that
$d^2$ satisfies: $d^2 \circ d^1=0$.

Thus, by iteration, for each $n \in N$, $n \geq 3$ we can
construct the $R$-linear sheaf morphism $d^n: {\Omega^n({\mathbf
K} (A_C))} \rightarrow {\Omega^{n+1} ({\mathbf K} (A_C))}$,
satisfying: $$d^n(t \wedge r)=(-1)^{n-1} t \wedge d^1(r) + d^{n-1}
(t) \wedge r$$ where $t \in {\Omega^{n-1}({\mathbf K} (A_C))(U)}$,
$r \in {\Omega^1 ({\mathbf K} (A_C))(U)}$, $U \subseteq X$.

In the above framework we obtain the following relations: $$d^3
\circ d^2=d^4 \circ d^3=\ldots=d^{n+1} \circ d^n=\ldots=0$$ where
$n \in N$, $n \geq 2$. This fact allows the construction of the de
Rham complex in our case, as a complex of $R$-linear sheaf
morphisms as follows:

$$0 \rightarrow R \rightarrow {\mathbf K} (A_C) \rightarrow
{\Omega^1({\mathbf K} (A_C))} \rightarrow {\Omega^2({\mathbf K}
(A_C))} \rightarrow {\ldots}$$ Now, if we remind that ${\mathbf K}
(A_C)$ consists of elements of the form $\psi_{A_C} \otimes
\xi=(\psi_{A_C}, \xi)+ l_Q(A_{C})$, as well as, that the
differential structure is defined by means of $(\psi_{A_C}, \xi)+
l_Q(A_{C}) \mapsto (\psi_{A_C},d_{A_C} \xi)+ l_Q(A_{C})$, where
$d_{A_C} \xi$ corresponds to the usual differential of a smooth
observable $\xi$, in the case considered, it can be checked the
exactness of the de Rham complex above, by reduction to the
well-known classical case of smooth functions. In this sense,
there can be obtained a version of the Poincare Lemma
corresponding to ${\mathbf K} (A_C)$.

\subsection{Functoriality of the Differential Geometric Mechanism}

At this subtle point of the present discussion the major
conceptual innovation of ADG consists of the realization that the
{\it differential geometric mechanism} as it is explicated by the
functioning of differential triads {\it is not dependent on both,
the arithmetics employed, and the localization methodology
adopted}. Put differently, {\it the form of the mechanism
describing the propagation of information  is universally the
same}, irrespectively of the arithmetics employed for encoding and
decoding its content, as well as, the localization contexts
deviced for its qualification through observation. This
essentially means that {\it the nature of the differential
geometric mechanism is functorial}; therefore, the differential
equations based on it, as well [37].

In order to explain the claim presented above in the context of
our inquiry related with the transition from the classical to
quantum regime of observable structure we will make use of a
topos-theoretic argument. The argument is based on the observation
that in the functorial environment of the topos of presheaves over
the category of commutative arithmetics the difference between
classical and quantum observable behaviour is expressed as a
switch on the representable functors of the corresponding
arithmetics from ${\mathbf y}[{A_C}] {\otimes}_{{\mathcal A}_C}
\mathbf M$ to ${\mathbf R}({A_{Q}}) {\otimes}_{\mathcal A_C}
{\mathbf M}$. In the classical case, $$ {\mathbf y}[{A_C}]
{\otimes}_{{\mathcal A}_C} \mathbf M  \cong {\mathbf M \circ
{\mathbf G}_{{\mathbf y}[{A_C}]}}({A_C},1_{A_C})={\mathbf
M}({A_C})$$ and the modelling functor ${\mathbf M}$ is assumed to
be the identity functor. Under this identification in the
classical case, we may equivalently assume that the category of
commutative arithmetics may be endowed with a discrete
Grothendieck topology, such that, the representable presheaves of
commutative arithmetics ${\mathbf y}[{A_C}]$ are being transformed
into sheaves for this topology. In the quantum case, respectively,
$${\mathbf R}({A_{Q}}) {\otimes}_{\mathcal A_C} {\mathbf M} \cong
A_{Q}$$ by virtue of the counit isomorphism, and moreover,
${\mathbf R}({A_{Q}})$ becomes a sheaf for the Grothendieck
topology of epimorphic families from commutative domain
arithmetics. Furthermore, inspecting the unit of the established
adjunction as applied to the representable functors ${\mathbf
y}[{A_C}]$ and ${\mathbf R}({A_{Q}})$ we obtain the corresponding
isomorphisms: $${\delta}_{{\mathbf y}[{A_C}]} : {\mathbf y}[{A_C}]
\to {{Hom}_{\mathcal {A_Q}}}({\mathbf M}(\_), {\mathbf y}[{A_C}]
{\otimes}_{{\mathcal A}_C} \mathbf M)$$ $${\delta}_{{\mathbf
R}({A_{Q}})} : {{\mathbf R}({A_{Q}})} \to {{Hom}_{\mathcal
{A_Q}}}({\mathbf M}(\_), {{{\mathbf R}({A_{Q}})}
{\otimes}_{{\mathcal A}_C}} \mathbf M)$$ At this instance, if we
remind  the construction of differential triads, we realize the
following: In the classical case, a differential triad is
specified locally by the triple ${\Delta}_C=({A_C}, d_{A_C},
\Omega(A_C))$, whereas in the quantum case, a differential triad
is specified locally by the triple ${\Delta}_Q=({\mathbf K} (A_C),
d_{{\mathbf K} (A_C)}, \Omega({\mathbf K} (A_C)))$. Of course, the
notion of local is solely determined with respect to the imposed
Grothendieck topology in each case correspondingly. Thus, formally
the differential geometric mechanism is expressed in the classical
case by a category of differential triads attached to the category
of commutative arithmetics equipped with the discrete Grothendieck
topology, which is equivalent to considering disjoint differential
triads globally, whereas in the quantum case, by a category of
differential triads attached to the category of commutative
arithmetics equipped with the Grothendieck topology of epimorphic
families, which is equivalent, correspondingly, to considering
diagrams of interconnected differential triads globally.
Consequently, the mechanism expressed universally as a  morphism
from the employed arithmetics to the modules of variation of these
arithmetics is functorial with respect to its  domain and codomain
instantiations in each case.

{\it Conclusively, whereas the ``mechanism of differentials" can
be relativized with respect to different arithmetics and different
modules of variation, the form it assumes is covariant, and simply
expressed as an $\mathcal R$-linear Leibniz morphism from the
arithmetics to their corresponding modules of variation, provided
that the same localization procedure is respectively employed in
both the domain and the codomain of this morphism within the
categorical environments specified}.

\subsection{The Notions of Connection and Curvature}

Interesting things start to happen from a differential geometric
point of view when the assumed  localization of the domain
categorical environment is different from the one that is actually
applicable in the codomain. This is exactly the case when a single
classical commutative arithmetic $A_{C}$, in the environment of
${\mathcal {A_C}}$ equipped with the discrete Grothendieck
topology, attempts to describe a quantum system whose actual
variation is described by the module ${{\Omega}^{A_{Q}}(A_{C})}$,
by setting up a mechanism of propagation of information. Although
we have not defined strictly ${{\Omega}^{A_{Q}}(A_{C})}$ yet, for
the heuristic purposes of the intuitive discussion of this
section, we may assume that it exists and denotes the
${A_Q}$-module of differentials on ${A_C}$ corresponding to the
arrow $A_C \rightarrow A_Q$. We will see in the next section,
where ${{\Omega}^{A_{Q}}(A_{C})}$ is strictly defined, that it
actually stands for an abelian group object of differentials in
the comma category ${\mathcal {A_Q}} / {A_Q}$.

It is instructive to consider two observers using classical
commutative arithmetic $A_{C}$ and $\acute {A_{C}}$ respectively.
The observers may localize their arithmetics over a topological
measurement space $X$ and thus have at their disposal the
corresponding algebra sheaves over $X$. In the terminology of ADG
a locally finite open covering $U=(U_a)$ of $X$ constitutes a
local frame. From the perspective of the arithmetics of the
observers, within their operational categorical environment, viz.,
${\mathcal {A_C}}$ equipped with the discrete Grothendieck
topology, quantum observable behaviour is being inferred and
uniquely determined, up to isomorphism in the same categorical
environment, by the cocycle ${W}_{ \acute {A_{C}}, {A_{C}}}$,
provided that ${\acute {A_{C}}} \bigcap {A_{C}} \neq 0$, using the
suggestive notation of Section 7.1. Thus, essentially each
observer is equipped with an arrow ${{\acute {A_{C}}} \bigcap
{A_{C}}} \rightarrow {\acute {A_{C}}}$ and ${{\acute {A_{C}}}
\bigcap {A_{C}}} \rightarrow {A_{C}}$ respectively, that provides
information about quantum observable behaviour. Let us now
restrict our attention to the observer using the commutative
arithmetic $A_C$, and pose the following question: How should the
observer $A_C$ set up a differential geometric mechanism of
information propagation related with quantum observable behaviour?
First of all, it is obvious that the expression of the mechanism
should be constrained by the existence of the arrow ${{\acute
{A_{C}}} \bigcap {A_{C}}} \rightarrow {A_{C}}$ in the environment
of ${\mathcal {A_C}}$. This means that the observer should
relativize the mechanism with respect to information contained in
${{\acute {A_{C}}} \bigcap {A_{C}}}$. For this purpose the
observer restricts the arithmetic $A_C$ at the image of the
cocycle in $A_C$, viz., restricts the scalars from $A_C$ to
${{\acute {A_{C}}} \bigcap {A_{C}}}$. Thus, the observer becomes
capable of expressing the mechanism in terms of the $A_C$-module
sheaf $E( {A_{C}})$, written suggestively as $E( {A_{C}}):=
{{{[Res]}^{A_C}}}_{{{\acute {A_{C}}} \bigcap {A_{C}}}} A_C$,
meaning that ${{\acute {A_{C}}} \bigcap {A_{C}}}$ is understood as
the $A_C$-module sheaf $E( {A_{C}})$.  Furthermore, from the
perspective of the arithmetic $A_C$ the observer perceives
variation of information regarding quantum behaviour by
relativizing the $A_{C}$-module sheaf of differentials
${{\Omega}}(A_C)$ with respect to the arrow ${{\acute {A_{C}}}
\bigcap {A_{C}}} \rightarrow {A_{C}}$. Let us denote this
relativization by ${{{\Omega}^{A_{C}}}_{{\acute {A_{C}}} \bigcap
{A_{C}}}}={{\Omega}}(E( {A_{C}}))$, meaning the $A_C$-module of
differentials on ${{\acute {A_{C}}} \bigcap {A_{C}}}$. Thus, the
observer $A_C$ should be able to set up a differential geometric
mechanism of information propagation related with quantum
observable behaviour, by means of the following $\mathcal
R$-linear Leibniz sheaf morphism: $$ D_{A_{C}}:E( {A_{C}})
\rightarrow {{\Omega}}(E( {A_{C}}))$$

We will now explain that the sheaf morphism $ D_{A_{C}}$ is
actually a connection on the $A_C$-module sheaf $E( {A_{C}})$,
introduced by the observer $A_C$ in order to express the
relativization of the differential mechanism with respect to the
arrow ${{\acute {A_{C}}} \bigcap {A_{C}}} \rightarrow {A_{C}}$
that induces information about quantum behaviour in the
categorical environment ${\mathcal {A_C}}$. For this purpose, we
initially notice that to give a derivation $ d_{A_{C}}:{A_{C}}
\rightarrow {{\Omega}}( {A_{C}})$ is equivalent to giving a
$\mathcal R$-linear sheaf morphism of $\mathcal R$-algebras
$${\tilde{d}}_{A_{C}}:{A_{C}} \rightarrow {A_{C}} \bigoplus
{{\Omega}}( {A_{C}}) \cdot \epsilon$$ $$a \mapsto a + da \cdot
\epsilon$$ where ${A_{C}} \bigoplus {{\Omega}}( {A_{C}}) \cdot
\epsilon$, with $\epsilon^2=0$, is the ring of dual numbers over
$A_C$ with coefficients in ${{\Omega}}( {A_{C}})$. We note that as
an abelian group ${A_{C}} \bigoplus {{\Omega}}( {A_{C}}) \cdot
\epsilon$ is the direct sum ${A_{C}} \bigoplus {{\Omega}}(
{A_{C}})$, and the multiplication law is defined by $$(a + da
\cdot \epsilon) \bullet (\acute a + \acute {da} \cdot \epsilon)=
(a \cdot \acute a +(a \cdot {\acute {da}} + \acute a \cdot da)
\cdot \epsilon)$$ We further require that the composition of the
augmentation $${A_{C}} \bigoplus {{\Omega}}( {A_{C}}) \cdot
\epsilon \rightarrow {A_{C}}$$ with ${\tilde{d}}_{A_{C}}$ is the
identity.

At a next stage, if we use the functor of scalars extension,
referring to the sheaf morphism of $\mathcal R$-algebras
${\tilde{d}}_{A_{C}}:{A_{C}} \rightarrow {A_{C}} \bigoplus
{{\Omega}}( {A_{C}}) \cdot \epsilon$, we obtain: $$E( {A_{C}})
\mapsto E( {A_{C}}) {\bigotimes}_{A_C} [{A_{C}} \bigoplus
{{\Omega}}( {A_{C}}) \cdot \epsilon]$$ Notice that $E( {A_{C}})
{\bigotimes}_{A_C} [{A_{C}} \bigoplus {{\Omega}}( {A_{C}}) \cdot
\epsilon]$ is an ${A_{C}} \bigoplus {{\Omega}}( {A_{C}}) \cdot
\epsilon$-module. Hence, by restricting it to $A_C$, denoted
obviously by the same symbol, we obtain a comparison morphism of
$A_C$-modules as follows: $${\tilde{D}}_{A_{C}}: E( {A_{C}})
\rightarrow E( {A_{C}}) {\bigotimes}_{A_C} [{A_{C}} \bigoplus
{{\Omega}}( {A_{C}}) \cdot \epsilon]$$ Thus, the information
incorporated in the comparison morphism can be now expressed as a
connection on $E( {A_{C}})$, viz., as an $\mathcal R$-linear
Leibniz sheaf morphism: $$ D_{A_{C}}:E( {A_{C}}) \rightarrow
E({A_{C}}) {\bigotimes}_{A_C} {{\Omega}}( {A_{C}})$$ Hence the
$A_C$-module of differentials on ${{\acute {A_{C}}} \bigcap
{A_{C}}}$, i.e. ${{\Omega}}(E( {A_{C}}))$, is identified with the
tensor product of $A_C$-modules $E({A_{C}}) {\bigotimes}_{A_C}
{{\Omega}}( {A_{C}})$, that is: $${{\Omega}}(E( {A_{C}}))\equiv
E({A_{C}}) {\bigotimes}_{A_C} {{\Omega}}( {A_{C}})$$ Thus, the
differential geometric mechanism of information propagation,
related with quantum observable behaviour, that the observer $A_C$
sets up for this purpose, which is expressed in terms of the
$\mathcal R$-linear Leibniz sheaf morphism, $ D_{A_{C}}:E(
{A_{C}}) \rightarrow {{\Omega}}(E( {A_{C}}))$, is equivalent with
the introduction of a connection on the $A_C$-module $E({A_{C}})$
in order to account for that observable behaviour, defined by
means of the $R$-linear Leibniz sheaf  morphism, $ D_{A_{C}}:E(
{A_{C}}) \rightarrow {{{\Omega}}(A_C)} \otimes E( {A_{C}})$. We
conclude this discussion by realizing the following;

{\it  Whereas an observer using quantum arithmetics, expressed
locally in the Grothendieck topology of epimorphic families of its
categorical environment by means of the algebra ${\mathbf K}
(A_C):=[{\mathbf R}({A_{Q}}) ({A_{C}}) \times {{\mathbf
M}({A_{C}})}] \slash l_Q(A_{C})$, formulates the differential
geometric mechanism locally in terms of the Leibniz morphism
${d^0}_{{\mathbf K}({A_C})}: {\mathbf K} (A_C) \rightarrow
\Omega({\mathbf K} (A_C))$, an observer using classical
arithmetics, expressed locally in the atomic Grothendieck topology
of its corresponding categorical environment by means of the
algebra ${A_{C}}$ has to device the notion of connection in order
to express the same mechanism}.

In the latter case, from the viewpoint of a classical observer
using a commutative arithmetic, in a discretely topologized
categorical environment, not respecting the localization
properties holding in the codomain of variations of his
observations, and by virtue of invariance of the mechanism under
relativizations, the only way that the formed discrepancy can be
compensated is through the introduction of the notion of
connection. Furthermore, using a local frame of $E( {A_{C}})$, it
can be readily shown that the ${A_{C}}$-connection $ D_{A_{C}}$
can be locally expressed in the form $$D_{A_{C}}={{{{
d}}^{0}}}_{A_{C}} + \omega_{A_{C}}$$ Hence, $D_{A_{C}}$ is locally
determined uniquely by $ \omega_{A_{C}}$, called the local
${A_{C}}$-connection matrix of $D_{A_{C}}$. This means that the
${A_{C}}$-connection $ D_{A_{C}}$ of an observer using a
commutative arithmetic plays locally the role of {\it potential}.

The notion of ${A_{C}}$-connection is always accompanied by the
{\it notion of curvature}, that in the context of ADG is expressed
as another appropriately defined sheaf morphism, however, now,
respecting the arithmetic used, in contradistinction with what
happens with $D_{A_{C}}$. More concretely, algebraically is
possible to define the various exterior powers of the module of
variations ${ {\Omega}}(A_C)={{ \Omega}^1}(A_C)$, and furthermore,
assume the existence of a second $\mathcal R$-linear morphism $${{
d}^1}_{A_C}: {{ \Omega}^1}(A_C) \rightarrow {{ \Omega}^2}(A_C):=
{{ \Omega}^1}(A_C) \wedge {{ \Omega}^1}(A_C)$$ such that ${{
d}^1}_{A_C} \circ {{ d}^0}_{A_C}=0$, where ${{ d}^1}_{A_C}$ is
called the first exterior derivation. Moreover, it is possible to
define the 1st prolongation of $ D_{A_{C}}$ by $$ {D^1}_{A_{C}}:
{{ \Omega}^1}(A_C) \otimes E( {A_{C}}) \rightarrow {{
\Omega}^2}(A_C) \otimes E( {A_{C}})$$ Finally, we can define the
curvature of the given ${A_{C}}$-connection by the following
commutative diagram:
\begin{diagram}
¤ ¤ ¤{E( {A_{C}})}¤  ¤ ¤¤ ¤ ¤\Swar {R_{A_C}} ¤  ¤\Sear
{{D}_{A_{C}}}¤ ¤¤ ¤{ {{ \Omega}^2}(A_C) \otimes E( {A_{C}})}¤
¤\War{{D^1}_{A_{C}}} ¤ ¤{{{ \Omega}^1}(A_C) \otimes E( {A_{C}})}
¤¤
\end{diagram}
where $R_{A_C}:={D^1}_{A_{C}} \circ D_{A_{C}}$.

It is readily seen that the curvature $R_{A_C}$ of the given
${A_{C}}$-connection $D_{A_{C}}$ is an ${A_C}$-morphism of the
${A_C}$-modules involved, that is $$R_{A_C} \in Hom (E( {A_{C}}),
{{ \Omega}^2}(A_C) \otimes E( {A_{C}}))$$ The physical meaning of
the curvature $R_{A_C}$ refers to the detectable effect or {\it
strength of the potential} represented by the connection
$D_{A_{C}}$. From our prism of interpretation,  we emphasize that
the curvature $R_{A_C}$ is the effect detected by an observer
employing a commutative arithmetic in a discretely topologized
categorical environment, in the attempt to understand the quantum
localization properties in the codomain of variations of his
observations, after having introduced a potential in order to
reproduce the differential geometric mechanism.

\section{QUANTUM FUNCTORIAL DIFFERENTIAL GEOMETRIC MECHANISM}

\subsection{Relativization and Abelian group Objects}
In the previous Section we have noticed that in the functorial
environment of the topos of presheaves over the category of
commutative arithmetics the difference between classical and
quantum observable behaviour is expressed as a switch on the
representable functors of the corresponding arithmetics from
${\mathbf y}[{A_C}] {\otimes}_{{\mathcal A}_C} \mathbf M$ to
${\mathbf R}({A_{Q}}) {\otimes}_{\mathcal A_C} {\mathbf M}$. The
problem of establishing a well defined functorial differential
geometric mechanism suitable for quantum observables algebras,
based on the adjunction $$\mathbf L :  {{\bf Sets}^{{{\mathcal
{A_C}}}^{op}}} \adjar  { {\mathcal {A_Q}}} : \mathbf R$$
necessitates the construction of a cohomological scheme of
interpretation of these algebras. For this purpose, it is
indispensable to have well defined notions of cohomology modules
and derivations in the category ${ {\mathcal {A_Q}}}$, as it is
actually the case in the category ${ {\mathcal {A_C}}}$. In order
to accomplish this task we adopt the following strategy: Firstly,
we unfold the notions of modules and derivations in the
paradigmatic case of the category ${ {\mathcal {A_C}}}$ using the
method of relativization, and secondly, we adapt appropriately the
definition of these notions in the category ${ {\mathcal {A_Q}}}$.

The categorical method of relativization involves the passage to
comma categories. The initial problem that is posed in this
context of inquiry has to do with the possibility of representing
the information contained in an $A_C$-module, where $A_C$ is a
commutative arithmetic in ${ {\mathcal {A_C}}}$, with a suitable
object of the relativization of ${ {\mathcal {A_C}}}$ with respect
to $A_C$, viz., with an object of the comma category ${ {\mathcal
{A_C}}} / {A_C}$. For this purpose, we define the split extension
of the commutative arithmetic $A_C$, considered as a commutative
ring, by an $A_C$-module $M$, denoted by $A_C \bigoplus M$, as
follows: The underlying set of $A_C \bigoplus M$ is the cartesian
product $A_C \times M$, where the group and ring theoretic
operations are defined respectively as; $$(a,m) + (b,n):= (a+b,
m+n)$$ $$(a,m) \bullet (b,n):=(ab, a \cdot n + b \cdot m)$$ Notice
that the identity element of $A_C \bigoplus M$ is $(1_{A_C},0_M)$,
and also that, the split extension $A_C \bigoplus M$ contains an
ideal ${0_{A_C}} \times M:= \langle M \rangle$, that corresponds
naturally to the $A_C$-module $M$. Thus, given a commutative
arithmetic $A_C$ in ${ {\mathcal {A_C}}}$, the information of an
$A_C$-module $M$, consists of an object $\langle M \rangle$ (ideal
in $A_C \bigoplus M$), together with a split short exact sequence
in ${ {\mathcal {A_C}}}$; $$\langle M \rangle \hookrightarrow A_C
\bigoplus M \rightarrow {A_C}$$ We infer that the ideal $\langle M
\rangle$ is identified with the kernel of the epimorphism $A_C
\bigoplus M \rightarrow {A_C}$, viz., $$\langle M \rangle= {Ker}
(A_C \bigoplus M \rightarrow {A_C})$$ From now on we focus our
attention to the comma category ${ {\mathcal {A_C}}} / {A_C}$,
noticing that $id_{A_C}:A_C \rightarrow A_C$ is the terminal
object in this category. If we consider the  split extension of
the commutative arithmetic $A_C$, by an $A_C$-module $M$, that is
$A_C \bigoplus M$, then the morphism: $$\lambda :A_C \bigoplus M
\rightarrow {A_C}$$ $$(a,m) \mapsto a$$ is obviously an object of
${ {\mathcal {A_C}}} / {A_C}$. It is a matter of simple algebra to
realize that it is actually an abelian group object in the comma
category ${ {\mathcal {A_C}}} / {A_C}$. This equivalently means
that for every object $\xi$ in ${ {\mathcal {A_C}}} / {A_C}$ the
set of morphisms $Hom_{{ {\mathcal {A_C}}} / {A_C}} (\xi,
\lambda)$ is an abelian group in $\mathbf {Sets}$. Moreover, the
arrow $\gamma: \kappa \rightarrow \lambda$ is a morphism of
abelian groups in ${ {\mathcal {A_C}}} / {A_C}$ if and only if for
every $\xi$ in ${ {\mathcal {A_C}}} / {A_C}$ the morphism; $${\hat
\gamma}_{\xi}: Hom_{{ {\mathcal {A_C}}} / {A_C}} (\xi, \kappa)
\rightarrow Hom_{{ {\mathcal {A_C}}} / {A_C}} (\xi, \lambda)$$ is
a morphism of abelian groups in $\mathbf {Sets}$. We denote the
category of abelian group objects in ${ {\mathcal {A_C}}} / {A_C}$
by the suggestive symbol ${[{{ {\mathcal {A_C}}} /
{A_C}}]_{\mathbf {Ab}}}$.  Based on our previous remarks it is
straightforward to show that the category of abelian group objects
in ${ {\mathcal {A_C}}} / {A_C}$ is equivalent with the category
of $A_C$-modules, viz.: $${[{{ {\mathcal {A_C}}} /
{A_C}}]_{\mathbf {Ab}}} \cong \mathcal M^{(A_C)}$$ Thus, we have
managed to characterize intrinsically $A_C$-modules as abelian
group objects in the relativization of the category of commutative
arithmetics ${ {\mathcal {A_C}}}$ with respect to $A_C$, and
moreover, we have concretely identified them as kernels of split
extensions of $A_C$.

This characterization is particularly useful if we consider an
$A_C$-module $M$ as a cohomology module, or equivalently, as a
codomain for derivations of objects of ${ {\mathcal {A_C}}} /
{A_C}$. For this purpose, let us initially notice that if $ k: B_C
\rightarrow A_C$ is an arbitrary object in ${ {\mathcal {A_C}}} /
{A_C}$, then any $A_C$-module $M$ is also a $B_C$-module via the
map $k$. We define a derivations functor from the comma category
${ {\mathcal {A_C}}} / {A_C}$ to the category of abelian groups
$\mathbf {Ab}$: $${\mathbf {Der}} (-,M) : { {\mathcal {A_C}}} /
{A_C} \rightarrow \mathbf {Ab}$$ Then if we evaluate the
derivations functor at the commutative arithmetic $B_C$ we obtain:
$${\mathbf {Der}} (B_C ,M) \cong Hom_{{ {\mathcal {A_C}}} / {A_C}}
(B_C, A_C \bigoplus M)$$ This means that, given an object $ k: B_C
\rightarrow A_C$ in ${ {\mathcal {A_C}}} / {A_C}$, then a
derivation ${d_{B_C}}: B_C \rightarrow M$ is the same as the
following morphism in ${ {\mathcal {A_C}}} / {A_C}$:

\begin{diagram}
¤ ¤ ¤{A_C}¤  ¤ ¤¤ ¤ ¤\Near {k} ¤  ¤\Nwar {\lambda}¤ ¤¤ ¤{B_C}¤
¤\Ear[100] {{{\tilde{d}}_{B_C}}} ¤ ¤{A_C \bigoplus M} ¤¤
\end{diagram}
Now we notice that the morphism: $\lambda :A_C \bigoplus M
\rightarrow {A_C}$ is actually an object in ${[{{ {\mathcal
{A_C}}} / {A_C}}]_{\mathbf {Ab}}}$. Hence, we consider it as an
object of ${[{{ {\mathcal {A_C}}} / {A_C}}]}$ via the action of an
inclusion functor: $$\Upsilon_{A_C}: {[{{ {\mathcal {A_C}}} /
{A_C}}]_{\mathbf {Ab}}} \hookrightarrow {[{{ {\mathcal {A_C}}} /
{A_C}}]}$$ $$[\lambda :A_C \bigoplus M \rightarrow {A_C} ] \mapsto
[{\Upsilon_{A_C}}(\lambda) : {\Upsilon_{A_C}}(M) \rightarrow
{A_C}]$$ Thus we obtain the isomorphism: $${\mathbf {Der}} (B_C
,M) \cong Hom_{{ {\mathcal {A_C}}} / {A_C}} (B_C,
{\Upsilon_{A_C}}(M))$$ The inclusion functor $\Upsilon_{A_C}$ has
a left adjoint functor; $${\mathbf \Omega}^{A_C}: {[{{ {\mathcal
{A_C}}} / {A_C}}]} \rightarrow {[{{ {\mathcal {A_C}}} /
{A_C}}]_{\mathbf {Ab}}}$$ Consequently, if we further take into
account the equivalence of categories ${[{{ {\mathcal {A_C}}} /
{A_C}}]_{\mathbf {Ab}}} \cong \mathcal M^{(A_C)}$, the isomorphism
above takes the following final form: $${\mathbf {Der}} (B_C ,M)
\cong Hom_{ \mathcal M^{(A_C)}}  ({ {\mathbf \Omega}^{A_C}} (B_C)
, M)$$ We conclude that the derivations functor ${\mathbf {Der}}
(-,M) : { {\mathcal {A_C}}} / {A_C} \rightarrow \mathbf {Ab}$ is
being represented by the abelianization functor ${\mathbf
\Omega}^{A_C}: {[{{ {\mathcal {A_C}}} / {A_C}}]} \rightarrow {[{{
{\mathcal {A_C}}} / {A_C}}]_{\mathbf {Ab}}}$. Furthermore, the
evaluation of the abelianization functor ${\mathbf \Omega}^{A_C}$
at an object $ k: B_C \rightarrow A_C$ of ${ {\mathcal {A_C}}} /
{A_C}$, viz. ${ {\mathbf \Omega}^{A_C}} (B_C)$, is interpreted as
the $A_C$-module of differentials on $B_C$.

At this stage of development it is obvious that, for cohomological
purposes, we can easily adapt the previously established notions
of modules and derivations to the category of quantum observables
algebras ${ {\mathcal {A_Q}}}$. Firstly, we simply define the
category of $A_Q$-modules as the category of abelian group objects
in the comma category ${ {\mathcal {A_Q}}} / {A_Q}$, viz.;
$$\mathcal M^{(A_Q)}:= {[{{ {\mathcal {A_Q}}} / {A_Q}}]_{\mathbf
{Ab}}}$$ Secondly, we use the above definition in order to
introduce the notion of an $A_Q$-module for derivations in the
category ${ {\mathcal {A_Q}}}$. For this purpose we define the
derivations functor from the comma category ${ {\mathcal {A_Q}}} /
{A_Q}$ to the category of abelian groups $\mathbf {Ab}$:
$${\mathbf {Der}} (-,N) : { {\mathcal {A_Q}}} / {A_Q} \rightarrow
\mathbf {Ab}$$ where N is now an $A_Q$-module, or equivalently, an
abelian group object in ${ {\mathcal {A_Q}}} / {A_Q}$. Hence, if $
K: B_Q \rightarrow A_Q$ denotes an object of ${ {\mathcal {A_Q}}}
/ {A_Q}$ we obtain the isomorphism: $${\mathbf {Der}} (B_Q ,N)
\cong Hom_{{ {\mathcal {A_Q}}} / {A_Q}} (B_Q,
{\Upsilon_{A_Q}}(N))$$ where; $$\Upsilon_{A_Q}: {[{{ {\mathcal
{A_Q}}} / {A_Q}}]_{\mathbf {Ab}}} \hookrightarrow {[{{ {\mathcal
{A_Q}}} / {A_Q}}]}$$ denotes the corresponding inclusion functor,
having a left adjoint abelianization functor: $${\mathbf
\Omega}^{A_Q}: {[{{ {\mathcal {A_Q}}} / {A_Q}}]} \rightarrow {[{{
{\mathcal {A_Q}}} / {A_Q}}]_{\mathbf {Ab}}}$$ Consequently we
obtain again the following isomorphism: $${\mathbf {Der}} (B_Q ,N)
\cong Hom_{ \mathcal M^{(A_Q)}}  ({ {\mathbf \Omega}^{A_Q}} (B_Q)
, N)$$ We conclude that the derivations functor ${\mathbf {Der}}
(-,N) : { {\mathcal {A_Q}}} / {A_Q} \rightarrow \mathbf {Ab}$ is
being represented by the abelianization functor ${\mathbf
\Omega}^{A_Q}: {[{{ {\mathcal {A_Q}}} / {A_Q}}]} \rightarrow {[{{
{\mathcal {A_Q}}} / {A_Q}}]_{\mathbf {Ab}}}$. Furthermore, the
evaluation of the abelianization functor ${\mathbf \Omega}^{A_Q}$
at an object $ K: B_Q \rightarrow A_Q$ of ${ {\mathcal {A_Q}}} /
{A_Q}$, viz. ${ {\mathbf \Omega}^{A_Q}} (B_Q)$, is interpreted
correspondingly as the $A_Q$-module of differentials on $B_Q$.

\subsection{Cohomology of Quantum Observables Algebras}

The representation of quantum observables algebras ${A_Q}$ in
${\mathcal {A_Q}}$ in terms of sheaves over commutative
arithmetics ${{A}_C}$ in ${\mathcal {A_C}}$ for the Grothendieck
topology of epimorphic families on ${\mathcal {A_C}}$, is based on
the existence of the adjunctive correspondence ${\mathbf L} \dashv
{\mathbf R}$ as follows: $$\mathbf L :  {{\bf Sets}^{{{\mathcal
{A_C}}}^{op}}} \adjar  { {\mathcal {A_Q}}} : \mathbf R$$ which
says that the functor of points ${\mathbf R}$ defined by
$${\mathbf R}({{A_Q}}) : {{A_C}} {\mapsto} {{Hom}_{ {\mathcal
{A_Q}}}({\mathbf M}({{A_C}}), {{A_Q}})}$$ has a left adjoint
$\mathbf L : {{\bf Sets}^{{{\mathcal {A_C}}}^{op}}} \to {
{\mathcal {A_Q}}}$, which is defined for each presheaf $\mathbf P$
in ${{\bf Sets}^{{{\mathcal {A_C}}}^{op}}}$ as the colimit
$${\mathbf {L}}({\mathbf P})= {\it Colim} \{ \bf{G}({\mathbf
P},{{\mathcal {A_C}}}) \Ar {{\mathbf G}_{\mathbf P}} {{\mathcal
{A_C}}} \Ar {\mathbf M} { {\mathcal {A_Q}}} \}$$ Equivalently,
there exists a bijection, natural in $\mathbf P$ and ${{A_Q}}$ as
follows: $$ Nat({\mathbf P},{\mathbf R}({{A_Q}})) \cong
{{Hom}_{\mathcal {{A_Q}}}({\mathbf L \mathbf P}, {{A_Q}})}$$ The
adjunction can be characterized in terms of  the unit and the
counit categorical constructions. For any presheaf $\mathbf P \in
{\bf Sets}^{{A_C}^{op}}$, the unit is defined as:
$${\delta}_{\mathbf P} : \mathbf P \ar \mathbf R \mathbf L
{\mathbf P}$$ On the other side, for each object ${A_Q}$ of
$\mathcal {A_Q}$ the counit is: $${\epsilon}_{A_Q} : \mathbf L
{\mathbf R}({A_Q}) \ar {A_Q}$$ The composite endofunctor $\mathbf
G:=\mathbf L {\mathbf R}: \mathcal {A_Q} \rightarrow  \mathcal
{A_Q} $, together with the natural transformations $\delta:
\mathbf G \rightarrow \mathbf G \circ \mathbf G$, called
comultiplication, and also, ${\epsilon} : \mathbf G \rightarrow
\mathbf I$, called counit, where $\mathbf I$ is the identity
functor on $\mathcal {A_Q}$, is defined as a comonad $(\mathbf G,
\delta, \epsilon)$ on the category of quantum observables algebras
$\mathcal {A_Q}$, provided that the diagrams below commute for
each object ${A_Q}$ of $\mathcal {A_Q}$;
\begin{diagram} ¤{\mathbf G}{{A_Q}}  ¤\Ear[50]   {{\delta}_{{A_Q}}} ¤
{{\mathbf G}^2} {{A_Q}}¤¤ ¤\Sar {{\delta}_{{A_Q}}} ¤ ¤\saR
{{\delta}_{{\mathbf G}{A_Q}}}¤¤ ¤ {{\mathbf G}^2} {{A_Q}}¤\Ear[50]
{{\mathbf G} {{\delta}_{{{A_Q}}}}} ¤{{\mathbf G}^3}{{A_Q}}¤¤
\end{diagram}

\begin{diagram}
¤¤ ¤ ¤ ¤{\mathbf G}{{A_Q}}¤   ¤¤ ¤ ¤\neeql  ¤\Sar
{{\delta}_{{A_Q}}} ¤\nweql ¤¤ ¤{\mathbf G}{{A_Q}}¤ \War[47]
{{{\epsilon}_{{\mathbf G}{A_Q}}}} ¤ {{\mathbf G}^2}{{A_Q}}¤
\Ear[47] {{{\mathbf G} {{\epsilon}_{{{A_Q}}}}}} ¤{\mathbf
G}{{A_Q}} ¤¤
\end{diagram}
For a comonad $(\mathbf G, \delta, \epsilon)$ on $\mathcal {A_Q}$,
a $\mathbf G$-coalgebra is an object ${A_Q}$ of $\mathcal {A_Q}$,
being equipped with a structural map $\kappa: {A_Q} \rightarrow
{\mathbf G} {A_Q}$, such that the following conditions are
satisfied; $$1=\epsilon_{A_Q} \circ \kappa : {A_Q} \rightarrow
{\mathbf G} {A_Q}$$ $${\mathbf G}\kappa \circ \kappa =
\delta_{A_Q} \circ \kappa : {A_Q} \rightarrow {{\mathbf G}^2}
{A_Q}$$ With the above obvious notion of morphism, this gives a
category ${\mathcal {A_Q}}_{\mathbf G}$ of all $\mathbf
G$-coalgebras.

The counit of the comonad $(\mathbf G, \delta, \epsilon)$ on
$\mathcal {A_Q}$, that is: $${\epsilon}_{A_Q} : \mathbf G {A_Q}:=
\mathbf L {\mathbf R}({A_Q})={\mathbf R}({A_{Q}})
{\otimes}_{\mathcal A_C} {\mathbf M} \ar {A_Q}$$ is intuitively
the first step of a functorial free resolution of an object
${A_Q}$ in $\mathcal {A_Q}$. Thus, by iteration of $\mathbf G$, we
may extend ${\epsilon}_{A_Q}$ to a free simplicial resolution of
${A_Q}$ in $\mathcal {A_Q}$. Most importantly, we will consider
the case of defining cohomology groups ${\mathbf
{\tilde{H}}}^n({A_Q}, X_Q)$, $n \geq 0$, of a quantum observables
algebra ${A_Q}$ in $\mathcal {A_Q}$ with coefficients in an
$A_Q$-module $X_Q$, relative to the given underlying functor of
points ${\mathbf R}: { {\mathcal {A_Q}}} \rightarrow {{\bf
Sets}^{{{\mathcal {A_C}}}^{op}}}$, defined by ${\mathbf
R}({{A_Q}}) : {{A_C}} {\mapsto} {{Hom}_{ {\mathcal
{A_Q}}}({\mathbf M}({{A_C}}), {{A_Q}})}$, having  a left adjoint
$\mathbf L : {{\bf Sets}^{{{\mathcal {A_C}}}^{op}}} \to {
{\mathcal {A_Q}}}$.

Thus, let $(\mathbf G, \delta, \epsilon)$ be the comonad on
$\mathcal {A_Q}$, that is induced by the adjoint pair of functors
$\mathbf L :  {{\bf Sets}^{{{\mathcal {A_C}}}^{op}}}  \adjar  {
{\mathcal {A_Q}}} : \mathbf R$. The following simplicial object in
$\mathcal {A_Q}$ is called the free simplicial comonadic
resolution of a quantum observables algebra ${A_Q}$ in $\mathcal
{A_Q}$, denoted by ${{\mathbf G}_\star}{A_Q} \rightarrow {A_Q}$:
\begin{diagram}
¤¤{{A_Q}}¤\War {{{{\epsilon}_{0}}}:={\epsilon}} ¤{\mathbf G} {A_Q}
¤\Wbiar {{\epsilon}_{0}} {{\epsilon}_{1}}   ¤{{\mathbf G}^2} {A_Q}
¤\War {{\epsilon}_{0,1,2}} ¤{\ldots} ¤  \War
{{\epsilon}_{0,1,\ldots,n-1}}¤{{\mathbf G}^n} {A_Q} ¤\War
{{\epsilon}_{{0,1,\ldots,n}}} ¤{{\mathbf G}^{n+1}} {A_Q}
¤{\ldots}¤¤
\end{diagram}
In the simplicial resolution above, ${{\epsilon}_{0,1,2}}$ denotes
a triplet of arrows etc. Notice that, ${{\mathbf G}^{n+1}}$ is the
term of degree $n$, whereas the face operator $\epsilon_i:
{{\mathbf G}^{n+1}} \rightarrow {{\mathbf G}^{n}}$ is ${{\mathbf
G}^{i}}\circ \epsilon \circ {{\mathbf G}^{n-i}}$, where $0\leq i
\leq n$. We can verify the following simplicial identities;
$$\epsilon_i \circ \epsilon_j=\epsilon_{j+1} \circ
\epsilon_i$$where $i \leq j$. The comonadic resolution ${{\mathbf
G}_\star}{A_Q} \rightarrow {A_Q}$ induces clearly a comonadic
resolution in the comma category ${[{{ {\mathcal {A_Q}}} /
{A_Q}}]}$, which we still denote by ${{\mathbf G}_\star}{A_Q}
\rightarrow {A_Q}$.

An $n$-cochain of a quantum observables algebra ${A_Q}$ with
coefficients in an $A_Q$-module $X_Q$, where, by definition, $X_Q$
is an object in ${[{{ {\mathcal {A_Q}}} / {A_Q}}]_{\mathbf
{Ab}}}$, is defined as a map ${{\mathbf G}^{n+1}} {A_Q}
\rightarrow {{{\mathbf \Upsilon}_{A_Q}} ({X_Q})}$ in the comma
category ${[{{ {\mathcal {A_Q}}} / {A_Q}}]}$. We remind that,
since $X_Q$ is an abelian group object in ${[{{ {\mathcal {A_Q}}}
/ {A_Q}}]}$, the set ${{Hom}_{\mathcal {A_Q}}}({A_Q},{X_Q})$ has
an abelian group structure for every object ${A_Q}$ in $\mathcal
{A_Q}$, and moreover, for every arrow $\acute {A_Q} \rightarrow
{A_Q}$ in $\mathcal {A_Q}$, the induced map of sets
${{Hom}_{\mathcal {A_Q}}}({A_Q},{X_Q}) \rightarrow
{{Hom}_{\mathcal {A_Q}}}(\acute {A_Q},{X_Q})$ is an abelian groups
map. Then, we can identify the set of $n$-cochains with the
abelian group of derivations of ${{\mathbf G}^{n+1}}{A_Q}$ into
the abelian group object $X_Q$ in ${[{{ {\mathcal {A_Q}}} /
{A_Q}}]_{\mathbf {Ab}}}$. Hence, we consider an $n$-cochain as a
derivation map ${{\mathbf G}^{n+1}} {A_Q} \rightarrow X_Q$.

Consequently, the face operators $\epsilon_i$, induce abelian
group maps; $${Der}({\epsilon_i} {A_Q}, {X_Q}): {Der}({{\mathbf
G}^{n}} {A_Q}, {X_Q}) \rightarrow {Der}({{\mathbf G}^{n+1}} {A_Q},
{X_Q})$$ Thus, the cohomology can be established by application of
the contravariant functor ${\mathbf {Der}}(-, {X_Q})$ on the free
simplicial resolution of a quantum observables algebra ${A_Q}$ in
$\mathcal {A_Q}$, obtaining the following cochain complex of
abelian groups;
\begin{diagram}
¤¤{0}¤\Ear {d^{0}} ¤{{Der}({{\mathbf G}}{A_Q}, {X_Q})}¤ ¤\Ear
{{d}^{1}} ¤ ¤{{Der}({{\mathbf G}^2}{A_Q}, {X_Q})} ¤¤ \Ear
{{d}^{2}}¤{\ldots} ¤\Ear {d^{n}}¤ ¤ ¤{{Der}({{\mathbf G}^{n+1}}
{A_Q}, {X_Q})}¤ ¤ ¤\Ear {{d}^{n+1}} ¤{\ldots}¤¤
\end{diagram}
where, because of the aforementioned simplicial identities we
have: $$d^{n+1}=\sum_i {(-1)^i {Der}({\epsilon_i} {A_Q}, {X_Q})}$$
where $0\leq i \leq {n+1}$, and also; $$d^{n+1} \circ d^n=0$$
written symbolically as; $$d^{\mathbf 2}=0$$

Finally we may also make use of the following isomorphism: $$
{{Der}({{{{\mathbf G}}{A_Q}, {X_Q})} \simeq {{Hom}({{\mathbf {
\Omega}}^{A_Q}} (\mathbf G}}{A_Q}), {X_Q})}$$ where the
abelinazation functor ${\mathbf \Omega}^{A_Q}: {[{{ {\mathcal
{A_Q}}} / {A_Q}}]} \rightarrow {[{{ {\mathcal {A_Q}}} /
{A_Q}}]_{\mathbf {Ab}}}$ represents the derivations functor
${\mathbf {Der}} (-,{X_Q}) : { {\mathcal {A_Q}}} / {A_Q}
\rightarrow \mathbf {Ab}$. In this precise sense, ${{\mathbf {
\Omega}}^{A_Q}} ({\mathbf G} {A_Q}):={{\mathbf {\hat
\Omega}}({{\mathbf G}}{A_Q})}$ is identified with the $A_Q$-module
of first order differentials or $1$-forms on ${{\mathbf G}}{A_Q}$.
Thus, equivalently, we obtain the following cochain complex of
abelian groups;
\begin{diagram}
¤¤ {0} ¤\Ear[33] {d^{0}}¤ ¤{{Hom}({{\mathbf {\hat
\Omega}}({{\mathbf G}}{A_Q}), {X_Q})}}¤  ¤\Ear[33] {{d}^{1}}¤ ¤
{{Hom}({{\mathbf {\hat \Omega}}({{\mathbf G}^2}{A_Q}), {X_Q})}}¤¤
\Ear {{d}^{2}}¤{\ldots} ¤\Ear[33] {d^{n}}¤ ¤ ¤ {{Hom}({{\mathbf
{\hat \Omega}}({{\mathbf G}^{n+1}}{A_Q}), {X_Q})}} ¤ ¤ ¤\Ear[33]
{{d}^{n+1}} ¤{\ldots}¤¤
\end{diagram}
Now, we define the cohomology groups ${\mathbf
{\tilde{H}}}^n({A_Q}, X_Q)$, $n \geq 0$, of a quantum observables
algebra ${A_Q}$ in $\mathcal {A_Q}$ with coefficients in an
$A_Q$-module $X_Q$ as follows: $${\mathbf {\tilde{H}}}^n({A_Q},
X_Q):={\mathbf H}^n [{\mathbf {Der}}({{{\mathbf G}_\star}{A_Q}},
{X_Q} )]=\frac{Ker ({d^{n+1}})}{Im ({d^n})}$$ According to the
above, a $1$-cocycle is a derivation map $\omega: {{\mathbf
G}^{2}} {A_Q} \rightarrow {X_Q}$, such that: $$\epsilon_2 \omega
\circ \epsilon_0 \omega=\epsilon_1 \circ \omega$$ where
$\epsilon_i: {{\mathbf G}^{3}} {A_Q} \rightarrow {{\mathbf G}^{2}}
{A_Q}$, $i=0,1,2$. Correspondingly, a $1$-coboundary is a
derivation map $\upsilon: {{\mathbf G}} {A_Q} \rightarrow {X_Q}$,
which can be presented as a mapping of $1$-cocycles $\upsilon:
\omega \rightarrow  \acute {\omega}$ modulo the conditions:
$$\omega \circ \epsilon_0 \upsilon=\epsilon_1 \upsilon \circ
\acute {\omega}$$

\subsection{Functorial Connection and Curvature Equation}

In this Section we are going to introduce the notion of a
functorial quantum connection, together with, the associated
curvature of that connection. The connection is intentionally
termed functorial because it is precisely induced by the functor
of points of a quantum observables algebra ${A_Q}$ in $\mathcal
{A_Q}$, restricted as usual to commutative arithmetics. For this
purpose it is necessary to define appropriately an $A_Q$-module,
denoted by ${{\Omega}}_{A_Q}$, that is going to play the role of a
universal object of quantum differential $1$-forms in analogy to
the classical case. At this stage, it is instructive to remind
briefly the analogous construction of the classical universal
object of differential $1$-forms ${{\Omega}}_{A_C}$, corresponding
to a commutative arithmetic $A_C$. According to Kahler, the free
$\mathcal A$-module $\mathbf \Omega$ can be constructed explicitly
form the fundamental form of scalars extension of $A_C$, that is
$\iota: A_C \hookrightarrow A_C {\bigotimes}_{\mathcal R} A_C$ by
considering the diagonal morphism: $$\delta: A_C
{\bigotimes}_{\mathcal R} A_C \rightarrow A_C$$ $$f_1 \otimes f_2
\mapsto f_1 \cdot f_2$$ where $f_1$, $f_2$ $\in$ $A_C$. Then by
taking the kernel of this morphism of algebras, that is the ideal;
$$ I= \{f_1 \otimes f_2 \in A_C {\bigotimes}_{\mathcal R} A_C :
\delta (f_1 \otimes f_2)=0 \} \subset A_C {\bigotimes}_{\mathcal
R} A_C$$ it can be easily proved that the morphism of
$A_C$-modules $$\Sigma : {{\Omega}}_{A_C} \rightarrow \frac{ I}{{
I}^2}$$ $$df \mapsto 1 \otimes f - f \otimes 1$$ is an
isomorphism. Thus the free $A_C$-module ${{\Omega}}_{A_C}$ of
$1$-forms is isomorphic with the free $A_C$-module
$\frac{I}{{I}^2}$ of Kahler differentials of the commutative
arithmetic $A_C$ over $\mathcal R$, according to the following
split short exact sequence: $${{\Omega}}_{A_C} \hookrightarrow A_C
\oplus {{\Omega}}_{A_C} \cdot \epsilon \rightarrow A_C$$ where
$\epsilon^2=0$, formulated equivalently as follows: $$0
\rightarrow {{\Omega}}_{A_C} \rightarrow A_C
{\bigotimes}_{\mathcal R} A_C \rightarrow A_C$$

In the quantum case, as we have explained in detain in Section
9.2, the counit of the adjunction ${\epsilon}_{A_Q} : \mathbf L
{\mathbf R}({A_Q}) \rightarrow {A_Q}$, defined by the composite
endofunctor $\mathbf G:=\mathbf L {\mathbf R}: \mathcal {A_Q}
\rightarrow \mathcal {A_Q} $, constitutes the first step of a
functorial free resolution of a quantum observables algebra
${A_Q}$ in $\mathcal {A_Q}$, generated by iterating the
endofunctor $\mathbf G$. In this setting, and in analogy to the
classical case, we define the $A_Q$-module ${{\Omega}}_{A_Q}$ of
quantum differential $1$-forms, by means of the following split
short exact sequence: $$0 \rightarrow {{J}}_{A_Q} \rightarrow
{\mathbf R}({A_{Q}}) {\otimes}_{\mathcal A_C} {\mathbf M}
\rightarrow {A_Q} $$ According to the above, we obtain that $
{{\Omega}_{A_Q}}=\frac {J_{A_Q}} {{J^2}_{A_Q}}$, where
${{J}}_{A_Q}= {\mathbf {Ker}}({\epsilon}_{A_Q})$ denotes the
kernel of the counit of the adjunction. Subsequently, we may apply
the algebraic construction, for each $n \in N$, $n \geq 2$, of the
$n$-fold exterior product ${{{{\Omega}^n}_{A_Q}}}={{\bigwedge}^n}
{{{{\Omega}^1}_{A_Q}}}$. Thus, we may now set up the algebraic de
Rham complex of ${A_Q}$ as follows: $${A_Q} \rightarrow
{{\Omega}_{A_Q}} \rightarrow {\ldots} \rightarrow
{{{{\Omega}^n}_{A_Q}}} \rightarrow {\ldots}$$ For the purpose of
introducing the notion of a functorial quantum connection, the
crucial idea comes from the realization that the functor of points
of a quantum observables algebra restricted to commutative
arithmetics, viz.,  ${\mathbf R}({{A_Q}})$, is a left exact
functor, because it is the right adjoint functor of the
established adjunction. Thus, it preserves the short exact
sequence defining the object of quantum differential $1$-forms, in
the following form: $$0 \rightarrow {\mathbf R} ({{\Omega}}_{A_Q})
\rightarrow {\mathbf R} ({\mathbf G}({A_{Q}})) \rightarrow
{\mathbf R}({{A_Q}})$$ Hence, we immediately obtain that:
${\mathbf R} ({{\Omega}}_{A_Q})= \frac{Z}{{Z}^2}$, where ${Z}=
{\mathbf {Ker}}({\mathbf R} ( {\epsilon}_{A_Q}))$.

Then, in analogy to the paradigmatic classical algebraic
situation, we define the notion of a functorial quantum
connection, denoted by ${{\mathcal {{\nabla}}}}_{{\mathbf
R}({{A_Q}})}$, in terms of the following Leibniz natural
transformation: $${{\mathcal {{\nabla}}}}_{{\mathbf R}({{A_Q}})}:
{\mathbf R} ({A_Q}) \rightarrow {\mathbf R} ({{\Omega}}_{A_Q})$$
Thus, the quantum connection ${{\mathcal {{\nabla}}}}_{{\mathbf
R}({{A_Q}})}$ induces a sequence of functorial morphisms, or
equivalently, natural transformations as follows: $${{\mathbf
R}({{A_Q}})} \rightarrow {\mathbf R} ({{\Omega}}_{A_Q})
\rightarrow {\ldots} \rightarrow {\mathbf R}
({{{\Omega}}^n}_{A_Q}) \rightarrow {\ldots}$$ Let us denote by;
$${\mathbf R}_\nabla : {{\mathbf R}({{A_Q}})} \rightarrow {\mathbf
R} ({{{\Omega}}^2}_{A_Q})$$ the composition ${{\mathcal
{{\nabla}}}}^1 \circ {{\mathcal {{\nabla}}}}^0$ in the obvious
notation, where ${{\mathcal {{\nabla}}}}^0:={{\mathcal
{{\nabla}}}}_{{\mathbf R}({{A_Q}})}$. The natural transformation
${\mathbf R}_\nabla$ is called  the curvature of the functorial
quantum connection ${{\mathcal {{\nabla}}}}_{{\mathbf
R}({{A_Q}})}$. Furthermore, the latter sequence of functorial
morphisms, is actually a complex if and only if ${\mathbf
R}_\nabla=0$. We say that the quantum connection ${{\mathcal
{{\nabla}}}}_{{\mathbf R}({{A_Q}})}$ is integrable or flat if
${\mathbf R}_\nabla=0$, referring to the above complex as the
functorial de Rham complex of the integrable connection
${{\mathcal {{\nabla}}}}_{{\mathbf R}({{A_Q}})}$ in that case.
Thus we arrive at the following conclusion: The vanishing of the
curvature of the functorial quantum connection, viz.: $${\mathbf
R}_\nabla=0$$  can be interpreted as the transposition of
Einstein's equations in the quantum regime, that is inside the
topos $\mathbf {Shv}(\mathcal {A_C})$ of sheaves of algebras over
the base category of commutative algebraic contexts, in the
absence of cohomological obstructions. We may explain the
curvature of the quantum connection as the effect of non-trivial
interlocking of commutative arithmetics, in some underlying
diagram of a quantum observables algebras being formed by such
localizing commutative arithmetics. The non-trivial gluing of
commutative arithmetics in localization systems of a quantum
observables algebra is caused by topological obstructions. These
obstructions can be associated with the elements of the
non-trivial cohomology groups of a quantum observables algebra
${A_Q}$, in $\mathcal {A_Q}$. From a physical viewpoint, these
obstructions can be understood as geometric phases related with
the monodromy of the quantum connection, being evaluated at points
$A_C$ of the functor of points of a quantum observables algebra
$A_Q$ restricted to commutative arithmetics, which, in turn, has
been respectively interpreted as a prelocalization system of
$A_Q$. Intuitively, a non-vanishing curvature may be understood as
the non-local attribute detected by an observer employing a
commutative arithmetic in a discretely topologized categorical
environment, in the attempt to understand the quantum localization
properties, after having introduced a potential, or equivalently,
a connection, in order to account for these properties by means of
a differential geometric mechanism. Thus, the physical meaning of
curvature is associated with the apparent existence of non-local
correlations from the restricted spatial perspective of disjoint
classical commutative arithmetics $A_{C}$.

\section{EPILOGUE}
The representation of quantum observables algebras, $A_Q$ in
${\mathcal A}_Q$, as sheaves, with respect to the Grothendieck
topology of epimorphic families on ${\mathcal A}_C$, is of a
remarkable physical significance. If we remind the discussion of
the physical meaning of the adjunction, expressed in terms of the
information content, communicated between commutative arithmetics
and quantum arithmetics, we arrive to the following conclusion:
the totality of the content of information, included in the
quantum species of observables structure remains invariant, under
commutative algebras decodings, corresponding to local arithmetics
for measurement of observables, in covering sieves of quantum
observables algebras, if, and only if, the counit of the
fundamental adjunction is a quantum algebraic isomorphism. In this
manner, the fundamental adjunction is being restricted to an
equivalence of categories ${\bf Sh}({{\mathcal A}_C}, {\mathbf J})
\cong { {\mathcal A}_Q}$; making thus, in effect, ${ {\mathcal
A}_Q}$ a Grothendieck topos, equivalent with the topos of sheaves
on the site $({\mathcal A}_C,{\mathbf J})$. The above
correspondence, that can be understood as a topos-theoretic
generalization of Bohr's correspondence principle, essentially
shows that the process of quantization is categorically equivalent
with the process of subcanonical localization and sheafification
of information in commutative terms, appropriately formulated in a
generalized topological fashion, {\it $\grave{a}$ la
Grothendieck}.

We also claim that the sheaf-theoretic representation of a quantum
observables algebra reveals that its deep conceptual significance
is related not to its global non-commutative character, but, on
the precise manner that distinct local contexts of observation,
understood as commutative arithmetics, are being interconnected
together, so as its informational content is preserved in the
totality of its operational commutative decodings. By the latter,
we precisely mean contextual operational procedures for probing
the quantum regime of observable structure, which categorically
give rise to covering sieves, substantiated as interconnected
epimorphic families of the generalized elements of the sheafified
functor of points of a quantum observables algebra $\mathbf
R({A_Q})$.  The sheaf-theoretic representation expresses exactly
the compatibility of these  commutative algebras of observables on
their overlaps in such a way as to leave invariant the amount of
information contained in a quantum system. We may adopt the term
reference frames of commutative arithmetics for a geometric
characterization of these local contexts of encoding the
information related to a quantum system, emphasizing their
prominent role in the organization of meaning associated with a
quantum algebra of observables. Moreover this terminology
signifies the intrinsic contextuality of algebras of quantum
observables, as filtered through the base commutative localizing
category, and is suggestive of the introduction of a relativity
principle in the quantum level of observable structure, as a
categorical extension of Takeuti's and Davis's research program
[38, 39], related, in the present embodiment,  with the invariance
of the informational content with respect to commutative
arithmetics reference frames contained in covering sieves of
quantum observables algebras.

Furthermore, the sheaf-theoretic representation of quantum
observables algebras makes possible the extension of the mechanism
of Differential Geometry in the quantum regime by a proper
adaptation of the methodology of ADG in a topos-theoretic
environment. More concretely, in the terminology of ADG the
differential geometric mechanism is incorporated in the
functioning of differential triads consisting of commutative
localized arithmetics, modules of variations of arithmetics and
Leibniz sheaf morphisms from the domains of the former to the
codomains of the latter, instantiating appropriate differentials.
Most importantly the mechanism itself is functorial in nature, or
equivalently, is always in force irrespectively of the
relativizations pertaining the domains and codomains of the
differentials introduced, provided that the same localization
properties are respected in the corresponding categorical
environments of the domains and codomains of differentials.

The differential geometric mechanism is expressed in the classical
case by a category of differential triads attached to the category
of commutative observable algebras equipped with the discrete
Grothendieck topology, whereas in the quantum case, by a category
of differential triads attached to the category of commutative
observables algebras, being a generating subcategory of the
category of quantum observables algebras, equipped with the
Grothendieck topology of epimorphic families. As a consequence of
the difference in the categorical localization properties
classical arithmetics are different from quantum arithmetics. In a
local cover belonging to a covering epimorphic sieve, a quantum
arithmetic appears as a quotient of a commutative algebra over an
ideal, incorporating information about all the other covers being
compatible with it in a localization system of the former. Despite
the difference in the corresponging arithmetics and modules of
variations of them, the mechanism expressed universally as a
morphism from the employed arithmetics to the modules of
variations of these arithmetics is functorial with respect to its
domain and codomain instantiations localized categorically in the
same fashion in each case. Realization of this subtle point has
subsequently forced us to argue that the real power of the
abstract differential geometric mechanism, referring to
propagation of information related to observation, is
substantiated in cases where the categorical localization of the
arithmetics used for observation is different from the categorical
localization that is actually applicable in the modules containing
variations of observations. As we have seen this is exactly the
case when a disjoint classical commutative arithmetic $A_{C}$, in
the environment of ${\mathcal {A_C}}$ equipped with the discrete
Grothendieck topology, is used for description of a quantum system
whose actual variation is described by the $A_Q$-module
${{\Omega}^{A_{Q}}(A_{C})}$, denoting the ${A_Q}$-module of
differentials on ${A_C}$ corresponding to the arrow $A_C
\rightarrow A_Q$. From the perspective of classical arithmetics in
discretely topologized categorical environments the explication of
the differential geometric mechanism necessitates the introduction
of a connection, termed quantum potential, for the explanation of
the -peculiar from their viewpoint- categorical localization rules
respected by variations of observations in the quantum regime. The
detectable effect emanating from the introduction of this
potential for the description of the mechanism of information
propagation in the resources offered by their arithmetics, is the
appearance of the strength of the employed potential, expressed
geometrically as the curvature of the associated connection. It is
instructive to make clear that, in the present scheme, the
geometric notion of curvature does not refer to an underlying
background manifold, since such a structure has neither been
postulated nor has it been required at all in the development of
the differential geometric mechanism according to ADG. The
physical meaning of curvature is associated with the apparent
existence of non-local correlations from the perspective of
disjoint classical commutative arithmetics $A_{C}$. Put
differently, curvature is the detectable effect on a locus
associated with a classical commutative arithmetic in a discretely
topologized categorical environment, when observation of quantum
behavior takes place, constituting the denotator of non-local
correlations, stemming exclusively from the restricted sense of
spatial locality that the locus shares in its discrete classical
categorical environment. On the contrary, the form of quantum
arithmetics,  constructed by epimorphic covering families of
sieves from interlocking commutative domain reference frames,
incorporate a generalized notion of localization, not associated
with its former restricted spatial connotation, but being defined
only in the relational local terms of compatible information
collation among those frames.

Considering the above scheme of interpretation seriously, we may
assert, characteristically, that the transition in the meaning of
generalized localization and its observable effects, as related
with the formulation of a dynamical model that reflects the
transition from the classical to the quantum regime, can be
understood in terms of the observational locoi, corresponding to
the respective arithmetics, as a conceptual transformation from a
space of unrelated points endowed with a classically conceived
point-localization structure, to a category of interconnected
generalized points, being themselves localizing morphisms in
covering sieves, and ultimately constituting a Grothendieck topos.

Thus, essentially the transition reflects a shift in the semantics
of localization schemes, from a set-theoretic to a topos-theoretic
one. Put differently, the notion of space of the classical theory
is replaced by that one of a Grothendieck topos, equivalent with
the topos of sheaves on the site $({\mathcal A}_C,{\mathbf J})$,
where the latter is simply understood as a generalized spatial
framework of interrelation of experimentally gathered information,
referring to quantum observable behaviour, being expressed in
reference frames of interlocking commutative arithmetics.
Remarkably, the algebraic sheaf-theoretic framework of ADG,
conceived via the categorical and topos-theoretic adaptation
attempted in this work, vindicates further the possibility of
extending the ``mechanism of differentials" in the quantum regime.
The latter is being effectuated by the realization that the
character of the mechanism is functorial with respect to the kind
of arithmetics used for description of observable physical
behaviour. Put differently, this means that the differential
geometric mechanism of description of information propagation in
physical terms, is covariant with respect to the arithmetics
employed, denoting reference frames of a topos-theoretic nature,
for decoding its content. Consequently, we are naturally directed
towards a functorial formulation of the differential geometric
mechanism, characterizing the dynamics of information propagation,
through observable attributes, localized over commutative
reference frames of variable form, thus, affording a language of
dynamics suited to localization schemes of a topos theoretical
nature, suitable for the transcription of dynamics from the
classical to the quantum regime of observable behaviour.

More concretely, the process of gluing information along the loci
of overlapping commutative arithmetics, interpreted in this
generalized topos-theoretic localization environment, generates
dynamics, involving the transition from the classical to the
quantum regime, by means of the notion of a functorial connection
and its associated curvature natural transformation, conceived in
a precise cohomological manner. In this sense, the vanishing of
the curvature of the functorial quantum connection, viz.,
${\mathbf R}_\nabla=0$,  can be interpreted as the transposition
of Einstein's equations in the quantum regime, that is inside the
topos $\mathbf {Shv}(\mathcal {A_C})$ of sheaves of algebras over
the base category of commutative algebraic contexts, in the
absence of cohomological obstructions.

\vspace{6mm}
\newpage
{\bf{Acknowledgments:}}  I would gratefully like to acknowledge A.
Mallios  for his continuous and enthousiastic interest in this
work, as well as, for numerous enlightening and really
instrumental discussions we have had in several occasions; the
present form of the paper owes much to all these. The author is
supported by an I.K.Y. Hellenic State Research Fellowship
Programme.

\end{document}